%% file: Takakura_PhD.tex
\documentclass[12pt,a4paper,oneside]{extbook}
\usepackage{layout}
\usepackage[top=31truemm,bottom=31truemm,inner=37truemm,outer=37truemm]{geometry}

\usepackage{color}
\usepackage{amsmath}
\usepackage{amssymb}
\usepackage{mathrsfs}
\usepackage{braket}
\usepackage{bm}
\usepackage{multirow}
\usepackage{mathabx}
\usepackage{bbm}
\usepackage{comment}
\usepackage{graphicx}
\makeatletter
\renewcommand{\theequation}
{\arabic{chapter}.\arabic{equation}}
\@addtoreset{equation}{chapter}
\makeatother

\usepackage{latexsym}
\def\qed{\hfill $\Box$}
\usepackage{ntheorem}
\theoremstyle{break}
\newtheorem{thm}{Theorem}[chapter]
\newtheorem{lem}[thm]{Lemma}
\newtheorem{cor}[thm]{Corollary}
\newtheorem{prop}[thm]{Proposition}
\newtheorem*{thm**}{Theorem}
\newtheorem*{lem**}{Lemma}
\newtheorem*{cor**}{Corollary}
\newtheorem*{prop**}{Proposition}

\newtheorem{lemapp}{Lemma}[section]

\newtheorem{propapp}{Proposition}[section]
\newtheorem*{propref}{Proposition \ref{prop_ortho repr}}
\newtheorem*{propref1}{Proposition \ref{prop_transitive self-dual}}

\newtheorem{axiom}{Axiom}
\newtheorem{assum}{Mathematical assumption}

\theorembodyfont{\normalfont}
\newtheorem{defi}[thm]{Definition}
\newtheorem*{defi**}{Definition}
\newtheorem*{pf}{Proof}

\theoremheaderfont{\itshape}
\newtheorem{eg}[thm]{Example}
\newtheorem*{eg**}{Example}
\newtheorem{rmk}[thm]{Remark}
\newtheorem*{rmk**}{Remark}

\usepackage{appendix}
\usepackage{authblk}
\usepackage{setspace}

\usepackage{lineno}
\if
\newcommand*\patchAmsMathEnvironmentForLineno[1]{%
	\expandafter\let\csname old#1\expandafter\endcsname\csname #1\endcsname
	\expandafter\let\csname oldend#1\expandafter\endcsname\csname end#1\endcsname
	\renewenvironment{#1}%
	{\linenomath\csname old#1\endcsname}%
	{\csname oldend#1\endcsname\endlinenomath}}%
\newcommand*\patchBothAmsMathEnvironmentsForLineno[1]{%
	\patchAmsMathEnvironmentForLineno{#1}%
	\patchAmsMathEnvironmentForLineno{#1*}}%
\AtBeginDocument{%
	\patchBothAmsMathEnvironmentsForLineno{equation}%
	\patchBothAmsMathEnvironmentsForLineno{align}%
	\patchBothAmsMathEnvironmentsForLineno{flalign}%
	\patchBothAmsMathEnvironmentsForLineno{alignat}%
	\patchBothAmsMathEnvironmentsForLineno{gather}%
	\patchBothAmsMathEnvironmentsForLineno{multline}%
}
\fi

\usepackage{hyperref}
\hypersetup{%
 bookmarksnumbered=true,%
 colorlinks=false,%
 setpagesize=false,%
 pdftitle={},%
 pdfauthor={Ryo Takakura},%
 pdfsubject={},%
 pdfkeywords={}
}

\newcommand{\N}{\mathbb{N}}

\newcommand{\R}{\mathbb{R}}
\newcommand{\C}{\mathbb{C}}

\newcommand{\Tr}{\mathrm{Tr}}

\newcommand{\ketbra}[2]{\ket{#1}\hspace{-0.25em}\bra{#2}}

\newcommand{\HH}{\mathcal{H}}
\newcommand{\LL}{\mathcal{L}}

\newcommand{\ang}[1]{\left\langle{#1}\right\rangle}

\newcommand{\ext}{\mathrm{ext}}

\newcommand{\hi}{\mathcal{H}} 
\newcommand{\hik}{\mathcal{K}} 
\newcommand{\id}{\mathbbm{1}}

\newcommand{\A}{\mathsf{A}}
\newcommand{\B}{\mathsf{B}}
\newcommand{\CC}{\mathsf{C}}
\newcommand{\D}{\mathsf{D}}
\newcommand{\E}{\mathsf{E}}
\newcommand{\F}{\mathsf{F}}
\newcommand{\G}{\mathsf{G}}
\newcommand{\M}{\mathsf{M}}
\newcommand{\vsigma}{\mathbf{\sigma}} 
\newcommand{\va}{\mathbf{a}} 
\newcommand{\vb}{\mathbf{b}} 
\newcommand{\vr}{\mathbf{r}} 
\newcommand{\vx}{\mathbf{x}} 
\newcommand{\vy}{\mathbf{y}}

\newcommand{\hin}{\hi_{in}} 
\newcommand{\state}{\mathcal{S}} 
\newcommand{\vz}{\mathbf{z}}
\newcommand{\x}{\mathrm{x}}
\newcommand{\y}{\mathrm{y}}
\newcommand{\z}{\mathrm{z}}
\newcommand{\half}{\tfrac{1}{2}}
\newcommand{\pTr}[2]{\mathrm{Tr}_{#1}[#2]}

\usepackage{xcolor}

\usepackage{multirow}

\usepackage[small,bf]{caption} 
\usepackage[subrefformat=parens]{subcaption}
\usepackage{here}
\usepackage {anyfontsize}
\newcommand{\kB}{k_{\mathrm{B}}}
\begin{document} 
\title{\vspace{-10mm}\textbf{{\fontsize{23.7pt}{23.7pt}\selectfont Convexity and uncertainty in\\
			\vspace{3mm} operational quantum foundations}}}

\author{\vspace{20mm}\large{A thesis presented by}\\
	\vspace{13mm}
{\fontsize{20pt}{20pt}\selectfont Ryo Takakura}\\
\vspace{12mm}

\large{in partial fulfillment of the requirements for the degree of Doctor of Philosophy (Engineering)} in the}

\affil{\vspace{6mm}\Large{Department of Nuclear Engineering \\Kyoto University}
	\vspace{6mm}
\begin{figure}[h]
\centering
\includegraphics[bb=0 0 500 500, scale=0.23]{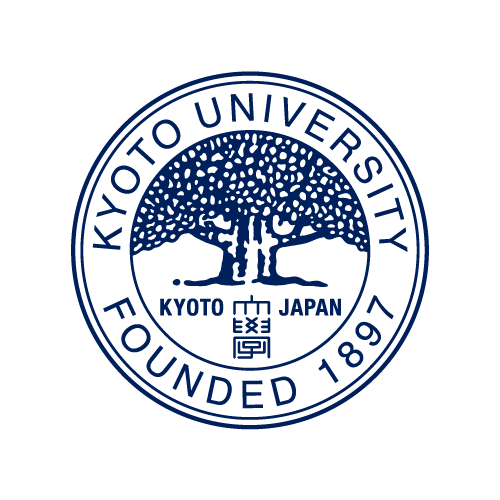}
\end{figure}}
\date{\vspace{-2mm}\Large{January \ 2022}}
\maketitle
\setcounter{page}{2}

\chapter*{\centering Abstract}
\addcontentsline{toc}{chapter}{Abstract}
To find the essential nature of quantum theory has been an important problem for not only theoretical interest but also applications to quantum technologies.
In those studies on quantum foundations, the notion of uncertainty, which appears in many situations, plays a primary role among several stunning features of quantum theory.
The purpose of this thesis is to investigate fundamental aspects of uncertainty.
In particular, we address this problem  focusing on convexity, which has an operational origin.

We first try to reveal why in quantum theory similar bounds are often obtained for two types of uncertainty relations, namely, preparation and measurement uncertainty relations.
In order to do this, we consider uncertainty relations in the most general framework of physics called generalized probabilistic theories (GPTs).
It is proven that some geometric structures of states connect those two types of uncertainty relations in GPTs in terms of several expressions such as entropic one.
From this result, we can find what is essential for the close relation between those uncertainty relations.

Then we consider a broader expression of uncertainty in quantum theory called quantum incompatibility.
Motivated by an operational intuition, we propose and investigate new quantifications of incompatibility which are related directly to the convexity of states.
It is also demonstrated that there can be observed a notable phenomenon for those quantities even in the simplest incompatibility, i.e., incompatibility for a pair of mutually unbiased qubit observables.

Finally, we study thermodynamical entropy of mixing in quantum theory, which also can be seen as a quantification of uncertainty.
Similarly to the previous approach, we consider its operationally natural extension to GPTs, and then try to characterize how specific the entropy in quantum theory is.
It is shown that the operationally natural entropy is allowed to exist only in classical and quantum-like theories among a class of GPTs called regular polygon theories.

\chapter*{\centering List of papers}
\addcontentsline{toc}{chapter}{List of papers}
This thesis is based on the following papers:\\
\begin{enumerate}
\item (Reproduced from \cite{Takakura2020}, with the permission of AIP Publishing)\\
Ryo Takakura, Takayuki Miyadera, ``Preparation Uncertainty Implies Measurement Uncertainty in a Class of Generalized Probabilistic Theories'', {\it Journal of Mathematical Physics}, {\bf 61}, 082203 (2020);\\

\item (\cite{Takakura_entropic_2021})\\
Ryo Takakura, Takayuki Miyadera, ``Entropic uncertainty relations in a class of generalized probabilistic theories'', {\it Journal of Physics A: Mathematical and Theoretical}, {\bf 54}, 315302 (2021);\\

\item (\cite{PhysRevA.104.032228})\\
Teiko Heinosaari, Takayuki Miyadera, Ryo Takakura, ``Testing incompatibility of quantum devices with few states'', {\it Physical Review A}, {\bf 104}, 032228 (2021);\\

\item (\cite{Takakura_2019})\\
Ryo Takakura, ``Entropy of mixing exists only for classical and quantum-like theories among the regular polygon theories'', {\it Journal of Physics A: Mathematical and Theoretical}, {\bf 52}, 465302 (2019).
\end{enumerate}

\tableofcontents
\pagenumbering{arabic}
\markright{Contents}


\setcounter{page}{5}
\chapter{Introduction}
\markright{}
Since its birth about a hundred years ago, quantum theory has been crucial in modern physics
because of its more accurate description of nature than classical theory; in addition, it was particularly revealed that there are many differences between the mathematical formulations of classical and quantum theories \cite{von1955mathematical}.
Then it is natural to ask the following questions.
What is physically the most significant difference between them?
Why is nature described by quantum theory? 
Since the dawn of quantum theory, they have remained central questions, and much effort has been devoted to finding an answer  to form the frontier of physics called quantum foundations \cite{Hardy_Spekken_foundation,Chiribella_Spekken_2016}.
Many significant results have been obtained in that field, and for results of particular importance such as uncertainty relations \cite{Heisenberg1927} and the violation of Bell inequality \cite{PhysRev.47.777,Bell_inequality}, active studies are still ongoing.
While studies on quantum foundations are motivated by the theoretical interest of exploring the root of nature, it should be emphasized that pursuing fundamental aspects of quantum theory also contributes to the development of its applications, i.e., quantum technologies.
For example, the original ideas of quantum cryptography (quantum key distribution) were derived using uncertainty relations and Bell nonlocality \cite{BB84,E91}.
Quantum foundations are valuable research objects from both theoretical and practical perspectives.

In this thesis, we are engaged in further developing of quantum foundations.
To elucidate how ``special'' quantum theory is, we focus on its convexity.
In quantum theory, convexity is one of the most fundamental ingredients, and appears in many situations.
A basic example that exhibits convexity is the set of all states (the state space) for some quantum system, which is in fact closed under operationally natural convex combinations \cite{von1955mathematical}.
There is one noteworthy approach to quantum foundations concentrating on this primitive convexity, which we call the convexity approach \cite{Lahti1985}.
The main aim of the convexity approach is to find what is needed to derive quantum theory besides the convexity, i.e., to distinguish quantum theory from other convex theories.
Its mathematical formulation and physical motivation are today succeeded to the framework called {\itshape generalized probabilistic theories} ({\itshape GPTs}).
As was seen above or will be seen in detail in subsequent chapters, GPTs are operationally the broadest framework to describe nature, and have been studied actively in recent years in the context of quantum foundations, followed by the intuition that seeing quantum theory from a broader perspective will contribute to elucidating its essence.
While this primitive convexity for states is focused in the study of GPTs, there are studies about quantum foundations based on other types of convexity such as convexity for separable states \cite{HORODECKI19961,TERHAL2000319} or compatibility \cite{doi:10.1063/1.5126496,PhysRevLett.122.130402}.
Considering the above facts, in this thesis we regard convexity as a significant concept for the research on quantum foundations, and demonstrate the results of several attempts to capture the essential nature of quantum theory via convexity.
In particular, we focus on ``uncertainty'', which is one of the most critical features in quantum theory, and try to reveal its essence.
We have to mention that all results are obtained for {\itshape operational convexity}, which means that every type of convexity considered in this thesis has an operational origin.
By means of the operational descriptions, our results are easier to understand physically, and thus may contribute more to the theoretical insights of quantum theory and technological applications.  

In Chapter \ref{chap:GPT}, we review the mathematical foundations of GPTs.
In recent studies, GPTs are usually introduced in a mathematically refined manner such as ``a state space is a compact convex set in a finite-dimensional Euclidean space.''
We try to give a detailed explanation of how those expressions are derived from physically abstract notions.
More precisely, we demonstrate how the operational convexity associated with probability mixtures of states or effects (observables) is expressed in terms of ordered Banach spaces.
There are also introduced additional topics for GPTs with physical or mathematical motivations such as the descriptions of composite system and transformations or the notions of transitivity and self-duality.

Based on the mathematical foundations of GPTs, in Chapter \ref{chap:URs in GPT} we extend the concept of uncertainty relations, which is one of the most astonishing consequences in quantum theory, to GPTs, and investigate how specific the quantum uncertainty is.
It is explained that two types of uncertainty, {\itshape preparation uncertainty} and {\itshape measurement uncertainty}, can also be naturally considered in GPTs, and how they are related is examined under various expressions such as entropic uncertainty relations.
Following the quantum results \cite{doi:10.1063/1.3614503,PhysRevLett.112.050401}, we prove that there is a quantitatively close connection between the two types of uncertainty in GPTs with the assumptions of transitivity and self-duality.
We also present numerical evaluations of uncertainty for GPTs called {\itshape regular polygon theories} from which we can observe how quantum uncertainty for a single qubit system is specific in regular polygon theories.

In Chapter \ref{chap:incomp dim}, we focus on another fundamental concept for quantum foundations called {\itshape quantum incompatibility}.
It is known that many astonishing results in quantum theory, such as the no-cloning theorem \cite{Wootters_Zurek_no-cloning} and uncertainty relations, are examples of quantum incompatibility \cite{Heinosaari_2016}.
In this way, quantum incompatibility provides such a unified framework to describe what is impossible or what becomes uncertain in quantum theory that it plays an essential role in the field of quantum foundations.
Further, we consider the operational convexity of quantum incompatibility, which is derived from that of states and effects.
There are introduced new quantifications of incompatibility called {\itshape compatibility dimension} and {\itshape incompatibility dimension} from a very operational perspective, and properties of those quantities  are examined for several cases.
In particular, for a pair of incompatible qubit observables, we demonstrate that there is a difference of interest between these quantities.
We note that similar quantities can also be defined in GPTs because they are introduced based on the convexity for states and effects, but we only concentrate on quantum incompatibility.

Finally, in Chapter \ref{chap:TD entropy in GPT}, we revisit GPTs, and consider thermodynamical entropy there.
We introduce operationally natural entropy which can be defined in every theory of GPTs but is required to satisfy some operational convexity for families of perfectly distinguishable states.
Then it is proven that the only theories that admit the existence of the natural entropy are classical and a quantum-like theories among regular polygon theories.

\input{chap2.tex}

\input{chap3.tex}
\input{chap4.tex}

\input{chap5.tex}

\chapter{Summary}
\label{chap:summary}
In this thesis, we have studied the notion of uncertainty in quantum theory via convexity.
We focused on three expressions of uncertainty: uncertainty relations, incompatibility, and thermodynamical entropy.
Our establishments were based on the idea that to see quantum uncertainty from a broader point of view makes it possible to understand its essence.
In fact, in each chapter, we considered uncertainty relations and thermodynamical entropy in a broader class of theories than quantum theory, and investigated quantum incompatibility, which is a broader notion than uncertainty.

In Chapter \ref{chap:GPT}, we introduced the mathematical framework of GPTs.
We saw that GPTs are constructed by requiring only primitive convexity originating from probability mixtures, and thus in this sense, they are the most general description of nature (in particular, broader than quantum theory).
There were proven that GPTs can be expressed mathematically in terms of ordered Banach spaces, and that it indeed reproduce the usual formulation of quantum theory.

In Chapter \ref{chap:URs in GPT}, several generalizations of uncertainty relations such as entropic uncertainty relations were considered in a class of GPTs which can be considered as generalized theories of quantum theory.
It was revealed that similar quantitative relations between preparation and measurement uncertainty to quantum case hold also in GPTs although only theories with transitivity and self-duality with respect to a certain inner product were considered.
We also gave concrete calculations of our results for regular polygon theories.
What is also specific to theorems is that they were obtained without considering entanglement or even composite systems while the quantum results of the previous studies were based on the ``ricochet'' property of maximally entangled states.
It may be indicated that some of the characteristics of quantum theory can be obtained without entanglement.
Future research should reveal the relations between the maximal entanglement and self-duality, which will be a key to generalizing our theorems to infinite-dimensional cases (remember that the maximally entangled states cannot be defined in infinite-dimensional quantum theories such as $\HH=L^{2}(\R)$).
To find information-theoretic applications of our results is also left for future work.

In Chapter \ref{chap:incomp dim}, we focused on incompatibility in quantum theory, and introduced the notions of compatibility and incompatibility dimensions for collections of quantum devices.
They describe the minimum number of states which are needed to detect incompatibility and the maximum number of states on which incompatibility vanishes, respectively.
We have not only presented general properties of those quantities but also examined concrete behaviors of them for a pair of unbiased qubit observables.
We have proven that even for this simple pair of incompatible observables there exist two types of incompatibility with different incompatibility dimensions which cannot be observed if we focus only on robustness of incompatibility under noise.
We expect that it is possible to apply this difference to some quantum protocols such as quantum cryptography.
Future work will be needed to investigate whether similar results can be obtained for observables in higher dimensional Hilbert space or other quantum devices.
As the definitions apply to devices in GPTs, an interesting task is further to see how quantum incompatibility dimension differs from incompatibility dimension in general.

In Chapter \ref{chap:TD entropy in GPT}, we returned to GPTs and considered theremodynamical entropy in regular polygon theories.
We showed that only classical and quantum-like theories (i.e. the triangle and disc theories respectively) allow the operationally natural entropy to be consistent. 
Further research is required to reveal if we can obtain the same results in higher dimensional cases.
Moreover, the proof of our main theorem indicates that the entropy discussed above is defined successfully in other theories where the probability coefficients obtained when a state is decomposed into perfectly distinguishable states are unique even though the state space is neither classical nor quantum. This means that we need to impose additional conditions on the entropy to remove those``unreasonable" theories, which is also a future problem.

\chapter*{Acknowledgments}
\addcontentsline{toc}{chapter}{Acknowledgments}
I would like to, first and foremost, thank my supervisor Takayuki Miyadera.
Since I became his student in 2016, he has taught me many interesting things on quantum theory to lead me to the study of quantum foundations.
Topics such as quantum uncertainty or incompatibility are examples of those things, and they are now the main interest of my research.
His insightful guidance, not only as a great researcher but also as a great person, has helped and will help me a lot in my research and daily life.
Nothing in my nine-year college experience makes me more proud of than being his student and able to learn from him.
I also would like to thank my collaborator Teiko Heinosaari.
I feel very honored to have worked with such a great person and exhibit our results in my PhD thesis.
I cannot miss thanking all members of my laboratory, especially Kenzo Ogure, Ikko Hamamura, and Kazuki Yamaga, for their fruitful help in my study.
This research has been supported by JSPS KAKENHI Grant Number JP21J10096, and I also acknowledge the JSPS Research Fellowship for Young Scientists.

Finally, I would like to express my thanks to my family and friends.
I am grateful to my parents for giving me birth with so good a constitution that I can stay healthy in everyday life.
I also thank my friends for their encouragement and support.
They are all great persons, and examples for me.
I have learned from them what are the most important things in life.
Those things are always within my mind, and are the principles that guide my life.
My deepest gratitude is expressed to my grandparents.
The greatest thing in my life is that they are my grandparents.
We live so far that we cannot often see each other, but I always feel their love supporting and encouraging me in my daily life.
I hope that my accomplishment of this PhD study will be their pleasure.

\chapter*{Appendix}
\addcontentsline{toc}{chapter}{Appendix}
\setcounter{section}{0}
\renewcommand{\thesection}{\Alph{section}}
\section{Proof of Proposition \ref{prop_ortho repr}}
\label{appA}
\renewcommand{\theequation}{A.\arabic{equation}}
\setcounter{equation}{0}
\renewcommand{\thesection}{\Alph{section}}
\setcounter{subsection}{0}

In this part, we give a proof of Proposition \ref{prop_ortho repr}. 
We need the following proposition, which holds without the assumption of the transitivity of $\Omega$.

\begin{propapp}
	\label{prop_generalized ortho repr}
	For a state space $\Omega$, define a linear map $P_{M}\colon V\rightarrow V$ by 
	\[
	P_{M}x=\int_{GL(\Omega)} Tx\ d\mu(T).
	\]
	Then $P_{M}$ is an orthogonal projection with respect to the inner product $\langle\cdot ,\cdot \rangle_{GL(\Omega)}$, i.e.
	\[
	P_{M}=P_{M}^{2}\quad\mbox{and}\quad\langle P_{M}x,\ y\rangle_{GL(\Omega)}=\langle x,\ P_{M}y\rangle_{GL(\Omega)}\ \ \mbox{for all}\ x, y\in V.
	\]
\end{propapp}
\begin{pf}
	We denote the inner product $\langle\cdot ,\cdot \rangle_{GL(\Omega)}$ simply by $\langle\cdot ,\cdot \rangle$ in this proof. 
	
	Let $V_{M}:=\{x\in V\mid Tx=x\ \mbox{for all}\ T\in GL(\Omega)\}$ be the set of all fixed points with respect to $GL(\Omega)$. 
	Then it is easy to see that $P_{M}x_{M}=x_{M}$ for any $x_{M}\in V_{M}$ and $V_{M}=ImP_{M}$ (in particular $V_{M}$ is a subspace of $V$). Therefore,
	\[
	P_{M}^{2}x=P_{M}(P_{M}x)=P_{M}x
	\]
	holds for any $x\in V$, and thus $P_{M}^{2}=P_{M}$. On the other hand, we can observe
	\begin{align}
		\langle P_{M}x,\ y\rangle
		&=\int_{GL(\Omega)}d\mu(T)\  (TP_{M}x,\ Ty)_{E}\notag\\
		&=\int_{GL(\Omega)}d\mu(T)\  (P_{M}x,\ Ty)_{E}\notag\\
		&=\int_{GL(\Omega)}d\mu(T)\  \left(\int_{GL(\Omega)}d\mu(S)Sx,\ Ty\right)_{E}.\label{pf_app_1}
	\end{align}
	Let us fix an orthonormal basis $\{w_{i}\}_{i=1}^{N+1}$ of $V$ compatible with the standard Euclidean inner product of $V$, i.e.
	\[
	(w_{i},\ w_{j})_{E}=\delta_{ij}.
	\]
	We can consider representing the vector $\int_{GL(\Omega)}d\mu(S)Sx\in V$ by means of the orthonormal basis $\{w_{i}\}_{i}$ as
	\[
	\int_{GL(\Omega)}d\mu(S)Sx=\sum_{i}\left(w_{i},\ \int_{GL(\Omega)}d\mu(S)Sx\right)_{E}w_{i}.
	\]
	In fact, the ``$i$th-element'' $\left(w_{i},\ \int_{GL(\Omega)}d\mu(S)Sx\right)_{E}$ is given by (see \cite{KIMURA20141} for more details)
	\[
	\left(w_{i},\ \int_{GL(\Omega)}d\mu(S)Sx\right)_{E}=\int_{GL(\Omega)}d\mu(S)\ (w_{i},\ Sx)_{E}.
	\]
	It results in
	\begin{align*}
		\left(\int_{GL(\Omega)}d\mu(S)Sx,\ Ty\right)_{E}
		&=\sum_{i}\left[\int_{GL(\Omega)}d\mu(S) (w_{i},\ Sx)_{E}\right](w_{i},Ty)_{E}\\
		&=\int_{GL(\Omega)}d\mu(S) \left[\sum_{i}(Sx,\ w_{i})_{E}(w_{i},Ty)_{E}\right]\\
		&=\int_{GL(\Omega)}d\mu(S)(Sx,\ Ty)_{E}.
	\end{align*}
	Therefore, we obtain
	\begin{align*}
		&\int_{GL(\Omega)}d\mu(T)\  \left(\int_{GL(\Omega)}d\mu(S)Sx,\ Ty\right)_{E}\\
		&\qquad\qquad\qquad\qquad=\int_{GL(\Omega)}d\mu(T)\  \left[\int_{GL(\Omega)}d\mu(S)\ (Sx,\ Ty)_{E}\right]\\
		&\qquad\qquad\qquad\qquad=\int_{GL(\Omega)}d\mu(S)\  \left[\int_{GL(\Omega)}d\mu(T)\ (Sx,\ Ty)_{E}\right]\\
		&\qquad\qquad\qquad\qquad=\int_{GL(\Omega)}d\mu(S)\  \left(Sx,\ \int_{GL(\Omega)}d\mu(T)Ty\right)_{E},
	\end{align*}
	where we use Fubini's theorem for the finite Haar measure $\mu$ on $GL(\Omega)$.
	\if
	Since the vector $\int_{GL(\Omega)}d\mu(S)Sx \in V$ is constructed with its $i$th element
	\[
	\left(w_{i},\ \int_{GL(\Omega)}d\mu(S)Sx\right)_{E}
	\]
	in terms of the Euclidean orthonormal basis $\{w_{i}\}_{i=1}^{N+1}$ of $V$ given by 
	\[
	\int_{GL(\Omega)}d\mu(S)\ (w_{i},\ Sx)_{E},
	\]
	\fi
	We can conclude together with \eqref{pf_app_1} that
	\begin{equation*}
		\langle P_{M}x,\ y\rangle=\langle x,\ P_{M}y\rangle
	\end{equation*}
	holds.
	\hfill $\Box$
\end{pf}

Proposition \ref{prop_generalized ortho repr} enables us to give an orthogonal decomposition of a vector $x\in V$ such that
\begin{equation}
	\label{def_general ortho repr}
	x=(\id-P_{M})x+P_{M}x,
\end{equation}
where $(\id-P_{M})x\in V_{M}^{\perp}$ and $P_{M}x\in V_{M}$. When the transitivity of $\Omega$ is assumed, \eqref{def_general ortho repr} is reduced to Proposition \ref{prop_ortho repr}.

\begin{propref}
	For a transitive state space $\Omega$, there exists a basis $\{v_{l}\}_{l=1}^{N+1}$ of $V$ orthonormal with respect to the inner product  $\langle\cdot ,\cdot  \rangle_{GL(\Omega)}$ such that $v_{N+1}=\omega_{M}$ and 
	\[
	x\in\mathit{aff}(\Omega)\iff x=\sum_{l=1}^{N}a_{l}v_{l}+v_{N+1}=\sum_{l=1}^{N}a_{l}v_{l}+\omega_{M}\ (a_{1}, \cdots, a_{N}\in\R).
	\]
\end{propref}
\begin{pf}
	Since we set $\mathrm{dim}\mathit{aff}(\Omega)=N$, there exists a set of $N$ linear independent vectors $\{v_{l}\}_{l=1}^{N}\subset[\mathit{aff}(\Omega)-\omega_{M}]$ which forms a basis of the $N$-dimensional vector subspace $[\mathit{aff}(\Omega)-\omega_{M}]\subset V$, and we can assume by taking an orthonormalization that they are orthonormal with respect to the inner product $\langle\cdot ,\cdot \rangle$. Hence $x\in\mathit{aff}(\Omega)$ if and only if it is represented as
	\begin{equation}
		\label{def_repr of aff}
		x=\sum_{l=1}^{N}a_{l}v_{l}+\omega_{M}\quad(a_{1}, \cdots, a_{N}\in\R).
	\end{equation}
	Moreover, because of the definition of $\mathit{aff}(\Omega)$, for every $v_{l}\in[\mathit{aff}(\Omega)-\omega_{M}]$ there exist $k\in\mathbb{N}$, real numbers $\{b_{i}\}_{i=1}^{k}$ satisfying $\sum_{i=1}^{k}b_{i}=1$, and states $\{\omega_{i}\}_{i=1}^{k}$ such that $v_{l}=\sum_{i=1}^{k}b_{i}\omega_{i}-\omega_{M}$. By means of  Proposition \ref{def_max mixed state}, we obtain for all $l=1, 2, \cdots, N$
	\begin{align}
		P_{M}v_{l}
		&=\sum_{i=1}^{k}b_{i}P_{M}\omega_{i}-P_{M}\omega_{M}\notag\\
		&=\sum_{i=1}^{k}b_{i}\omega_{M}-\omega_{M}=0.\label{app_pf0}
	\end{align}
	Therefore, because of Proposition \ref{prop_generalized ortho repr}
	\begin{align*}
		\langle\omega_{M},\ v_{l}\rangle
		&=\langle P_{M}\omega_{M},\ v_{l}\rangle\\
		&=\langle \omega_{M},\ P_{M}v_{l}\rangle\\
		&=0
	\end{align*}
	holds for all $l=1, 2, \cdots, N$, and we can conclude together with the unit norm of $\omega_{M}$ that $\{v_{1}, \cdots, v_{N}, \omega_{M}\}$ in \eqref{def_repr of aff} forms an orthonormal basis of the $(N+1)$-dimensional vector space $V$ with respect to $\langle\cdot ,\cdot \rangle$ and Proposition \ref{prop_ortho repr} is proven (we can also find that \eqref{def_repr of aff} corresponds to \eqref{def_general ortho repr}).\qed
\end{pf}

\appendix
\setcounter{section}{1}
\section{Proof of Proposition \ref{prop_transitive self-dual}}
\label{appB}
\renewcommand{\theequation}{B.\arabic{equation}}
\setcounter{equation}{0}
\renewcommand{\thesection}{\Alph{section}}
\setcounter{subsection}{0}

In this part, we prove Proposition \ref{prop_transitive self-dual}. As we have so far, we let $\Omega$ be a state space, $V_{+}$ be the positive cone generated by $\Omega$, and $GL(\Omega)$ be the set of all state automorphisms on $\Omega$ in the following.
\begin{lemapp}
	\label{lemma0}
	$V_{+\langle\cdot, \cdot\rangle_{GL(\Omega)}}^{*int}$ is a $GL(\Omega)$-invariant set. 
	That is, $T V_{+\langle \cdot, \cdot\rangle_{GL(\Omega)}}^{*int}=V_{+\langle \cdot, \cdot\rangle_{GL(\Omega)}}^{*int}$ for all $T\in GL(\Omega)$.
\end{lemapp}
\begin{pf}
	Let $w\in V_{+\langle\cdot, \cdot\rangle_{GL(\Omega)}}^{*int}$. 
	It holds that $\langle w, v\rangle_{GL(\Omega)}\geq 0$
	for all $v\in V_+$. 
	Because any $T\in GL(\Omega)$ is an orthogonal transformation with respect to $\langle \cdot, \cdot\rangle_{GL(\Omega)}$, we obtain 
	\[
	\langle Tw, v\rangle_{GL(\Omega)}
	= \langle w, T^{-1}v\rangle_{GL(\Omega)} \geq 0
	\]
	for all $v\in V_{+}$. Therefore, $T V_{+\langle \cdot, \cdot\rangle_{GL(\Omega)}}^{*int}\subset V_{+\langle \cdot, \cdot\rangle_{GL(\Omega)}}^{*int}$holds, and a similar argument for $T^{-1}\in GL(\Omega)$ proves the lemma.
	\qed
\end{pf}
\begin{lemapp}
	\label{lemma1}
	Let $(\cdot ,\cdot )$ be an arbitrary inner product on $V$. $V_+$ is self-dual if and only if there exists a linear map $J\colon V\to V$ such that $J$ is strictly positive with respect to $(\cdot ,\cdot )$, i.e. $(x, Jy)=(Jx, y)$ for all $x, y\in V$ and $(x, Jx)>0$ for all $x\in V$, and $J(V_+) = V^{*int}_{+(\cdot ,\cdot )}$. 
\end{lemapp}
\begin{pf}
	If part: We introduce an inner product 
	$(\cdot, \cdot)_{J}= (\cdot, J\cdot)$.  $V_{+(\cdot,\cdot)_{J}}^{*int}$ 
	is written as 
	\begin{align*}
		V_{+(\cdot,\cdot)_{J}}^{*int}
		&=\{v\mid(v, w)_{J} \geq 0, \ ^{\forall} w
		\in V_+\}\\
		&=\{v\mid(v, Jw)\geq 0, \ ^{\forall} w
		\in V_+\}
		\\
		&= \{v\mid(Jv, w)\geq 0, \ ^{\forall} w\in V_+\}. 
	\end{align*}
	Thus $v\in V_{+(\cdot , \cdot )_{J}}^{*int}$ 
	is equivalent to $Jv\in V_{+(\cdot, \cdot )}^{*int}$. It concludes $V_{+(\cdot,\cdot)_{J}}^{*int}
	=J^{-1}(
	V_{+(\cdot , \cdot )}^{*int}
	)=V_+$. \\
	Only if part: Let $V_+$ be self-dual with respect to 
	an inner product $\langle\cdot ,\cdot \rangle$. There exists some $K\colon V\to V$ strictly positive with respect to $(\cdot ,\cdot )$ such that $\langle\cdot ,\cdot\rangle=(\cdot, K\cdot)$. We obtain 
	\begin{align*}
		V_{+}=V_{+\langle\cdot ,\cdot \rangle}^{*int}
		&=\{v|\ \langle v, w\rangle \geq 0, \ ^{\forall} w
		\in V_+\}\\
		&=\{v|\ (v, Kw)\geq 0, \ ^{\forall} w
		\in V_+\}
		\\
		&= \{v|\ (Kv, w)\geq 0, \ ^{\forall} w\in V_+\}
	\end{align*}
	Thus $v\in V_+=V_{+\langle\cdot ,\cdot \rangle}^{*int}$ is 
	equivalent to $Kv\in V^{*int}_{+(\cdot ,\cdot )}$, i.e.
	$KV_{+}=V^{*int}_{+(\cdot ,\cdot )}$. Define $J=K$.
	\qed
\end{pf}

In Lemma \ref{lemma1}, we gave a necessary and 
sufficient condition for $V_+$ with an inner product 
$(\cdot ,\cdot )$ to be self-dual. 
The condition was the existence of a strictly positive map 
$J$ satisfying $J(V_+) = 
V_{+(\cdot ,\cdot )}^{*int}$. 
This map $J$ may not be unique. For instance, let us consider 
a classical system in $\R^{2}$ whose extreme points are 
two points $(1,1)$ and $(1,-1)$. The positive cone is a 
``forward lightcone'' $V_+=\{(x_0, x_1)| \ x_0\geq 0, x_0^2 - x_1^2 \geq 0\}$. 
It is easy to see that $V_{+}=V^{*int}_{+(\cdot ,\cdot )_{E}}$ with the standard Euclidean inner product $(\cdot ,\cdot )_{E}$.
However, 
if we choose an orthogonal basis $\{v_{0}, v_{1}\}$ of $\R^{2}$ given by $v_{0}=(1, 1)$ and $v_{1}=(1, -1)$, then
every linear map of the form 
\[
\left(
\begin{aligned}
	&v_{0}\\
	&v_{1}
\end{aligned}
\right)
\mapsto
\left(
\begin{aligned}
	&\lambda_0v_{0}\\
	&\lambda_1v_{1}
\end{aligned}
\right)
\]
for $\lambda_0, \lambda_1>0$ (which contains ``Lorentz transformations'' in $1+1$ dimension) is strictly positive and makes $V_{+}$ invariant.
Nevertheless, when $|\Omega^{\mathrm{ext}}|<\infty$, we can demonstrate that such strictly positive maps are ``equivalent'' to each other .

\begin{lemapp}\label{lemma5}
	Let $|\Omega^{\mathrm{ext}}|<\infty$.
	If a linear map 
	$J:V\to V$ is strictly positive with respect to an inner product $(\cdot ,\cdot )$, i.e. $(x, Jy)=(Jx, y)$ for all $x, y\in V$ and $(x, Jx)>0$ for all $x\in V$, and 
	satisfies $J(V_+)=V_{+}$,   
	then for each $\omega_{}^\mathrm{ext}\in \Omega^\mathrm{ext}$ there exists $\mu(\omega_{}^\mathrm{ext})>0$ such that $J(\omega_{}^\mathrm{ext}) =\mu(\omega_{}^\mathrm{ext}) \omega_{}^\mathrm{ext}$. 
\end{lemapp}
\begin{pf}
	Any $\omega^{\mathrm{ext}}\in\Omega^\mathrm{ext}$ is represented as 
	$\omega^{\mathrm{ext}}=c(\omega^{\mathrm{ext}})w$ with $c(\omega^{\mathrm{ext}}):=\|\omega^{\mathrm{ext}}\|=(\omega^{\mathrm{ext}}, \omega^{\mathrm{ext}})^{1/2}$ and $w$ satisfying $\|w\|=1$. 
	Suppose that there exists a family 
	\[
	\{\omega_k^{\mathrm{ext}}\}_{k=1}^{Z}=\{ c(\omega_k^{\mathrm{ext}})w_{k}\}_{k=1}^{Z}
	\subset \Omega^\mathrm{ext}
	\] 
	such that there is no $\mu(\omega_k^{\mathrm{ext}})>0$ for every $k=1, 2, \cdots, Z$ satisfying 
	$J(\omega_k^{\mathrm{ext}})=\mu(\omega_k^{\mathrm{ext}}) \omega_k^{\mathrm{ext}}$, and define $W:=\{w_{k}\}_{k=1}^{Z}$.
	Since $J$ maps each extreme ray of $V_{+}$ to an extreme ray of $V_{+}$, $J(w_k)$ with $w_{k}\in W$ is proportional to some $\omega^{\mathrm{ext}}\in\Omega^{\mathrm{ext}}$ (remember that an extreme ray of $V_{+}$ is the set of positive scalar multiples of an extreme point of $\Omega$). We can see that $J(w_{k})$ is proportional to some $w_p\in W$ with $p\neq k$ considering that $J(J(w_{k}))=\mu J(w_{k})$ holds if and only if $J(w_{k})=\mu w_{k}$ holds.

	We shall show in the following that there is a $w_{q}\in W$ such that $J(w_{q})\notin W$ despite of the argument above.
	To prove the claim, let us diagonalize $J$. 
	It is written as
	$J=\sum_{n=1}^M \tau_n R_n$, where $\tau_1 
	> \tau_2 >\cdots >\tau_M>0$ and $\{R_n\}_{n=1}^{M}$ are orthogonal projections. 
	We choose $w_1$ so that  
	$0 \neq (w_1, R_1 w_1)
	\geq (w_k, R_1 w_k)$
	for all $w_{k}\in W$. 
	Although such $w_1$ may not be unique, 
	the following argument does not depend on the choice.  
	If it happens that 
	$(w_k, R_1 w_k)=0$
	for all 
	$w_{k}\in W$, 
	we choose $w_1$ so that 
	$0 \neq (w_1, R_2 w_1)
	\geq (w_k, R_2 w_k)$
	for all 
	$w_{k}\in W$. If still $(w_k, R_2 w_k)=0$ for all $w_{k}\in W$, 
	we repeat the argument for $R_3, R_4, \cdots$. 
	For simplicity, we assume hereafter that 
	$(w_1, R_1 w_1)\neq 0$ holds. The general cases can be treated similarly. 
	Let $r_1:=R_1 w_1/\Vert R_1 w_1\Vert\neq0$, then $J$ is written as 
	\begin{align*}
		J= \tau_1\ketbra{r_{1}}{r_{1}}
		+ \tau_1 (R_1 -\ketbra{r_{1}}{r_{1}}) 
		+ \sum_{n\geq 2}\tau_n E_n
		=\tau_1 \hat{R}_0+\tau_1 \hat{R}_1
		+ \sum_{n\geq 2} \tau_n \hat{R}_n,  
	\end{align*} 
	where we define $\hat{R}_0:= \ketbra{r_{1}}{r_{1}}$, $\hat{R}_1:= 
	R_1-\ketbra{r_{1}}{r_{1}}$ and 
	$\hat{R}_n := R_n$ for $n\geq 2$ satisfying $\hat{R}_a\hat{R}_b=\delta_{ab}\hat{R}_a$ for $a,b=0,1,\cdots, M$. 
	Now we consider a vector 
	\begin{align*}
		\frac{J(w_1)}{\Vert J(w_1)\Vert}
		=
		\frac{\tau_1 \hat{R}_{0}w_1
			+ \tau_1 \hat{R}_1 w_1 
			+ \sum_{n\geq 2} \tau_n \hat{R}_n w_1}
		{
			\left(
			\tau_1^2 (w_1, \hat{R}_0w_1)
			+\tau_1^2 (w_1, \hat{R}_1w_1)
			+ \sum_{n \geq 2} \tau_n^2 
			(w_1, \hat{R}_n w_1)
			\right)^{1/2}
		},
	\end{align*}
	which must coincide with some $w_p\in W$. Its ``$\hat{R}_{0}$ -element'' can be calculated as
	\begin{align}
		&\left( \frac{J(w_1)}{\Vert J(w_1)\Vert}, 
		\hat{R}_0 
		\frac{J(w_1)}{\Vert J(w_1)\Vert}\right)\notag
		\\
		&\qquad\qquad
		=\frac{\tau_1^2 ( w_1, \hat{R}_0 w_1)}
		{
			\tau_1^2 ( w_1, \hat{R}_0 w_1)
			+\tau_1^2 ( w_1, \hat{R}_1w_1)
			+ \sum_{n \geq 2} \tau_n^2 
			( w_1, \hat{R}_n w_1)
		}\notag
		\\&\qquad\qquad
		= \frac{( w_1, \hat{R}_0 w_1)}
		{( w_1, \hat{R}_0 w_1)
			+ ( w_1, \hat{R}_1 w_1 )
			+ 
			\sum_{n=2}^M
			\frac{\tau_n^2}{\tau_1^2}
			( w_1, \hat{R}_n w_1)}.\label{eq_app_R0 element}
	\end{align} 
	On the other hand, we can obtain that
	\begin{align*}
		&( w_1, \hat{R}_0 w_1)
		+ ( w_1, \hat{R}_1 w_1 )
		+ 
		\sum_{n=2}^M
		\frac{\tau_n^2}{\tau_1^2}
		( w_1, \hat{R}_n w_1)
		\\
		&\qquad\qquad
		<  
		( w_1, \hat{R}_0 w_1)
		+ ( w_1, \hat{R}_1 w_1 )
		+ 
		\sum_{n=2}^M
		( w_1, \hat{R}_n w_1)=1 
	\end{align*}
	because there exists a
	$n\geq 2$ such that $( w_1, \hat{R}_n w_1 )
	\neq 0$ (otherwise $w_{1}=(\hat{R}_{0}+\hat{R}_{1})w_{1}=R_{1}w_{1}$ and thus $J(w_{1})=\tau_{1}w_{1}$ hold, which contradicts $w_{1}\in W$). 
	Therefore, \eqref{eq_app_R0 element} results in
	\begin{align*}
		\label{eq_app_R0 element ineq}
		\left( \frac{J(w_1)}{\Vert J(w_1)\Vert}, 
		\hat{R}_0 
		\frac{J(w_1)}{\Vert J(w_1)\Vert}\right)>( w_1, \hat{R}_0 w_1).
	\end{align*}
	This observation concludes a contradiction to $J(w_{1})/\|J(w_{1})\|=w_{p}\in W$ because $w_{1}$ satisfies $(w_1, \hat{R}_0 w_1)\geq(w_k, \hat{R}_0 w_k)$ for all $w_{k}\in W$. Overall, we find that 
	every $\omega^{\mathrm{ext}}\in \Omega^{\mathrm{ext}}$ has some
	$\mu(\omega^{\mathrm{ext}})>0$ such that 
	$J(\omega^{\mathrm{ext}})= \mu(\omega^{\mathrm{ext}})\omega^{\mathrm{ext}}$. 
	\qed
\end{pf}

\begin{lemapp}\label{prop1}
	Let $|\Omega^{\mathrm{ext}}|<\infty$, and
	suppose that linear maps 
	$J$ and $K$ strictly positive with respect to an inner product 
	$(\cdot ,\cdot )$ satisfy 
	$J(V_+) = 
	K(V_+)=
	V_{+(\cdot ,\cdot )}^{*int}$ (in particular, $V_{+}$ is self-dual). 
	Then there exists a $\mu(\omega^{\mathrm{ext}})>0$ 
	for each $\omega^{\mathrm{ext}}\in \Omega^{\mathrm{ext}}$ 
	such that $K(\omega^{\mathrm{ext}}) =\mu(\omega^{\mathrm{ext}}) J(\omega^{\mathrm{ext}})$ holds. 
\end{lemapp}
\begin{pf}
	As was seen in Lemma \ref{lemma1}, 
	the inner products $(\cdot, \cdot)_{J}:= 
	(\cdot, J\cdot)$ and $(\cdot, \cdot)_{K}:= 
	(\cdot, K\cdot)$ satisfy 
	$V_{+(\cdot, \cdot)_J}^{*int} = V_+$ and $V_{+(\cdot, \cdot)_K}^{*int} = V_+$ respectively. 
	Because $(\cdot, \cdot)_K$ 
	is represented as 
	$(\cdot, \cdot)_K=(\cdot, L\cdot)_J$ 
	with some linear map $L$ strictly positive with 
	respect to $(\cdot, \cdot)_J$, we have for arbitrary $v, w\in V$
	\begin{align*}
		(v, w)_K=(v, Kw) 
		= (v, Lw)_J= (v, JLw),
	\end{align*}
	and thus $L=J^{-1}\circ K$ holds. On the other hand, $L$ satisfies 
	\begin{align*}
		V_{+(\cdot, \cdot)_K}^{*int}
		&= \{v\mid(v, w)_K \geq 0, \ ^{\forall} w\in V_+\}
		\\
		&=\{v\mid(v, Lw)_J \geq 0, \ ^{\forall} w\in V_+\}
		\\
		&=\{v\mid(Lv, w)_J\geq 0, \ ^{\forall} w\in V_+\}
		=L^{-1}(V^{*int}_{+(\cdot, \cdot)_{J}}).
	\end{align*}
	That is, $L(V_+) =V_+$ holds. Therefore, we can apply Lemma \ref{lemma5} to $L$, and conclude that 
	\begin{align*}
		L(\omega^{\mathrm{ext}})=\mu(\omega^{\mathrm{ext}})\omega^{\mathrm{ext}}=J^{-1}(K(\omega^{\mathrm{ext}})),
	\end{align*}
	i.e. $K(\omega^{\mathrm{ext}})=\mu(\omega^{\mathrm{ext}})J(\omega^{\mathrm{ext}})$ holds.
	\qed
\end{pf}

\begin{propref1}
	Let $\Omega$ be transitive with $|\Omega^{\mathrm{ext}}|<\infty$ and $V_+$ be self-dual with respect to some inner product. There exists a linear bijection $\Xi\colon V\to V$ such that $\Omega':=\Xi\Omega$ is transitive and the generating positive cone $V'_{+}$ is self-dual with respect to $\langle\cdot,\cdot\rangle_{GL(\Omega')}$, i.e.
	$V^{'}_+ = V_{+\langle\cdot ,\cdot \rangle_{GL(\Omega')}}^{'*int}$.
\end{propref1}

\begin{pf}
	Because of the transitivity of $\Omega$, we can adopt the orthogonal coordinate system of $V$ introduced in Proposition \ref{prop_ortho repr}.
	Since $V_+$ is self-dual, there exists a linear map 
	$J\colon V\to V$ strictly positive with respect to $\langle\cdot, \cdot\rangle_{GL(\Omega)}$ such that 
	$J(V_+) = V_{+\langle \cdot, \cdot\rangle_{GL(\Omega)}}^{*int}$ (Lemma \ref{lemma1}). We can assume without loss of generality that $J$ satisfies $\langle\omega_{M}, J\omega_{M}\rangle_{GL(\Omega)}=1$.
	\if
	In fact, for any $v\in V_+$, as $V_+$ is 
	$GL(\Omega)$-invariant $T(v) \in V_+$ follows. 
	Thus $N\circ T(v) \in 
	V_{+, \langle \cdot, \cdot\rangle_{GL(\Omega)}}^{
		*int}$. As $V_{+, \langle \cdot, \cdot\rangle_{GL(\Omega)}}^{
		*int}$ is also $GL(\Omega)$-invariant, $N_T(v)\in 
	V_{+, \langle \cdot, \cdot\rangle_{GL(\Omega)}}^{
		*int}$ holds. Thus we find $N_T(V_+) \subseteq 
	V_{+, \langle \cdot, \cdot\rangle_{GL(\Omega)}}^{
		*int}$. 
	In addition, for any $w \in 
	V_{+, \langle \cdot, \cdot\rangle_{GL(\Omega)}}^{
		*int}$, $v:= T^{-1} \circ N^{-1} 
	\circ T(w) \in V_+$ satisfies 
	$N_T (v)=w$. Thus $N_T(V_+) 
	= V_{+, \langle \cdot, \cdot\rangle_{GL(\Omega)}}^{
		*int}$ is concluded. 
	\fi
	Let us introduce 
	\[
	\Omega^{*}:=V^{*int}_{+\langle \cdot, \cdot\rangle_{GL(\Omega)}}\cap[z=1]=\{v\in V^{*int}_{+\langle \cdot, \cdot\rangle_{GL(\Omega)}}\mid\langle v, \omega_{M}\rangle_{GL(\Omega)}=1\},
	\]
	where we identify the ``$\omega_{M}$-coordinate'' with ``$z$-coordinate'' in $V$ and define $[z=1]:=\{x\in V\mid\langle x, \omega_{M}\rangle_{GL(\Omega)}=1\}(=\mathit{aff}(\Omega))$ (see Proposition \ref{prop_ortho repr}). 
	Note that since both $V^{*int}_{+\langle \cdot, \cdot\rangle_{GL(\Omega)}}$ and $[z=1]$
	are $GL(\Omega)$-invariant, $\Omega^{*}$ is 
	also $GL(\Omega)$-invariant. 
	It is easy to demonstrate that $\Omega^{*}$ is convex (and compact), and we denote by $\Omega^{*\mathrm{ext}}$ the set of all extreme points of $\Omega^{*}$. We can also see that $\Omega^{*\mathrm{ext}}$ generates the extreme rays of $V^{*int}_{+\langle \cdot, \cdot\rangle_{GL(\Omega)}}$. Because $J$ satisfying $J(V_{+})=V_{+\langle \cdot, \cdot\rangle_{GL(\Omega)}}^{*int}$ is bijective and maps extreme rays of $V_{+}$ to extreme rays of $V_{+\langle \cdot, \cdot\rangle_{GL(\Omega)}}^{*int}$, it holds that $|\Omega^{*\mathrm{ext}}|=|\Omega^{\mathrm{ext}}|$.
	Thus there exists 
	a bijection $f\colon\Omega^{\mathrm{ext}} \to \Omega^{*\mathrm{ext}}$ and 
	$\beta(\omega^{\mathrm{ext}})>0$ for each $\omega^{\mathrm{ext}}\in \Omega^{\mathrm{ext}}$
	satisfying $J(\omega^{\mathrm{ext}})=\beta(\omega^{\mathrm{ext}}) f(\omega^{\mathrm{ext}})$.

	For each $T \in GL(\Omega)$, we introduce 
	$J_T:= T^{-1} \circ J \circ T$. It is easy to see that $J_T$ satisfies $J_T(V_+) = V_{+\langle \cdot, 
		\cdot\rangle_{GL(\Omega)}}^
	{*int}$ by virtue of Lemma \ref{lemma0}. Furthermore, $J_T$ is shown to be strictly positive with respect to $\langle \cdot, \cdot\rangle_{GL(
		\Omega)}$ because $T\in GL(\Omega)$ is an orthogonal transformation with respect to $\langle \cdot, \cdot\rangle_{GL(
		\Omega)}$.
	Therefore, applying Lemma \ref{prop1} to $J$ and $J_{T}$, 
	there exists $\mu_T:
	\Omega^{\mathrm{ext}} \to \mathbf{R}_{>0}$ 
	such that $J_T(\omega^{\mathrm{ext}})=\mu_{T}(\omega^{\mathrm{ext}})J(\omega^{\mathrm{ext}})$ for $\omega^{\mathrm{ext}}\in \Omega^{\mathrm{ext}}$, that is,
	\begin{align*}
		J_T(\omega^{\mathrm{ext}})&= \mu_T (\omega^{\mathrm{ext}}) J(\omega^{\mathrm{ext}})\\
		&= \mu_T(\omega^{\mathrm{ext}}) \beta(\omega^{\mathrm{ext}}) f(\omega^{\mathrm{ext}})\\
		&=: \beta_T(\omega^{\mathrm{ext}}) f(\omega^{\mathrm{ext}}),
	\end{align*} 
	where we define $\beta_T(\omega^{\mathrm{ext}}):=\mu_T(\omega^{\mathrm{ext}}) \beta(\omega^{\mathrm{ext}})$.
	We calculate this $\beta_T(\omega^{\mathrm{ext}})$.  
	It holds that 
	\begin{align*}
		J_T(\omega^{\mathrm{ext}}) 
		&=T^{-1}\circ J(T\omega^{\mathrm{ext}})\\
		&= T^{-1} (\beta(T\omega^{\mathrm{ext}}) f(T \omega^{\mathrm{ext}}))\\
		&=\beta(T\omega^{\mathrm{ext}}) T^{-1} f(T\omega^{\mathrm{ext}})\\
		&=\beta_T(\omega^{\mathrm{ext}}) f(\omega^{\mathrm{ext}}).  
	\end{align*}
	This relation shows that $T^{-1} f(T\omega^{\mathrm{ext}})$ is proportional to $f(\omega^{\mathrm{ext}})$. 
	Considering that the $z$-coordinates of $f(T\omega^{\mathrm{ext}})$ and $f(\omega^{\mathrm{ext}})$ are $1$ and that 
	$T^{-1}$ preserves $z$-coordinates, 
	we find that $T^{-1} f(T\omega^{\mathrm{ext}}) =f(\omega^{\mathrm{ext}})$ (equivalently, $f(T\omega^{\mathrm{ext}})= T f(\omega^{\mathrm{ext}})$) holds.
	Consequently, we obtain 
	\begin{align*}
		J_T(\omega^{\mathrm{ext}}) = \beta(T\omega^{\mathrm{ext}}) f(\omega^{\mathrm{ext}}). 
	\end{align*}

	Now we introduce 
	\begin{align*}
		J_{av}:= \frac{1}{|GL(\Omega)|}\sum_{
			T \in GL(\Omega)} J_T. 
	\end{align*}
	We note that $|GL(\Omega)|<\infty$ when $|\Omega^{\mathrm{ext}}|<\infty$ because $|GL(\Omega)|\le|\Omega^{\mathrm{ext}}|\hspace{0.1em}!$.
	$J_{av}$ acts on $\omega^{\mathrm{ext}}\in \Omega^{\mathrm{ext}}$ as 
	\begin{align*}
		J_{av}(\omega^{\mathrm{ext}}) = \frac{1}{|GL(\Omega)|}
		\sum_{T \in GL(\Omega)}\beta(T\omega^{\mathrm{ext}}) \cdot f(\omega^{\mathrm{ext}}) 
		=: C f(\omega^{\mathrm{ext}}),
	\end{align*}
	where $C:=\frac{1}{|GL(\Omega)|}
	\sum_{T \in GL(\Omega)}\beta(T\omega^{\mathrm{ext}})$ is a positive constant which does not depend on the choice of $\omega^{\mathrm{ext}}\in \Omega^{\mathrm{ext}}$ because of the transitivity of $\Omega$.
	Thus the map satisfies 
	$J_{av}(V_+) = 
	V_{+\langle \cdot, \cdot\rangle_{GL(\Omega)}}^{
		*int}$ 
	since $J_{av}(\Omega^{\mathrm{ext}})
	= C \Omega^{*\mathrm{ext}}$, and is strictly positive with respect to $\langle \cdot, \cdot\rangle_{GL(\Omega)}$ since it is a summation of the strictly positive operators $\{J_{T}\}_{T\in GL(\Omega)}$. 
	Moreover, it satisfies  
	\begin{align*}
		J_{av} \circ T = T \circ J_{av}. 
	\end{align*}
	for any $T \in GL(\Omega)$.
	We thus find that $J_{av} \circ P_{M} = P_{M} \circ J_{av}$ holds for the orthogonal projection $P_{M}$ introduced in Proposition \ref{prop_generalized ortho repr}. In fact, 
	\begin{align*}
		J_{av}(P_{M}x)
		&=\frac{1}{|GL(\Omega)|}J_{av}\left(
		\sum_{T \in GL(\Omega)}Tx\right)\\
		&=\frac{1}{|GL(\Omega)|}
		\sum_{T \in GL(\Omega)}T(J_{av}x)\\
		&=P_{M}(J_{av}x)
	\end{align*}
	holds for all $x\in V$.
	Therefore, $J_{av}$ is decomposed into two parts as 
	\begin{align}
		\label{eq_ortho_decomp}
		J_{av}= P_{M}\circ J_{av} \circ P_{M} + 
		P_{M}^{\perp} \circ J_{av} \circ P_{M}^{\perp},
	\end{align}
	where $P_{M}^{\perp}=\id-P_{M}$.
	We note that $V_{M}^{\perp}=ImP_{M}^{\perp}=[\mathit{aff}(\Omega)-\omega_{M}]=\R^{N}$ and $\mathrm{dim}\ V_{M}=\mathrm{dim}\ ImP_{M}=1$ hold by virtue of Proposition \ref{prop_ortho repr}.
	Therefore, the first part of \eqref{eq_ortho_decomp} is proportional to $\id_{V_{M}}=\id_{z}=P_{M}$, and because we set $\langle\omega_{M}, J\omega_{M}\rangle_{GL(\Omega)}=1$ and thus
	\begin{align*}
		\left\langle \omega_{M},\ P_{M}\circ J_{av} \circ P_{M}\omega_{M}\right\rangle_{GL(\Omega)}
		&=\langle \omega_{M}, J_{av} \omega_{M}\rangle_{GL(\Omega)}\\
		&=\langle \omega_{M}, P_{M}J \omega_{M}\rangle_{GL(\Omega)}\\
		&=\langle \omega_{M}, J \omega_{M}\rangle_{GL(\Omega)}\\
		&=1\\
		&=\langle \omega_{M}, P_{M}\omega_{M}\rangle_{GL(\Omega)}
	\end{align*}
	holds, it is proven that 
	\[
	P_{M}\circ J_{av} \circ P_{M}=P_{M}.
	\]
	Let us examine the second part. 
	Suppose that there exists a nonzero $x\in V_{M}^{\perp}$ such that $Tx=x$ for all $T\in GL(\Omega)$. 
	Then $P_{M}x=x\neq0$ holds, and it contradicts \eqref{app_pf0}. Thus we can find that $GL(\Omega)$ acts irreducibly on $V_{M}^{\perp}$, that is, only $\{0\}$ and $V_{M}^{\perp}=\R^N$ itself are 
	invariant subspaces. It concludes that $P_{M}^{\perp} J_{av} P_{M}^{\perp}$,
	which commutes with every element in $GL(\Omega)$,
	is proportional to $\id_{V_{M}^{\perp}}=\id_{\R^{N}}=P_{M}^{\perp}$ due to Schur's lemma. 
	Consequently, we obtain for some $\xi >0$
	\begin{align*}
		J_{av} = P_{M} + \xi P_{M}^{\perp},
	\end{align*}
	and thus
	\begin{align}
		\label{app_pf1}
		J_{av}(V_{+})
		&=(P_{M} + \xi P_{M}^{\perp})(V_{+})
		=V_{+\langle \cdot, \cdot\rangle_{GL(\Omega)}}^{*int}.
	\end{align}	
	
	Let us introduce a linear bijection 
	\[
	\Xi:=\sqrt{J_{av}}=P_{M}+\sqrt{\xi}P_{M}^{\perp},
	\] 
	strictly positive with respect to $\langle \cdot, \cdot\rangle_{GL(\Omega)}$, and define $\Omega':=\Xi\Omega$. It is easy to check that the positive cone $V_{+}'$ generated by $\Omega'$ is given by $V_{+}'=\Xi V_{+}$, and $GL(\Omega')=\Xi GL(\Omega)\Xi^{-1}=GL(\Omega)$ (moreover, the unique maximally mixed state of $\Omega'$ is still $\omega_{M}$). In addition, we can find that
	\begin{align*}
		V_{+\langle \cdot, \cdot\rangle_{GL(\Omega')}}^{'*int}
		&=\{v\mid \langle v, w'\rangle_{GL(\Omega)}\ge 0,\ ^{\forall}w'\in V'_{+}
		\}\\
		&=\{v\mid \langle v, \Xi w\rangle_{GL(\Omega)}\ge 0,\ ^{\forall}w\in V_{+}
		\}\\
		&=\Xi^{-1}V_{+\langle \cdot, \cdot\rangle_{GL(\Omega)}}^{*int}.
	\end{align*}
	holds.
	Since \eqref{app_pf1} can be rewritten as
	\[
	\Xi V_{+}=\Xi^{-1}V_{+\langle\cdot ,\cdot \rangle_{GL(\Omega)}}^{*int},
	\]
	we can conclude
	\[
	V^{'}_+ = V_{+\langle\cdot ,\cdot \rangle_{GL(\Omega')}}^{'*int}.
	\]
	\qed
\end{pf}

\if
We employ R-representation. We may rescale 
the coordinate so that every
pair $v,w \in \Omega$ satisfies
$\langle v, w\rangle_{GL(\Omega)}>0$. 
Let us introduce 
$\Omega^{*}:= V^{*int}_{+, \langle \cdot, \cdot\rangle_{GL(\Omega)}}\cap [z=1]$. 
As $V^{*int}_{+, \langle \cdot, \cdot\rangle_{GL(\Omega)}}$ 
is $GL(\Omega)$-invariant, $\Omega^{*}$ is 
also $GL(\Omega)$-invariant. 
Thus $\Omega^{*ext}=\{Tw_0| T\in GL(\Omega)\}$ holds, 
where $w_0$ is a point on a hyperplane $[z=1]$. 
Moreover, because $N$ is bijective, 
$|\Omega^{*ext}|=|\Omega^{ext}|$ follows.  
Each point in $\Omega^{ext}$ is mapped by $N$ 
to a point which is proportional to 
a point in $\Omega^{*ext}$. 
As $N$ is strictly positive it can be diagonalized as 
$N= \sum_{n} \lambda_n E_n$, where $\lambda_n >0$ 
and $E_n$ is a projection. 
Each $v_j \in \Omega^{ext}$ is mapped as 
\begin{eqnarray*}
	Nv_j = \sum_n \lambda_n E_n v_j,  
\end{eqnarray*}
which must be proportional to some $w_j\in \Omega^{*ext}$. 
That is, there exists $\mu_j>0$ such that 
\begin{eqnarray*}
	\sum_n \lambda_n E_n v_j = \mu_j w_j 
	= \sum_n \mu_j E_n w_j. 
\end{eqnarray*}
Operating $E_m$ to the above equation, 
we obtain 
\begin{eqnarray*}
	\lambda_m E_m v_j = \mu_j E_m w_j.  
\end{eqnarray*}
It implies that for $m$ with $E_m w_j =0$, 
$E_m v_j=0$ holds as $\lambda_m, \mu>0$. 
It implies that 
for $m$ with $E_m w_j \neq 0$, 
both $
\lambda_m = \mu_j >0$ and 
$E_m v_j = E_m w_j$ holds. 
For $m$ with $E_m w_j =0$, 
either $\lambda_m =0$ or $E_m v_j =0$ must hold. 
Thus we obtain 
\begin{eqnarray*}
	v_j = w_j + \sum_{m: E_m w_j =0} 
	E_m v_j.
\end{eqnarray*}
Thus we can conclude that 
$v_j = w_j$ and $\lambda_m =\mu$ for all $m$. 
\fi

\begin{rmk**}
	In the case of $|\Omega^{\mathrm{ext}}|=\infty$, 
	there exists a counterexample of Lemma \ref{lemma5}. 
	\if
	Let us consider $\Omega=\ ^{t}\{(\boldsymbol{x}, 1)\in \R^4\mid |\boldsymbol{x}| \leq 1\}$ (Bloch ball),  
	whose extreme point set is 
	$\Omega^{\mathrm{ext}}=\{\boldsymbol{x}\mid
	\boldsymbol{x}\in \mathbf{R}^3, |\mathbf{x}|=1\}$. 
	This $\Omega^{ext}$ is transitive 
	with respect to $GL(\Omega) =O(3)$.  
	Let us embed $\Omega$ into $\mathbf{R}^4$ 
	so that $\Omega \subset [x_0=1]$. 
	\fi
	Let us consider a state space
	\[
	\Omega=\{\ ^t(
	1,\boldsymbol{x})=\ ^t(
	1,x_1,x_2,x_3)\in\R^{4}\mid |\boldsymbol{x}|^{2}=x_1^2 + x_2^2 + x_3^2 \leq 1\}
	\]
	(the Bloch ball).
	$\Omega$ defines a corresponding positive cone $V_+$ 
	as
	\[
	V_+=\{x\in\R^{4}\mid x_0^2 - |\boldsymbol{x}|^2 \geq 0, 
	x_0\geq 0\},
	\]
	which can be identified with a forward light cone
	of a Minkowski spacetime.
	We examine a pure Lorentz transformation 
	$\Lambda$ defined for $\lambda \in \R$
	as
	\begin{eqnarray*}
		\Lambda= \left[
		\begin{array}{cccc}
			\cosh \lambda& \sinh \lambda& 0& 0\\
			\sinh \lambda& \cosh \lambda& 0&0\\
			0&0&1&0\\
			0&0&0&1
		\end{array}
		\right].
	\end{eqnarray*}
	It is easy to prove that this $\Lambda$ is strictly positive.
	\if
	as it holds that for any $x
	\in \mathbf{R}^4$ 
	\begin{eqnarray*}
		(x, \Lambda x) = \frac{1}{2}
		(e^{\lambda} (x_0+x_1)^2 
		+2^{-\lambda} (x_0-x_1)^2) 
		+x_2^2 + x_3^2 \geq 0. 
	\end{eqnarray*}
	\fi
	Since the pure Lorentz transformation 
	preserves the Minkowski metric, 
	it satisfies $\Lambda(V_+)=V_+$. 
	However, $\Lambda$ transforms an extreme point 
	$x=\ ^{t}(1,0,1,0)$ to 
	\[
	\Lambda(x)=\ ^{t}(\cosh \lambda, 
	\sinh \lambda, 1,0),
	\]
	which is not proportional to 
	$x$. Investigating whether Proposition \ref{prop_transitive self-dual} still holds when $|\Omega^{\mathrm{ext}}|=\infty$ is a future problem. 
\end{rmk**}

\cleardoublepage
\phantomsection
\addcontentsline{toc}{chapter}{\bibname}
\bibliographystyle{hieeetr_url} 
\bibliography{ref_PhD_0125}
\end{document}

%% file: chap2.tex
\chapter{Generalized Probabilistic Theories}
\label{chap:GPT}
Quantum theory is the most successful theory that describes nature:
it does explain phenomena that cannot be recognized if we live in the classical world.
The existence of superposition or entanglement is an instance of those remarkable phenomena, but probably the most drastic one is that nature is probabilistic: even if we conduct a ``perfect'' preparation of a physical system and measurement, we do not always obtain one determined outcome.
Generalized probabilistic theories (GPTs) are the framework that focuses on those probabilistic behaviors of nature.
The only requirement for GPTs is the convexity for primitive notions of states and effects, and there are in general not assumed any Hilbert space structures or operator algebraic properties.
In this sense, GPTs are a more general framework than quantum theory and classical theory, and play an active role in the study of quantum foundations \cite{hardy2001quantum,Barnum_nocloning,PhysRevA.75.032304,PhysRevLett.99.240501,PhysRevA.81.062348,PhysRevA.84.012311,Masanes_Muller_2011,barnum2012teleportation,Lami_PhD,Plavala_2021_GPTs}\footnote{Recent results on GPTs are summarized briefly in \cite{Lami_PhD,Plavala_2021_GPTs}.} after their initial proposition and development in the 1960s and 1970s \cite{Mackey1963,Ludwig1964,Ludwig1967,Davies1970,cmp/1103858551,Hartkaemper_Neumann_1974}.\footnote{For historical review of GPTs, we recommend \cite{Lami_PhD,Cassinelli2016}.}
In this chapter, we explore the mathematical foundations of GPTs in detail to show how they give the most intuitive and fundamental description of nature.

This chapter is organized as follows.
In Section \ref{2sec:fundamental}, we give the two most fundamental notions of GPTs, namely, states and effects.
They are introduced in a conceptual and operational way, and mathematically embedded into a vector space and its dual (more generally, a Banach space and its Banach dual) respectively.
These embeddings form the mathematical foundations of GPTs.
In fact, thanks to this embedding theorem, studies on GPTs usually begin with the assumption that a state space is a compact convex set in a finite-dimensional vector space (more generally, a closed base of a base norm Banach space).
After giving the descriptions of states and effects, we explain other basic but somewhat more advanced topics, composite systems and transformations in GPTs, in Section \ref{2sec:composites} and Section \ref{2sec:channels}.
It is found that the previously introduced embeddings into vector spaces make it mathematically convenient to discuss those concepts.
In Section \ref{2sec:additional}, we introduce the notions of transitivity and self-duality.
These additional notions often appear in the field of GPTs, and our main results in the following chapter are also obtained based on them.
In Section \ref{2sec:example}, we illustrate some examples of GPTs including classical and quantum theories with finite levels and other important theories often considered in the study of quantum foundations.
Throughout this chapter, explicit proofs of mathematical matters are given in principle, but some of them are omitted when they are too technical or lengthy.

\section{States and effects}
\label{2sec:fundamental}
A physical experiment is described by three procedures: to prepare an object system, to perform a measurement, and to obtain an outcome.
However, in general, even if the same preparations are conducted and the same measurements are performed, each outcome obtained is different, and we can only predict from the preparation and measurement how frequently each outcome is obtained, i.e., the probabilities  \cite{Kraus1983,Araki_mathematical, Busch_quantummeasurement,Holevo1982}.\footnote{In \cite{Cassinelli2016}, this primitive assumption of physics is called the {\itshape statistical causality}.}
Let us give a concrete description.
For a preparation procedure $\mathsf{P}$, measurement apparatus $\mathsf{A}$, and a measurable set $(X, \mathcal{A})$, where $X$ is the nonempty set of outcomes associated with $\mathsf{A}$ and $\mathcal{A}$ is a $\sigma$-algebra of subsets of $X$, we denote by $\mu(\mathsf{A}, \mathsf{P})(U)$ the probability of obtaining an outcome in $U\in \mathcal{A}$ when measuring $\mathsf{A}$ on $\mathsf{P}$.
Then each pair $(\mathsf{A}, U)$ reflects whether a measurement of $\mathsf{A}$ yields a result in the set $U$ or not.
We regard such ``yes-no measurements'' as a more fundamental notion than the original measurement apparatus because the latter is an assemblage of the former.
\begin{figure}[h]
	\centering
	\includegraphics[bb=0.000000 0.000000 898.000000 352.000000, scale=0.38]{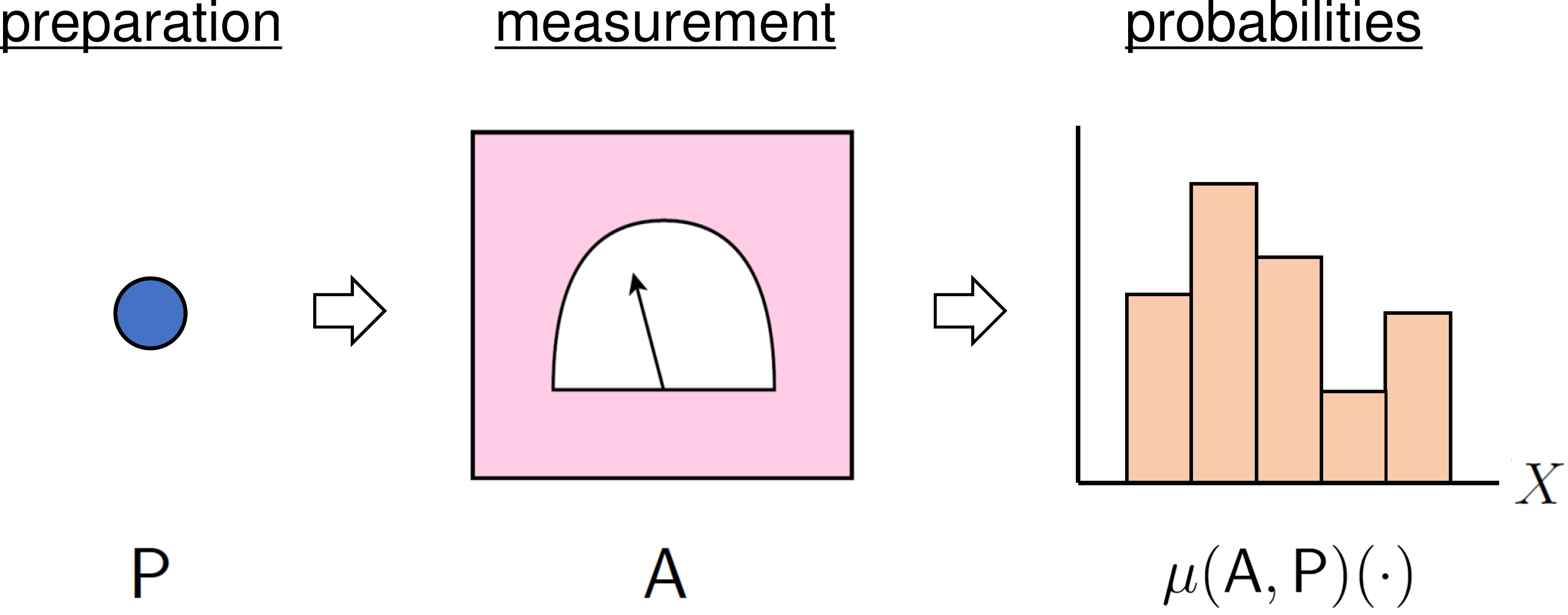}
	\caption{Description of physical experiments.}
	\label{fig:experiment}
\end{figure}

In this section, we shall demonstrate how to describe two fundamental concepts of physics, preparations and measurements, in mathematical language.
As explained above, we focus mainly on yes-no measurements, and write a yes-no measurement and the probability $\mu(\mathsf{A}, \mathsf{P})(U)$ simply as $\mathsf{M}$ and $\mu(\mathsf{M}, \mathsf{P})$ respectively.
It will be shown that they are reduced to the notions of {\itshape states} and {\itshape effects}, and are embedded naturally into some vector space and its dual space respectively.
The embedding theorem enables us to treat abstract concepts of preparations and measurements as mathematically well-defined objects, which is the very starting point for GPTs.
After their investigations, we will go back to descriptions of general measurement apparatuses to obtain the notion of {\itshape observables}.
This section is mainly in accord with \cite{Lami_PhD,Plavala_2021_GPTs,Busch_quantummeasurement,Gudder_stochastic,Kuramochi_compactconvex_2020,kimura2010physical}.

\subsection{Axiomatic description}
\label{2subsec:states_effects}
\begin{figure}[h]
	\centering
	\includegraphics[bb=0.000000 0.000000 884.000000 355.000000, scale=0.402]{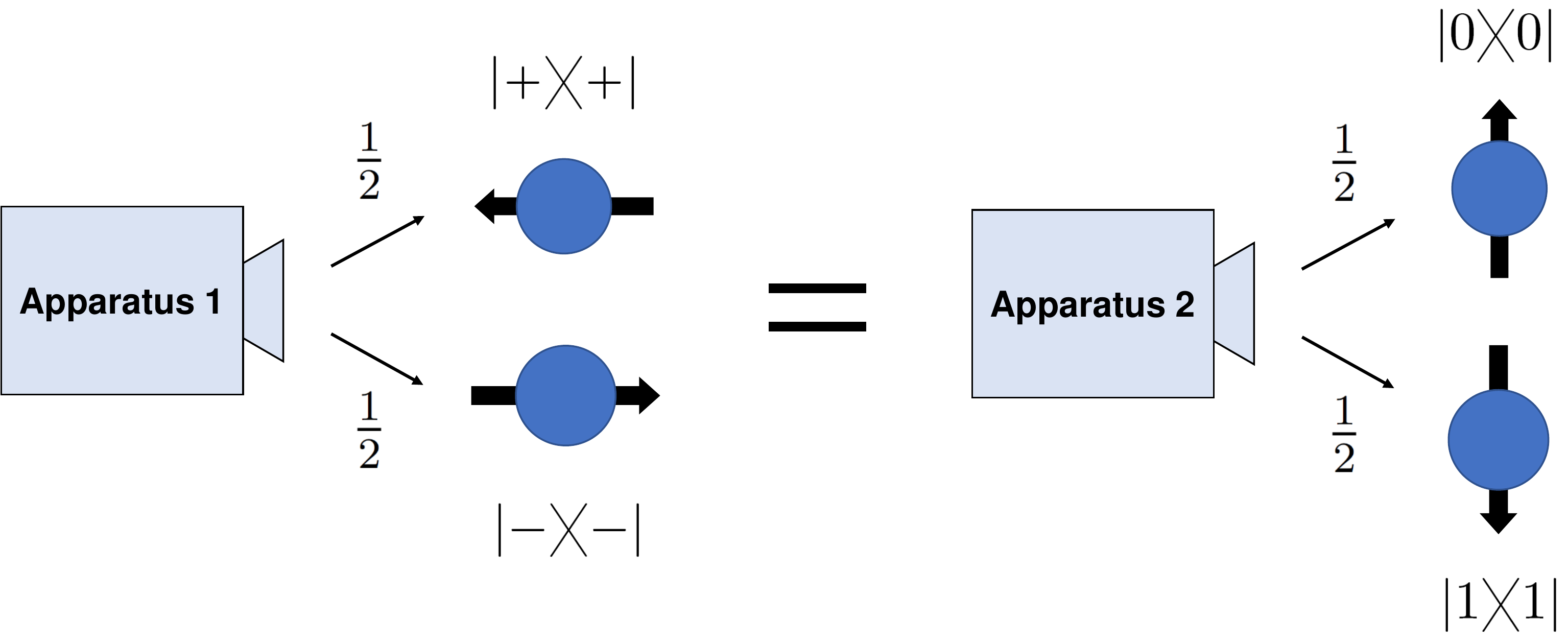}
	\caption{We identify apparatuses that have different ``contexts'' but generate the same statistics.}
	\label{fig:context}
\end{figure}

Let {\sffamily Prep} and {\sffamily Meas} be the set of all procedures of preparations and yes-no measurements for some physical experiment respectively.
For example, in the experiment of detecting the spin of an electron, each element of {\sffamily Prep} represents an apparatus that emits an electron, and each element of {\sffamily Meas} represents a value of the meter of some measurement apparatus or the corresponding yes-no apparatus itself.
What is specific to this description is that apparatuses with different physical implementations are distinguished.
In the previous example, an apparatus that emits randomly (i.e., with probabilities $\half$ and $\half$) electrons with $x+$ spin and $x-$ spin, and apparatus that emits randomly electrons with $z+$ spin and $z-$ spin are different elements of {\sffamily Prep}, even though they describe the same quantum state $\frac{\id}{2}$.
In the field of GPTs, we do not pay attention to those differences of ``context'' \cite{PhysRevA.71.052108} for both preparations and measurements, but only focus on the statistics: if we have two apparatuses that are different but output the same statistics, then we identify those two apparatuses in our framework (see Figure \ref{fig:context}).

Let us present its mathematical expression. 
Preparation procedures $\mathsf{P_1, P_2}\in\mathsf{Prep}$ are called {\itshape operationally equivalent} (denoted by $\mathsf{P_1}\sim_P \mathsf{P_2}$) if $\mu(\mathsf{M}, \mathsf{P_1})=\mu(\mathsf{M}, \mathsf{P_2})$ holds for all $\mathsf{M}\in\mathsf{Meas}$.
In a similar way, measurement procedures $\mathsf{M_1, M_2}\in\mathsf{Meas}$ are called {\itshape operationally equivalent} (denoted by $\mathsf{M_1}\sim_M \mathsf{M_2}$) if $\mu(\mathsf{M_1}, \mathsf{P})=\mu(\mathsf{M_2}, \mathsf{P})$ for all $\mathsf{P}\in\mathsf{Prep}$.
The binary relations $\sim_P$ and $\sim_M$ define equivalence relations, and thus we can introduce the corresponding quotient sets $\tilde{\Omega}:=\mathsf{Prep}/\hspace{-1.5mm}\sim_P$ and $\tilde{\mathcal{E}}:=\mathsf{Meas}/\hspace{-1.5mm}\sim_M$.
These two sets $\tilde{\Omega}$ and $\tilde{\mathcal{E}}$ are called the {\itshape state space} and {\itshape effect space} respectively, and each element of $\tilde{\Omega}$ and $\tilde{\mathcal{E}}$ are called a {\itshape state} and an {\itshape effect} respectively \cite{Ludwig1964,Ludwig1967,Kraus1983,Gudder_stochastic}.
Here, we express those descriptions above as an axiom.
\begin{axiom}[Separation principle]
	\label{axiom:separation}
States and effects separate each other.
That is, for any distinct $\tilde{\omega}_1, \tilde{\omega}_2\in\tilde{\Omega}$, there exists an effect $\tilde{e}\in\tilde{\mathcal{E}}$ such that $\mu(\tilde{e}, \tilde{\omega}_1)\neq\mu(\tilde{e}, \tilde{\omega}_2)$, and also,
for any distinct $\tilde{e}_1, \tilde{e}_2\in\tilde{\mathcal{E}}$, there exists a state $\tilde{\omega}\in\tilde{\Omega}$ such that $\mu(\tilde{e}_1, \tilde{\omega})\neq\mu(\tilde{e}_2, \tilde{\omega})$.
\end{axiom}
We note that in the statement above we regard the function $\mu(\cdot, \cdot)$ on $\mathsf{Meas}\times\mathsf{Prep}$ as on $\tilde{\mathcal{E}}\times\tilde{\Omega}$ in an well-defined way.
States and effects are two primitive notions in GPTs.

Next, we focus on another fundamental concept, {\itshape probabilistic mixtures}.
It is operationally natural to assume that if we can prepare states $\tilde{\omega}_1, \tilde{\omega}_2, \ldots, \tilde{\omega}_n$, then we can also prepare a state through the probabilistic mixture of $\tilde{\omega}_1, \tilde{\omega}_2,$ $\ldots, \tilde{\omega}_n$ with respective probabilities $\lambda_{1}, \lambda_{2}, \ldots, \lambda_{n}$, where $\lambda_{i}\ge0$ and $\sum_{i=1}^{n}\lambda_i=1$.\footnote{From an operational viewpoint, it seems unnatural to consider mixtures with irrational ratios because we can only conduct a finite number of experiments. However, in this thesis, we focus on theories with the completeness assumption (see Mathematical assumption \ref{math_assume:completeness}), so at this point admit those irrational mixtures.}
We denote the newly introduced state by $\ang{\lambda_{1},  \lambda_{2}, \ldots, \lambda_{n};\tilde{\omega}_1, \tilde{\omega}_2, \ldots, \tilde{\omega}_n}_{\tilde{\Omega}}$.
The notion of probabilistic mixtures should be considered also for effects, and we denote the effect obtained through the mixture of effects $\{\tilde{e}_j\}_{j=1}^{m}\subset\tilde{\mathcal{E}}$ with a probability weight $\{\sigma_j\}_{j=1}^{m}$ by $\ang{\sigma_{1},  \sigma_{2}, \ldots, \sigma_{m};\tilde{e}_1, \tilde{e}_2, \ldots, \tilde{e}_m}_{\tilde{\mathcal{E}}}$.
Then the nature of probabilistic mixtures motivates us to give the following axiom.
\begin{axiom}[Probabilistic mixtures]
	\label{axiom:prob mix}
For any finite set of states $\{\tilde{\omega}_i\}_{i=1}^{n}\subset\tilde{\Omega}$ and probability weight $\{\lambda_{i}\}_{i=1}^{n}$ ($\lambda_{i}\ge0$ and $\sum_i\lambda_{i}=1$), there exists a state $\ang{\lambda_{1},  \lambda_{2}, \ldots, \lambda_{n};\tilde{\omega}_1, \tilde{\omega}_2, \ldots, \tilde{\omega}_n}_{\tilde{\Omega}}\in\tilde{\Omega}$ satisfying
\begin{equation}
\label{eq1:convexity}
\mu\left(\tilde{e}, \ang{\lambda_{1},  \lambda_{2}, \ldots, \lambda_{n};\tilde{\omega}_1, \tilde{\omega}_2, \ldots, \tilde{\omega}_n}_{\tilde{\Omega}}\right)=\sum_{i=1}^{n}\lambda_{i}\mu(\tilde{e}, \tilde{\omega}_i)
\end{equation}
for all $\tilde{e}\in\tilde{\mathcal{E}}$.
Similarly, for any finite set of effects $\{\tilde{e}_j\}_{j=1}^{m}\subset\tilde{\mathcal{E}}$ and probability weight $\{\sigma_{j}\}_{j=1}^{m}$, 
there exists an effect $\ang{\sigma_{1},  \sigma_{2}, \ldots, \sigma_{m};\tilde{e}_1, \tilde{e}_2, \ldots, \tilde{e}_m}_{\tilde{\mathcal{E}}}\in\tilde{\mathcal{E}}$ satisfying
\begin{equation}
	\label{eq1:convexity2}
	\mu\left(\ang{\sigma_{1},  \sigma_{2}, \ldots, \sigma_{m};\tilde{e}_1, \tilde{e}_2, \ldots, \tilde{e}_m}_{\tilde{\mathcal{E}}}, \tilde{\omega}\right)=\sum_{j=1}^{m}\sigma_{j}\mu(\tilde{e}_j, \tilde{\omega})
\end{equation}
for all $\tilde{\omega}\in\tilde{\Omega}$.
From Axiom \ref{axiom:separation}, they are uniquely determined.
\end{axiom}
Axiom \ref{axiom:separation} and Axiom \ref{axiom:prob mix} ensure that, in addition to \eqref{eq1:convexity}, several properties that probabilistic mixtures should satisfy hold successfully for the state $\ang{\lambda_{1},  \lambda_{2}, \ldots, \lambda_{n};\tilde{\omega}_1, \tilde{\omega}_2, \ldots, \tilde{\omega}_n}_{\tilde{\Omega}}$.
For example, we can derive easily that 
\[
\ang{\lambda_{1},  \lambda_{2}, \ldots, \lambda_{n};\tilde{\omega}_1, \tilde{\omega}_2, \ldots, \tilde{\omega}_n}_{\tilde{\Omega}}
=
\ang{\lambda_{2},  \lambda_{1}, \ldots, \lambda_{n};\tilde{\omega}_2, \tilde{\omega}_1, \ldots, \tilde{\omega}_n}_{\tilde{\Omega}}
\]
holds, i.e., the mixture does not depend on the ``order'' of the states and probabilities (similar observations also can be obtained for effects).

We require additional conditions for $\tilde{\mathcal{E}}$
according to \cite{Lami_PhD,Filippov2020_restriction,PhysRevA.87.052131}.
The first requirement is that $\tilde{\mathcal{E}}$ includes the {\itshape unit effect} $\tilde{u}$ satisfying $\mu(\tilde{u}, \tilde{\omega})=1$ for all $\tilde{\omega}\in\tilde{\Omega}$.
In other words, we suppose the existence of a yes-no measurement apparatus that always outputs ``yes'', and this seems to be an operationally natural condition.
We note that such $\tilde{u}$ is unique due to Axiom \ref{axiom:separation}.
The second one is that if $\tilde{e}$ is an element of $\tilde{\mathcal{E}}$, then the complement effect $\tilde{e}^{\perp}$ such that $\mu(\tilde{e}^{\perp}, \tilde{\omega})=1-\mu(\tilde{e}, \tilde{\omega})$ for all $\tilde{\omega}\in\tilde{\Omega}$ is also an element of $\tilde{\mathcal{E}}$.
This condition comes from an operationally natural intuition that if we admit a certain yes-no measurement apparatus, then we should also admit the apparatus constituted by exchanging the ``yes'' and ``no'' of the original one.
We remark similarly that such $\tilde{e}^{\perp}$ is unique.
For the complement of the unit effect $\tilde{u}$, we sometimes denote it by $\tilde{0}$ in this thesis.
These conditions are summarized as follows.
\begin{axiom}[Existence of unit and complement effects]
	\label{axiom:unit-complement}
(i) There exists the unit effect $\tilde{u}$ in $\tilde{\mathcal{E}}$ such that $\mu(\tilde{u}, \tilde{\omega})=1$ for all $\tilde{\omega}\in\tilde{\Omega}$.\\
(ii) If $\tilde{e}\in\tilde{\mathcal{E}}$, then its complement $\tilde{e}^{\perp}\in\tilde{\mathcal{E}}$ such that $\mu(\tilde{e}^{\perp}, \tilde{\omega})=1-\mu(\tilde{e}, \tilde{\omega})$ for all $\tilde{\omega}\in\tilde{\Omega}$.
\end{axiom}
We note that the effects $\tilde{u}$ and $\tilde{e}^{\perp}$ in Axiom \ref{axiom:unit-complement} are consistent with Axiom \ref{axiom:prob mix}.

Now we can give the definition of a GPT.
\begin{defi}[Generalized probabilistic theories]
	\label{2def:GPTs}
A triple $(\tilde{\Omega}, \tilde{\mathcal{E}}, \mu)$ of two sets $\tilde{\Omega}$ and $\tilde{\mathcal{E}}$, and a function $\mu\colon\tilde{\Omega}\times\tilde{\mathcal{E}}\to[0,1]$ satisfying Axiom \ref{axiom:separation}, Axiom \ref{axiom:prob mix}, and Axiom \ref{axiom:unit-complement} is called a {\itshape generalized probabilistic theory} (a {\itshape GPT} for short).
The set $\tilde{\Omega}$ and its element are called the {\itshape state space} and a {\itshape state} of the theory, and 
$\tilde{\mathcal{E}}$ and its element are called the {\itshape effect space} and an {\itshape effect} of the theory respectively.
\end{defi}

Let us consider infinite countable mixtures for states.\footnote{For infinite countable mixtures of effects, see footnote \ref{2footnote:completeness of effects}.}
In the following, we denote mixtures of two states $\ang{\lambda, 1-\lambda;\tilde{\omega}_1, \tilde{\omega}_2}_{\tilde{\Omega}}$ simply by $\ang{\lambda;\tilde{\omega}_1, \tilde{\omega}_2}$.
In order to treat infinite limits, some topological structure should be introduced into $\tilde{\Omega}$.
Here we define a topology on $\tilde{\Omega}$ in line with Gudder \cite{Gudder_stochastic}.
We suppose that if states $\tilde{\omega}_1$ and $\tilde{\omega}_2$ are ``close'', then 
\[
\ang{\lambda;\tilde{\omega}_1', \tilde{\omega}_1}=\ang{\lambda;\tilde{\omega}_2', \tilde{\omega}_2}
\]
with small $\lambda$ holds for some $\tilde{\omega}_1', \tilde{\omega}_2'\in\tilde{\Omega}$.
That is, the closeness between $\tilde{\omega}_1$ and $\tilde{\omega}_2$ should be evaluated by 
\begin{equation}
	\label{2eq:distance}
	\begin{aligned}
	\tilde{d}(\tilde{\omega}_1,  \tilde{\omega}_2):=
	\inf\{0<\lambda\le1\mid&\ang{\lambda;\tilde{\omega}_1', \tilde{\omega}_1}=\ang{\lambda;\tilde{\omega}_2', \tilde{\omega}_2}
	\\
	&\qquad\qquad\qquad\quad\mbox{for some $\tilde{\omega}_1', \tilde{\omega}_2'\in\tilde{\Omega}$}\}.
	\end{aligned}
\end{equation}
We note that \eqref{2eq:distance} always can be defined since $\ang{\half;\tilde{\omega}_2, \tilde{\omega}_1}=\ang{\half;\tilde{\omega}_1, \tilde{\omega}_2}$ holds due to Axiom \ref{axiom:separation}.
We assume that infinite countable mixtures are allowed in our framework.
It is described in the following form.
\begin{assum}[Completeness]
	\label{math_assume:completeness}
If $\tilde{d}$ defined in \eqref{2eq:distance} satisfies $\lim_{n, 
m\to\infty}\tilde{d}(\tilde{\omega}_n,  \tilde{\omega}_m)=0$ for a family of states $\{\tilde{\omega}_n\}_{n=1}^{\infty}\subset\tilde{\Omega}$, then there exists a unique $\tilde{\omega}\in\tilde{\Omega}$ such that $\lim_{n	\to\infty}\tilde{d}(\tilde{\omega}_n,  \tilde{\omega})=0$.
\end{assum}
There are two things to remark on Mathematical assumption \ref{math_assume:completeness}.
The first one is about the notion of completeness.
In fact, we can prove that the function $\tilde{d}$ is a metric function on $\tilde{\Omega}$ (see Subsection \ref{2subsec:embeddings}), and thus Mathematical assumption \ref{math_assume:completeness} is equivalent to the requirement that $(\tilde{\Omega}, \tilde{d})$ is a complete metric space, which especially admits infinite countable mixtures.
The other remark is about the terminology ``Mathematical assumption''.
In the field of GPTs, the assumption of closedness or completeness for a state space with respect to some physically natural topology is a common one \cite{Plavala_2021_GPTs}.
That is, if we can prepare states that are very ``close'' to some fixed state, then it is usually assumed that the fixed state can also be prepared.
This seems to be a natural, but at the same time more artificial assumption than the previous ones, so in this thesis we regard it as a mathematical assumption rather than an axiom.

\subsection{Convex structures and embedding theorems}
\label{2subsec:embeddings}
In the previous section, we presented the primitive descriptions of states and effects from a physical perspective.
We can rephrase them via the mathematical notion of {\itshape convex structures} \cite{cmp/1103858551,Gudder_stochastic}.
\begin{defi}
	\label{2def:convex structure}
	(i) A set $S$ with a map $\ang{\cdot;\cdot}$ such that
	\begin{enumerate}
		\item $\ang{\lambda_{1},  \lambda_{2}, \ldots, \lambda_{n}; s_1, s_2, \ldots, s_n}$ defines a unique element of $S$ for any finite $s_1, s_2, \ldots, s_n\in S$ and probability weight $\{\lambda_{1},  \lambda_{2}, \ldots, \lambda_{n}\}$ (i.e., each $\lambda_{i}\ge0$ and $\sum_{i=1}^{n}\lambda_i=1$);
		\item $\ang{\lambda_{1},  \lambda_{2}, \ldots, \lambda_{n}; s, s, \ldots, s}=s$
	\end{enumerate}
	is called a {\itshape convex (pre-)structure}.
	Elements of the form $\ang{\lambda, 1-\lambda; s, t}$ are denoted simply by $\ang{\lambda; s, t}$.\\
	(ii) Let $S$ and $T$ be convex structures.
	A map $F\colon S\to T$ is called {\itshape affine} if 
\begin{equation}
\label{eq:def of affine}
\begin{aligned}
&F\left(\ang{\lambda_{1},  \lambda_{2}, \ldots, \lambda_{n}; s_1, s_2, \ldots, s_n}\right)\\
&\qquad\qquad\qquad\qquad=\ang{\lambda_{1},  \lambda_{2}, \ldots, \lambda_{n}; F(s_1), F(s_2), \ldots, F(s_n)},
\end{aligned}
\end{equation}
and the set of all affine maps from $S$ to $T$ is denoted by $\mathit{Aff}(S, T)$.
If there exists an affine bijection $J\colon S\to T$, then $S$ and $T$ are called {\itshape affinely isomorphic}, and $J$ is called an {\itshape affine isomorphism}.
\\
(iii) Because a convex subset of a vector space is naturally a convex structure with usual convex combinations\footnote{A subset $A$ of a vector space $L$ is called {\itshape convex} if $\lambda x +(1-\lambda)y\in A$ whenever $x, y\in A$ and $\lambda\in(0, 1)$, and a vector sum $\sum_{i=1}^{n}\lambda_{i}x_i$ for $x_1, \ldots, x_n\in A$ is called a {\itshape convex combination} if $\{\lambda_{1}, \ldots, \lambda_{n}\}$ is a probability weight.
	For a more detailed description of convex sets, see \cite{Rockafellar+2015,schaefer_TopVecSp}}:
$\ang{\lambda_{1},  \lambda_{2}, \ldots, \lambda_{n}; s_1, s_2, \ldots, s_n}=\sum_{i=1}^{n}\lambda_i s_i$, we can define successfully the set $\mathit{Aff}(S, \R)$ for a convex structure $S$, and call its element an {\itshape affine functional} on $S$.
In particular, the set of all $f\in \mathit{Aff}(S, \R)$ such that $f(s)\in [0,1]$ for all $s\in S$ is denoted by $\tilde{\mathcal{E}}_S$.
We regard $\mathit{Aff}(S, \R)$ as a real vector space in a natural way.
\\
(iv) A convex structure $(S, \ang{\cdot;\cdot})$ is called a {\itshape total convex structure} if 
	\begin{enumerate}
	\item $S$ is equipped with a function $\tilde{d}\colon S\times S\to \R$ defined as
	\begin{equation}
		\label{2eq:distance1}
		\begin{aligned}
			\tilde{d}(s_1,  s_2):=
			\inf\{0<\lambda\le1\mid&\ang{\lambda;s_1', s_1}=\ang{\lambda;s_2', s_2}
			\\
			&\qquad\qquad\qquad\quad\mbox{for some $s_1', s_2'\in S$}\},
		\end{aligned}
	\end{equation}
and for every family $\{s_n\}_n\subset S$ satisfying $\lim_{n,m\to\infty}\tilde{d}(s_n, s_m)=0$, there exists a unique $s\in S$ such that $\lim_{n\to\infty}\tilde{d}(s_n, s)=0$;
	\item $f(s)=f(t)$ for every $f\in\tilde{\mathcal{E}}_{S}$ implies $s=t$.
	\end{enumerate}
\end{defi}
Let us consider a GPT with a state space $\tilde{\Omega}$ and effect space $\tilde{\mathcal{E}}$.
Clearly, $\tilde{\Omega}$ satisfies conditions (i)-1, (i)-2, (iv)-1, and (iv)-2 in Definition \ref{2def:convex structure}, and thus is a total convex structure.
On the other hand, it is easy to see that the functional $\tilde{e}^{\circ}$ defined for $\tilde{e}\in\tilde{\mathcal{E}}$ as $\tilde{e}^{\circ}\colon\tilde{\omega}\mapsto\mu(\tilde{e}, \tilde{\omega})$ is an affine functional on $\tilde{\Omega}$ due to Axiom \ref{axiom:prob mix}.
Because we are interested only in probabilities, it is not problematic to identify the effect $\tilde{e}$ representing the associated yes-no apparatus with the affine functional $\tilde{e}^{\circ}$, and we also call the latter an {\itshape effect}.\footnote{In \cite{Cassinelli2016}, $\tilde{e}$ is called an {\itshape experimental proposition}, while the term ``effect'' (also called {\itshape experimental function}) is used for the induced affine functional $\tilde{e}^{\circ}$.}
In other words, if we define the map $\circ\colon\tilde{e}\mapsto\tilde{e}^{\circ}$, then it is an injection from $\tilde{\mathcal{E}}$ to $\tilde{\mathcal{E}}_{\tilde{\Omega}}$ because of Axiom \ref{axiom:separation}, and thus $\tilde{\mathcal{E}}$ and $\tilde{\mathcal{E}}^{\circ}\subset\tilde{\mathcal{E}}_{\tilde{\Omega}}$ can be identified with each other.
Moreover, we can observe from Axiom \ref{axiom:prob mix} that the notion of mixtures is represented mathematically as
\begin{equation}
\label{2eq:conv structure to conv comb}
\ang{\sigma_{1},  \sigma_{2}, \ldots, \sigma_{m};\tilde{e}_1, \tilde{e}_2, \ldots, \tilde{e}_m}_{\tilde{\mathcal{E}}}^{\circ}=\sum_{j=1}^{m}\sigma_{j}\tilde{e}^{\circ}_{j},
\end{equation}
and from Axiom \ref{axiom:unit-complement} that $\tilde{\mathcal{E}}^{\circ}$ includes a special effect $\tilde{u}^{\circ}$ such that $\tilde{u}^{\circ}(\tilde{\omega})=1$ for all $\tilde{\omega}\in\tilde{\Omega}$ and  $\tilde{e}^{\perp}\hspace{0.1mm}^{\circ}=\tilde{u}^{\circ}-\tilde{e}^{\circ}\in\tilde{\mathcal{E}}^{\circ}$ holds whenever $\tilde{e}^{\circ}\in\tilde{\mathcal{E}}^{\circ}$.
We note that $\tilde{\mathcal{E}}^{\circ}$ is a convex subset of the vector space $\mathit{Aff}(\tilde{\Omega}, \R)$ due to \eqref{2eq:conv structure to conv comb}.
In this way, we regard the effect space $\tilde{\mathcal{E}}^{\circ}$ as a convex subset of $\tilde{\mathcal{E}}_{\tilde{\Omega}}$: $\tilde{\mathcal{E}}^{\circ}\subset \tilde{\mathcal{E}}_{\tilde{\Omega}}$.
In this thesis, we require that the converse inclusion also holds, which is called the {\itshape no-restriction hypothesis} \cite{PhysRevA.81.062348}.
\begin{assum}[No-restriction hypothesis]
	\label{math_assume:no-restriction}
Any affine functional $\tilde{e}^{\circ}$ on $\tilde{\Omega}$ with $\tilde{e}^{\circ}(\tilde{\omega})\in[0,1]$ for all $\tilde{\omega}\in\tilde{\Omega}$ is an effect.
That is, $\tilde{\mathcal{E}}^{\circ}= \tilde{\mathcal{E}}_{\tilde{\Omega}}$.
\end{assum}
The no-restriction hypothesis means that any mathematically valid affine functional is also physically valid.
There is no physical background for this assumption, and GPTs without assuming it were investigated for example in \cite{Filippov2020_restriction,PhysRevA.87.052131,Barnum_2014,PhysRevLett.120.200402}.
However, in this thesis, we suppose that all theories satisfy the no-restriction hypothesis based on the fact that it is satisfied both in classical and quantum theory.
Now we can conclude the following.
\begin{prop}
\label{2prop:state sp as convex structure}
A GPT is identified with $(\tilde{\Omega}, \tilde{\mathcal{E}}_{\tilde{\Omega}})$, where $\tilde{\Omega}$ is a total convex structure and $\tilde{\mathcal{E}}_{\tilde{\Omega}}$ is the set of all affine functionals on it whose values lie in $[0,1]$.
\end{prop}
\begin{eg}[Examples of convex structures]
	\label{2eg:completeness}
	(i) Let $S$ be the convex structure of the closed interval $[0,1]$ of $\R$.
	If we consider its elements $s_1=0$ and $s_2=k$ $(0<k\le1)$, then an easy calculation shows $\tilde{d}(s_1, s_2)=1-\frac{1}{1+k}$, which is an increasing function of $k$.
	This observation indicates that the function $\tilde{d}$ is a valid measure to represent how close two states are.
	We can also prove that $S$ is a total convex structure.\\
	(ii) Let $S=\R$, which is naturally a convex structure.
	We can find easily that $\tilde{d}(s_1, s_2)=0$ for all $s_1, s_2\in S$, and thus this $S$ is not a total convex structure.
\end{eg}
The above examples show that under Mathematical assumption \ref{math_assume:completeness}, the state space $\tilde{\Omega}$ is ``closed'' and ``bounded'', and the function $\tilde{d}$ defined in \eqref{2eq:distance} represents properly the closeness between two states in $\tilde{\Omega}$.
In subsequent parts, we will give the mathematically rigorous verification of these observations.

It is known that a total convex structure can be embedded into a certain Banach space.
In order to show this, we need the following lemma.
\begin{lem}
	\label{2lem:continuity}
Let $(S, \ang{\cdot;\cdot})$ be a total convex structure with a ``metric'' $\tilde{d}$ defined in \eqref{2eq:distance1}.\\
(i) If a family of elements $\{s_n\}_{n=1}^{\infty}$ satisfies $\lim_{n\to\infty}\tilde{d}(s_n, s)=0$ with some $s\in S$, then $\lim_{n\to\infty}f(s_n)=f(s)$ holds for all $f\in\tilde{\mathcal{E}}_S$.\\
(ii) Let $(T, \ang{\cdot;\cdot}_T)$ be another total convex structure equipped with a similar ``metric'' $\tilde{d}_T$.
For all $s_1, s_2\in S$ and $F\in\mathit{Aff}(S, T)$, it holds that $\tilde{d}_T(F(s_1), F(s_2))\le\tilde{d}(s_1, s_2)$. 
If $F$ is bijective, then $\tilde{d}_T(F(s_1), F(s_2))=\tilde{d}(s_1, s_2)$.
\end{lem}
\begin{pf}
(i) Because $\lim_{n\to\infty}\tilde{d}(s_n, s)=0$ holds, there exists $N\in\N$ for any $\varepsilon>0$ such that $\tilde{d}(s_n, s)<\frac{\varepsilon}{\varepsilon+2}$ holds whenever $n>N$.
It implies that there are $\lambda\in(0, \frac{\varepsilon}{\varepsilon+2})$ and $t_1, t_2\in S$ satisfying $\ang{\lambda; t_1, s_n}=\ang{\lambda; t_2, s}$, which results in
\[
\lambda f(t_1)+(1-\lambda) f(s_n)=\lambda f(t_2)+(1-\lambda) f(s)
\]
for $f\in\tilde{\mathcal{E}}_S$.
It follows that
\[
|f(s_n)-f(s)|=\frac{\lambda}{1-\lambda}|f(t_2)-f(t_1)|\le\frac{2\lambda}{1-\lambda},
\]
and thus $|f(s_n)-f(s)|<\varepsilon$ holds because
\[
\frac{2\lambda}{1-\lambda}<\left.\frac{2\lambda}{1-\lambda}\right|_{\lambda\to\frac{\varepsilon}{\varepsilon+2}}=\varepsilon.
\]
(ii) It holds from the definition of $\tilde{d}$ that
\begin{align*}
	\tilde{d}_T(F(s_1), F(s_2))
	&=\inf\{0<\lambda\le1\mid
	\\
	&\qquad\qquad\ang{\lambda;t_1, F(s_1)}_T=\ang{\lambda;t_2, F(s_2)}_T,\ t_1, t_2\in T\}\\
	&\le\inf\{0<\lambda\le1\mid
	\\
	&\qquad\qquad\ang{\lambda;F(s), F(s_1)}_T=\ang{\lambda;F(s'), F(s_2)}_T,\ s, s'\in S\}\\
	&=\inf\{0<\lambda\le1\mid F(\ang{\lambda;s, s_1})=F(\ang{\lambda;s', s_2}),\ s, s'\in S\}\\
	&\le\inf\{0<\lambda\le1\mid \ang{\lambda;s, s_1}=\ang{\lambda;s', s_2},\ s, s'\in S\}\\
	&=\tilde{d}(s_1, s_2).
\end{align*}
If $F$ is bijective, then the two ``$\le$'' in the above consideration become ``$=$'', and thus $\tilde{d}_T(F(s_1), F(s_2))=\tilde{d}(s_1, s_2)$ holds.
\qed
\end{pf}
We remember that $\mathit{Aff}(S, \R)$ is a real vector space for a convex structure $S$.
The set $\mathit{Aff}(S, \R)':=\{\alpha\mid\alpha\colon\mathit{Aff}(S, \R)\to\R,\ \mbox{linear}\}$ is naturally a vector space called the {\itshape algebraic dual} of $\mathit{Aff}(S, \R)$.
Then there is a standard embedding $J$ of $S$ into $\mathit{Aff}(S, \R)'$ such that each element $J(s)\in\mathit{Aff}(S, \R)'\ (s\in S)$ is defined as
\begin{align}
	\label{eq:standard J}
[J(s)](f)=f(s)\quad(f\in\mathit{Aff}(S, \R)).
\end{align}
We can prove the following proposition.
\begin{prop}
	\label{2prop:canonical embedding}
Let $(S, \ang{\cdot;\cdot})$ be a total convex structure with a ``metric'' $\tilde{d}$ defined in \eqref{2eq:distance}.\\
(i) The standard embedding $J\colon S\to \mathit{Aff}(S, \R)'$ defined via \eqref{eq:standard J} is an affine isomorphism between $S$ and the convex subset $J(S)$ of $\mathit{Aff}(S, \R)'$.\\
(ii) If there is an affine isomorphism $\eta$ between $S$ and a convex subset $S_{0}$ of some real vector space $V_0$ such that $\mathit{aff}(S_{0})$ does not include the origin $0$ of $V_0$, then there is a linear bijection $\Phi\colon \mathit{span}(S_0)\to\mathit{span}(J(S))$ satisfying $\Phi(S_0)=J(S)$.\footnote{For a subset $A$ of a vector space $W$, its {\itshape affine hull} $\mathit{aff}(A)$ and {\itshape linear span} $\mathit{span}(A)$ are defined as $\mathit{aff}(A):=\{\sum_{i=1}^{n}\lambda_{i}a_{i}\mid a_i\in A,\ \lambda_{i}\in\R,\ \sum_{i}\lambda_{i}=1,\ \mbox{$n$: finite}\}$ and $\mathit{span}(A):=\{\sum_{i=1}^{n}\lambda_{i}a_{i}\mid a_i\in A,\ \lambda_{i}\in\R,\ \mbox{$n$: finite}\}$ respectively.}
\\
(iii) $(S, \tilde{d})$ is a complete metric space.
\end{prop}
The claims (i) and (ii) demonstrate that the total convex structure $S$ can be identified with a convex set in some vector space in an essentially unique way via the standard embedding $J$.
We note that the functional $0\in\mathit{Aff}(S, \R)'$ defined as $0(f)=0$ for all $f\in\mathit{Aff}(S, \R)$, which is the origin of the vector space $\mathit{Aff}(S, \R)'$, does not belong to $J(S)$ because $0\in J(S)$ contradicts the existence of the unit effect.
On the other hand, the claim (iii) shows that $\tilde{d}$ is indeed a metric (see Mathematical assumption \ref{math_assume:completeness}).
\begin{pf}[proof of Proposition \ref{2prop:canonical embedding}]
(i) It is easy to see that $J$ is an affine map from $S$ to $\mathit{Aff}(S, \R)'$ and $J(S)$ is a convex set in $\mathit{Aff}(S, \R)'$.
Since $S$ is total (see (iv)-2 in Definition \ref{2def:convex structure}), for $s, t\in S$ with $s\neq t$, there exists an affine functional $f\in\mathit{Aff}(S, \R)$ such that $f(s)\neq f(t)$, i.e., $[J(s)](f)\neq [J(t)](f)$.
This implies $J(s)\neq J(t)$.\\
(ii) Let us introduce a subset $K:=\{\sum_{i=1}^{n}\lambda_{i}x_{i}\mid\lambda_{i}\ge0,\ x_{i}\in S_0,\ n:\mbox{finite}\}$, i.e., the {\itshape conic hull} of $S_0$ (see Definition \ref{2def:cone}).
Then any $y\in K\setminus\{0\}$ can be represented as $y=\lambda x$ with $\lambda>0$ and $x\in S_0$ in a unique way.
To see this, assume that $y\in K\setminus\{0\}$ satisfies $y=\lambda x=\lambda' x'$ with $\lambda, \lambda'>0$ and $x, x'\in J(S)$.
If $\lambda\neq\lambda'$, then it holds that 
\[
0=\lambda x-\lambda' x'=(\lambda-\lambda')
\left(
\frac{\lambda}{\lambda-\lambda'}x-\frac{\lambda'}{\lambda-\lambda'}x'
\right).
\]
Because $0\notin\mathit{aff}(S_0)$, the above equation implies $\lambda=\lambda'$, which is a contradiction.
Thus we can conclude $\lambda=\lambda'$ and $x=x'$.
Now let us construct the linear bijection $\Phi$ from the affine isomorphism $\eta$.
First, we define an affine bijection $\phi_0\colon S_0\to J(S)$ by $\phi_0=J\circ\eta^{-1}$ (note that $J$ is a bijection between $S$ and $J(S)$).
From the above consideration, we can extend this $\phi_0$ successfully to a bijection $\phi$ from $K$ to the conic hull of $J(S)$: $\phi(y)=\lambda \phi_0(x)$ for $y=\lambda x$ with $y\in K$, $x\in S_0$, and $\lambda\ge0$.
It is easy to verify that $\phi(\lambda y+\mu z)=\lambda\phi(y)+\mu\phi(z)$ holds for $y, z\in K$ and $\lambda, \mu\ge0$. 
Since any $y\in\mathit{span}(S_0)$ such that $y=\sum_{i=1}^{n}\lambda_{i}x_{i}$ with $x_i\in S_0$, $\lambda_{i}\in\R$, and a finite $n$ can be expressed as $y=u-v$, where $u, v\in K$, we can consider the extension of $\phi$ to a map $\Phi$ from $\mathit{span}(S_0)$ to $\mathit{span}(J(S))$ by $\Phi(y)=\phi(u)-\phi(v)$ for $y=u-v$ with $y\in\mathit{span}(S_0)$ and $u, v\in K$.
We note that this $\Phi$ is well-defined: if $y=u_1-v_1=u_2-v_2$ with $u_1, u_2, v_1, v_2\in K$ holds, then $u_1+v_2=u_2+v_1$ holds, and thus $\Phi(u_1+v_2)=\Phi(u_2+v_1)$, i.e., $\Phi(u_1)+\Phi(v_2)=\Phi(u_2)+\Phi(v_1)$ follows, which implies $\Phi(u_1)-\Phi(u_2)=\Phi(v_1)-\Phi(v_2)$.
It is easy to confirm that $\Phi\colon\mathit{span}(S_0)\to\mathit{span}(J(S))$ is linear and bijective.\\
(iii) It is trivial that $\tilde{d}(s, t)\ge0$ and $\tilde{d}(s, t)=\tilde{d}(t, s)$ holds for all $s, t\in S$.
Let $\tilde{d}(s, t)=0$.
Then there exist a family of positive numbers $\{\lambda_i\}_i$ with $\lim_{i\to\infty}\lambda_i=0$ and families $\{s_i\}_i$ and $\{t_i\}_i$ of elements of $S$ such that $\ang{\lambda_{i}; s_i, s}=\ang{\lambda_{i}; t_i, t}$.
It follows that
\[
\lambda_i f(s_i)+(1-\lambda_i) f(s)=\lambda_i f(t_i)+(1-\lambda_i) f(t)
\]
holds for all $f\in\tilde{\mathcal{E}}_S$.
Because $0\le f(s_i), f(t_i)\le1$ holds, taking $i\to\infty$ in the above equation, we obtain $f(s)=f(t)$ for all $f\in\tilde{\mathcal{E}}_S$.
By the assumption of totality, we can conclude $s=t$.
To verify the triangle inequality for $\tilde{d}$, it is enough to prove that $\tilde{d}'\colon J(S)\times J(S)\to\R$ defined on $J(S)$ in a similar way to $\tilde{d}$ satisfies it.
This is because $\tilde{d}'(J(s), J(t))=\tilde{d}(s, t)$ holds for all $s, t\in S$ as we have seen in Lemma \ref{2lem:continuity}.
For the evaluation of $\tilde{d}'(p, r)+\tilde{d}'(r, q)$ with $p, q, r\in J(S)$, let us assume that $\lambda_{1}, \lambda_2\in(0, 1)$ satisfy 
\begin{align*}
&\lambda_1p_1+(1-\lambda_1)p=\lambda_1r_1+(1-\lambda_1)r,\\
&\lambda_2r_2+(1-\lambda_2)r=\lambda_2q_1+(1-\lambda_2)q
\end{align*}
for $p_1, q_1, r_1, r_2\in J(S)$.
We obtain from these equations 
\begin{align*}
\lambda_1(1-\lambda_2)p_1&+\lambda_2(1-\lambda_1)r_2+(1-\lambda_1)(1-\lambda_2)p\\
&=\lambda_2(1-\lambda_1)q_1+\lambda_1(1-\lambda_2)r_1+(1-\lambda_1)(1-\lambda_2)q.
\end{align*}
It can be rewritten as
\begin{align}
\label{2eq:lambdas}
\lambda_0p_2+(1-\lambda_0)p=\lambda_0q_2+(1-\lambda_0)q,
\end{align}
where
\begin{align*}
\lambda_0
=
\frac{\lambda_1(1-\lambda_2)+\lambda_2(1-\lambda_1)}{\lambda_1(1-\lambda_2)+\lambda_2(1-\lambda_1)+(1-\lambda_1)(1-\lambda_2)}
=\frac{\lambda_1+\lambda_2-2\lambda_1\lambda_2}{1-\lambda_1\lambda_2}
\end{align*}
and
\begin{align*}
&p_2=\frac{\lambda_1(1-\lambda_2)}{\lambda_1(1-\lambda_2)+\lambda_2(1-\lambda_1)}p_1+\frac{\lambda_2(1-\lambda_1)}{\lambda_1(1-\lambda_2)+\lambda_2(1-\lambda_1)}r_2,\\
&q_2=\frac{\lambda_2(1-\lambda_1)}{\lambda_1(1-\lambda_2)+\lambda_2(1-\lambda_1)}q_1+\frac{\lambda_1(1-\lambda_2)}{\lambda_1(1-\lambda_2)+\lambda_2(1-\lambda_1)}r_1.
\end{align*}
Because $\lambda_0\le\lambda_1+\lambda_2$, we can see from \eqref{2eq:lambdas} that 
\[
\tilde{d}'(p,q)\le\tilde{d}'(p, r)+\tilde{d}'(r, q) 
\]
holds, and thus we can conclude that $(S, \tilde{d})$ is a metric space.
The completeness clearly holds due to (iv)-1 in Definition \ref{2def:convex structure}.\qed
\end{pf}
We note that we can prove the same claim as (iii) also for the function $\tilde{d}_{0}$ defined as
\begin{align}
	\label{2eq:metric2}
\tilde{d}_{0}(s, t)=\frac{\tilde{d}(s, t)}{1-\tilde{d}(s, t)}.
\end{align}
In fact, it was shown in \cite{cmp/1103858551} that this $\tilde{d}_{0}$ is a metric on $S$, and the completeness holds similarly.
Before proceeding to the main theorem of this section, we introduce the notion of convex cones \cite{schaefer_TopVecSp,Bourbaki_topvec, boyd_vandenberghe_2004}.
\begin{defi}
	\label{2def:cone}
Let $L$ be a vector space and $0\in L$ be its origin.\\
(i) A subset $C$ of $L$ is called a {\itshape cone} of vertex $0$ if $\lambda C\subset C$ for all $\lambda>0$.
A cone of vertex $x_0$ is a set of the form $x_0+C$, where $C$ is a cone of
vertex 0.
In this thesis, the vertex of a cone is always assumed to be $0$.\\
(ii) A cone $C\subset L$ is called 
\begin{enumerate}
\item{\itshape convex} if it is convex, i.e., satisfies $C+C\subset C$;
\item{\itshape pointed} if $C \cap -C=\{0\}$;
\item{\itshape generating} (or {\itshape spanning}) if $\mathit{span} (C)=L$, i.e., $C-C=L$.
\end{enumerate}
(iii) The {\itshape conic hull} of a subset $A$ of $L$ is defined as $\mathit{cone}(A):=\{\sum_{i=1}^{n}\lambda_{i}a_{i}\mid\lambda_{i}\ge0,\ a_{i}\in A,\ n:\mbox{finite}\}$. It is easy to see that $\mathit{cone}(A)$ is a convex cone.
\end{defi}
Let us write $\mathit{cone}(J(S))$ and $\mathit{span}(J(S))$ generated by $J(S)$ simply as $K$ and $V$ respectively.
It is easy to see that $K$ is a convex, pointed, and generating cone for $V$, and thus any $v\in V$ is written in the form $v=k_+-k_-=\alpha p-\beta q$, where $k_{\pm}\in K$, $p, q\in J(S)$, and $\alpha, \beta\ge0$.
It follows that we can introduce the following quantity for $v\in V$:
\begin{equation}
	\label{2eq:intro base norm}
\|v\|=\inf\{\alpha+\beta\mid v=\alpha p-\beta q,\ \alpha, \beta\ge0,\ p, q\in J(S)\}.
\end{equation}
Now we can present an embedding theorem for a total convex structure as follows.
We shall omit the proof, but it is given in \cite{Gudder_stochastic} (see the proofs of Theorem 4.11 and Theorem 4.12 there).
\begin{thm}
\label{2thm:embedding}
Let $S$ be a total convex structure, and $K$ and $V$ be the cone and the real vector space generated by the standard embedding $J(S)$ of $S$ into $\mathit{Aff}(S, \R)'$ (see \eqref{eq:standard J}) respectively.\\
(i) The function $\|\cdot\|$ on $V$ defined in \eqref{2eq:intro base norm} is a norm on $V$ satisfying $\|J(s)-J(t)\|=2\tilde{d}_{0}(s, t)$ for all $s, t\in S$ and $\|J(s)\|=1$ for all $s\in S$.
Moreover, $(V, \|\cdot\|)$ is a real Banach space, and $K$ is closed.\\
(ii) Let $f\in\mathit{Aff}(S, \R)$.
Then the affine functional $f\circ J^{-1}\colon J(S)\to\R$ on $J(S)$ has a unique linear extension $\check{f}\colon V\to\R$.\\
(iii) If we let $e\colon V\to\R$ be the unique linear extension of $\tilde{e}^\circ\in\tilde{\mathcal{E}}_{S}\subset\mathit{Aff}(S, \R)$ described in (ii) above, then $e$ is continuous, and thus belongs to the Banach dual $V^{*}:=\{f\mid f\colon V\to\R,\ \mbox{linear, bounded (continuous)}\}$ of $V$.
In particular, the linear extension $u$ of the unit effect $\tilde{u}^\circ$ such that $u(J(s))=1$ for all $J(s)\in J(S)$ satisfies $u\in V^{*}$.
\end{thm}
Let us consider a GPT $(\tilde{\Omega}, \tilde{\mathcal{E}}_{\tilde{\Omega}})$ (see Proposition \ref{2prop:state sp as convex structure}).
By setting $S=\tilde{\Omega}$ in Theorem \ref{2thm:embedding}, we can identify the state space $\tilde{\Omega}$ with a convex set $\Omega:=J(\tilde{\Omega})$\footnote{It will be shown in the following part that $\Omega$ is in fact a closed convex set in $V$ inheriting the closedness of $K$.} in a Banach space $V=\mathit{span}(\Omega)$ equipped with the norm $\|\cdot\|$ in \eqref{2eq:intro base norm} called the {\itshape base norm}, and the effect space $\tilde{\mathcal{E}}_{\tilde{\Omega}}$ with a subset $\mathcal{E}_{\Omega}:=\{e\in V^{*}\mid e(\omega)\in[0,1]\ \mbox{for all $\omega\in\Omega$}\}$ of the Banach dual $V^{*}$. 
We also call $\Omega$ and $\mathcal{E}_{\Omega}$ the {\itshape state space} and the {\itshape effect space} of the GPT respectively.
In the next part, we give further explanations about the Banach space $V$ and its Banach dual $V^{*}$.

\subsection{Ordered Banach spaces}
\label{2subsec:order vector space}
The vector spaces $V$ and $V^{*}$ introduced in the previous part are equipped with both order and Banach space structures, that is, they are ordered Banach spaces.
In this subsection, we make a brief review of ordered Banach spaces.
Mathematical terms shown in this subsection are according mainly to \cite{Lami_PhD,Plavala_2021_GPTs,Kuramochi_compactconvex_2020,schaefer_TopVecSp,Bourbaki_topvec,Alfsen_compact_convex,Ellis_duality_1964}.
Also, there can be found the technical proofs of some theorems which we omit.
We begin with the definition of an ordered vector space.
\begin{defi}
\label{2def:OLS}
A real vector space $L$ equipped with a partial ordering\footnote{
A binary relation $\le$ on a set $X$ is called a {\itshape preorder} if it is reflexive, i.e., $x\le x\ (x\in X)$, and transitive, i.e., $x\le y$ and $y\le z$ implies $x\le z\ (x, y, z\in X)$.
A preorder $\le$ is called a {\itshape partial order} if it is antisymmetric, i.e., $x\le y$ and $y\le x$ implies $x=y$ ($x, y\in X$).
We remark that some authors use the term ``partial order'' to represent a preorder here \cite{Kelley_Namioka_topological}.} $\le$ is called an {\itshape ordered vector space} if it satisfies\\
(i) $x\le y$ implies $x+z\le y+z$ for all $x, y, z\in L$;\\
(ii) $x\le y$ implies $\lambda x\le \lambda y$ for all $x, y\in L$ and $\lambda\ge0$.
\end{defi}
We can prove easily the following (recall Definition \ref{2def:cone}).
\begin{prop}
\label{2prop:OVS to cone}
Let $L$ be an ordered vector space and $\le$ be its ordering.\\
(i) $L_+:=\{x\in L\mid x\ge0\}$ is a convex and pointed cone.\\
(ii) If $(L, \le)$ is directed, i,e, for every $x, y\in L$ there is $z\in L$ such that $x\le z, y\le z$, then $L_+$ in (i) is also generating.
\end{prop}
\begin{pf}
(i) For $x\ge 0$, it holds clearly that $\lambda x\ge 0$ ($\lambda\ge0$), and thus $L_+$ is a cone.
Because, for $x, y\ge 0$, both $px\ge0$ and $(1-p)y\ge 0$ ($0\le p\le 1$) hold, $px+(1-p)y \ge0$ follows, which implies $L_+$ is convex. 
The claim that $L_+$ is pointed follows from the observation that $x\ge0$ and $x\le0$ implies $x=0$.\\
(ii) Because $L$ is directed, for any $x\in L$, there exists $z\in L$ such that $x\le z$ and $-x\le z$, equivalently, $z-x\ge0$ and $z+x\ge0$ hold.
Because $\half(z-x)\ge0$ and $\half(z+x)\ge0$,  the expression $x=\half(z+x)-\half(z-x)$ implies that $L_+$ is generating.\qed
\end{pf}
\begin{defi}
\label{def:order dual}
Let $L$ be an ordered vector space and $\le$ be its ordering.\\
(i)  The cone $L_+:=\{x\in L\mid x\ge0\}$ is called the {\itshape positive cone} of $L$.\\
(ii) For the positive cone $L_+$ of $L$, its {\itshape order dual cone} $L^{\Diamond}_+$ is defined as the set of all ``positive'' functionals on $L_+$, i.e., $L^{\Diamond}_+:=\{f\in L'\mid f(x)\ge0\ \mbox{for all $x\in L_+$}\}$.
It is clear that $L^{\Diamond}_+$ is a convex cone in the algebraic dual $L'$ of $L$ and in the subspace $L^{\Diamond}:=L^{\Diamond}_+-L^{\Diamond}_+=\mathit{span}(L^{\Diamond}_+)$ called the {\itshape order dual} of $L$.
Moreover, we can find that $L^{\Diamond}_+$ is pointed in $L'$ and $L^{\Diamond}$ if $L_+$ is generating.
\end{defi}

We have proven in Proposition \ref{2prop:OVS to cone} that a positive cone can be introduced through an order vector space.
Conversely, we can construct an order structure for a vector space when there is a convex cone.
\begin{prop}
\label{2prop:cone to OVS}
Let $C$ be a convex and pointed cone in a real vector space $L$.\\
(i) If we define a binary relation $\le$ as $x\le y\iff y-x\in C$ for $x, y\in V$, then the relation $\le$ is a partial ordering, and $L$ is an ordered vector space with its ordering given by $\le$.\\
(ii) The positive cone $L_+$ for $L$ defined via the order $\le$ in (i) is identical to $C$, i.e., $L_+=C$.\\
(iii) If $C$ is in addition generating, then $(L, \le)$ is directed.
\end{prop}
\begin{pf}
(i) Because $C$ is pointed, $x-x=0\in C$, and $y-x\in C$ and $x-y\in C$ imply $y-x=0$, i.e., $x=y$ for $x, y\in L$.
Moreover, if $y-x\in C$ and $z-y\in C$ ($z, y, z\in L$), then $z-x=(z-y)+(y-x)\in C$.
Therefore, we can conclude that $\le$ is a partial ordering.
On the other hand, because $y-x=(y+z)-(x+z)$ ($x, y, z\in L$), $x+z\le y+z$ holds when $x\le y$.
Since $C$ is a cone, $y-x\in C$ ($x, y\in L$) implies $\lambda (y-x)=\lambda y-\lambda x\in C$ ($\lambda\ge0$), i.e., $\lambda x\le\lambda y$ when $x\le y$.\\
(ii) The claim $L_+=C$ is trivial since $x\ge0$ is equivalent to $x\in C$.\\
(iii) For $x, y\in L$, because $C$ is generating, there exist $x_1, x_2, y_1, y_2\in C$ such that $x=x_1-x_2$ and $y=y_1-y_2$.
Defining $z=x_1+y_1$, we have $z-x=y_1+x_2\in C$ and $z-y=x_1+y_2\in C$, which means that $(L, \le)$ is directed.\qed
\end{pf}
It follows from these propositions that a positive cone and a convex and pointed cone can be identified naturally with each other.

Next, we give descriptions of ordered Banach spaces.
An ordered vector space $L$ is called an ordered Banach space if $L$ is also a Banach space (see \cite{Conway_functionalanalysis} for a review of Banach space).
There are two important types of ordered Banach space in the field of GPTs: {\itshape base norm Banach spaces} and {\itshape order unit Banach spaces}, which are related with state spaces and effect spaces respectively.
Let us first introduce base norm Banach spaces.
\begin{defi}
\label{def:base for positive cone}
Let $L$ be an ordered vector space with its positive cone $L_+$.
A convex subset $B\subset L_+$ is called a {\itshape base} of $L_+$ if for any $x\in L_+$ there exists a unique $\lambda\ge0$ such that $x\in\lambda B$.
\end{defi}
The following lemma is important.
\begin{lem}
\label{2lem:0 notin aff base}
Let $L$ be an ordered vector space with its positive cone $L_+$, and let $B$ be its base.
Then $\mathit{aff}(B)$ does not contain the origin $0$ of $L$.
\end{lem}
\begin{pf}
Suppose $0\in\mathit{aff}(B)$.
Then there exist real numbers $\{\lambda_i\}_{i=1}^n$ with $\sum_{i=1}^n\lambda_i=1$ and elements $\{x_i\}_{i=1}^n$ of $B$ such that $\sum_{i=1}^n\lambda_i x_i=0$.
Dividing $\{\lambda_i\}_{i=1}^n$ into positive and negative parts, we obtain
\[
\sum_j\lambda_{j}^+ x_j^+=\sum_k\lambda_{k}^- x_k^-,
\]
where $\{x_{j}^+\}_j$ and $\{x_{k}^-\}_k$ are subsets of $\{x_i\}_{i=1}^n$, and $\{\lambda_{j}^+\}_j$ and $\{\lambda_{k}^-\}_k$ are positive numbers satisfying $\sum_j\lambda_{j}^+ -\sum_k\lambda_{k}^-=1$.
If we suppose $K:=\sum_k\lambda_{k}^-\neq0$, then we can rewrite the above equation as
\[
\frac{K+1}{K}\cdot\frac{1}{K+1}\sum_j\lambda_{j}^+ x_j^+=\frac{1}{K}\sum_k\lambda_{k}^- x_k^-.
\]
Because $y:=\frac{1}{K+1}\sum_j\lambda_{j}^+ x_j^+$ and $y':=\frac{1}{K}\sum_k\lambda_{k}^- x_k^-$ are convex combinations of elements of $B$, they belong to $B$.
Then the above equation $\frac{K+1}{K}y=y'$ contradicts the uniqueness condition in the definition of the base $B$, and thus we obtain $K=0$.
It implies $0\in B$, but this also contradicts the uniqueness condition because any positive number $\lambda$ satisfy $\lambda0=0$.\qed
\end{pf}
By means of this lemma, we can associate a base of a positive cone with a linear functional in the following way \cite{Lami_PhD,Ellis_duality_1964}. 
\begin{prop}
\label{2prop:base to unit}
Let $L$ and $L_+$ be an ordered vector space and its positive cone respectively.
The positive cone $L_+$ has a base $B$ if and only if there exists a strictly positive functional $e_B$ (i.e., $e_B\in L^{\Diamond}$ and satisfies $e_B(x)>0$ for all nonzero $x\in L_+$) such that 
\begin{equation}
	\label{2eq:def of intensity}
B=\{x\in L_+\mid e_B(x)=1\}.
\end{equation}
\end{prop}
\begin{pf}
The if part is easy, so we prove the only if part.
Let $B$ be a base of $L_+$.
Applying Zorn's lemma to the set $A$ of all affine sets that include $\mathit{aff}(B)$ but not $\{0\}$, we obtain the maximal affine set $H$ in $A$.
It can be shown \cite{Guler_optimization_2010} that this $H$ is a hyperplane in $L$, and thus there exists a linear functional $e_B$ such that $e_B(x)=1$ for all $x\in H$. 
This functional $e_B$ is easily found to be strictly positive because $B$ is a base.\qed
\end{pf}
We call the functional $e_B$ the {\itshape intensity functional} for the base $B$ \cite{Cassinelli2016}.
\begin{lem}
	\label{2lem_minkowski}
Let $L$ be an ordered vector space and $L_+$ be its positive cone, and assume that $L_+$ is generating.
For a base $B\subset L_+$ of $L_+$, the set $D:=\mathit{conv}(B\cup -B)$ is a radial, circled, and convex subset of $L$.\footnote{A subset $U$ of a vector space $L$ (assumed to be on the field $F=\R$ or $\C$) is {\itshape radial} if for any $x\in L$ there exists $\lambda_0\in F$ such that $|\lambda|\ge|\lambda_0|$ implies $x\in \lambda U$, and is {\itshape circled} if $\lambda U\subset U$ for any $\lambda$ with $|\lambda|\le1$ \cite{schaefer_TopVecSp}.}
\end{lem}
\begin{pf}
The convexity is clear.
It is easy to see $0\in D$, and thus $D$ is circled.
Because $L_+$ is generating, any $x\in L$ can be written as $x=\lambda_+ x_+ +\lambda_- x_-$ with $\lambda_{\pm}\ge0$ and $x_{+}\in B, x_-\in -B$.
Let $\lambda_0=\lambda_++\lambda_-$.
For $\lambda\ge\lambda_0$, the vector $x$ can be rewritten as 
\[
x=\lambda\cdot\frac{\lambda_+ +\lambda_-}{\lambda}\left(\frac{\lambda_+}{\lambda_++\lambda_-}x_++\frac{\lambda_-}{\lambda_++\lambda_-}x_-
\right).
\]
Because $D$ is circled, $\frac{\lambda_+ +\lambda_-}{\lambda}\left(\frac{\lambda_+}{\lambda_++\lambda_-}x_++\frac{\lambda_-}{\lambda_++\lambda_-}x_-
\right)\in D$ is obtained. 
It implies $x\in \lambda D$, and thus $D$ is radial.\qed
\end{pf}
According to Lemma \ref{2lem_minkowski}, if $L_+$ is generating, then the Minkowski functional of $D=\mathit{conv}(B\cup -B)$ defined as
\begin{align}
\label{2eq:minkowski}
p_D(x):=\inf\{\lambda>0\mid x\in \lambda D\}\quad(x\in L)
\end{align}
is a seminorm on $L$ \cite{schaefer_TopVecSp}.
It is not difficult to see that, with $e_B$ introduced in Proposition \ref{2prop:base to unit}, the function $p_D$ satisfies
\begin{align}
\label{2eq:minkowski2}
p_D(x)=\inf\{e_B(x_+)+e_B(x_-)\mid x=x_+-x_-,\ x_\pm\in L_+\}\quad(x\in L),
\end{align}
or equivalently
\begin{align}
\label{2eq:minkowski3}
p_D(x)=\inf\{\alpha+\beta\mid x=\alpha b_+-\beta b_-,\  \alpha, \beta\ge0,\ b_\pm\in B\}\quad(x\in L)
\end{align}
since it holds that $p_D(x_+)=e_B(x_+)$ for all $x_+\in L_+$.
Now we can give the definition of a base norm space.
\begin{defi}
\label{def:base norm sp}
Let $L$ be an ordered vector space with its positive cone $L_+$ generating, and let $B$ be a base of $L_+$.
If the function $p_D$ defined in \eqref{2eq:minkowski}-\eqref{2eq:minkowski3} through the base $B$ is a norm on $L$, then $(L, B)$ is called a {\itshape base norm space}.
In this case, we write $p_D(\cdot)$ as $\|\cdot\|_B$ and call it the {\itshape base norm}.
A base norm space $(L, B)$ is called a {\itshape base norm Banach space} if $L$ is complete with respect to the base norm $\|\cdot\|_B$.
\end{defi}
\begin{rmk}
If we set $L=\R^2$ and $L_+=\{(u, v)\mid v>0\}\cup (0,0)$ with a base $B=\{(u, v)\mid v=1\}$, then the function $p_D$ satisfies $p_D((u,0))=0$ for all $u\in \R$, and thus it is not a norm in $L$.
In fact, it can be shown that $p_D$ is a norm if and only if $D=\mathit{conv}(B\cup -B)$ is linearly bounded, i.e., $M\cap D$ is a bounded subset of $L$ whenever $M$ is a one-dimensional subspace \cite{Ellis_duality_1964} (in the example, $M\cap D$ is not bounded for $M=\{(u,0)\mid u\in\R\}$).
\end{rmk}
In this thesis, for a Banach space $X$, we denote its \textit{Banach dual} by $X^*=\{f\mid f\colon X\to \R,\ \mbox{linear, bounded}\}$.
When $X$ is in addition an ordered vector space (i.e., an \textit{ordered Banach space}) and $X_+$ is its positive cone, we define a subset $X^*_+$ of $X^*$ as $X^*_+:=\{f\in X^*\mid f(x)\ge0\ \mbox{for all $x\in X_+$}\}$, and call it the \textit{Banach dual cone} for $X_+$.
It is verified easily that $X^{*}_+$ is a convex and closed (in the weak*\footnote{For a Banach space $X$ and its Banach dual $X^{*}$, the {\itshape weak topology} of $X$ often dented by $\sigma(X, X^{*})$ is the weakest topology on $X$ which makes all $f\in X^{*}$ continuous, and the {\itshape weak* topology} of $X^{*}$ often dented by $\sigma(X^{*}, X)$ is the weakest topology on $X^{*}$ which makes all $x\in X\subset X^{**}$ continuous \cite{schaefer_TopVecSp,Conway_functionalanalysis}.} and norm topologies) cone in $X^*$,\footnote{\label{closed dual cone}Clearly, $X^*_+$ satisfies $X^*_+=\bigcap_{x\in X_+}\{f\in X^{*}\mid f(x)\ge0\}$, and thus is weakly* and norm closed.} and is in addition pointed if $X_+$ is generating.

We present miscellaneous facts about base norm Banach spaces. 
\begin{prop}
\label{2prop:base norm spaces}
Let $(L, B)$ be a base norm Banach space, and $L_+$ be the positive cone of $L$.
For a subset $A$ of $L$, we denote its norm closure by $\overline{A}$.\\
(i) The intensity functional $e_B$ for the base $B$ (see Proposition \ref{2prop:base to unit}) is continuous, i.e., $e_B\in L^{*}$.\\
(ii) $B$ is closed if and only if $L_+$ is closed.\\
(iii) The closed unit ball of $L$ is given by $\overline{D}=\overline{\mathit{conv}(B\cup -B)}$.\\
(iv) The dual norm $\|\cdot\|^*$ on the Banach dual $L^{*}$ defined as $\|f\|^*:=\sup\{|f(x)|\mid \|x\|_B\le1\}$ satisfies $\|f\|^*=\sup\{|f(x)|\mid x\in B\}$.\\
(v) $\overline{L_+}$ is a convex, pointed, and generating cone in $L$, and $\overline{B}$ is a base of $\overline{L_+}$ with its intensity functional identical with that of the original base $B$: $\overline{B}=\overline{L_+}\cap e_B^{-1}(1)$.
Moreover, the base norm induced by $\overline{B}$ coincides with the original one by $B$.
\\
(vi) If $L_+$ is closed, then the Banach dual and order dual coincide with each other: $L^*=L^{\Diamond}$.
\end{prop}
\begin{pf}	
(i) Representing $x\in L$ as $x=x_+-x_-$ ($x_{\pm}\in L_+$), we have 
\begin{align*}
|e_B(x)|=|e_B(x_+)-e_B(x_-)|\le e_B(x_+)+e_B(x_-).
\end{align*}
It implies $|e_B(x)|\le\|x\|_B$, i.e., $e_B$ is bounded.\\
(ii) Let $e_B$ be the intensity functional for $B$, which is continuous.
When $L_+$ is closed, its base $B=L_+\cap\{x\in L\mid e_B(x)=1\}$ is also closed.
Assume conversely that $B$ is closed.
Since $L$ is complete, for a Cauchy sequence $\{\alpha_i x_i\}_i$ in $L_+$ such that $\alpha_i\ge0$ and $x_i\in B$, there exists $v_*\in L$ to which $\{\alpha_i x_i\}_i$ converges.
From the continuity of $e_B$, we obtain $\alpha_i=e_B(\alpha_i x_i)\underset{i\to\infty}{\longrightarrow} e_B(v_*)$ (remember that $e_B(x_i)=1$ holds for every $x_i\in B$).
If $e_B(v_*)=0$, then $\alpha_i\underset{i\to\infty}{\longrightarrow}0$ holds.
Since each $\alpha_i x_i$ is an element of $L_+$, we have $\alpha_i=e_B(\alpha_i x_i)=\|\alpha_i x_i\|_B$, and thus $\|\alpha_i x_i\|_B\underset{i\to\infty}{\longrightarrow}0$, i.e., $v_*=\lim_i\alpha_i x_i=0$.
This observation implies $v_*\in L_+$ because $L_+$ is pointed and thus $0\in L_+$ (see Proposition \ref{2prop:OVS to cone}).
If $e_B(v_*)\neq0$, then 
\begin{align*}
e_B(v_*)\|x_i-x_j\|_B
&\le \|e_B(v_*)x_i-v_*\|_B+\|v_*-e_B(v_*)x_j\|_B\\
&\le \|e_B(v_*)x_i-\alpha_i x_i\|_B+\|\alpha_i x_i-v_*\|_B\\
&\qquad\qquad\quad+\|v_*-\alpha_j x_j\|_B+\|\alpha_j x_j-e_B(v_*)x_j\|_B\\
&=|e_B(v_*)-\alpha_i|+\|\alpha_i x_i-v_*\|_B\\
&\qquad\qquad\quad+\|v_*-\alpha_j x_j\|_B+|\alpha_j -e_B(v_*)|.
\end{align*}
The last equation converges to $0$ as $i,j\to\infty$, and thus $\{x_i\}_i$ is a Cauchy sequence.
Because $B$ is closed, $\{x_i\}_i$ converges to $x_*\in B$.
Therefore, we obtain $\lim_i\alpha_ix_i=e_B(v_*)x_*\in L_+$.\\
(iii) This claim follows directly from the definition of $\|\cdot\|_B$ as the Minkowski functional of $D$.\\
(iv) It can be found that 
\begin{align*}
\|f\|^*
&=\sup\{|f(x)|\mid \|x\|_B\le1\}\\
&=\sup\{|f(x)|\mid x\in \overline{D}\}\\
&=\sup\{|f(x)|\mid x\in D\}\\
&=\sup\{|f(x)|\mid x\in B\}.
\end{align*}
For the proofs of (v) and (vi), see Proposition 1.40 in \cite{Lami_PhD}.\qed
\end{pf}
Roughly speaking, the base norm and the intensity functional considered above correspond to the trace norm and the identity operator in the usual formulation of quantum theory respectively.
In fact, if we let $L$ be the set $\LL_S(\HH)$ of all self-adjoint operators on a finite-dimensional Hilbert space $\HH$, then any $x\in L$ is decomposed as $x=x_+-x_-$ with $x_\pm\ge0$ in the usual ordering for self-adjoint operators, and thus the trace norm $\|x\|_{\Tr}$ of $x$ is given via the identity operator $\id$ by $\|x\|_{\Tr}=\Tr[x_+]+\Tr[x_-]=\Tr[\id x_+]+\Tr[\id x_-]$, which corresponds to \eqref{2eq:minkowski2}.

Let us move to the introduction of order unit Banach spaces.
\begin{defi}
	\label{def:order unit}
	Let $L$ be an ordered vector space equipped with an ordering $\le$.\\
	(i) $L$ is called {\itshape Archimedean} if $x\le 0$ whenever there exists $y\in L$ such that $nx\le y$ for all $n\in \N$.\\
	(ii) $L$ is called {\itshape almost Archimedean} if $x=0$ whenever there exists $y\in L$ such that $-y\le nx\le y$ for all $n\in \N$.\\
	(iii) A positive element $u$ of $L$ is called an {\itshape order unit} if for any $x\in L$ there exists some $n\in\N$ such that $-nu\le x \le nu$.
\end{defi}
It is clear that if $L$ is Archimedean, then it is almost Archimedean. 
For $a, b\in L$, we define the order interval $[a, b]$ as $[a, b]:=\{x\in L\mid a\le x\le b\}$.
The following lemma is important.
\begin{lem}
\label{2lem:order unit norm}
Let $L$ be an ordered vector space with an ordering $\le$, and let $u$ be an order unit associated with the ordering $\le$.\\
(i) The order interval $\Delta:=[-u, u]$ is a radial, circled, and convex subset of $L$.\\
(ii) The Minkowski functional of $\Delta$ defined as
\begin{align}
	\label{2eq:minkowski order unit}
p_\Delta(x)=\inf\{\lambda>0\mid x\in\lambda\Delta\}\quad(x\in L)
\end{align}
is a norm on $L$ if and only if $L$ is almost Archimedean.
\end{lem}
\begin{pf}
It is easy to see that (i) holds due to the definition of the order unit $u$, and thus the Minkowski functional $p_\Delta$ is a seminorm on $L$.
Assume that $p_\Delta$ is a norm and $x\in L$ satisfies $-y\le nx\le y$ for all $n\in\N$ and some $y\in L$.
Since there exists $m\in\N$ such that $-mu\le y\le mu$, we obtain $-mu\le nx\le mu$, or $-\frac{m}{n}u\le x\le \frac{m}{n}u$ for all $n\in\N$.
Thus $\inf\{\lambda>0\mid-\lambda u\le x\le \lambda u\}=0$ holds, and we can conclude $x=0$ because $\|\cdot\|_u$ is a norm.
Conversely, assume that $L$ is almost Archimedean and $x\in L$ satisfies $p_\Delta(x)=0$. 
Then $-u\le\frac{1}{\lambda}x\le u$ holds for arbitrary small $\lambda$, and thus $x=0$ follows from the assumption that $L$ is almost Archimedean, which concludes (ii).\qed
\end{pf}
We can give the definition of an order unit Banach space.
\begin{defi}
\label{def:order unit Banach}
Let $L$ be an ordered vector space with an order unit $u\in L$ associated with the ordering of $L$.
$(L, u)$ is called an {\itshape order unit Banach space} if $L$ is Archimedean and complete with respect to the norm $p_{\Delta}$ defined in \eqref{2eq:minkowski order unit}.
In this case, we write $p_{\Delta}(\cdot)$ as $\|\cdot\|_u$, and call it the {\itshape order unit norm}.
\end{defi}
We present miscellaneous facts about order unit Banach spaces according mainly to \cite{Lami_PhD,Alfsen_compact_convex}.
\begin{prop}
	\label{2prop:order unit spaces}
Let $(L, u)$ be an order unit Banach space and $\le$ be the ordering of $L$.\\
(i) The positive cone $L_+$ of $L$ is generating and closed.\\
(ii) The closed unit ball of $L$ is given by $\Delta=[-u, u]$.\\
(iii) If $f$ is a positive functional on $L$, then $f$ is bounded, and its dual norm $\|f\|^{*}$ on the Banach dual $L^*$ is given by $\|f\|^{*}=f(u)$.
Conversely, if a linear functional $f\colon L\to \R$ satisfies $\|f\|^{*}=f(u)$, then $f$ is positive.
\\
(iv) If we define $B_u:=\{f\in L^{*}_+\mid f(u)=1\}$, then $B_u$ is a base for the Banach dual cone $L^{*}_+$.
\\
(v) The Banach dual and order dual coincide with each other: $L^*=L^{\Diamond}$.
\end{prop}
\begin{pf}
(i) For $x\in L$, there exists $n\in \N$ such that $-nu\le x\le nu$.
Then $x=nu+(x-nu)$ shows $x\in L_+-L_+$, i.e., $L_+$ is generating.
Let $\{x_i\}_i$ be a Cauchy sequence in $L_+$ and converge to $x_\star\in L$.
For any $n\in\N$, we have $\|x_\star-x_i\|_u\le\frac{1}{n}$ for sufficiently large $i$.
It implies $-\frac{1}{n}u\le x_\star-x_i\le \frac{1}{n}u$, and thus $-nx_\star\le u$. 
Since this holds for all $n\in \N$ and $L$ is Archimedean, we obtain $-x_\star\le0$, i.e., $x\in L_+$.\\
(ii) Because $\Delta=[-u, u]=(u-L_+)\cap(-u+L_+)$ and $L_+$ is closed, we can observe that $\Delta$ is closed.
Then the definition of $\|\cdot\|_u$ as the Minkowski functional of $\Delta$ proves the claim.\\
(iii) Assume that $f$ is positive.
For $x\in \Delta$, we have $-f(u)\le f(x)\le f(u)$, i.e., $\|f\|^*\le f(u)$.
The equality clearly holds for $x=u$, and thus we obtain $\|f\|^*=f(u)$ (in particular, $f$ is bounded).
Assume conversely that $\|f\|^*=f(u)$.
For $x\in L_+$ with $\|x\|_u=1$, we have $0\le x\le  u$, or $0\le u-x\le u$.
It follows that $\|u-x\|_u\le1$, and because $\|f\|^*=f(u)$, we obtain $|f(u)-f(x)|\le f(u)$, which implies $f(x)\ge0$.\\
(iv) It can be seen from (iii) that every $f\in L_+^{*}$ satisfies $\|f\|^*=f(u)$, and thus, when considered as an element of $L^{**}:=(L^*)^*$, the functional $u$ is strictly positive on $L_+^*$.
Then, applying Proposition \ref{2prop:base to unit}, we obtain the claim.\\
(v) See Proposition 1.29 in \cite{Lami_PhD}.\qed
\end{pf}
It can be verified easily that the order unit norm corresponds to the usual operator norm in the formulation of quantum theory.

Now we can give the most general description of GPTs in terms of base norm Banach spaces and order unit Banach spaces.
We present first of all a fundamental theorem for our description on a close relationship between base norm Banach spaces and order unit Banach spaces (see \cite{Lami_PhD,Alfsen_compact_convex,Ellis_duality_1964} for the proof).
\begin{thm}
	\label{2thm:duality thm for base and order unit}
(i) Let $(L, B)$ be a base norm Banach space whose positive cone is $L_+$, and let $e_B$ be the intensity functional for $B$ satisfying $B=\{x \in L_+\mid e_B(x)=1\}$.
Then $(L^*, e_B)$ is an order unit Banach space, and $L^*_+:=\{f\in L^*\mid f(x)\ge0\ \mbox{for all $x\in L_+$}\}$ is its positive cone.
Moreover, the order unit norm coincides with the usual Banach dual norm in $L^*$.\\
(ii) Let $(L, u)$ be an order unit Banach space whose positive cone is $L_+$, and let $B_u:=\{f\in L^{*}_+\mid f(u)=1\}$.
Then $(L^*, B_u)$ is a base norm Banach space, and $L^*_+:=\{f\in L^*\mid f(x)\ge0\ \mbox{for all $x\in L_+$}\}$ is its positive cone.
Moreover, the base norm coincides with the usual Banach dual norm in $L^*$, and $B_u$ is a weakly* compact subset of $L^*$.
\end{thm}
We can also find that the converse of Theorem \ref{2thm:duality thm for base and order unit} holds (see \cite{Lami_PhD,Ellis_duality_1964,Olubummo1999} for the proof)
\begin{thm}
	\label{2thm:converse thm}
	Let $L$ be a Banach space that has a predual $L_*$.\footnote{Let $X$ be a Banach space.
		If there exists a Banach space $X_*$ such that its Banach dual $(X_*)^*$ satisfies $(X_*)^*=X$, then $X_*$ is called a \textit{predual} of $X$ \cite{Kuramochi_compactconvex_2020}.}\\
	(i) If $L$ is an order unit Banach space with $L_+$ its positive cone and $u\in L_+$ its order unit, then $L_*$ is a base norm Banach space whose positive cone and base are given by $L_{*+}=\{x\in L_*\mid f(x)\ge0\ \mbox{for all $f\in L_+$}\}$ and $B_{*u}=\{x\in L_{*+}\mid u(x)=1\}$ respectively.
	Moreover, the base norm coincides with the original Banach norm in $L_*$.\\
	(ii) If $L$ is a base norm Banach space with $L_+$ its positive cone and $B$ an weakly* compact base of $L_+$, then there exists $e_{*B}\in L_*$ such that $f(e_{*B})=1$ for all $f\in B$, and $L_*$ is an order unit Banach space whose positive cone and order unit are given by $L_{*+}=\{x\in L_*\mid f(x)\ge0\ \mbox{for all $f\in L_+$}\}$ and $e_{*B}$ respectively.
	Moreover, the order unit norm coincides with the original Banach norm in $L_*$.
\end{thm}
In the next subsection, we interpret these theorems in the language of GPTs and present the most standard formulation of GPTs based on them.

\subsection{Standard formulations of GPTs}
\label{2subsec:formulatoin of GPT}
We adopt Theorem \ref{2thm:duality thm for base and order unit} (i) to our expression of GPTs.
To do this, we recall that in Subsection \ref{2subsec:embeddings} (Theorem \ref{2thm:embedding}) the state space of a GPT was shown to be represented as a convex subset $\Omega$ of some Banach space $V$ (note that $0\notin\Omega$ by its construction).
We presented that the embedding vector space $V$ is constructed by $V=\mathit{span}(\Omega)$, and there is a convex, pointed, and generating cone $K$ in $V$ given by $K=\mathit{cone}(\Omega)$.
Moreover, we defined a norm in $V$ by 
\begin{equation*}
	\|v\|=\inf\{\alpha+\beta\mid v=\alpha p-\beta q,\ \alpha, \beta\ge0,\ p, q\in \Omega\}\quad(v\in V)
\end{equation*}
(see \eqref{2eq:intro base norm}), and found that $V$ is a Banach space and $K$ is closed with respect to the norm.
These observations can be interpreted in the language of ordered Banach spaces.
That is, $V$ is a base norm Banach space whose positive cone and base are given by $V_+=K=\mathit{cone}(\Omega)$ and $\Omega$ respectively.
The positive cone $V_+$ is closed and generating, and the base $\Omega$ is closed (see Proposition \ref{2prop:base norm spaces} (ii)).
On the other hand, it follows from Proposition \ref{2prop:base to unit} that there exists a strictly positive functional $e_{\Omega}$ such that $e_{\Omega}(\omega)=1$ for all $\omega\in\Omega$.
Then Proposition \ref{2prop:base norm spaces} (i) and Theorem \ref{2thm:duality thm for base and order unit} (i) result in that this $e_{\Omega}$ is an element of the Banach dual $V^{*}$, and in fact is an order unit of $V^*$ ordered via the Banach dual cone $V_+^*$.
Since $V=\mathit{span}(\Omega)$, we can find that the order unit $e_B$ coincides with the unit effect $u\in V^*$ (see Theorem \ref{2thm:embedding} (iii)).
Overall, we have obtained the following observation.
\begin{thm}
\label{2thm:GPTs for Banach}
A GPT is given by $(\Omega, \mathcal{E}_\Omega)$, where
\begin{enumerate}
\item the state space $\Omega$ is a closed base of the closed positive cone $V_+$ in a base norm Banach space $V$ such that $V_+=\mathit{cone}(\Omega)$ and $V=\mathit{span}(\Omega)$;
\item the effect space $\mathcal{E}_\Omega$ is a subset $[0, u]=\{e\in V^{*}\mid0\le e\le u\}$ of the order unit Banach space $V^{*}$ dual to $V$ with $V^{*}_+:=\{f\in V^{*}\mid f(x)\ge 0\ \mbox{for all $x\in L_+$}\}$ its positive cone and $u\in V^*$ its order unit determined by $u(\omega)=1$ for all $\omega\in\Omega$.
\end{enumerate}
\end{thm}
The contents of Theorem \ref{2thm:GPTs for Banach} are the most general formulation of GPTs.
In this thesis, the vector space $V$ in the theorem is called the \textit{standard embedding vector space} of the state space $\Omega$.
We remark that the positive cone $V_+$ represents the set of all ``unnormalized'' states, which are not necessarily mapped to 1 by the unit effect $u$, and that $\mathcal{E}_\Omega$ spans $V^{*}$ because $V_+^*$ is generating.
We define another primitive notion of {\itshape observables} based on this representation.\footnote{Observables can be introduced also in terms of the abstract description of convex structures \cite{Gudder_stochastic}, but in this thesis we present the definition of observables after embedding them into vector spaces for simplicity.}
\begin{defi}
\label{2def:obs}
Let $(\Omega, \mathcal{E}_\Omega)$ be a GPT.
An {\itshape observable} whose outcome space is given by a measurable space $(X, \mathcal{A})$ is defined as a normalized effect-valued measure $E$ on $(X, \mathcal{A})$, i.e., $E\colon\mathcal{A}\to\mathcal{E}_\Omega$ such that\\
(i) $E(X)=u$;\\
(ii) $E(\bigcup_i U_i)=\sum_i E(U_i)$ for any countable family $\{U_i\}_i$ of pairwise disjoint sets in $\mathcal{A}$ (the sum converges in the weak* topology on $V^*$).
\end{defi}
When the outcome set $X$ of an observable $E$ is finite, we often describe it as $E=\{e_{x}\}_{x\in X}$ with $e_x=E(\{x\})$ representing the yes-no measurement corresponding to the outcome $x\in X$.
We also use the notation $E=\{e_{i}\}_{i=1}^{l}$ when $|X|=l$ $(l<\infty)$, where $e_i$ represents the $i$th yes-no measurement. 
We note that $\sum_{x\in X}e_x=u$ and $\sum_{i=1}^{l}e_{i}=u$ hold.
In this thesis, we assume that observables are composed of a finite number of nonzero effects, and the trivial observable $\{u\}$ is not considered.

Although those descriptions above are of the most general form including theories with $\dim V=\infty$, we are interested only in finite-dimensional cases in this thesis.
We present explicitly this assumption as follows.
\begin{assum}[Finite dimensionality]
\label{math_assum:finite dim} 
For a GPT $(\Omega, \mathcal{E}_\Omega)$, the standard embedding vector space $V$ of $\Omega$ is a finite-dimensional Euclidean space.
\end{assum}
We note that any Hausdorff topological vector space of finite dimension is isomorphic linearly and topologically to the Euclidean space with the same dimension, and the norm, weak, and weak* topologies on a Banach space and its dual are Hausdorff (thus these topologies coincide with each other to be Euclidean in finite-dimensional cases) \cite{schaefer_TopVecSp,Conway_functionalanalysis}.
It should be also noted that a finite-dimensional vector space is isomorphic to its dual.
If a GPT satisfies Mathematical assumption \ref{math_assum:finite dim}, then we call it a finite-dimensional GPT. 
Let us develop how we can simplify the formulation of GPTs shown in Theorem \ref{2thm:GPTs for Banach} when dealing with finite-dimensional theories.
The following facts derived for the standard Euclidean topology are useful \cite{Lami_PhD,schaefer_TopVecSp,Aliprantis_cone_2007}.
\begin{prop}
	\label{2prop:finite dim_properties}
Let $L=\R^{d}$ be a finite-dimensional ordered vector space (in particular, an ordered Banach space with respect to the Euclidean norm) whose positive cone $L_+$ is generating.\\
(i) The condition that $L_+$ is generating is equivalent to the condition that $L_+$ has an interior point.\\
(ii) $L_+$ is closed if and only if $L$ is Archimedean.\\
(iii) If $L_+$ is closed, then the following statements for $e\in L^*_+$ are equivalent (remember that $L^*_+$ is defined as $L^*_+=\{f\in L^*\mid f(x)\ge0\ \mbox{for all $x\in L_+$}\}$, and the Banach dual $L^*$ of $L$ is an ordered Banach space with $L_+^*$ its positive cone because $L_+$ is generating):
\begin{enumerate}
	\item $e$ is strictly positive, i.e., $e(x)>0$ for all $x\in L_+\backslash\{0\}$;
	\item $e$ is an interior point of $L_+^*$;
	\item $e$ is an order unit in $L^*$.
\end{enumerate}
(iv) If $L_+$ is closed and $B$ is a base of $L_+$, then $B$ is bounded.\\
(v) If $L_+$ is closed, then $L_+$ admits a bounded base, i.e., there exists a bounded base for $L_+$.\\
(vi) If $L_+$ is closed, then all types of dual $L'$, $L^{\Diamond}$, and $L^{*}$ coincide with each other.
\end{prop}
\begin{pf}
In this proof, we denote the ordering of $L$ by $\le$ (thus, $x\ge0$ if and only if $x\in L_+$).\\
(i) Let $u$ be an interior point of $L_+$.
Then there exists an open ball $C$ in $L$ such that $u+C\subset L_+$.
For $v\in C$, because $C$ is a ball and thus $-v\in C$, we have $u\pm v\ge0$, i.e., $-u\le v \le u$.
Thus we obtain $C\subset [-u, u]$, i.e., $u$ is an order unit, which implies that $L_+$ is generating (see the proof of Proposition \ref{2prop:order unit spaces} (i)).
Assume conversely that $L_+$ is generating.
It is not difficult to see that the maximal set $\{v_i\}_{i=1}^k$ of linearly independent elements in $L_+$ is a basis of $L$ (and thus $k=d$).
Let us consider a subset $U:=\{v\in L\mid v=\sum_{i=1}^d \lambda_i v_i \ \mbox{with}\ \sum_{i=1}^d|\lambda_i|<1\}$ of $L$.
Because a map $\|\cdot\|'$ on $L$ given by $\|\sum_{i=1}^d \lambda_i v_i\|'=\sum_{i=1}^d |\lambda_i|$ defines a norm on $L$, the above $U$ is an open subset in $L$ (remember that all norm topologies are equivalent to each other in finite-dimensional cases).
Defining $v_\star:=\sum_{i=1}^d v_i\in L_+$, we can see that for any $v=\sum_{i=1}^d \lambda_i v_i\in U$, it holds that $v_\star+v=\sum_{i=1}^d (1+\lambda_i)v_i\in L_+$ because $v_i\in L_+$ and $1+\lambda_i>0$.
This implies $v_\star+U\subset L_+$, and thus $v_\star$ is an interior point of $L_+$.\\
(ii) Suppose that $L_+$ is closed.
If $x, y\in L$ satisfy $nx\le y$ for all $n\in \N$, then a sequence $\{\frac{1}{n}y-x\}_n$ in $L_+$ converges to $-x\in L_+$, and thus we have $x\le0$.\\
Conversely, suppose that $L$ is Archimedean and consider $x\in\overline{L_+}$, where $\overline{L_+}$ is the norm closure of $L_+$.
Because the interior of $L_+$ denoted by $\mathrm{int}(L_+)$ is nonempty (see (i)), there exists $y\in \mathrm{int}(L_+)$, and we can see that  $\frac{1}{n+1}y+(1-\frac{1}{n+1})x\in\mathrm{int}(L_+)$ holds for any $n\in \N$ \cite{schaefer_TopVecSp}.
It follows that $y+nx\in\mathrm{int}(L_+)\subset L_+$, and thus $-nx\le y$ for all $n\in\N$.
Since $L$ is Archimedean, we obtain $x\ge0$, which means $\overline{L_+}\subset L_+$.\\
(iii)
\textit{(1$\rightarrow$2)} Let $e\in L^*$ be strictly positive, and consider a closed unit ball $C:=\{x\in L\mid\|x\|\le1\}$ and a unit sphere $D:=\{f\in L\mid\|x\|=1\}$ in $L$, where $\|\cdot\|$ is the Euclidean norm.
Because $L_+$ is closed and $L=\R^d$ is finite-dimensional, $S:=L_+\cap D$ is a compact subset of $L$.
It implies that there exists a minimum value $M>0$ for the strictly positive and continuous functional $e$ on $S$.
On the other hand, if we define a closed unit ball $C^*:=\{f\in L^*\mid\|f\|^*\le1\}$ in $L^*$ with the Banach dual norm $\|\cdot\|^*$ (which is equivalent to Euclidean norm in this finite-dimensional case), then, for $f\in C^*$, we have $\|f\|^*=\sup_{y\in D}|f(y)|$ \cite{Conway_functionalanalysis}, and thus $-1\le f(y)\le 1$ holds for all $y\in S$.
It follows that if we take $0<\varepsilon<M$, then the functional $e+\varepsilon f$ satisfies $(e+\varepsilon f)(y)>0$ for all $y\in S$.
Since this holds for every $f\in C^*$ and any $x\in L_+$ can be represented as $x=\lambda y$ with $\lambda=\|x\|\ge0$ and $y\in S$, we can conclude that $e+\varepsilon C^*\subset L_+^*$, i.e., $e$ is an internal point of $L_+^*$.\\
\textit{(2$\rightarrow$3)} Because $e$ is an interior point of $L_+^*$, there exist $\alpha, \beta>0$ for every $f\in L^*$ such that $e+\alpha f\in L_+^*$ and $e+\beta(-f)\in L_+^*$.
It can be rewritten as $-\frac{1}{\alpha}e\le f\le \frac{1}{\beta}e$, and thus we can conclude that $e$ is an order unit in $L^*$.\\
\textit{(3$\rightarrow$1)} Suppose that there exists $x_0\in L_+\backslash\{0\}$ such that $e(x_0)=0$.
Since $e$ is an order unit, for $f\in L^*$, there exists $n\in\N$ such that $-ne\le f\le ne$, i.e., $f(x_0)=0$.
Because this holds for all $f\in L^*$, we obtain $x_0=0$, which is a contradiction.\\
(iv) 
Let $e_B$ be the intensity functional for $B$, which is strictly positive according to Proposition \ref{2prop:base to unit}.
Since any linear functional is continuous in a finite dimensional topological vector space (see Theorem 3.4 in \cite{schaefer_TopVecSp}), we obtain $e_B\in L^*_+$.
It follows from (iii) that $e_B$ is an order unit in $L^*$, and thus, for $f\in L^*$, there exists $n\in\N$ such that $-ne_B\le f\le ne_B$.
We obtain $|f(x)|\le n$ for all $x\in B$, and because $f\in L^*$ is arbitrary, we can conclude that $B$ is bounded.\\
(v) For the unit sphere $D$ in $L$ introduced above, consider $T:=L_+\cap D$ and its convex hull $T':=\mathit{conv}(T)$.
Clearly, $T'$ does not include $0$, and we can find that $T'$ is compact because $T$ is compact (see Theorem 10.2 in \cite{schaefer_TopVecSp}).
Thus there exists $x_0\in T'$ such that the continuous norm function $\|\cdot\|$ takes its minimum in $T'$.
It follows that any $x'\in T'$ satisfies $\|x_0\|\le\|x_0-t(x_0-x')\|$ for $0\le t\le1$ because $x_0-t(x_0-x')=(1-t)x_0+tx'\in T'$.
It can be rewritten as $t^2\|x_0-x'\|^2-2t(x_0, x_0-x')_E\ge0$, where $(\cdot, \cdot)_E$ is the Euclidean inner product in $L=\R^d$.
Since this holds for all $0\le t\le1$, it must hold that $(x_0, x_0-x')_E\le0$, that is, any $x'\in T'$ satisfies $(x_0, x')_E\ge (x_0, x_0)_E>0$.
On the other hand, any $x\in L_+$ can be written as $x=\|x\|y$ with $y\in D$ (in particular, $y\in T'$).
Hence we obtain $(x_0, x)_E>0$ for all $x\in L_+\backslash\{0\}$.
By means of the Riesz representation theorem \cite{Conway_functionalanalysis}, we can identify the inner product $(x_0, \cdot)_E$ as an element $f_0\in L^*$ such that $f_0(x)=(x_0, x)_E$.
This $f_0$ is a strictly positive functional for $L_+$, and thus defines a base, which is bounded as shown in (iv).\\ 
(vi) As we have seen in (iv) above, any linear functional on $L=\R^d$ is continuous, and thus we obtain $L^*=L'$ (and $L_+^*=L^\Diamond_+$).
On the other hand, it follows from (v) above that there are a base $B$ in $L$ and a strictly positive functional $e_B\in L_+^*$ associated with $B$.
Then (iii) and (i) imply that the Banach dual cone $L_+^*$ generates the Banach dual $L^*$, and because $L_+^*=L^\Diamond_+$, we can conclude the claim (remember that the order dual $L^\Diamond$ is given by $L^\Diamond=\mathit{span}(L^\Diamond_+)$).
\qed
\end{pf}
\begin{rmk}
The claim (iii)-(vi) in Proposition \ref{2prop:finite dim_properties} do not necessarily hold when $L_+$ is not closed.
To confirm this, let us consider the case where $L=\R^2$ and $L_+=\{(x, y)\in\R^2\mid y>0\}\cup (0, 0)$.
It is easy to see that $L_+$ defines a convex, pointed, and generating cone, but we cannot find a bounded base for this $L_+$ or verify $L^*=L^\Diamond$. 
\end{rmk}
Theorem \ref{2thm:GPTs for Banach} now can be rewritten as follows.
\begin{cor}
	\label{2cor:GPTs finite dim}
	A GPT is given by $(\Omega, \mathcal{E}_\Omega)$, where
	\begin{enumerate}
		\item the state space $\Omega$ is a compact convex set of some finite-dimensional Euclidean space $V=\R^{N+1}$ $(N<\infty)$ such that $\mathit{span}(\Omega)=V$ and $0\notin\mathit{aff}(\Omega)$ (in particular,
		$\dim\mathit{aff}(\Omega)=N$ holds\footnote{For an affine set $A$ of a finite-dimensional vector space $L$, its dimension $\dim A$ is defined as the dimension of the set $A-a_0 (a_0\in A)$ as a vector subspace of $L$.});
		\item the effect space $\mathcal{E}_\Omega$ is a subset $[0, u]=\{e\in V^{*}\mid0\le e\le u\}$ of the dual space $V^{*}$ of $V$ with $u\in V^*$ satisfying $u(\omega)=1$ for all $\omega\in\Omega$.\footnote{Although the dual space $V^{*}$ of $V$ is isomorphic to $\R^{N+1}$, we do not identify them here (see Subsection \ref{2subsec:self dual}).}
	\end{enumerate}
\end{cor}
The mathematical expression given in Corollary \ref{2cor:GPTs finite dim} is the standard formulation of GPTs in this thesis, and all observations on GPTs are based on this description.
We note that order structures similar to the ones described in Theorem \ref{2thm:GPTs for Banach} can be introduced for these finite-dimensional $V$ and $V^*$.
In fact, in Corollary \ref{2cor:GPTs finite dim}, we can verify easily that an order structure can be introduced for $V$ by a generating cone $V_+:=\mathit{cone}(\Omega)$, and $\Omega$ is a compact (thus closed) base for $V_+$ with which $V$ is a base norm Banach space.
There we can also find that $V^*$ can be ordered via a generating cone $V^{*}_+:=\{f\in V^{*}\mid f(x)\ge 0\ \mbox{for all $x\in L_+$}\}$, and the functional $u$, which is the intensity functional for the base $\Omega$, is an order unit with which $V^*$ is an order unit Banach space.\footnote{A triple $(V, V_+, u)$, where $V$ is a finite-dimensional ordered vector space with a closed positive cone $V_+$ and $u\in V^{*}$ is a strictly positive functional on $V$, is sometimes called an {\itshape abstract state space} \cite{barnum2012teleportation}
	The subset $V_+\cap u^{-1}(1)$ in this formulation corresponds to a state space in our formulation.}

Let us further introduce several notions about finite-dimensional GPTs.
For a state space $\Omega$, we can consider its extreme points,\footnote{For a convex subset $C$ in a vector space, $x\in C$ is called an {\itshape extreme point} of $C$ if $x=\lambda y+(1-\lambda)z$ with $y, z\in C$ and $0<\lambda<1$ implies $y=z=x$.} and denote the set of all extreme points of $\Omega$ by $\Omega^{\ext}=\{\omega_{i}^{\mathrm{ext}}\}_{i\in\mathcal{I}}$, where $\mathcal{I}$ is an index set.
Because $\Omega$ is a compact convex set in $\R^{N+1}$, thanks to the Krein-Milman theorem, $\Omega^{\ext}$ is not empty and $\Omega=\mathit{conv}(\Omega^\ext)$ \cite{Rockafellar+2015,schaefer_TopVecSp,Conway_functionalanalysis}.
Similar arguments also hold for the corresponding effect space $\mathcal{E}_{\Omega}$ since $\mathcal{E}_{\Omega}=V_+^{*}\cap (u-V_+^{*})$ and Proposition \ref{2prop:order unit spaces} (ii) imply $\mathcal{E}_{\Omega}$ is closed and bounded, i.e., compact.\footnote{\label{2footnote:completeness of effects}Therefore, the effect space $\mathcal{E}_{\Omega}$ is closed under infinite countable mixtures.}
\begin{defi}
\label{2def:pure}
(i) An extreme point of $\Omega$ is called a {\itshape pure state}, and a state that is not pure is called a {\itshape mixed state}.\\
(ii) An extreme point of $\mathcal{E}_\Omega$ is called a {\itshape pure effect}, and an effect that is not pure is called a {\itshape mixed effect}.\\
(iii) An effect $e$ is called {\it indecomposable} if $e\neq0$ and a decomposition $e=e_{1}+e_{2}$, where $e_{1}, e_{2}\in\mathcal{E}_\Omega$, implies that both $e_{1}$ and $e_{2}$ are scalar multiples of $e$.
We denote the set of all pure and indecomposable effects (shown to be nonempty \cite{KIMURA2010175}) by $\mathcal{E}^{\mathrm{ext}}_\Omega=\{e_{j}^{\mathrm{ext}}\}_{j\in \mathcal{J}}$, where $\mathcal{J}$ is an index set.
\end{defi}
It is easy to see that the unit effect $u$ is pure and $e^{\perp}:=u-e\in\mathcal{E}_\Omega$ is pure whenever $e\in\mathcal{E}_\Omega$ is pure.
It can be also observed that pure and indecomposable effects correspond to rank-1 projections in quantum theory (see Subsection \ref{eg_QT}), and that $e\in\mathcal{E}_\Omega$ is indecomposable if and only if $e$ is on an extremal ray of $V^{*}_{+}$.\footnote{A ray $P\subset V_{+}^{*}$ is called an {\it extremal ray} of $V^{*}_{+}$ if $x\in P$ and $x=y+z$ with $y, z\in V^{*}_{+}$ imply $y, z\in P$.} 
We call two GPTs $(\Omega_{1}, \mathcal{E}_{\Omega_1})$ and $(\Omega_{2}, \mathcal{E}_{\Omega_2})$ {\itshape equivalent} if there exists an affine bijection (affine isomorphism) $\psi$ such that $\psi(\Omega_{1})=\Omega_{2}$.
In this case, we can find easily that $\mathcal{E}_{\Omega_{2}}=\mathcal{E}_{\Omega_{1}}\circ\psi^{-1}$, and thus physical predictions are covariant (equivalent), which can be regarded as a physical expression of Proposition \ref{2prop:canonical embedding} (ii).
We remark that the affine isomorphism $\psi$ is indeed a linear isomorphism on the underlying vector spaces $V_1=\mathit{span}(\Omega_1)$ and $V_2=\mathit{span}(\Omega_2)$ (see the proof of Proposition \ref{2prop:canonical embedding} (ii)).
A set of $m$ states $\{\omega_{1}, \omega_{2}, \cdots, \omega_{m}\}$ is called ${\it perfectly}$ ${\it distinguishable}$ if there exists an observable $\{e_{1}, e_{2}, \cdots, e_{m}\}$ such that $e_{i}(\omega_{j})=\delta_{ij}$ $(i,j=1,\ 2,\ \cdots,\ m)$. 
In general, we can not identify the state of a system by a single measurement. 
However, for perfectly distinguishable states, there exists a measurement by which we can detect perfectly in which state the system is prepared.

\begin{rmk}
	\label{2rmk:Hardy's formulation}
There is a physical interpretation for the mathematical assumption of finite dimensionality.
In \cite{hardy2001quantum}, Hardy assumed that any state is determined by a finite set of effects named {\itshape fiducial measurements}.
If we denote those fiducial measurements by $\{e^{\mathrm{fid}}_i\}_{i=0}^N$ $(N<\infty)$, then a state $\omega$ can be identified with a vector
\begin{align}
\label{2eq:Hardy vector}
\omega=\left(
\begin{array}{c}
a_0 \\
a_1 \\
\vdots \\
a_{N}
\end{array}
\right),
\end{align}
where the $i$th row $a_i$ represents the probability $e^{\mathrm{fid}}_i(\omega)$.
It is easy to see that Hardy's formulation is consistent with ours: the state space $\Omega$ composed by $\omega$ of the form \eqref{2eq:Hardy vector} is a compact (or closed and bounded) convex set in $\R^{N+1}$ (by requiring completeness), and the normalization $u(\omega)=1$ for the unit effect $u$ yields the condition $\dim\mathit{aff}(\Omega)=N$.
We note that similar formulations for infinite-dimensional cases are given in \cite{Araki_mathematical}.
That is, a state $\omega$ is regarded as an element of the product set $[0,1]^{\mathcal{E}}$ with a set of effects $\mathcal{E}$ similarly to \eqref{2eq:Hardy vector}, and the state space is a subset of $[0,1]^{\mathcal{E}}$ which is compact with respect to the pointwise convergence topology corresponding to the weak* topology
(see also Theorem \ref{2thm:duality thm for base and order unit} (ii)).
\end{rmk}
\begin{rmk}
\label{2rmk:duality}
In our formulation, effects are constructed from states in the way how a state space is given first as a closed base of a base norm Banach space and then effects are given in its dual (see Theorem \ref{2thm:GPTs for Banach} and Corollary \ref{2cor:GPTs finite dim}).
On the other hand, as in the operator algebraic formulation of quantum theory \cite{Araki_mathematical,Bratteli_Robinson,Haag_localquantum_1996,Fewster_algebraic_2019}, it should be allowed to construct theories starting with effects.
In fact, for a finite-dimensional GPT $(\Omega, \mathcal{E}_{\Omega})$, if we consider the set $\Theta:=\{x\in V^{**}_+\mid x(u)=1\}$ in $V^{**}$, where $V^{**}$ is the double Banach dual of $V$ or the Banach dual of $V^{*}$, i.e., $V^{**}=(V^*)^*$ , then by means of the canonical identification of $V$ with $V^{**}$ it holds that $\Theta=\Omega$. 
This can be proven in a similar way to Proposition \ref{2prop:canonical embedding} (i) by just regarding $\Omega$ as $S$ (an explicit proof is given in \cite{Plavala_2021_GPTs}).
The equation $\Theta=\Omega$ holds also in an infinite-dimensional case\footnote{
It may be useful to understand the present descriptions from the perspective of the operator algebraic quantum theory.
Consider a concrete von Neumann algebra $\mathfrak{M}$ as representing observables (for the review of operator algebras, see \cite{Bratteli_Robinson,Takesaki_operatoralgebra,Sakai1998}).
Then the sets $\Omega$ and $\Theta$ given here represent respectively the set of all normal states, which are equivalent to the usual quantum states represented by density operators, and the set of all states on $\mathfrak{M}$.
In particular, $\Omega$ is a subset of the predual $\mathfrak{M}_*$ of $\mathfrak{M}$ while $\Theta$ is a subset of the Banach dual $\mathfrak{M}^*$ of $\mathfrak{M}$ (see also Theorem \ref{2thm:duality thm for base and order unit} and Theorem \ref{2thm:converse thm}).} when $\Omega$ is weakly compact, which is identical to the reflexivity of the underlying base norm Banach space $V$ (Lemma 8.71 in \cite{AlfsenShultz2003}).

There is also an axiomatic way of deriving our expression of GPTs from effects.  
As was proven that states represented by a total convex structure can be embedded into a base norm Banach space, one can show that an abstract expression of effects called a {\itshape convex effect algebra} (with some completeness) can be embedded into an order unit Banach space \cite{doi:10.1063/1.532031,Gudder_Pulmannova_1998,Gudder_EffectAlgebras}.
Then, due to Theorem \ref{2thm:duality thm for base and order unit} and the above argument, we can obtain successfully the corresponding state space in a base norm Banach space.
\end{rmk}

\section{Composite systems}
\label{2sec:composites}
In the previous section, we have presented the mathematical formulation of single systems in GPTs.
Then it is natural to ask how a system composed of several single systems, a {\itshape composite system}, is described mathematically in GPTs.
This is also motivated by another physical reason that it is in general difficult to isolate perfectly a system from environments: a composite system of the target system and its environments emerges naturally \cite{Busch_quantummeasurement}.
In this part, we establish the mathematical formulation of composite systems in GPTs based on that of single systems.
We note that we only study theories for bipartite systems in this thesis.
Our description may seem to be only for limited cases and not general, but it is in fact an essential one also for multipartite cases,\footnote{For the description of multipartite systems, see \cite{barnum2012teleportation,Plavala_2021_GPTs}.} and we can develop sufficiently interesting observations for this simplest scenario.

Let us consider a composite system composed of two single systems characterized by GPTs $(\Omega_A, \mathcal{E}_{\Omega_A})$ and $(\Omega_B, \mathcal{E}_{\Omega_B})$.
By convention, we suppose that the two subsystems are controlled by Alice and Bob respectively.
A fundamental assumption that is usually assumed implicitly is that the total system is also expressed by a GPT.
In the following, we follow this assumption, and denote the GPT for the total system by $(\Omega_{AB}, \mathcal{E}_{\Omega_{AB}})$.
Similarly to the previous section, we write the standard embedding vector spaces of $\Omega_A$, $\Omega_B$, and $\Omega_{AB}$ as $V_A$, $V_B$, and $V_{AB}$ respectively (thus $\mathcal{E}_{\Omega_A}$ is embedded into the dual vector space $V_A^{*}$, for example).
For the joint system, it is natural to require that every individual and independent preparation or measurement by Alice and Bob is a valid preparation or measurement in the bipartite system respectively.
It is also reasonable to assume that if such an independent preparation by Alice or Bob is probabilistic with some probability weight, then the total preparation is also probabilistic with the same probability weight (similarly for independent measurements).
Its mathematical expression is given as follows \cite{Lami_PhD}.
\begin{axiom}[Validity of individual preparations and measurements]
	\label{axiom:product states}
	There exist biaffine maps\footnote{Let $X, Y, Z$ be convex sets. A map $f\colon X\times Y\to Z$ is called \textit{biaffine} if $f(x, \cdot)$ is an affine map from $Y$ to $Z$ for every $x\in X$ and $f(\cdot, y)$ is an affine map from $X$ to $Z$ for every $y\in Y$.} $\phi\colon\Omega_A\times\Omega_B\to\Omega_{AB}$ and $\psi\colon\mathcal{E}_{\Omega_{A}}\times\mathcal{E}_{\Omega_{B}}\to\mathcal{E}_{\Omega_{AB}}$ such that
	\begin{align}
		[\psi(e_A, e_B)]\left(\phi(\omega_A, \omega_B)\right)=e_A(\omega_A)\cdot e_B(\omega_B)
	\end{align}
	for all $\omega_A\in \Omega_A$, $\omega_B\in \Omega_B$ and $e_A\in\mathcal{E}_{\Omega_A}$, $e_B\in\mathcal{E}_{\Omega_B}$.
	Each $\phi(\omega_A, \omega_B)$ and $\psi(e_A, e_B)$ are called a product state and product effect respectively.
\end{axiom} 
In the assumption, each product state $\phi(\omega_A, \omega_B)$ represents the individual preparation of $\omega_A$ and $\omega_B$ by Alice and Bob, and the individual convexity is reflected via the notion of biaffinity of the map $\phi$ (similarly for each product effect $\psi(e_A, e_B)$ and the map $\psi$).
We also require that if Alice and Bob measure their respective unit effects $u_A$ and $u_B$ individually on any joint state (not necessarily a product state), then the observed probability is 1.
In other words, the unit effect of the total system is $\psi(u_A, u_B)$.
\begin{axiom}[Unit effect of the total system]
	\label{axiom:total unit effect}
The unit effect $u_{AB}$ of the joint system is given by the product effect $\psi(u_A, u_B)$ of each unit effect $u_A$ and $u_B$ of Alice and Bob respectively.
\end{axiom}
Let us give an easy consequence of these axioms according mainly to \cite{Lami_PhD}.
\begin{lem}
	\label{2lem:injections}
Assume Axiom \ref{axiom:product states} and Axiom \ref{axiom:total unit effect}.
There are linear injections $\Phi\colon V_A\otimes V_B\to V_{AB}$ and $\Psi\colon V_A^{*}\otimes V_B^{*}\to V_{AB}^{*}$ such that\\
(i) $\Phi(\omega_A\otimes\omega_B)=\phi(\omega_A, \omega_B)$ for all $\omega_A\in\Omega_{A}$ and $\omega_B\in\Omega_{B}$;\\
(ii) $\Psi(e_A\otimes e_B)=\psi(e_A, e_B)$ for all $e_A\in\mathcal{E}_{\Omega_A}$ and $e_B\in\mathcal{E}_{\Omega_B}$;\\
(iii) $u_{AB}=\Psi(u_A\otimes u_B)$.
\end{lem}
\begin{pf}
Let us first construct a bilinear extension $\Phi'$ on $V_A\times V_B$ of the biaffine map $\phi$ on $\Omega_A\times\Omega_B$.
Due to the assumption of the biaffinity, $\phi(\omega_A, \cdot)$ defines an affine map from $\Omega_B$ to $\Omega_{AB}$ for a fixed $\omega_A\in\Omega_A$, and it can be extended (uniquely) to a linear map $[\phi'({\omega_A})](\cdot)$ from $\mathit{span}(\Omega_B)=V_B$ to $\mathit{span}(\Omega_{AB})=V_{AB}$ such that $[\phi'({\omega_A})](\omega_B)=\phi(\omega_A, \omega_B)$ for all $\omega_B\in\Omega_B$ (see the proof of Proposition \ref{2prop:canonical embedding} (ii)).
In this way, we obtain a map $P\colon\Omega_A\to\mathcal{L}(V_B, V_{AB})$, where $\mathcal{L}(V_B, V_{AB})$ is the set of all linear operators from $V_B$ to $V_{AB}$.
It is easy to see that $P$ is affine, and thus, similarly to the above argument, it has a unique linear extension $P\colon V_A\to\mathcal{L}(V_B, V_{AB})$ such that $[P(\omega_A)](\cdot)=[\phi'({\omega_A})](\cdot)$ for all $\omega_A\in\Omega_A$.
The bilinear extension $\Phi'$ of $\phi$ is now obtained by $\Phi'(v_A, v_B)=[P(v_A)](v_B)$ for $v_A\in V_A, v_B\in V_B$.
Then the existence of the linear map $\Phi\colon V_A\otimes V_B\to V_{AB}$ satisfying $\Phi(v_A\otimes v_B)=\Phi'(v_A, v_B)$ for all $v_A\in V_A, v_B\in V_B$ (in particular (i)) follows immediately from the universal property of tensor product \cite{Roman_advanced_linear_alge}.
The existence of $\Psi$ satisfying (ii) is proved similarly, and (iii) is an easy consequence of Axiom \ref{axiom:total unit effect}.

The remaining problem is to show the injectivity of $\Phi$.
Because $\Omega_A$ and $\Omega_B$ span $V_A$ and $V_B$ respectively, any $v_{A}\otimes v_B\in V_{A}\otimes V_B$ with $v_A\in V_A, v_B\in V_B$ is expressed as $v_{A}\otimes v_B\in V_{AB}=\sum_{i, j}a_{ij}\omega_A^i\otimes\omega_B^j$ with $a_{ij}\in\R$ and $\omega_A^i\in\Omega_A, \omega_B^j\in\Omega_B$.
Similarly, any $w_{A}\otimes w_B\in V^*_{A}\otimes V^*_{B}$ with $w_A\in V_A^*, w_B\in V^*_B$ is expressed as $w_{A}\otimes w_{B}=\sum_{k,l}b_{kl}e_A^{k}\otimes e_B^{l}$ with $b_{kl}\in\R$ and $e_A^k\in\mathcal{E}_{\Omega_A}, e_B^l\in\mathcal{E}_{\Omega_B}$.
Thus we can observe from the linearity of $\Phi$ and $\Psi$ that 
\[
[\Psi(w_{A}\otimes w_{B})](\Phi(v_{A}\otimes v_B))=w_{A}(v_A)\cdot w_B(v_B).
\]
Let $v_{AB}\in V_{A}\otimes V_B$ satisfy $\Phi(v_{AB})=0$.
Since $v_{AB}$ is expressed by $v_{AB}=\sum_{i}v_A^i\otimes v_B^i$ with $v_A^i\in V_A, v_B^i\in V_B$, it holds for all $w_A\in V_A^*, w_B\in V^*_B$ that 
\begin{align*}
[w_{A}\otimes w_{B}](v_{AB})
&=\sum_{i}w_{A}(v_A^i)\cdot w_B(v_B^i)=[\Psi(w_A\otimes w_B)](\Phi(v_{AB}))=0.
\end{align*}
Because $\{w_{A}\otimes w_{B}\mid w_A\in V_A^*, w_B\in V^*_B\}$ spans $V_{A}^*\otimes V_B^*=(V_{A}\otimes V_B)^*$, we can conclude $v_{AB}=0$, which means that $\Phi$ is injective.
The injectivity of $\Psi$ can be proved similarly.
\qed
\end{pf}

\begin{rmk}
It seems to be assumed implicitly in Axiom \ref{axiom:product states} and Axiom \ref{axiom:total unit effect} that Alice's actions do not influence Bob, and vice versa.
For example, there we require that Alice and Bob can prepare individually their states and effects without influencing each other,
or we can see from the biaffinity (bilinearity) of $\psi$ that the statistics observed by Alice alone are independent of Bob's measurements: 
for any joint state $\omega_{AB}$, the probability of Alice observing $e_A\in\mathcal{E}_{\Omega_A}$ does not depend on Bob's observable $\{f_B^{\prime\hspace{0.5mm} i}\}_i$ because it holds that
\[
\sum_i[\psi(e_A, f_B^{\prime\hspace{0.5mm} i})](\omega_{AB})=[\psi(e_A, u_B)](\omega_{AB}).
\]
In fact, Axiom \ref{axiom:product states} and Axiom \ref{axiom:total unit effect} can be rephrased in terms of the so-called {\itshape no-signaling principle} \cite{PhysRevA.75.032304,PR-box,Barnum_post-classical},\footnote{How the no-signaling principle is formulated in GPTs is explained in detail in \cite{Barnum_post-classical}.} or the requirement of {\itshape causality} \cite{PhysRevA.81.062348,PhysRevA.84.012311}.
\end{rmk}
There is another important requirement for bipartite systems.
We require that every joint state can be determined by local measurements.
This claim called the {\itshape tomographic locality for states} \cite{hardy2001quantum,PhysRevA.75.032304,Hardy2011_quantum} is described mathematically as follows.
\begin{axiom}[Tomographic locality for states]
	\label{axiom:tomographic locality}
	If $\omega_{AB}, \omega'_{AB}\in\Omega_{AB}$ satisfy $[\psi(e_A, e_B)](\omega_{AB})=[\psi(e_A, e_B)](\omega'_{AB})$ for all $e_A\in\mathcal{E}_{\Omega_A}$ and $e_B\in\mathcal{E}_{\Omega_B}$, then $\omega_{AB}=\omega'_{AB}$.
\end{axiom}
\begin{lem}
	\label{2lem:tensor product}
Assume Axiom \ref{axiom:product states}, Axiom \ref{axiom:total unit effect}, and Axiom \ref{axiom:tomographic locality}.
The linear injections $\Phi$ and $\Psi$ in Lemma \ref{2lem:injections} are also surjective, that is, $\Phi$ is a linear bijection between $V_A\otimes V_B$ and $V_{AB}$, and $\Psi$ between $V_A^{*}\otimes V_B^{*}$ and $V_{AB}^{*}$.
\end{lem}
\begin{pf}
Suppose that $V_{AB}^{*}\backslash\Psi(V_A^*\otimes V_B^*)$ is nonempty, and $w_{AB}'\in V_{AB}^{*}\backslash\Psi(V_A^*\otimes V_B^*)$.
Because $w_{AB}'$ and a basis of $\Psi(V_A^*\otimes V_B^*)$ are linearly independent, we can construct an element $v_{AB}'$ of $V_{AB}^{**}$ such that $v_{AB}'(w_{AB}')=1$ and $v_{AB}'(w_{AB})=0$ for all $w_{AB}\in\Psi(V_A^*\otimes V_B^*)$.
We note that $V_{AB}^{**}=V_{AB}$ holds due to the assumption of finite dimensionality, and thus $v_{AB}'$ above can be regarded as an element of $V_{AB}$.
It follows that if we define $M:=\{v_{AB}\in V_{AB}\mid w_{AB}(v_{AB})=0\ \mbox{for all $w_{AB}\in\Psi(V_A^*\otimes V_B^*)$}\}$, then $M\backslash\{0\}$ is nonempty.
In the following, we prove that $M=\{0\}$, which implies $V_{AB}^{*}=\Psi(V_A^*\otimes V_B^*)$.
Let $v_{AB}^{\star}\in M$.
For a state $\omega_{AB}\in\mathit{int}(V_{AB+})$, where $\mathit{int}(V_{AB+})$ is the interior of the positive cone $V_{AB+}$ of $V_{AB}$ generated by $\Omega_{AB}$ (see Proposition \ref{2prop:finite dim_properties}), we can make $\omega_{AB}^{\star}:=\omega_{AB}+\varepsilon v_{AB}^{\star}$ belong to $V_{AB+}$ if we take sufficiently small $\varepsilon>0$.
Because $u_{AB}=\Psi(u_A\otimes u_B)$, it holds from the definition of $M$ that $u_{AB}(\omega_{AB}^{\star})=u_{AB}(\omega_{AB})=1$, i.e., $\omega_{AB}^{\star}\in\Omega_{AB}$.
Moreover, we can find in a similar way that $[\Psi(e_A\otimes e_B)](\omega_{AB}^{\star})=[\Psi(e_A\otimes e_B)](\omega_{AB})$ holds for all $e_A\in\mathcal{E}_{\Omega_A}, e_B\in\mathcal{E}_{\Omega_B}$, and thus, from Axiom \ref{axiom:tomographic locality}, $\omega_{AB}^{\star}=\omega_{AB}$ holds.
This implies $v_{AB}^{\star}=0$, which means $M=\{0\}$ and $V_{AB}^{*}=\Psi(V_A^*\otimes V_B^*)$.
Therefore, we can conclude $\Psi$ is surjective (i.e., bijective).
Then it is easy to derive $\dim V_{AB}=\dim V_{A}\otimes V_B=\dim V_{A}\cdot\dim V_{B}$, and the surjectivity (bijectivity) of $\Phi$ follows from this observation.
\qed
\end{pf}
We assume Axiom \ref{axiom:product states}, Axiom \ref{axiom:total unit effect}, and Axiom \ref{axiom:tomographic locality} (thus Lemma \ref{2lem:tensor product}) in this thesis.
Then it does not cause any problem to identify the subsets $\Phi^{-1}(\Omega_{AB})$ and $\Psi^{-1}(\mathcal{E}_{\Omega_{AB}})$ of $V_A\otimes V_B$ and $V_A^{*}\otimes V_B^{*}=(V_A\otimes V_B)^{*}$ with the state space and effect space of the joint system respectively (see the argument above Remark \ref{2rmk:Hardy's formulation}).
We hereafter write $\Phi^{-1}(\Omega_{AB})$ simply as $\Omega_{AB}$, and $\Psi^{-1}(\mathcal{E}_{\Omega_{AB}})$ as $\mathcal{E}_{\Omega_{AB}}$, and work with these expressions of states and effects, where product states and effects are represented as $\omega_A\otimes \omega_B$ and $e_A\otimes e_B$ ($\omega_A\in\Omega_A, \omega_B\in\Omega_B$ and $e_A\in\mathcal{E}_{\Omega_A}, e_B\in\mathcal{E}_{\Omega_B}$) respectively.
\begin{rmk}
One may consider Axiom \ref{axiom:total unit effect} to be more artificial when compared to the other axioms.
In \cite{Lami_PhD}, it was explained that $u_{AB}=u_A\otimes u_B$ holds if the {\itshape tomographic locality for effects} is imposed together with Axiom \ref{axiom:product states} and Axiom \ref{axiom:tomographic locality}. 
\end{rmk}

Let us give more detailed specifications of bipartite systems.
For GPTs $(\Omega_A, \mathcal{E}_{\Omega_A})$ and $(\Omega_B, \mathcal{E}_{\Omega_B})$ of local systems, we define the following classes of convex sets \cite{HULANICKI1968177,https://doi.org/10.1112/plms/s3-19.1.177}.
\begin{defi}
\label{2def:min max tensor}
Let $(\Omega_A, \mathcal{E}_{\Omega_A})$ and $(\Omega_B, \mathcal{E}_{\Omega_B})$ be GPTs.\\
(i) The convex subset $\Omega_A \otimes_{min}\Omega_{B}$ of $V_A\otimes V_B$ defined as 
\[
\Omega_A \otimes_{min}\Omega_{B}:=\left\{\sum_{i}p_i\omega_A^{i}\otimes\omega_B^{i}\mid p_i\ge0,\ \sum_{i}p_i=1,\ \omega_A^{i}\in\Omega_A,\ \omega_B^{i}\in\Omega_B\right\}
\]
is called the {\itshape minimal tensor product} of $\Omega_A$ and $\Omega_B$.
The minimal tensor product $\mathcal{E}_{\Omega_A} \otimes_{min}\mathcal{E}_{\Omega_B}$ of the effect spaces $\mathcal{E}_{\Omega_A}$ and $\mathcal{E}_{\Omega_B}$ is defined in the same way.\\
(ii) The convex subset $\Omega_A \otimes_{max}\Omega_{B}$ of $V_A\otimes V_B$ defined as 
\begin{align*}
\Omega_A\otimes_{max}\Omega_{B}:=\{\omega_{AB}\in V_A\otimes &V_B\mid (e_{A}\otimes e_B)(\omega_{AB})\in[0,1],\\
&\qquad e_{A}\in\mathcal{E}_{\Omega_A},\  e_{B}\in\mathcal{E}_{\Omega_B},\ (u_A\otimes u_{B})(\omega_{AB})=1\}
\end{align*}
is called the {\itshape maximal tensor product} of $\Omega_A$ and $\Omega_B$.
The maximal tensor product $\mathcal{E}_{\Omega_A} \otimes_{max}\mathcal{E}_{\Omega_B}$ of the effect spaces $\mathcal{E}_{\Omega_A}$ and $\mathcal{E}_{\Omega_B}$ is defined in the same way.
\end{defi}
It is verified easily that the minimal and maximal tensor products are dual to each other in the sense that $\mathcal{E}_{\Omega_A \otimes_{min}\Omega_{B}}=\mathcal{E}_{\Omega_A}\otimes_{max}\mathcal{E}_{\Omega_B}$ and $\mathcal{E}_{\Omega_A \otimes_{max}\Omega_{B}}=\mathcal{E}_{\Omega_A}\otimes_{min}\mathcal{E}_{\Omega_B}$ hold.
A similar observation can be obtained if we start from effects (see Remark \ref{2rmk:duality}).
We also note that $\Omega_A \otimes_{min}\Omega_{B}\subset\Omega_A \otimes_{max}\Omega_{B}$ clearly holds.

By means of the axioms introduced so far, we can specify the joint state space $\Omega_{AB}$ in the following way.
First, it can be found that $\Omega_{AB}$ must include $\Omega_A \otimes_{min}\Omega_B$ because product states and probabilistic mixtures are required to exist.
Similarly, the existence of product effects are imposed, and it follows that $\Omega_{AB}$ is included in $\Omega_A \otimes_{max}\Omega_{B}$.
We have now obtained the following description for bipartite systems.
\begin{thm}
\label{2thm:joint state space}
Let $(\Omega_{AB}, \mathcal{E}_{\Omega_{AB}})$ be a GPT describing a bipartite system composed of two subsystems $(\Omega_A, \mathcal{E}_{\Omega_A})$ and $(\Omega_B, \mathcal{E}_{\Omega_B})$.
Then 
\begin{align}
	\label{2eq:composite states}
\Omega_A \otimes_{min}\Omega_{B}\subset\Omega_{AB}\subset\Omega_A \otimes_{max}\Omega_{B}
\end{align}
holds.
Dually, 
\begin{align}
	\label{2eq:composite effects}
\mathcal{E}_{\Omega_A} \otimes_{min}\mathcal{E}_{\Omega_B}\subset\mathcal{E}_{\Omega_{AB}}\subset\mathcal{E}_{\Omega_A} \otimes_{max}\mathcal{E}_{\Omega_B}
\end{align}
holds.
\end{thm}
It can be found that when a bipartite system $(\Omega_{AB}, \mathcal{E}_{\Omega_{AB}})$ composed of $(\Omega_{A}, \mathcal{E}_{\Omega_{A}})$ and $(\Omega_{B}, \mathcal{E}_{\Omega_{B}})$ satisfies both \eqref{2eq:composite states} and \eqref{2eq:composite effects}, then Axiom \ref{axiom:product states}, Axiom \ref{axiom:total unit effect}, and Axiom \ref{axiom:tomographic locality} hold conversely.
In fact, Axiom \ref{axiom:product states} and Axiom \ref{axiom:total unit effect} clearly hold, and because any element of $V_{AB}^*$ can be written as a linear combination of effects of the form $e_A\otimes e_B$ (remember that $\mathcal{E}_{\Omega_{A}}$ and $\mathcal{E}_{\Omega_{B}}$ span $V_A^*$ and $V_B^*$ respectively), Axiom \ref{axiom:tomographic locality} also can be verified.
\begin{defi}
Each element of $\Omega_A \otimes_{min}\Omega_{B}$ is called a {\itshape separable state}, and an element of the form $\omega_A\otimes\omega_B$ is particularly called a {\itshape product state}.
Each element of $\Omega_A \otimes_{max}\Omega_{B}\backslash\Omega_A \otimes_{min}\Omega_{B}$ is called an {\itshape entangled state}.
Separable effects, product effects, and entangled effects are defined in the same way.
\end{defi}
It should be noted that entangled states exist unless either theory is classical. 
More precisely, it was shown in \cite{Aubrun2021} that $\Omega_A \otimes_{min}\Omega_{B}=\Omega_A \otimes_{min}\Omega_{B}$ holds if and only if either $\Omega_A$ or $\Omega_B$ is a simplex (i.e., a classical theory).
\begin{eg}[Quantum theory over a real Hilbert space]
Let $\mathcal{K}=\R^{d}$ ($d<\infty$) be a finite-dimensional real Hilbert space.
We can consider a GPT whose state space is given by $\Omega_{\mathrm{rQT}}(\mathcal{K})=\{\rho\in\LL_{S}(\mathcal{K})\mid\rho\ge0,\ \Tr[\rho]=1\}$ with $\LL_{S}(\mathcal{K})$ the set of all self-adjoint operators on $\mathcal{K}$.
The real quantum theory described by $\Omega_{\mathrm{rQT}}(\mathcal{K})$ often appears in the field of GPTs when deriving the standard quantum theory (i.e., complex quantum theory) from physical principles \cite{hardy2001quantum,Barnum_2014}.
It is easy to see that $\mathit{aff}(\Omega_{\mathrm{rQT}}(\mathcal{K}))$ and the standard embedding vector space $V(\mathcal{K})$ are given by $\mathit{aff}(\Omega_{\mathrm{rQT}}(\mathcal{K}))=\{\rho\in\LL_{S}(\mathcal{K})\mid\Tr[\rho]=1\}$ and $V(\mathcal{K})=\LL_{S}(\mathcal{K})$ respectively.
We can also observe that $\dim\mathit{aff}(\Omega_{\mathrm{rQT}}(\mathcal{K}))=\half(d^2+d)-1$ and $\dim V(\mathcal{K})=\half(d^2+d)$ hold (in particular, $\dim V(\mathcal{K})=\dim\mathit{aff}(\Omega_{\mathrm{rQT}}(\mathcal{K}))+1$ holds).
Suppose in analogy with the formulation of a finite-dimensional quantum theory over a complex Hilbert space that the state space of the composite system composed of two identical state spaces $\Omega_{\mathrm{rQT}}(\mathcal{K})$ is given by $\Omega_{\mathrm{rQT}}(\mathcal{K}\otimes\mathcal{K})
=\{\rho\in\LL_{S}(\mathcal{K}\otimes\mathcal{K})\mid\rho\ge0,\ \Tr[\rho]=1\}$.
Then we can derive
\[
\dim V_A\cdot\dim V_B=[\half\left(d^2+d\right)]^2,\ \ \dim V_{AB}=\half\left(d^4+d^2\right),
\]
where $V_A=V_B=V(\mathcal{K})$ and $V_{AB}=V(\mathcal{K}\otimes\mathcal{K})$ are the standard embedding vector spaces of the individual and total state spaces respectively.
The equations imply $\dim V_A\cdot\dim V_B<\dim V_{AB}$, i.e., $V_A\otimes V_B=V_{AB}$ does not hold.
Thus we can conclude that the tomographic locality is not satisfied in a finite-dimensional quantum theory over a real Hilbert space (it is not difficult to see that Axiom \ref{axiom:product states} and Axiom \ref{axiom:total unit effect} hold in this case).\footnote{We can also eliminate a finite-dimensional quantum theory over a quaternionic Hilbert space by a similar observation \cite{Araki_composite_1980}.}
\end{eg}

\section{Transformations}
\label{2sec:channels}
In this section, we explain how transformations between systems are formulated in GPTs, which completes our review for basic notions on GPTs.
It is found that not only state changes such as time evolution but also measurements can be described in terms of transformations or their more refined form {\itshape channels}.
We also introduce the notions of compatibility and incompatibility for channels, which play a key role in the following chapters.

\subsection{Channels in GPTs}
\label{subsec:channel}
In quantum theory, transformations of systems are described via the notion of channels \cite{Busch_quantummeasurement,heinosaari_ziman_2011}.
In this part, we explain how channels are generalized in GPTs according mainly to \cite{Plavala_2021_GPTs,Heinosaari2019nofreeinformation}.
\begin{defi}
Let $(\Omega_1, \mathcal{E}_{\Omega_1})$ and $(\Omega_2, \mathcal{E}_{\Omega_2})$ be GPTs.
An affine map $T\colon\Omega_{1}\to\Omega_2$ is called a {\itshape channel} from $\Omega_{1}$ to $\Omega_{2}$.
A linear map $T\colon V_1\to V_2$, where $V_1$ and $V_2$ are the embedding vector spaces of $\Omega_{1}$ and $\Omega_2$ respectively, is equivalently called a channel from $\Omega_{1}$ to $\Omega_{2}$ if $T(\Omega_{1})\subset\Omega_2$ (thus it is positive in the sense that $T((V_1)_+)\subset T((V_2)_+)$\footnote{It is sometimes more convenient to consider a linear map $T\colon V_1\to V_2$ satisfying $T((V_1)_+)\subset T((V_2)_+)$ and $u_2(T(\omega))\le1$ ($\omega\in\Omega_1$), where $u_2$ is the unit effect for $\Omega_2$, as representing a transformation of states.
Such positive and normalization-nonincreasing maps in GPTs correspond to the notion of \textit{operations} in quantum theory \cite{Haag_Kastler_1964}, although operations in quantum theory are sometimes assumed also to be completely positive \cite{heinosaari_ziman_2011}.}).
We denote the set of all channels from $\Omega_{1}$ to $\Omega_{2}$ by $\mathcal{C}(\Omega_{1}, \Omega_{2})$, and denote the set $\mathcal{C}(\Omega_{1}, \Omega_{1})$ simply by $\mathcal{C}(\Omega_{1})$
\end{defi}
A channel $T\colon V_1\to V_2$ in the above definition induces a map $T'\colon V_2^{*}\to V_1^{*}$ such that $[T'e](\omega)=e(T\omega)$ for all $e\in\mathcal{E}_{\Omega_2}$ and $\omega\in \Omega_1$.
In this way, we can focus on transformations between effects instead of transformations between states.
However, in this thesis, when channels are considered, they always represent transformations between states, that is, the \textit{Schrodinger picture} is adopted  although similar arguments can be developed with channels considered as transformations between effects (the \textit{Heisenberg picture}).

It is easy to obtain the following observations.
\begin{prop}
Let $(\Omega_1, \mathcal{E}_{\Omega_1})$, $(\Omega_2, \mathcal{E}_{\Omega_2})$, and $(\Omega_3, \mathcal{E}_{\Omega_3})$ be GPTs.\\
(i) For $T, T'\in \mathcal{C}(\Omega_{1}, \Omega_{2})$, if we define $\lambda T+(1-\lambda)T'$ as $[\lambda T+(1-\lambda)T'](\omega_1)=\lambda T(\omega_1)+(1-\lambda)T'(\omega_1)$ $(0\le\lambda\le1)$, then $\lambda T+(1-\lambda)T'\in \mathcal{C}(\Omega_{1}, \Omega_{2})$\\
(ii) If $S\in\mathcal{C}(\Omega_{1}, \Omega_{2})$ and $T\in\mathcal{C}(\Omega_{2}, \Omega_{3})$, then $T\circ S\in \mathcal{C}(\Omega_{1}, \Omega_{3})$.
\end{prop}
Let us give several examples of channels.
\begin{eg}[Basic examples of channels]
	\label{eq:basic examples of channels}
Let $(\Omega_1, \mathcal{E}_{\Omega_1})$, $(\Omega_2, \mathcal{E}_{\Omega_2})$ be GPTs, and $V_1$ and $V_2$ be the standard embedding vector spaces of $\Omega_{1}$ and $\Omega_2$ respectively.\\
(i) If we define a map $id_{\Omega_1}\colon\Omega_1\to\Omega_1$ by $id_{\Omega_1}(\omega_1)=\omega_1$ for all $\omega_1\in \Omega_1$, then $id_{\Omega_1}\in\mathcal{C}(\Omega_1)$.
We call $id_{\Omega_1}$ the {\itshape identity channel} on $\Omega_1$.\\
(ii) Let $\omega^*\in \Omega_2$.
If we define a map $T_{\omega^*} \colon\Omega_1\to\Omega_2$ by $T_{\omega^*}(\omega_1)=\omega^{*}$ for all $\omega_1\in \Omega_1$, then $T_{\omega^*}\in\mathcal{C}(\Omega_1, \Omega_2)$.\\
(iii) Consider a bipartite system $(\Omega_{12}, \mathcal{E}_{\Omega_{12}})$ composed of $(\Omega_1, \mathcal{E}_{\Omega_1})$, $(\Omega_2, \mathcal{E}_{\Omega_2})$.
For the linear maps $id_{\Omega_1}\colon V_1\to V_1$ and $u_2\colon V_2\to \R$, where $id_{\Omega_1}$ is the identity channel on $\Omega_1$ and $u_2$ is the unit effect on $\Omega_2$, we define their tensor product $id_{\Omega_1}\otimes u_2$.
Then $id_{\Omega_1}\otimes u_2$ as a linear map from $V_{1}\otimes V_2$ to $V_1$ is a channel from $\Omega_{12}$ to $\Omega_1$, and called the {\itshape partial trace}.
\end{eg}
We can demonstrate that even the fundamental notions of states and observables can be represented in terms of channels.
To show this, we need to define the following convex sets.
\begin{defi}
	\label{2def:simplex}
Let $\{x_i\}_{i=1}^{n+1}$ be a set of affinely independent\footnote{Vectors $v_0,\ v_1,\ \ldots,\ v_{n}$ in a vector space $V$ are called \textit{affinely independent} if the vectors $v_1-v_0,\ \ldots,\ v_n-v_0$ are linearly independent.} vectors in $\R^{d}$ $(n\le d)$.
The convex set $conv(\{x_i\}_{i=1}^{n+1})$ is called an {\itshape $n$-dimensional simplex} \cite{Rockafellar+2015}.
In particular, we denote the simplex generated by orthonormal vectors $\{p_i\}_{i=1}^{n+1}$ with $p_1=(1, 0, \ldots, 0)$, $p_2=(0, 1, 0, \ldots, 0),\ldots$ simply by $\Delta_n$, and call it the {\itshape $n$-dimensinoal standard simplex}.
It is trivial that any $n$-dimensional simplex is isomorphic to $\Delta_n$.
\end{defi}
\begin{eg}[States, observables, and instruments as channels]
	\label{2eg: obs as channels}
Let $(\Omega_1, \mathcal{E}_{\Omega_1}), (\Omega_2, \mathcal{E}_{\Omega_2})$ be GPTs, and let us follow similar notations in Definition \ref{2def:simplex} above.\\
(i) 
A state $\omega\in\Omega_1$ is equivalent to a channel from $\Delta_1$ to $\Omega_1$ by the identification of $\omega$ with a channel $P_{\omega}\colon\Delta_1\to\Omega_1$ defined as $P_{\omega}(p_1)=\omega$.
Similarly, we can introduce a conditional preparation channel $P_{\{\omega_{i}\}_{i=1}^{n+1}}\in \mathcal{C}(\Delta_n, \Omega_1)$ by $P_{\{\omega_i\}_{i=1}^{n+1}}(v)=\sum_{i=1}^{n+1}v_i\omega_i$, where $v_i$ is the $i$th element of the vector $v\in \R^{n+1}$.
The channel $P_{\{\omega_{i}\}_{i=1}^{n+1}}$ represents an apparatus that outputs the states $\{\omega_{i}\}_{i=1}^{n+1}$ according to the proportion determined by a classical input $v=(v_1, \ldots, v_{n+1})$.\\
(ii) An observable $E=\{e_i\}_{i=1}^{n+1}$ on $\Omega_1$ with $(n+1)$ outcomes is equivalent to a channel from $\Omega_1$ to $\Delta_n$ by the identification of $E$ with a channel $M_E\colon\Omega_1\to\Delta_n$ defined as 
$M_E(\omega)=(e_1(\omega), \ldots, e_{n+1}(\omega))=\sum_{i=1}^{n+1}e_i(\omega)p_i$.\\
(iii) For a conditional preparation channel $P_{\{\omega_{i}\}_{i=1}^{n+1}}\in \mathcal{C}(\Delta_n, \Omega_2)$ and a measurement channel $M_E\colon\Omega_1\to\Delta_n$, the composition $P_{\{\omega_{i}\}_{i=1}^{n+1}}\circ M_E\in \mathcal{C}(\Omega_1, \Omega_2)$ is called a {\itshape measure-and prepare channel}.
Preparation channels or measurement channels in (i) or (ii) above respectively are examples of measure-and prepare channels (see \cite{Plavala_2021_GPTs} for other examples).\\
(iv) A channel from $\Omega_1$ to $\Omega_1\otimes_{min}\Delta_n$ is called an {\itshape instrument}.
It outputs the measurement outcomes of an observable and the ensemble of the post measurement states.
\end{eg}
\begin{rmk}
\label{2rmk:CP}
In this part, we introduce channels in GPTs as positive and normalization-preserving maps, while in quantum theory channels are defined as trace-preserving (normalization-preserving) and completely positive maps \cite{Busch_quantummeasurement,heinosaari_ziman_2011,Kraus_CP_1971}.
The notion of complete positivity can be introduced also in GPTs based on the above formulation of bipartite systems \cite{Plavala_2021_GPTs}.
However, completely positive maps do not always correspond to physical processes in GPTs.
This is because, while in quantum theory all completely positive maps are physically valid transformations in the sense that their physical implementations exist via the Steinspring's theorem \cite{Hayashi_etal_2015},
there is in general not ensured the existence of such physical implementations in GPTs.
\end{rmk}

\subsection{Compatibility and incompatibility for channels}
\label{2subsec:compatibility}
In quantum theory, we cannot always obtain simultaneously statistics for a pair of observables such as position and momentum, or cannot always duplicate a family of states \cite{Wootters_Zurek_no-cloning}.
These impossibilities are essential ingredients of quantum theory: for example, without them, the violation of Bell inequality or the security of quantum cryptography never occurs.
Those impossibilities can be described by the notion of {\itshape incompatibility} in a unified way \cite{Heinosaari_2016}.
In this part, we demonstrate that the notion of incompatibility can be introduced successfully also in GPTs. 
\begin{defi}
	\label{2def:compatibility}
Let $(\Omega_1, \mathcal{E}_{\Omega_1})$, $(\Omega_2, \mathcal{E}_{\Omega_2})$, and $(\Omega_3, \mathcal{E}_{\Omega_3})$ be GPTs, and $(\Omega_{23}, \mathcal{E}_{\Omega_{23}})$ be a GPT that describes a joint system of $(\Omega_2, \mathcal{E}_{\Omega_2})$ and $(\Omega_3, \mathcal{E}_{\Omega_3})$.
Channels $S\in\mathcal{C}(\Omega_1, \Omega_2)$ and $T\in\mathcal{C}(\Omega_1, \Omega_3)$ are called {\itshape compatible} if there exists a channel $R\in\mathcal{C}(\Omega_1, \Omega_{23})$ called a {\itshape joint channel} of $S$ and $T$ such that the marginal actions of $R$ reproduce each action of $S$ and $T$, that is, 
\begin{align*}
&(id_{\Omega_2}\otimes u_3)\circ R=S,\\
&(u_2\otimes id_{\Omega_3})\circ R=T,
\end{align*}
where $id_{\Omega_2}\otimes u_3$ and $u_2\otimes id_{\Omega_3}$ are the partial traces in $\Omega_{23}$ (see Example \ref{eq:basic examples of channels}).
If $S$ and $T$ are not compatible, then they are called {\itshape incompatible}
\end{defi}
This definition of incompatibility applies to cases when three or more channels are considered.
For incompatibility of observables, we can derive a simpler expression.
\begin{prop}
\label{2prop:compatibility for obs}
Let $(\Omega, \mathcal{E}_{\Omega})$ be a GPT, and $M_E$ and $M_F$ be the measurement channels associated with observables $E=\{e_i\}_{i=1}^{l}$ and $F=\{f_j\}_{j=1}^m$ on $\Omega$ respectively (see Example \ref{2eg: obs as channels} (ii)).
Then $M_E$ and $M_F$ are compatible if and only if there exists an observable (called a {\itshape joint observable}) $G=\{g_{ij}\}_{i=1,j=1}^{l,m}$ on $\Omega$ such that 
\[
\sum_{j=1}^m g_{ij}=e_i,\qquad \sum_{i=1}^l g_{ij}=f_j.
\] 
\end{prop}
\begin{pf}
If there exists an observable $G=\{g_{ij}\}_{i=1,j=1}^{l,m}$ on $\Omega$ such that 
\[
\sum_{j=1}^m g_{ij}=e_i,\qquad \sum_{i=1}^l g_{ij}=f_j, 
\]
then it is easy to see that the measurement channel $M_G\in\mathcal{C}(\Omega, \Delta_{l-1}\otimes_{min}\Delta_{m-1})$ defined as $M_G(\omega)
=(m_{11}(\omega), \ldots, m_{lm}(\omega))
=\sum_{i, j}m_{ij}(\omega)p_i\otimes p_j$, where $p_1=(1, 0, 0, \ldots)$, $p_2=(0, 1, 0, \ldots)$, (see Definition \ref{2def:simplex}), is a joint channel of $M_E$ and $M_F$.
We note that the composite of two simplices is always given by their minimal tensor product.
Conversely, if there exists a joint channel $M\in\mathcal{C}(\Omega, \Delta_{l-1}\otimes_{min}\Delta_{m-1})$ of $M_E$ and $M_F$, then, representing $M(\omega)\in\Delta_{l-1}\otimes_{min}\Delta_{m-1}$ as $M(\omega)=\sum M(\omega)_{ij} p_i\otimes p_j$ ($M(\omega)_{ij}\in [0,1]$), we obtain $\sum_jM(\omega)_{ij}=e_i(\omega)$ and $\sum_iM(\omega)_{ij}=f_j(\omega)$.
We can naturally introduce effects $m_{ij}\colon\Omega\to [0,1]$ by $m_{ij}(\omega)=M(\omega)_{ij}$, and it is easy to verify that $\sum_jm_{ij}=e_i$ and $\sum_im_{ij}=f_j$ (and thus $\sum_{i,j}m_{ij}=u$, i.e., $\{m_{ij}\}_{i,j}$ is an observable).
\qed
\end{pf}
In \cite{PhysRevA.94.042108}, it was shown that there exists an incompatible pair of observables in every finite-dimensional GPT unless it is classical.
We can present the existence of another type of incompatibility. 
\begin{eg}[Generalized no-broadcasting theorem]
Let $(\Omega, \mathcal{E}_{\Omega})$ be a GPT, and let $(\Omega_{12}, \mathcal{E}_{\Omega_{12}})$ be a GPT describing a composite system of $(\Omega_1, \mathcal{E}_{\Omega_1})$ and $(\Omega_2, \mathcal{E}_{\Omega_2})$, where $\Omega_1=\Omega_2=\Omega$.
A set of states $\{\omega_i\}_i\subset\Omega$ is called {\it broadcastable} if there exists a channel $T\in\mathcal{C}(\Omega, \Omega_{12})$ such that $(id_{\Omega_1}\otimes u_2)(T(\omega_i))=\omega_i$ and $(u_1\otimes id_{\Omega_2})(T(\omega_i))=\omega_i$ hold for all $i$.
It was shown in \cite{Barnum_nocloning,PhysRevLett.99.240501} (see also \cite{Plavala_2021_GPTs}) that $\{\omega_i\}_i\subset\Omega$ is broadcastable if and only if it lies in a simplex.
In other words, the identity channels $id_{\Omega_1}$ and $id_{\Omega_2}$ are compatible if and only if $\Omega_1=\Omega_{2}=\Omega$ is a simplex (i.e., the theory is classical).
\end{eg}
These results on GPTs manifest interesting facts that properties once thought to be specific to quantum theory are in fact more universal ones.

\section{Additional notions}
\label{2sec:additional}
So far we have reviewed fundamental notions in GPTs especially focusing on states and effects.
It was shown that states and effects are represented in terms of ordered Banach spaces, and under the assumption of finite dimensionality, they are reduced to elements of Euclidean spaces.
In this part, based on those descriptions, we develop additional notions on states and effects that will play significant roles in demonstrating several results of this thesis.
To do this, we follow the notations that have been used so far.
That is, a GPT is given by a pair $(\Omega, \mathcal{E}_{\Omega})$ of a state space and the corresponding effect space such that $\Omega\subset V=\R^{N+1}$ with $\mathit{span}(\Omega)=V$ and $0\notin \Omega$ and $\mathcal{E}_\Omega\subset V^{*}$.
We should also recall that the set of all pure states is denoted by $\Omega^{\ext}$, and the set of all pure and indecomposable effects by $\mathcal{E}_{\Omega}^{\ext}$.

\subsection{Physical equivalence of pure states}
It is known that in quantum theory all pure states are physically equivalent via unitary (and antiunitary) transformations \cite{Busch_quantummeasurement}. 
A similar notion to this physical equivalence of pure states can be introduced also in GPTs.

Let $\Omega$ be a state space. A map $T\colon\Omega\to\Omega$ is called a {\it state automorphism} on $\Omega$ if $T$ is an affine bijection. We denote the set of all state automorphisms on $\Omega$ by $GL(\Omega)$, and say that a state $\omega_{1}\in\Omega$ is {\it physically equivalent} to a state $\omega_{2}\in\Omega$ if there exists a $T\in GL(\Omega)$ such that $T\omega_{1}=\omega_{2}$. It was shown in \cite{kimura2010physical} that the physical equivalence of $\omega_{1}, \omega_{2}\in\Omega$ is equal to the existence of some unit-preserving affine bijection $T'\colon\mathcal{E}_\Omega\to\mathcal{E}_\Omega$ satisfying $e(\omega_{1})=T'(e)(\omega_{2})$ for all $e\in\mathcal{E}_\Omega$, which means that $\omega_{1}$ and $\omega_{2}$ have the same physical contents on measurements. 
Because any affine map on $\Omega$ can be extended uniquely to a linear map on $V$, it holds that $GL(\Omega)=\{T\colon V\to V\mid T:\mbox{linear, bijective},\ T(\Omega)=\Omega\}$. It is clear that $GL(\Omega)$ forms a group, and we can represent the notion of physical equivalence of pure states by means of the transitive action of $GL(\Omega)$ on $\Omega^{\mathrm{ext}}$.
\begin{defi}[Transitive state space]
	\label{def_transitive state sp}
	A state space $\Omega$ is called {\it transitive} if $GL(\Omega)$ acts transitively on $\Omega^{\mathrm{ext}}$, that is, for any pair of pure states $\omega_{i}^{\mathrm{ext}}, \omega_{j}^{\mathrm{ext}}\in\Omega^{\mathrm{ext}}$ there exists an affine bijection $T_{ji}\in GL(\Omega)$ such that $\omega_{j}^{\mathrm{ext}}=T_{ji}\omega_{i}^{\mathrm{ext}}$.
\end{defi}
We remark that the equivalence of pure states does not depend on how the theory is expressed. In fact, when $\Omega$ is a transitive state space and $\Omega':=\psi(\Omega)$ is equivalent to $\Omega$ with a linear bijection $\psi$, it is easy to check that $GL(\Omega')=\psi\circ GL(\Omega)\circ\psi^{-1}$ and $\Omega'$ is also transitive.

In the remaining of this subsection, we let $\Omega$ be a transitive state space. In a transitive state space, we can introduce successfully the maximally mixed state as a unique invariant state with respect to every state automorphism \cite{Davies_compactconvex}.
\begin{prop}
	\label{def_max mixed state}
	For a transitive state space $\Omega$, there exists a unique state $\omega_{M}\in\Omega$ (which we call the {\it maximally mixed state}) such that $T\omega_{M}=\omega_{M}$ for all $T\in GL(\Omega)$. The unique maximally mixed state $\omega_{M}$ is given by 
	\[
	\omega_{M}=\int_{GL(\Omega)} T\omega^{\mathrm{ext}}\ d\mu(T),
	\]
	where $\omega^{\mathrm{ext}}$ is an arbitrary pure state and $\mu$ is the normalized two-sided invariant Haar measure on $GL(\Omega)$.
\end{prop}
Note in Proposition \ref{def_max mixed state} that the transitivity of $\Omega$ guarantees the independence of $\omega_{M}$ on the choice of $\omega^{\mathrm{ext}}$. 
When $\Omega^{\mathrm{ext}}$ is finite and $\Omega^{\mathrm{ext}}=\{\omega_{i}^{\mathrm{ext}}\}_{i=1}^{n}$, the maximally mixed state $\omega_{M}$ has a simpler form
\[
\omega_{M}=\frac{1}{n}\sum_{i=1}^{n} \omega_{i}^{\mathrm{ext}}.
\]
We should recall that the action of the linear bijection $\eta:=\frac{1}{\|\omega_{M}\|_{E}}\id_{V}$ on $\Omega$ does not change the theory, where $\|\omega_{M}\|_{E}=(\omega_{M}, \omega_{M})^{1/2}_{E}$ with the standard Euclidean inner product $(\cdot, \cdot)_E$ and $\id_{V}$ is the identity map on $V$. 
Since $\eta T \eta^{-1}=T$ holds for all $T\in GL(\Omega)$, the set $GL(\Omega)$ is invariant under the rescaling of $\Omega$ by $\eta$, i.e., $GL(\eta(\Omega))=GL(\Omega)$. 
It follows that the unique maximally mixed state of the rescaled state space $\eta(\Omega)$ is $\frac{1}{\|\omega_{M}\|_{E}}\omega_{M}$. 
In the remaining of this thesis, when a transitive state space is discussed, we apply this rescaling and assume that $\|\omega_{M}\|_{E}=1$ holds. 
This assumption makes it easy to prove our main theorems in Chapter \ref{chap:URs in GPT} via Proposition \ref{prop_ortho repr} introduced in the following.

The Haar measure $\mu$ on $GL(\Omega)$ makes it possible for us to construct a convenient representation of the theory. First of all, we define an inner product $\langle\cdot,\cdot \rangle_{GL(\Omega)}$ on $V$ as
\[
\langle x, y\rangle_{GL(\Omega)}:=\int_{GL(\Omega)}(Tx, Ty)_{E}\ d\mu(T)\quad (x, y\in V).
\]
Remark that in this thesis we adopt $(\cdot,\cdot)_{E}$ as the reference inner product of $\langle\cdot,\cdot \rangle_{GL(\Omega)}$ although the following discussion still holds even if it is not $(\cdot,\cdot)_{E}$. 
Thanks to the properties of the Haar measure $\mu$, it holds that 
\[
\langle Tx, Ty\rangle_{GL(\Omega)}=\langle x, y\rangle_{GL(\Omega)} \quad \ ^{\forall} T\in GL(\Omega),
\]
which proves that any $T\in GL(\Omega)$ to be an orthogonal transformation on $V$ with respect to the inner product $\langle\cdot,\cdot \rangle_{GL(\Omega)}$. Therefore, together with the transitivity of $\Omega$, we can see that all pure states of $\Omega$ are of equal norm, that is,
\begin{equation}
	\label{eq_equal norm}
	\begin{aligned}
		\|\omega_{i}^{\mathrm{ext}}\|_{GL(\Omega)}
		&=\langle\omega_{i}^{\mathrm{ext}}, \omega_{i}^{\mathrm{ext}}\rangle^{1/2}_{GL(\Omega)}\\
		&=\langle T_{i0}\omega_{0}^{\mathrm{ext}}, T_{i0}\omega_{0}^{\mathrm{ext}}\rangle^{1/2}_{GL(\Omega)}\\
		&=\langle\omega_{0}^{\mathrm{ext}}, \omega_{0}^{\mathrm{ext}}\rangle^{1/2}_{GL(\Omega)}\\
		&=\|\omega_{0}^{\mathrm{ext}}\|_{GL(\Omega)}
	\end{aligned}
\end{equation}
holds for all $\omega_{i}^{\mathrm{ext}}\in\Omega^{\mathrm{ext}}$, where $\omega_{0}^{\mathrm{ext}}$ is an arbitrary reference pure state. We remark that when $\|\omega_{M}\|_{E}=1$, we can obtain from the invariance of $\omega_{M}$ for $GL(\Omega)$
\begin{align*}
	\|\omega_{M}\|_{GL(\Omega)}^{2}
	&=\int_{GL(\Omega)}(T\omega_{M}, T\omega_{M})_{E}\ d\mu(T)\\
	&=\int_{GL(\Omega)}(\omega_{M}, \omega_{M})_{E}\ d\mu(T)\\
	&=\|\omega_{M}\|_{E}^{2}\int_{GL(\Omega)}\ d\mu(T)\\
	&=\|\omega_{M}\|_{E}^{2},
\end{align*}
and thus $\|\omega_{M}\|_{GL(\Omega)}=1$ . The next proposition allows us to give a useful representation of the theory (the proof is given in Appendix \ref{appA}).
\begin{prop}
	\label{prop_ortho repr}
	For a transitive state space $\Omega$, there exists a basis $\{v_{l}\}_{l=1}^{N+1}$ of $V$ orthonormal with respect to the inner product  $\langle\cdot,\cdot\rangle_{GL(\Omega)}$ such that $v_{N+1}=\omega_{M}$ and 
	\[
	x\in\mathit{aff}(\Omega)\iff x=\sum_{l=1}^{N}a_{l}v_{l}+v_{N+1}=\sum_{l=1}^{N}a_{l}v_{l}+\omega_{M}\ (a_{1}, \cdots, a_{N}\in\R).
	\]
\end{prop}
By employing the representation shown in Proposition \ref{prop_ortho repr}, an arbitrary $x\in\mathit{aff}(\Omega)$ can be written as a vector form that
\begin{equation}
	\label{eq_vec repr}
	x=
	\left(
	\begin{array}{c}
		\bm{x} \\
		1
	\end{array}
	\right)
	\quad
	\mbox{with}
	\quad
	\omega_{M}=
	\left(
	\begin{array}{c}
		\bm{0} \\
		1
	\end{array}
	\right),
\end{equation}
where the vector $\bm{x}$ is sometimes called the {\it Bloch vector} \cite{muller2012unifying,PhysRevLett.108.130401} corresponding to $x$.

\subsection{Self-duality}
\label{2subsec:self dual}
In this part, we introduce the notion of self-duality, which also plays an important role in our work. 

Let $V_{+}$ be the positive cone generated by a state space $\Omega$. We define the {\it internal dual cone} of $V_{+}$ relative to an inner product $(\cdot ,\cdot )$ on $V$ as $V^{*int}_{+(\cdot ,\cdot )}:=\{y\in V\mid(x, y)\ge0,\ ^{\forall} x\in V_{+}  \}$, which is isomorphic to the dual cone $V^{*}_{+}$ because of the Riesz representation theorem \cite{Conway_functionalanalysis}.\footnote{In the field of GPTs, effects are often defined as elements of $V=\R^{N+1}$ through the identification $V^*=V^{*int}_{+(\cdot ,\cdot )}$, and the action of effects on states is represented via the inner product $(\cdot ,\cdot)$.}
The self-duality of $V_{+}$ can be defined as follows.
\begin{defi}[Self-duality]
	\label{def_self-duality}
	$V_{+}$ is called {\it self-dual} if there exists an inner product $(\cdot ,\cdot )$ on $V$ such that $V_{+}=V^{*int}_{+(\cdot ,\cdot )}$.
\end{defi}
We remark similarly to Definition \ref{def_transitive state sp} that if $V_{+}$ generated by a state space $\Omega$ is self-dual, then the cone $V'_{+}$ generated by $\Omega':=\psi(\Omega)$ with a linear bijection $\psi$ (i.e. $V'_{+}=\psi(V_{+})$) is also self-dual. In fact, we can confirm that if $V_{+}=V^{*int}_{+(\cdot,\cdot)}$ holds for some inner product $(\cdot,\cdot)$, then $V'_{+}=V^{'*int}_{+(\cdot,\cdot)'}$ holds, where the inner product $(\cdot,\cdot)'$ is defined as $(x,y)'=(\psi^{-1}x,\  \psi^{-1}y)\ \ (x, y\in V)$.

Let us consider the case where $\Omega$ is transitive and $V_{+}$ is self-dual with respect to the inner product $\langle\cdot ,\cdot \rangle_{GL(\Omega)}$. Since $V_{+}=V^{*int}_{+\langle\cdot ,\cdot \rangle_{GL(\Omega)}}$, we can regard $V_{+}$ also as the set of unnormalized effects. In particular, every pure state $\omega_{i}^{\mathrm{ext}}\in\Omega^{\mathrm{ext}}$ can be considered as an unnormalized effect, and if we define
\begin{equation}
	\label{def_indecomp effect0}
	e_{i}:=\frac{\omega_{i}^{\mathrm{ext}}}{\|\omega_{i}^{\mathrm{ext}}\|_{GL(\Omega)}^{2}}=\frac{\omega_{i}^{\mathrm{ext}}}{\|\omega_{0}^{\mathrm{ext}}\|_{GL(\Omega)}^{2}},
\end{equation}
then from Cauchy-Schwarz inequality
\begin{align*}
	\langle e_{i},\omega_{k}^{\mathrm{ext}}\rangle_{GL(\Omega)}
	&\le\|e_{i}\|_{GL(\Omega)}\|\omega_{k}^{\mathrm{ext}}\|_{GL(\Omega)}=1
\end{align*}
holds for any pure state $\omega_{k}^{\mathrm{ext}}\in\Omega^{\mathrm{ext}}$ (thus $e_{i}$ is indeed an effect). 
The equality holds if and only if $\omega_{k}^{\mathrm{ext}}$ is parallel to $e_{i}$, i.e. $\omega_{k}^{\mathrm{ext}}=\omega_{i}^{\mathrm{ext}}$, and we can also conclude that an effect is pure and indecomposable if and only if it is of the form defined as \eqref{def_indecomp effect0} together with the fact that effects on the extremal rays of $V^{*int}_{+\langle\cdot ,\cdot \rangle_{GL(\Omega)}}=V_{+}$ are indecomposable (for more details see \cite{KIMURA2010175}):
\begin{equation}
	\label{def_indecomp effect}
	e_{i}=\frac{\omega_{i}^{\mathrm{ext}}}{\|\omega_{i}^{\mathrm{ext}}\|_{GL(\Omega)}^{2}}=\frac{\omega_{i}^{\mathrm{ext}}}{\|\omega_{0}^{\mathrm{ext}}\|_{GL(\Omega)}^{2}}\equiv e_{i}^{\mathrm{ext}}\in\mathcal{E}^{\mathrm{ext}}(\Omega).
\end{equation}
When $|\Omega^{\mathrm{ext}}|<\infty$, it is sufficient for the discussion above that $\Omega$ is transitive and self-dual with respect to an arbitrary inner product.
\begin{prop}
	\label{prop_transitive self-dual}
	Let $\Omega$ be transitive with $|\Omega^{\mathrm{ext}}|<\infty$ and $V_+$ be self-dual with respect to some inner product. There exists a linear bijection $\Xi\colon V\to V$ such that $\Omega':=\Xi\Omega$ is transitive and the generating positive cone $V'_{+}$ is self-dual with respect to $\langle\cdot,\cdot\rangle_{GL(\Omega')}$, i.e.
	$V^{'}_+ = V_{+\langle\cdot ,\cdot \rangle_{GL(\Omega')}}^{'*int}$.
\end{prop}
The proof is given in Appendix \ref{appB}. Proposition \ref{prop_transitive self-dual} reveals that if a theory with finite pure states is transitive and self-dual, then the theory can be expressed in the way it is self-dual with respect to $\langle\cdot,\cdot \rangle_{GL(\Omega)}$.

\section{Examples of GPTs}
\label{2sec:example}
In this section, we present some examples of GPTs with relevant structures to transitivity or self-duality.
\subsection{Classical theories with finite levels}
\label{eg_CT}
Let us denote by $\Omega_{\mathrm{CT}}$ the state space of a classical system with a finite level.
$\Omega_{\mathrm{CT}}$ can be represented by means of some finite $N\in\mathbb{N}$ as the set of all probability distributions (probability vectors) $\{\mathbf{p}=(p_{1},\ \cdots,\ p_{N+1})\}\subset V=\R^{N+1}$ on some sample space $\{a_{1},\ \cdots,\ a_{N+1}\}$, i.e., $\Omega_{\mathrm{CT}}$ is the $N$-dimensional standard  simplex $\Delta_N$. 
It is easy to justify that the set of all pure states $\Omega_{\mathrm{CT}}^{\mathrm{ext}}$ is given by $\Omega_{\mathrm{CT}}^{\mathrm{ext}}=\{\mathbf{p}_{i}^{\mathrm{ext}}\}_{i=1}^{N+1}$, where $\mathbf{p}_{i}^{\mathrm{ext}}$ is the probability distribution satisfying $(\mathbf{p}_{i}^{\mathrm{ext}})_{j}=\delta_{ij}$, and the positive cone $V_{+}$ by $V_{+}=\{\sigma=(\sigma_{1},\cdots,\sigma_{N+1})\in V\mid \sigma_{i}\ge0,\ ^{\forall} i\}$. Remark that the set
	\[
	\{\mathbf{p}_{i}^{\mathrm{ext}}\}_{i=1}^{N+1}=\{(1, 0, \cdots, 0), (0, 1, \cdots, 0), \cdots, (0, 0, \cdots, 1)\}
	\]
	forms a standard orthonormal basis of $V$.
	Since any state automorphism maps pure states to pure states, it can be seen that the set $GL(\Omega_{\mathrm{CT}})$ of all state automorphisms on $\Omega_{\mathrm{CT}}$ is exactly the set of all permutation matrices with respect to the orthonormal basis $\{\mathbf{p}_{i}^{\mathrm{ext}}\}_{i=1}^{N+1}$ of $V$. Therefore, $\Omega_{\mathrm{CT}}$ is a transitive state space, and any $T\in GL(\Omega_{\mathrm{CT}})$ is orthogonal, which results in
	\begin{align}
		\label{eg_eq_CT_inner product}
		\langle x, y\rangle_{GL(\Omega_{\mathrm{CT}})}
		&=\int_{GL(\Omega_{\mathrm{CT}})}(Tx, Ty)_{E}\ d\mu(T)\notag\\
		&=\int_{GL(\Omega_{\mathrm{CT}})}(x, y)_{E}\ d\mu(T)\notag\\
		&=(x, y)_{E}\int_{GL(\Omega_{\mathrm{CT}})}d\mu(T)\notag\\
		&=(x, y)_{E}.
	\end{align}
	The set of all positive linear functionals on $\Omega_{\mathrm{CT}}$ can be identified with the internal dual cone  $V^{*int}_{+(\cdot ,\cdot )_{E}}$, and every $h\in V^{*int}_{+(\cdot ,\cdot )_{E}}$ can be identified with $h=(h(\mathbf{p}_{1}^{\mathrm{ext}}),\ \cdots,\ h(\mathbf{p}_{N+1}^{\mathrm{ext}}))$ with all entries nonnegative since
	\[
	h(\mathbf{p}_{i}^{\mathrm{ext}})=(h, \mathbf{p}_{i}^{\mathrm{ext}})_{E}=(h)_{i}\ge0
	\]
	holds for all $i$. Therefore, we can conclude together with \eqref{eg_eq_CT_inner product} $V_{+}=V^{*int}_{+(\cdot ,\cdot )_{E}}=V^{*int}_{+\langle\cdot ,\cdot \rangle_{GL(\Omega_{\mathrm{CT}})}}$. Note that we can find the representation \eqref{eq_vec repr} to be valid for this situation
	by taking a proper basis of $V=\R^{N+1}$ and normalization.

\subsection{Quantum theories with finite levels}
	\label{eg_QT}
	The state space of a quantum system with a finite level denoted by $\Omega_{\mathrm{QT}}$ is the set of all density operators on $N$-dimensional Hilbert space $\HH$ ($N<\infty$), that is, $\Omega_{\mathrm{QT}}:=\{\rho\in\LL_{S}(\HH)\mid\rho\ge0, \Tr[\rho]=1\}$, where $\LL_{S}(\HH)$ is the set of all self-adjoint operators on $\HH$. The set of all pure states $\Omega_{\mathrm{QT}}^{\mathrm{ext}}$ is given by the rank-1 projections: $\Omega_{\mathrm{QT}}^{\mathrm{ext}}=\{\ketbra{\psi}{\psi}\mid\ket{\psi}\in\HH, \braket{\psi|\psi}=1\}$. It has been demonstrated in \cite{KIMURA2003339} that with the identity operator $\id_{N}$ on $\HH$ and the generators $\{\sigma_{i}\}_{i=1}^{N^{2}-1}$ of $SU(N)$ satisfying
	\begin{equation}
		\label{eq_Pauli}
		\sigma_{i}\in\LL_{S}(\HH),\ \ \Tr[\sigma_{i}]=0,\ \ \Tr[\sigma_{i}\sigma_{j}]=2\delta_{ij},
	\end{equation}
	any $A\in\LL_{S}(\HH)$ can be represented as
	\begin{equation}
		\label{eq_Bloch repr1}
		A=c_{0}\id_{N}+\sum_{i=1}^{N^{2}-1}c_{i}\sigma_{i}\quad(c_{0}, c_{1}, \cdots, c_{N^{2}-1}\in\R)
	\end{equation}
	and any $B\in\mathit{aff}(\Omega_{\mathrm{QT}})$ as
	\begin{equation}
		\label{eq_Bloch repr2}
		B=\frac{1}{N}\id_{N}+\sum_{i=1}^{N^{2}-1}c_{i}\sigma_{i}\quad(c_{1}, \cdots, c_{N^{2}-1}\in\R).
	\end{equation}
	Since \eqref{eq_Pauli} implies that $\{\id_{N}, \sigma_{1}, \cdots, \sigma_{N^{2}-1}\}$ forms an orthogonal basis of $\LL_{S}(\HH)$ with respect to the Hilbert-Schmidt inner product $(\cdot ,\cdot )_{HS}$ defined by
	\[
	(X,Y)_{HS}=\Tr[X^{\dagger}Y],
	\]
	and \eqref{eq_Bloch repr1} and \eqref{eq_Bloch repr2} prove $\mathrm{dim}(\LL_{S}(\HH))=\mathrm{dim}(\mathit{aff}(\Omega_{\mathrm{QT}}))+1$, it seems natural to consider $\Omega_{\mathrm{QT}}$ to be embedded in $V=\LL_{S}(\HH)$ equipped with $(\cdot ,\cdot )_{HS}$. Because it holds that 
	\begin{align*}
		\mathcal{E}(\Omega_{\mathrm{QT}})
		&=\{E\in\LL_{S}(\HH)\mid0\le\Tr[E\rho]\le1,\ ^{\forall}\rho\in\Omega_{\mathrm{QT}}\}\\
		&=\{E\in\LL_{S}(\HH)\mid0\le E\le\id_{N}\},
	\end{align*}
	we can see $V_{+}=V^{*int}_{+(\cdot ,\cdot )_{HS}}=\{A\in\LL_{S}(\HH)\mid A\ge0\}$, and rank-1 projections are pure and indecomposable effects in quantum theories.
	We note that while higher dimensional classical theories are represented by simplices as shown in the previous example, higher dimensional quantum theories have more complicated structures \cite{KIMURA2003339, Bengtsson2013}:
	we cannot represent them with higher dimensional balls just generalizing the three dimensional ball for the qubit case (the Bloch ball).

	On the other hand, it is known that in quantum theory any state automorphism is either a unitary or antiunitary transformation \cite{Busch_quantummeasurement}, and for any pair of pure states one can find a unitary operator that links them. Thus, $\Omega_{\mathrm{QT}}$ is transitive, and any state automorphism is of the form
	\[
	\rho\mapsto U\rho U^{\dagger}\quad\ ^{\forall}\rho\in\Omega_{\mathrm{QT}},
	\]
	where $U$ is unitary or antiunitary. Considering that 
	\begin{align*}
		(UXU^{\dagger}, UYU^{\dagger})_{HS}
		&=\Tr\left[UX^{\dagger}U^{\dagger}UYU^{\dagger}
		\right]\\
		&=\Tr[X^{\dagger}Y]\\
		&=(X,Y)_{HS}
	\end{align*}
	holds for any unitary or antiunitary operator $U$, we can obtain in a similar way to \eqref{eg_eq_CT_inner product}
	\begin{align}
		\label{eg_eq_QT_inner product}
		\langle X, Y\rangle_{GL(\Omega_{\mathrm{QT}})}=(X, Y)_{HS}.
	\end{align}
	Therefore, we can conclude $V_{+}=V^{*int}_{+(\cdot ,\cdot )_{HS}}=V^{*int}_{+\langle \cdot ,\cdot \rangle_{GL(\Omega_{\mathrm{QT}})}}$. We remark similarly to the classical cases that we may rewrite \eqref{eq_Bloch repr2} as \eqref{eq_vec repr} by taking a suitable normalization and considering that $\omega_{M}=\id_{N}/N$.

\subsection{Regular polygon theories}
	\label{eg_polygon}
	If the state space of a GPT is in the shape of a regular polygon with $n(\ge3)$ sides, then we call it a {\it regular polygon theory} and denote the state space by $\Omega_{n}$. 
	We set $V=\R^{3}$ when considering regular polygon theories, and it can be seen in \cite{1367-2630-13-6-063024} that the pure states of $\Omega_{n}$ are described as
	\[
	\Omega^{\mathrm{ext}}_{n}=\{\omega^{n}_{i}\}_{i=0}
	^{n-1}
	\]
	with
	\begin{align}
		\label{def_polygon pure state}
		\omega^{n}_{i}=
		\left(
		\begin{array}{c}
			r_{n}\cos({\frac{2\pi i}{n}})\\
			r_{n}\sin({\frac{2\pi i}{n}})\\
			1
		\end{array}
		\right),\ \ &r_{n}=\sqrt{\frac{1}{\cos({\frac{\pi}{n}})}}
	\end{align}
	when $n$ is finite, and when $n=\infty$ (the state space $\Omega_{\infty}$ is a disc), 
	\[
	\Omega_{\infty}^{\mathrm{ext}}=\{\omega^{\infty}_{\theta}\}_{\theta\in[0, 2\pi)}
	\]
	with
	\begin{align}
		\label{def_disc pure state}
		\omega^{\infty}_{\theta}=
		\left(
		\begin{array}{c}
			\cos\theta\\
			\sin\theta\\
			1
		\end{array}
		\right).
	\end{align}
	The state space $\Omega_{3}$ represents a classical trit system (the 2-dimensional standard simplex), while $\Omega_{\infty}$ represents a qubit system with real coefficients (the unit disc can be considered as an equatorial plane of the Bloch ball). 
	Regular polygon theories can be regarded as intermediate theories of those theories.
	
	The state space of the regular polygon theory with $n$ sides (including $n=\infty$) defines its positive cone $V_{+}$, and it is also shown in \cite{1367-2630-13-6-063024} that the corresponding internal dual cone $V^{*int}_{+(\cdot ,\cdot )_{E}}\subset\R^{3}$ is given by the conic hull
	of the following extreme effects (in fact, those effects are also indecomposable)
	\begin{equation}
		\label{def_polygon pure effect}
		\begin{aligned}
			&e^{n}_{i}=\frac{1}{2}
			\left(
			\begin{array}{c}
				r_{n}\cos({\frac{(2i-1)\pi}{n}})\\
				r_{n}\sin({\frac{(2i-1)\pi}{n}})\\
				1
			\end{array}
			\right),\ \ i=0, 1, \cdots, n-1\ \ (n:\mbox{even})\ ;\\
			&e^{n}_{i}=\frac{1}{1+r_{n}^{2}}
			\left(
			\begin{array}{c}
				r_{n}\cos({\frac{2i\pi}{n}})\\
				r_{n}\sin({\frac{2i\pi}{n}})\\
				1
			\end{array}
			\right),\ \ i=0, 1, \cdots, n-1\ \ (n:\mbox{odd})\ ;\\
			&e^{\infty}_{\theta}=\frac{1}{2}
			\left(
			\begin{array}{c}
				\cos\theta\\
				\sin\theta\\
				1
			\end{array}
			\right),\ \ \theta\in[0, 2\pi)\ \ \ (n=\infty).
		\end{aligned}
	\end{equation}
	Moreover, for finite $n$, we can see that the group $GL(\Omega_{n})$ (named the {\it dihedral group}) is composed of orthogonal transformations with respect to $(\cdot ,\cdot )_{E}$ \cite{Dummit_abstractalgebra}, which also holds for $n=\infty$. Similar calculations to \eqref{eg_eq_CT_inner product} or \eqref{eg_eq_QT_inner product} demonstrate $(\cdot ,\cdot )_{E}=\langle\cdot ,\cdot \rangle_{GL(\Omega_{n})}$ for $n=3, 4, \cdots, \infty$. Therefore, from \eqref{def_polygon pure state} - \eqref{def_polygon pure effect}, we can conclude that $V_{+}$ is self-dual, i.e. $V_{+}=V^{*int}_{+(\cdot ,\cdot )_{E}}=V^{*int}_{+\langle \cdot ,\cdot \rangle_{GL(\Omega_{n})}}$, when $n$ is odd or $\infty$, while $V_{+}$ is not identical but only isomorphic to $V^{*int}_{+\langle \cdot ,\cdot \rangle_{GL(\Omega_{n})}}$ when $n$ is even (in that case, $V_{+}$ is called {\it weakly self-dual} \cite{barnum2012teleportation,1367-2630-13-6-063024}).

Among regular polygon theories, the square theory described by the state space $\Omega_4$ is physically of particular importance, and is often called a {\itshape gbit} (generalized bit) system \cite{PhysRevA.75.032304}.
It can be observed that the so-called PR-box \cite{PR-box} is represented by a pure entangled state of the composite system $\Omega_4\otimes_{max}\Omega_4$ \cite{PhysRevA.75.032304}, and thus can violate the CHSH inequality maximally in the sense that it attains the value 4 for that entangled state \cite{1367-2630-13-6-063024}.
The square theory is also known for its interesting behavior on incompatibility.
It was demonstrated in \cite{PhysRevA.96.022113} that a pair of two-outcome observables for $\Omega_4$ exhibits maximal incompatibility, which means that we need maximal noise to make them compatible (see also Example \ref{eg_deg of inc}).

%% file: chap3.tex
\chapter{Preparation uncertainty implies measurement uncertainty in a class of GPTs}
\label{chap:URs in GPT}
Since it was propounded by Heisenberg \cite{Heisenberg1927}, the existence of uncertainty relations, which is not observed in classical theory, has been regarded as one of the most significant features of quantum theory. 
The importance of uncertainty relations lies not only in their conceptual aspects but also in practical use such as the security proof of quantum key distribution \cite{BB84,Koashi_2006}. 
There have been researches to capture and formulate the notion of ``uncertainty'' in several ways. 
One of the most outstanding works was given by Robertson \cite{PhysRev.34.163}. 
There was shown an uncertainty relation in terms of standard derivation which stated that the probability distributions obtained by the measurements of a pair of noncommutative observables cannot be simultaneously sharp.
While this type of uncertainty (called {\it preparation uncertainty}) has been studied also in a more direct way \cite{Uffink_PhD,PhysRevA.71.052325,PhysRevA.76.062108} or the entropic way \cite{Hirschman1957_entropy,Beckner1975_entropy,UncertaintyRelationsForInformationEntropy,PhysRevLett.50.631,PhysRevLett.60.1103,10.2307/25051432}, another type of uncertainty called {\it measurement uncertainty} 
is known to exist in quantum theory \cite{Busch_quantummeasurement}. 
It describes that when we consider measuring jointly a pair of noncommutative observables, there must exist {\it measurement error} for the joint measurement, that is, we can only conduct their {\it approximate joint measurement}. 
There have been researches on measurement uncertainty with measurement error formulated in terms of standard derivation \cite{PhysRevLett.60.2447,doi:10.1002/j.1538-7305.1965.tb01684.x,PhysRevA.67.042105} or entropy \cite{PhysRevLett.112.050401}. 
Their measurement uncertainty relations were proven by using preparation uncertainty relations. 
It implies that there may be a close connection between those two kinds of uncertainty. 
From this perspective, in \cite{doi:10.1063/1.3614503}, simple inequalities were proven which demonstrate in a more explicit way than other previous studies that preparation uncertainty indicates measurement uncertainty and the bound derived from the former also bounds the latter. 
The main results of \cite{doi:10.1063/1.3614503} were obtained with preparation uncertainty quantified by overall widths and minimum localization error, and measurement uncertainty by error bar widths, Werner's measure, and $l_{\infty}$ distance \cite{doi:10.1063/1.2759831,PhysRevA.78.052119,10.5555/2017011.2017020,10.5555/2011593.2011606}. 
Concerning about uncertainty, both preparation and measurement uncertainty can be introduced naturally also in GPTs.
For example, both types of uncertainty for GPTs analogical with a qubit system were investigated in \cite{PhysRevA.101.052104}, and there are also researches on joint measurability of observables \cite{PhysRevA.94.042108,PhysRevLett.103.230402,Busch_2013,PhysRevA.89.022123,PhysRevA.98.012133}, which are related with measurement uncertainty, in GPTs.
It is of interest to give further research on how two types of generalized uncertainty are related with each other.

In this part, we study the relations between two kinds of uncertainty in GPTs. 
We focus on a class of GPTs that are transitive and self-dual including finite-dimensional classical and quantum theories, and demonstrate similar results to \cite{doi:10.1063/1.3614503} in the GPTs: preparation uncertainty relations indicate measurement uncertainty relations.
More precisely, it is proven in a certain class of GPTs that if a preparation uncertainty relation gives some bound, then it is also a bound on the corresponding measurement uncertainty relation with the quantifications of uncertainty in \cite{doi:10.1063/1.3614503} generalized to GPTs. 
We also prove its entropic expression by generalizing the quantum results in \cite{PhysRevLett.112.050401} to those GPTs.
Our results manifest that the close connections between two kinds of uncertainty exhibited in quantum theory are more universal ones.
We also present, as an illustration, concrete expressions of our uncertainty relations in regular polygon theories.

This part is organized as follows. 
In Section \ref{4sec:URs in GPTs}, we introduce measures that quantify the width of a probability distribution. 
These measures are used for considering whether it is possible to localize jointly two probability distributions obtained by two kinds of measurement on one certain state, that is, they are used for describing preparation uncertainty. We also introduce measures quantifying measurement error by means of which we can formulate measurement uncertainty resulting from approximate joint measurements of two incompatible observables.
After the introductions of those quantifications, we present the main theorems and their proofs. 
In Section \ref{4sec:entropic URs}, we demonstrate that similar contents of those theorems can be also expressed in an entropic way.
In Section \ref{4sec:eg UR}, we investigate uncertainty relations in regular polygon theories.

\section{Preparation uncertainty and measurement uncertainty in GPTs}
\label{4sec:URs in GPTs}
In this section, our main results on the relations between preparation uncertainty and measurement uncertainty are given in GPTs with transitivity and self-duality with respect to $\langle\cdot ,\cdot \rangle_{GL(\Omega)}$ (see Section \ref{2sec:additional}). 
Measures quantifying the width of a probability distribution or measurement error are also given to describe those results. 
Throughout this section, we consider observables whose sample spaces are finite metric spaces.
\subsection{Widths of probability distributions}
\label{sec3-1}
In this subsection, we give two kinds of measure to quantify how concentrated a probability distribution is. 

Let $A$ be a finite metric space equipped with a metric function $d_{A}$, and $O_{d_{A}}(a;\,w)$ be the ball defined by $O_{d_{A}}(a;\,w):=\{x\in A\mid d_{A}(x, a)\le w/2\}$. For $\epsilon\in[0,1]$ and a probability distribution $\mathbf{p}$ on $A$, we define the {\it overall width} (at confidence level $1-\epsilon$) \cite{doi:10.1063/1.3614503,doi:10.1063/1.2759831} as
\begin{equation}
\label{def_overall width original}
W_{\epsilon}(\mathbf{p}):=\inf\{w>0\mid\exists a\in A : \mathbf{p}(O_{d_{A}}(a;\,w))\ge1-\epsilon\}.
\end{equation}
We can give another formulation for the width of $\mathbf{p}$. We define the {\it minimum localization error} \cite{doi:10.1063/1.3614503} of $\mathbf{p}$ as
\begin{equation}
\label{def_localization error original}
LE(\mathbf{p}):=1-\underset{a\in A}{\max}\ p(a).
\end{equation}
Both \eqref{def_overall width original} and \eqref{def_localization error original} can be applied to probability distributions observed in physical experiments. Let us consider a GPT with $\Omega$ its state space. 
For a state $\omega\in\Omega$ and an observable $F=\{f_{a}\}_{a\in A}$ on $A$, we denote by $\omega^{F}$ the probability distribution obtained by the measurements of $F$ on $\omega$, i.e.
\[
\omega^{F}:=\{f_{a}(\omega)\}_{a\in A}.
\]
The overall width and minimum localization error for $\omega^{F}$ can be defined as
\begin{equation}
\label{def_overall width}
W_{\epsilon}(\omega^{F}):=\inf\{w>0\mid\exists a\in A : \sum_{a'\in O_{d_{A}}(a;\,w)}f_{a'}(\omega)\ge1-\epsilon\}
\end{equation}
and
\begin{equation}
\label{def_localization error}
LE(\omega^{F}):=1-\underset{a\in A}{\max}\ f_{a}(\omega)
\end{equation}
respectively. Note that as in \cite{doi:10.1063/1.3614503,doi:10.1063/1.2759831}, overall widths can be defined properly even if the sample spaces of probability distributions are infinite. For example, overall widths are considered in \cite{doi:10.1063/1.2759831} for probability measures on $\R$ derived from the measurement of position or momentum of a particle.

Those two measures above are used for the mathematical description of {\it preparation uncertainty relations (PURs)}. As a simple example, we consider a qubit system with Hilbert space $\HH=\C^{2}$. For two projection-valued measures (PVMs) $Z=\{\ketbra{0}{0}, \ketbra{1}{1}\}$ and $X=\{\ketbra{+}{+}, \ketbra{-}{-}\}$, where $\{\ket{0}, \ket{1}\}$ and $\{\ket{+}, \ket{-}\}=\{\frac{1}{\sqrt{2}}(\ket{0}+\ket{1}), \frac{1}{\sqrt{2}}(\ket{0}-\ket{1})\}$ are the $z$-basis and $x$-basis of $\HH$ respectively, it holds from \cite{PhysRevA.71.052325,PhysRevLett.60.1103} that 
\begin{equation}
\label{eq_landau pollak}
LE(\rho^{Z})+LE(\rho^{X})\ge1-\frac{1}{\sqrt{2}}>0
\end{equation}
for any state $\rho$ (see also \eqref{eq:quantum LP}).
The inequality \eqref{eq_landau pollak} shows that there is no state $\rho$ which makes both $LE(\rho^{Z})$ and $LE(\rho^{X})$ zero, that is, $\rho^{Z}$ and $\rho^{X}$ cannot be localized simultaneously even if the observables are ideal ones (PVMs). PURs in terms of overall widths were also discussed in \cite{doi:10.1063/1.2759831} for the position and momentum observables.

\subsection{Measurement error}
\label{sec3-2}
In this part, we introduce the concept of measurement error in GPTs, which derives from joint measurement problems, and describe how to quantify it. 

Let us consider a GPT with its state space $\Omega$, and two observables $F=\{f_{a}\}_{a\in A}$ and $G=\{g_{b}\}_{b\in B}$ on $\Omega$. 
Although general descriptions of (in)compatibility was already given in Subsection \ref{2subsec:compatibility}, here we show the definition again.
We call $F$ and $G$ are {\it compatible} or {\it jointly measurable} if there exists a {\it joint observable} $M^{FG}=\{m^{FG}_{ab}\}_{(a, b)\in A\times B}$ of $F$ and $G$ satisfying 
\begin{align*}
&\sum_{b\in B}m^{FG}_{ab}=f_{a}\ \ \mbox{for all $a\in A$},\\
&\sum_{a\in A}m^{FG}_{ab}=g_{b}\ \ \mbox{for all $b\in B$},
\end{align*}
and if $F$ and $G$ are not jointly measurable, then they are called {\it incompatible} \cite{Heinosaari_2016,Busch_2013}. 
As was mentioned in Subsection \ref{2subsec:compatibility}, 
there exist pairs of observables that are incompatible in all non-classical GPTs, but we can nevertheless conduct their {\it approximate joint measurements} allowing {\it measurement error}. 
Assume that $F$ and $G$ are incompatible.
It is known that one way to compose their approximate joint measurement is adding some trivial noise to them. 
To see this, we consider as a simple example the incompatible pair of observables $Z=\{\ketbra{0}{0}, \ketbra{1}{1}\}$ and $X=\{\ketbra{+}{+}, \ketbra{-}{-}\}$ in a qubit system described in the last subsection. 
It was demonstrated in \cite{PhysRevA.87.052125} that the observables
\begin{equation}
\label{def_appro Z and X}
\begin{aligned}
\widetilde{Z}^{\lambda}:&=\lambda Z+(1-\lambda)I\\
&=\left\{\lambda\ketbra{0}{0}+\frac{1-\lambda}{2}\id_{2},\  \lambda\ketbra{1}{1}+\frac{1-\lambda}{2}\id_{2}\right\},\\
\widetilde{X}^{\lambda}:&=\lambda X+(1-\lambda)I,\\
&=\left\{\lambda\ketbra{+}{+}+\frac{1-\lambda}{2}\id_{2},\  \lambda\ketbra{-}{-}+\frac{1-\lambda}{2}\id_{2}\right\}
\end{aligned}
\end{equation}
are jointly measurable for $0\le\lambda\le\frac{1}{\sqrt{2}}$, where $I:=\{\id_{2}/2, \id_{2}/2\}$ with $\id_2$ the identity operator on $\HH=\C^2$ is a trivial observable.
The joint measurablity of \eqref{def_appro Z and X} implies that the addition of trivial noise described by a trivial observable makes incompatible observables compatible in an approximate way. 
In fact, it is observed also in GPTs that adding trivial noise results in approximate joint measurements of incompatible observables \cite{Busch_2013,PhysRevA.89.022123,PhysRevA.87.052125}. 

Because the notion of measurement error derives from the difference between ideal and approximate observables as discussed above, we have to define ideal observables in GPTs to quantify measurement error. 
In this chapter, they are defined in an analogical way with the ones in finite-dimensional quantum theories, where PVMs are considered to be ideal \cite{Busch_quantummeasurement}. If we denote a PVM by $E=\{P_{a}\}_{a}$, then each effect is of the form 
\[
P_{a}=\sum_{i_{(a)}}\ketbra{\psi_{i_{(a)}}}{\psi_{i_{(a)}}}.
\]
In particular, every effect is a sum of pure and indecomposable effects, and we call in a similar way an observable $F=\{f_{a}\}_{a\in A}$ on $\Omega$ {\itshape ideal} if each effect $f_{a}$ satisfies
\begin{equation}
\label{def of ideal meas}
f_{a}=\sum_{i_{(a)}}e_{i_{(a)}}^{\mathrm{ext}},\quad\mbox{or}\quad f_{a}=u-\sum_{i_{(a)}}e_{i_{(a)}}^{\mathrm{ext}},
\end{equation}
where we should recall that the set of all pure and indecomposable effects is denoted by $\{e_{i}^{\mathrm{ext}}\}_{i}$ and we do not consider the trivial observable $F=\{u\}$. 
It is easy to see that observables defined as \eqref{def of ideal meas} result in PVMs in finite-dimensional quantum theories. This type of observable was considered also in \cite{Barnum_2014}.

The introduction of ideal observables makes it possible for us to quantify measurement error. 
Consider an ideal observable $F=\{f_{a}\}_{a}$ and a general observable $\widetilde{F}=\{\widetilde{f}_{a}\}_{a}$, and suppose similarly to the previous subsection that $A$ is a finite metric space with a metric $d_{A}$. 
$F$ may be understood as the measurement intended to be measured, while $\widetilde{F}$ as a measurement conducted actually.
Taking into consideration the fact that for each nonzero pure effect there exists at least one state which is mapped to 1 (an ``eigenstate'' \cite{KIMURA2010175}), we can define for $\epsilon\in[0,1]$ the {\it error bar width} of $\widetilde{F}$ relative to $F$ \cite{doi:10.1063/1.3614503,doi:10.1063/1.2759831} as
\begin{equation}
\label{def_error bar width}
\begin{aligned}
\mathcal{W}_{\epsilon}(\widetilde{F}, F)
&=\inf\{w>0\mid\ ^{\forall} a\in A, \ ^{\forall} \omega\in\Omega :\\ 
&\qquad\qquad\qquad\qquad
f_{a}(\omega)=1\Rightarrow\sum_{a'\in O_{d_{A}}(a;\,w)}\widetilde{f}_{a'}(\omega)\ge1-\epsilon\}.
\end{aligned}
\end{equation}
$\mathcal{W}_{\epsilon}(\widetilde{F}, F)$ represents the spread of probabilities around the ``eigenvalues'' of $F$ observed when the corresponding ``eigenstates'' of $F$ are measured by $\widetilde{F}$, and thus it can be thought to be one of the quantifications of measurement error. Note that although error bar widths in general (not necessarily finite) metric spaces were defined in \cite{doi:10.1063/1.2759831}, we consider only finite metric spaces in this chapter, so we employ their convenient forms \eqref{def_error bar width} in finite metric spaces shown in \cite{doi:10.1063/1.3614503}. Another measure is the one given by Werner \cite{10.5555/2011593.2011606} as the difference of expectation values of ``slowly varying functions'' on the probability distributions obtained when $F$ and $\widetilde{F}$ are measured. It is defined as
\begin{equation}
\label{def_Werner's measure}
D_{W}(\widetilde{F}, F):=\underset{\omega\in\Omega}{\sup}\ \underset{h\in\Lambda}{\sup}
\left|(\tilde{F}[h])(\omega)-(F[h])(\omega)\right|,
\end{equation}
where
\[
\Lambda:=\{h\colon A\to\R\mid |h(a_{1})-h(a_{2})|\le d_{A}(a_{1}, a_{2}),\ ^{\forall} a_{1}, a_{2}\in A\}
\]
is the set of all ``slowly varying functions'' (called the {\it Lipshitz ball} of $(A, d_{A})$) and
\[
F[h]:=\sum_{a\in A}h(a)f_{a}
\]
is a map which gives the expectation value of $h\in\Lambda$ when $F$ is measured on a state $\omega$ (similarly for $\widetilde{F}[h]$). 
There is known a simple relation between \eqref{def_error bar width} and \eqref{def_Werner's measure}.
\begin{prop}[{\rm \cite{doi:10.1063/1.3614503,doi:10.1063/1.2759831}}]
	\label{prop_error bar and Werner}
	Let $(A,d_{A})$ be a finite metric space, and $F=\{f_{a}\}_{a\in A}$ and $\widetilde{F}=\{\widetilde{f}_{a}\}_{a\in A}$ be an ideal and general observable respectively. 
	Then
	\[
	\mathcal{W}_{\epsilon}(\widetilde{F}, F)\le\frac{2}{\epsilon}D_{W}(\tilde{F}, F)
	\]
	holds for $\epsilon\in(0, 1].$
\end{prop}
\begin{pf}
	Let us define $n:=\frac{D_{W}(\tilde{F}, F)}{\epsilon}$ for $\epsilon\in(0, 1]$, and consider for $a\in A$ a state $\omega\in\Omega$ satisfying $f_{a}(\omega)=1$. 
	Remember that such state does exist for every $a\in A$ because $F$ is ideal.
	We also define a function $h_{n}$ on $A$ as
	\begin{align*}
	h_{n}(x):=
	\left\{
	\begin{aligned}
	&n-d_{A}(x, a) &&(d(x, a)\le n)\\
	&0 &&(d(x, a)>n).
	\end{aligned}
	\right.
	\end{align*}
	It can be seen that
	\[
	|h_{n}(x_{1})-h_{n}(x_{2})|\le d_{A}(x_{1}, x_{2})
	\]
	holds for $x_{1}, x_{2}\in A$, and thus we can obtain from the definition of $D_{W}(\tilde{F}, F)$ \eqref{def_Werner's measure}
	\[
	\left|(\tilde{F}[h_{n}])(\omega)-(F[h_{n}])(\omega)\right|\le D_{W}(\tilde{F}, F).
	\]
	It results in 
	\begin{equation}
	\label{eq_appC}
	\left|(\tilde{F}[g_{n}])(\omega)-(F[g_{n}])(\omega)\right|\le \frac{D_{W}(\tilde{F}, F)}{n}=\epsilon,
	\end{equation}
	where we set $g_{n}:=h_{n}/n$.
	Since it holds that $g_{n}(x)\le\chi_{O_{d_{A}}(a;\ 2n)}(x)\le 1$ for all $x\in A$, where $\chi_{O_{d_{A}}(a;\ 2n)}$ is the indicator function of the ball $O_{d_{A}}(a;\ 2n)=\{x\in A\mid d_{A}(x, a)\le n\}$, and 
	\[
	(F[g_{n}])(\omega)=\sum_{x\in A} g_{n}(x)f_{x}(\omega)=g_{n}(a)f_{a}(\omega)=1
	\]
	because $f_{a}(\omega)=1$, \eqref{eq_appC} can be rewritten as
	\[
	1-(\tilde{F}[\chi_{O_{d_{A}}(a;\ 2n)}])(\omega)\le \epsilon,
	\]
	that is,
	\begin{equation}
	\label{eq_appC1}
	\sum_{x\in O_{d_{A}}(a;\ 2n)} \widetilde{f}_{x}(\omega)\ge 1-\epsilon.
	\end{equation}
	\eqref{eq_appC1} holds for all $a\in A$ and all $\omega\in \Omega$ such that $f_{a}(\omega)=1$, and thus 
	\[
	2n=\frac{2}{\epsilon}D_{W}(\tilde{F}, F)\ge\mathcal{W}_{\epsilon}(\widetilde{F}, F)
	\] 
	is concluded (see the definition of $\mathcal{W}_{\epsilon}(\widetilde{F}, F)$ \eqref{def_error bar width}).
	\qed
\end{pf}
On the other hand, there can be introduced a more intuitive quantification of measurement error called {\it $l_{\infty}$ distance} \cite{10.5555/2017011.2017020}:
\begin{equation}
\label{def_l distance}
D_{\infty}(\widetilde{F}, F):=\underset{\omega\in\Omega}{\sup}\ \underset{a\in A}{\max}\left|\widetilde{f}_{a}(\omega)-f_{a}(\omega)\right|.
\end{equation}

By means of those quantifications of measurement error above, we can formulate {\it measurement uncertainty relations (MURs)}. As an illustration, we again consider the joint measurement problem of incompatible observables $Z$ and $X$ in a qubit system. Suppose that $\widetilde{M}^{ZX}$ is an approximate joint observable of $Z$ and $X$, and $\widetilde{M}^{Z}$ and $\widetilde{M}^{X}$ are its marginal observables corresponding to $Z$ and $X$ respectively. It was proven in \cite{10.5555/2017011.2017020} that 
\begin{equation}
\label{eq_MUR_ZX}
D_{\infty}(\widetilde{M}^{Z}, Z)+D_{\infty}(\widetilde{M}^{X}, X)\ge 1-\frac{1}{\sqrt{2}}>0.
\end{equation}
\eqref{eq_MUR_ZX} gives a quantitative representation of the incompatibility of $Z$ and $X$ that $D_{\infty}(\widetilde{M}^{Z}, Z)$ and $D_{\infty}(\widetilde{M}^{X}, X)$ cannot be simultaneously zero, that is, measurement error must occur when conducting any approximate joint measurement of $Z$ and $X$ (see \cite{PhysRevA.78.052119} for another inequality). MURs for the position and momentum observables were given in \cite{doi:10.1063/1.2759831} and \cite{10.5555/2011593.2011606} in terms of \eqref{def_error bar width} and \eqref{def_Werner's measure} respectively.

\subsection{Relations between preparation uncertainty and measurement uncertainty in a class of GPTs}
In the previous subsections, we have introduced several measures to review two kinds of uncertainty, preparation uncertainty and measurement uncertainty. 
In this part, we shall manifest as our main results how they are related with each other in GPTs, which is a generalization of the quantum ones in \cite{doi:10.1063/1.3614503}.

Before demonstrating our main theorems, we confirm the physical settings and mathematical assumptions to state them. In the following, we focus on a GPT with a state space $\Omega$, and suppose that $\Omega$ is transitive and the positive cone $V_{+}$ is self-dual with respect to $\langle\cdot ,\cdot \rangle_{GL(\Omega)}$ (see Section \ref{2sec:additional}).
While our assumptions may seem curious, it can be observed in \cite{PhysRevLett.108.130401} that those two conditions are satisfied simultaneously if the state space is {\it bit-symmetric}. There are also researches where they are derived from certain conditions possible to be interpreted physically \cite{Barnum_2014,1367-2630-19-4-043025}. 
In addition, we consider ideal observables $F=\{f_{a}\}_{a\in A}$ and $G=\{g_{b}\}_{b\in B}$ on $\Omega$, whose sample spaces are finite metric spaces $(A, d_{A})$ and $(B, d_{B})$ respectively, and consider an observable $\widetilde{M}^{FG}:=\{\widetilde{m}_{ab}^{FG}\}_{(a, b)\in A\times B}
$ as an approximate joint observable of $F$ and $G$, whose marginal observables are given by
\[
\begin{aligned}
&\widetilde{M}^{F}:=\{\widetilde{m}_{a}^{F}\}_{a},\quad \widetilde{m}_{a}^{F}:=\sum_{b\in B}\widetilde{m}_{ab}^{FG};\\
&\widetilde{M}^{G}:=\{\widetilde{m}_{b}^{G}\}_{b},\quad \widetilde{m}_{b}^{G}:=\sum_{a\in A}\widetilde{m}_{ab}^{FG}.
\end{aligned}
\]
Remember that, as shown in Subsection \ref{sec3-2}, the ideal observable $F=\{f_{a}\}_{a}$ satisfies
\begin{equation}
\label{eq_general PVM}
f_{a}=\sum_{i_{(a)}}e_{i_{(a)}}^{\mathrm{ext}},\quad\mbox{or}\quad f_{a}=u-\sum_{i_{(a)}}e_{i_{(a)}}^{\mathrm{ext}}
\end{equation}
in terms of the pure and indecomposable effects $\{e_{i}^{\mathrm{ext}}\}_{i}$ shown in \eqref{def_indecomp effect} (similarly for $G=\{g_{b}\}_{b}$).
The following lemmas are needed to prove our main results.
\begin{lem}
	\label{lem_u=omegaM}
	If $\Omega$ is transitive, then the unit effect $u\in V^{*int}_{+\langle\cdot ,\cdot \rangle_{GL(\Omega)}}\subset V$ is identical to the maximally mixed state $\omega_{M}$, i.e. $u=\omega_{M}$.
\end{lem}
\begin{pf}
	It is an easy consequence of Proposition \ref{prop_ortho repr}. 
	In fact, \eqref{eq_vec repr} gives
	\[
	u=\omega_{M}=
	\left(
	\begin{array}{c}
	\boldsymbol{0} \\
	1
	\end{array}
	\right).
	\]\qed
\end{pf}

\begin{lem}
	\label{lem_eigenstates}
	If $\Omega$ is a transitive state space and its positive cone $V_{+}$ is self-dual with respect to $\langle\cdot,\cdot\rangle_{GL(\Omega)}$, 
	then for any effect $e\in \mathcal{E}(\Omega)$ on $\Omega$ it holds that 
	\begin{equation}
	\label{eq_eigenstate0}
	\frac{e}{\langle u, e\rangle}_{GL(\Omega)}\in\Omega,
	\end{equation}
	and for any ideal observable $F=\{f_{a}\}_{a\in A}$ on $\Omega$ it holds that
	\begin{equation}
	\label{eq_eigenstate}
	\left\langle f_{a},\ \frac{f_{a}}{\langle u, f_{a}\rangle_{GL(\Omega)}}\right\rangle_{GL(\Omega)}=1
	\end{equation}
	for all $a\in A$. In particular, each $f_{a}/\langle u, f_{a}\rangle_{GL(\Omega)}$ is an ``eigenstate'' of $F$.
\end{lem}
\begin{pf}
	In this proof, we denote the inner product $\langle\cdot ,\cdot \rangle_{GL(\Omega)}$ and the norm $\|\cdot\|_{GL(\Omega)}$ simply by $\langle\cdot ,\cdot \rangle$ and $\|\cdot\|$ respectively.

	For any element $e\in V^{*\mathrm{int}}_{+\langle\cdot ,\cdot \rangle}$, the vector $e/\langle u, e\rangle$ defines a state
	because $\left\langle u,\ e/\langle u, e\rangle\right\rangle=1$ and $e\in V_{+}$ due to the the self-duality: $V_{+}=V^{*\mathrm{int}}_{+\langle\cdot ,\cdot \rangle}$, which proves \eqref{eq_eigenstate0}.
	To prove \eqref{eq_eigenstate}, we focus on the fact that $f_{a}$ in \eqref{eq_general PVM} is an effect (thus $u-f_{a}$ is also an effect), that is, $\sum_{i_{(a)}}e_{i_{(a)}}^{\mathrm{ext}}$ is an effect and it satisfies $0\le\langle\sum_{i_{(a)}}e_{i_{(a)}}^{\mathrm{ext}},\ \omega\rangle\le1$ for any state $\omega\in\Omega$. However, if we act $\sum_{i_{(a)}}e_{i_{(a)}}^{\mathrm{ext}}$ on the pure state $\omega_{j_{(a)}}^{\mathrm{ext}}$, then \eqref{def_indecomp effect} shows that $\langle e_{j_{(a)}}^{\mathrm{ext}},\ \omega_{j_{(a)}}^{\mathrm{ext}}\rangle=1$, and thus we have
	\[
	\langle e_{i_{(a)}}^{\mathrm{ext}},\ \omega_{j_{(a)}}^{\mathrm{ext}}\rangle=0\quad\mbox{for $i_{(a)}\neq j_{(a)}$},
	\]
	that is,
	\begin{equation}
	\label{eq_prf lemma}
	\langle e_{i_{(a)}}^{\mathrm{ext}},\ e_{j_{(a)}}^{\mathrm{ext}}\rangle=0\quad\mbox{for $i_{(a)}\neq j_{(a)}$}.
	\end{equation}
	Because
	\begin{align*}
	\langle e_{i_{(a)}}^{\mathrm{ext}},\ e_{i_{(a)}}^{\mathrm{ext}}\rangle=\frac{1}{\|\omega_{0}^{\mathrm{ext}}\|^{2}}\quad\mbox{and}\quad
	\langle u,\ e_{i_{(a)}}^{\mathrm{ext}}\rangle=\frac{1}{\|\omega_{0}^{\mathrm{ext}}\|^{2}}
	\end{align*}
	hold from \eqref{def_indecomp effect}, we obtain together with \eqref{eq_prf lemma}
	\begin{equation}
	\label{eq_values of indecomp effects}
	\begin{aligned}
	\langle\sum_{i_{(a)}}e_{i_{(a)}}^{\mathrm{ext}},\ \sum_{i_{(a)}}e_{i_{(a)}}^{\mathrm{ext}}\rangle=\frac{(\#i_{(a)})}{\|\omega_{0}^{\mathrm{ext}}\|^{2}},&\quad
	\langle u,\ \sum_{i_{(a)}}e_{i_{(a)}}^{\mathrm{ext}}\rangle=\frac{(\#i_{(a)})}{\|\omega_{0}^{\mathrm{ext}}\|^{2}},\\
	\langle u-\sum_{i_{(a)}}e_{i_{(a)}}^{\mathrm{ext}},\ u-\sum_{i_{(a)}}e_{i_{(a)}}^{\mathrm{ext}}\rangle=1-\frac{(\#i_{(a)})}{\|\omega_{0}^{\mathrm{ext}}\|^{2}},&\quad
	\langle u,\ u-\sum_{i_{(a)}}e_{i_{(a)}}^{\mathrm{ext}}\rangle=1-\frac{(\#i_{(a)})}{\|\omega_{0}^{\mathrm{ext}}\|^{2}},
	\end{aligned}
	\end{equation}
	where $(\#i_{(a)})$ is the number of elements of the index set $\{i_{(a)}\}$ and we use $\langle u, u\rangle=\langle u, \omega_{M}\rangle=1$ (Lemma \ref{lem_u=omegaM}). Therefore, we can conclude that every effect $f_{a}=\sum_{i_{(a)}}e_{i_{(a)}}^{\mathrm{ext}}\ \mbox{or}\  u-\sum_{i_{(a)}}e_{i_{(a)}}^{\mathrm{ext}}$ composing $F$ satisfies
	\begin{equation*}
	\left\langle f_{a},\ \frac{f_{a}}{\langle u, f_{a}\rangle}\right\rangle=1.
	\end{equation*}
\qed
\end{pf}

Now, we can state our main theorems connecting PURs and MURs. 
Similar results to ours were proven \cite{doi:10.1063/1.3614503} for finite-dimensional quantum theories. Because GPTs shown above include those theories, our theorems can be considered to demonstrate that the relations between PURs and MURs introduced in \cite{doi:10.1063/1.3614503} are more general ones.
\begin{thm}
	\label{thm_error bar}
	Let $\Omega$ be a transitive state space and its positive cone $V_{+}$ be self-dual with respect to $\langle\cdot,\cdot\rangle_{GL(\Omega)}$, and let $(F, G)$ be a pair of ideal observables on $\Omega$.
	For an arbitrary approximate joint observable $\widetilde{M}^{FG}$ of $(F, G)$ and $\epsilon_{1}, \epsilon_{2}\in[0,1]$ satisfying $\epsilon_{1}+\epsilon_{2}\le1$, there exists a state $\omega\in\Omega$ such that
	\[
	\begin{aligned}
	&\mathcal{W}_{\epsilon_{1}}(\widetilde{M}^{F}, F)\ge W_{\epsilon_{1}+\epsilon_{2}}(\omega^{F}),\\
	&\mathcal{W}_{\epsilon_{2}}(\widetilde{M}^{G}, G)\ge W_{\epsilon_{1}+\epsilon_{2}}(\omega^{G}).
	\end{aligned}
	\]
\end{thm}
Theorem \ref{thm_error bar} manifests that if one cannot make both $W_{\epsilon_{1}+\epsilon_{2}}(\omega^{F})$ and $W_{\epsilon_{1}+\epsilon_{2}}(\omega^{G})$ vanish, then one also cannot make both $\mathcal{W}_{\epsilon_{1}}(\widetilde{M}^{F}, F)$ and $\mathcal{W}_{\epsilon_{2}}(\widetilde{M}^{G}, G)$ vanish. That is, if there exists a PUR, then there also exists a MUR. Moreover, Theorem \ref{thm_error bar} also demonstrates that bounds for MURs in terms of error bar widths can be given by ones for PURs described by overall widths.

\begin{pf}[Proof of Theorem \ref{thm_error bar}]
	In this proof, we denote again the inner product $\langle\cdot ,\cdot \rangle_{GL(\Omega)}$ and the norm $\|\cdot\|_{GL(\Omega)}$ simply by $\langle\cdot ,\cdot \rangle$ and $\|\cdot\|$ respectively.

	\if
	Since we assume that $f_{a}$ in \eqref{eq_general PVM} is an effect (thus $u-f_{a}$ is also an effect),  $\sum_{i_{(a)}}e_{i_{(a)}}^{\mathrm{ext}}$ is an effect and it satisfies $0\le\langle\sum_{i_{(a)}}e_{i_{(a)}}^{\mathrm{ext}},\ \omega\rangle\le1$ for any state $\omega\in\Omega$. However, if we act $\sum_{i_{(a)}}e_{i_{(a)}}^{\mathrm{ext}}$ on the pure state $\omega_{j_{(a)}}^{\mathrm{ext}}$, then \eqref{def_indecomp effect} shows that $\langle e_{j_{(a)}}^{\mathrm{ext}},\ \omega_{j_{(a)}}^{\mathrm{ext}}\rangle=1$, and thus we have
	\[
	\langle e_{i_{(a)}}^{\mathrm{ext}},\ \omega_{j_{(a)}}^{\mathrm{ext}}\rangle=0\quad\mbox{for $i_{(a)}\neq j_{(a)}$},
	\]
	that is,
	\begin{equation}
	\langle e_{i_{(a)}}^{\mathrm{ext}},\ e_{j_{(a)}}^{\mathrm{ext}}\rangle=0\quad\mbox{for $i_{(a)}\neq j_{(a)}$}.
	\end{equation}
	Because
	\begin{align*}
	\langle e_{i_{(a)}}^{\mathrm{ext}},\ e_{i_{(a)}}^{\mathrm{ext}}\rangle=\frac{1}{\|\omega_{0}^{\mathrm{ext}}\|^{2}}\quad\mbox{and}\quad
	\langle u,\ e_{i_{(a)}}^{\mathrm{ext}}\rangle=\frac{1}{\|\omega_{0}^{\mathrm{ext}}\|^{2}}
	\end{align*}
	hold from \eqref{def_indecomp effect}, we obtain
	\begin{equation}
	\begin{aligned}
	\langle\sum_{i_{(a)}}e_{i_{(a)}}^{\mathrm{ext}},\ \sum_{i_{(a)}}e_{i_{(a)}}^{\mathrm{ext}}\rangle&=\frac{(\#i_{(a)})}{\|\omega_{0}^{\mathrm{ext}}\|^{2}}\\
	\langle u,\ \sum_{i_{(a)}}e_{i_{(a)}}^{\mathrm{ext}}\rangle&=\frac{(\#i_{(a)})}{\|\omega_{0}^{\mathrm{ext}}\|^{2}}\\
	\langle u-\sum_{i_{(a)}}e_{i_{(a)}}^{\mathrm{ext}},\ u-\sum_{i_{(a)}}e_{i_{(a)}}^{\mathrm{ext}}\rangle&=1-\frac{(\#i_{(a)})}{\|\omega_{0}^{\mathrm{ext}}\|^{2}}\\
	\langle u,\ u-\sum_{i_{(a)}}e_{i_{(a)}}^{\mathrm{ext}}\rangle&=1-\frac{(\#i_{(a)})}{\|\omega_{0}^{\mathrm{ext}}\|^{2}},
	\end{aligned}
	\end{equation}
	where $(\#i_{(a)})$ is the number of elements of the index set $\{i_{(a)}\}$ and we use $\langle u, u\rangle=\langle u, \omega_{M}\rangle=1$.
	
	On the other hand, for any element $e\in V^{*\mathrm{int}}_{+\langle\cdot ,\cdot \rangle}$, the vector $e/\langle u, e\rangle$ defines a state because $V_{+}=V^{*\mathrm{int}}_{+\langle\cdot ,\cdot \rangle}$. In particular, for the effect $f_{a}=\sum_{i_{(a)}}e_{i_{(a)}}^{\mathrm{ext}}\ \mbox{or}\  u-\sum_{i_{(a)}}e_{i_{(a)}}^{\mathrm{ext}}$, we can see from \eqref{eq_values of indecomp effects} that
	\begin{equation}
	\left\langle f_{a},\ \frac{f_{a}}{\langle u, f_{a}\rangle}\right\rangle=1
	\end{equation}
	holds, that is, the state $f_{a}/\langle u, f_{a}\rangle$ is an eigenstate of $f_{a}$ (similarly for $g_{b}$). 
	\fi

	From Lemma \ref{lem_eigenstates} and the definition of $\mathcal{W}_{\epsilon_{1}}(\widetilde{M}^{F}, F)$ \eqref{def_error bar width}, for any $w_{1}\ge\mathcal{W}_{\epsilon_{1}}(\widetilde{M}^{F}, F)$ we have
	\[
	\sum_{a'\in O_{d_{A}}(a;\,w_{1})}\left\langle \widetilde{m}_{a'}^{F},\ \frac{f_{a}}{\langle u, f_{a}\rangle}\right\rangle\ge1-\epsilon_{1},
	\]
	equivalently,
	\[
	\sum_{b'\in B}\sum_{a'\in O_{d_{A}}(a;\,w_{1})}\left\langle \widetilde{m}_{a'b'}^{FG},\ \frac{f_{a}}{\langle u, f_{a}\rangle}\right\rangle\ge1-\epsilon_{1}
	\]
	for all $a\in A$. 
	Multiplying both sides by $\langle u, f_{a}\rangle=\langle \omega_{M}, f_{a}\rangle (>0)$ (Lemma \ref{lem_u=omegaM}) and taking the summation over $a$ yield
	\begin{equation}
	\label{eq_1}
	\sum_{a\in A}\sum_{b'\in B}\sum_{a'\in O_{d_{A}}(a;\,w_{1})}\left\langle \widetilde{m}_{a'b'}^{FG},\ f_{a}\right\rangle\ge1-\epsilon_{1},
	\end{equation}
	where we use the relation $\sum_{a\in A}\langle u, f_{a}\rangle=\langle u, u\rangle=\langle u, \omega_{M}\rangle=1$.
	Defining a function $\chi_{[d_{A}, w_{1}]}$ on $A\times A$ such that
	\[
	\chi_{[d_{A}, w_{1}]}(a, a')=
	\left\{
	\begin{aligned}
	&\ 1\qquad(d_{A}(a, a')\le\frac{w_{1}}{2})\\
	&\ 0\qquad(d_{A}(a, a')>\frac{w_{1}}{2}),
	\end{aligned}
	\right.
	\]
	it holds that 
	\begin{align*}
	\sum_{a\in A}\sum_{a'\in O_{d_{A}}(a;\,w_{1})}\left\langle \widetilde{m}_{a'b'}^{FG},\ f_{a}\right\rangle
	&=\sum_{(a, a')\in A\times A}\chi_{[d_{A}, w_{1}]}(a, a')\left\langle \widetilde{m}_{a'b'}^{FG},\ f_{a}\right\rangle\\
	&=\sum_{a'\in A}\sum_{a\in O_{d_{A}}(a';\,w_{1})}\left\langle \widetilde{m}_{a'b'}^{FG},\ f_{a}\right\rangle
	\end{align*}
	because of the symmetric action of $\chi_{[d_{A}, w_{1}]}$ on $a$ and $a'$.
	Therefore, \eqref{eq_1} can be rewritten as
	\[
	\sum_{a'\in A}\sum_{b'\in B}\sum_{a\in O_{d_{A}}(a';\,w_{1})}\left\langle \widetilde{m}_{a'b'}^{FG},\ f_{a}\right\rangle\ge1-\epsilon_{1}.
	\]
	Overall, we obtain
	\begin{equation}
	\label{ineq_fa}
	\sum_{a'\in A}\sum_{b'\in B}\sum_{a\in O_{d_{A}}(a';\,w_{1})}\langle u, \widetilde{m}_{a'b'}^{FG}\rangle\left\langle f_{a},\ \frac{\widetilde{m}_{a'b'}^{FG}}{\langle u, \widetilde{m}_{a'b'}^{FG}\rangle}\right\rangle\ge1-\epsilon_{1}.
	\end{equation}
	Similar calculations show that for any $w_{2}\ge\mathcal{W}_{\epsilon_{2}}(\widetilde{M}^{G}, G)$
	\begin{equation}
	\label{ineq_gb}
	\sum_{a'\in A}\sum_{b'\in B}\sum_{b\in O_{d_{B}}(b';\,w_{2})}\langle u, \widetilde{m}_{a'b'}^{FG}\rangle\left\langle g_{b},\ \frac{\widetilde{m}_{a'b'}^{FG}}{\langle u, \widetilde{m}_{a'b'}^{FG}\rangle}\right\rangle\ge1-\epsilon_{2}
	\end{equation}
	holds.
	We obtain from \eqref{ineq_fa} and \eqref{ineq_gb}
	\begin{equation*}
	\begin{aligned}
	\sum_{a'\in A}\sum_{b'\in B}\langle u, \widetilde{m}_{a'b'}^{FG}\rangle
	\left[
	\left(
	\sum_{a\in O_{d_{A}}(a';\,w_{1})}\left\langle f_{a},\ \frac{\widetilde{m}_{a'b'}^{FG}}{\langle u, \widetilde{m}_{a'b'}^{FG}\rangle}\right\rangle
	\right)\right.\qquad\qquad\qquad\qquad\qquad\\
	+
	\left.\left(
	\sum_{b\in O_{d_{B}}(b';\,w_{2})}\left\langle g_{b},\ \frac{\widetilde{m}_{a'b'}^{FG}}{\langle u, \widetilde{m}_{a'b'}^{FG}\rangle}\right\rangle
	\right)
	\right]\ge 2-\epsilon_{1}-\epsilon_{2},
	\end{aligned}
	\end{equation*}
	which implies that there exists a $(a'_{0}, b'_{0})\in A\times B$ such that 
	\begin{equation}
	\label{ineq_fa gb}
	\begin{aligned}
	\left(
	\sum_{a\in O_{d_{A}}(a'_{0};\,w_{1})}\left\langle f_{a},\ \frac{\widetilde{m}_{a'_{0}b'_{0}}^{FG}}{\langle u, \widetilde{m}_{a'_{0}b'_{0}}^{FG}\rangle}\right\rangle
	\right)\quad\qquad\qquad\qquad\qquad\qquad\qquad\\
	+
	\left(
	\sum_{b\in O_{d_{B}}(b'_{0};\,w_{2})}\left\langle g_{b},\ \frac{\widetilde{m}_{a'_{0}b'_{0}}^{FG}}{\langle u, \widetilde{m}_{a'_{0}b'_{0}}^{FG}\rangle}\right\rangle
	\right)\ge 2-\epsilon_{1}-\epsilon_{2}
	\end{aligned}
	\end{equation}
	since $\sum_{a'\in A}\sum_{b'\in B}\langle u, \widetilde{m}_{a'b'}^{FG}\rangle=\langle u, u\rangle=1$ and $0\le\langle u, \widetilde{m}_{a'b'}^{FG}\rangle\le1$ for all $(a', b')\in A\times B$. 
	We can see from \eqref{ineq_fa gb} that
	\begin{align}
	\sum_{a\in O_{d_{A}}(a'_{0};\,w_{1})}\left\langle f_{a},\ \frac{\widetilde{m}_{a'_{0}b'_{0}}^{FG}}{\langle u, \widetilde{m}_{a'_{0}b'_{0}}^{FG}\rangle}\right\rangle
	&\ge
	1-\epsilon_{1}-\epsilon_{2}\notag\\
	&\qquad+\left(1-\sum_{b\in O_{d_{B}}(b'_{0};\,w_{2})}\left\langle g_{b},\ \frac{\widetilde{m}_{a'_{0}b'_{0}}^{FG}}{\langle u, \widetilde{m}_{a'_{0}b'_{0}}^{FG}\rangle}\right\rangle\right)\notag\\
	\label{ineq fa epsilon}
	&\ge 1-\epsilon_{1}-\epsilon_{2}
	\end{align}
	holds for an arbitrary $w_{1}\ge\mathcal{W}_{\epsilon_{1}}(\widetilde{M}^{F}, F)$, where we use 
	\[
	\sum_{b\in O_{d_{B}}(b'_{0};\,w_{2})}\left\langle g_{b},\ \frac{\widetilde{m}_{a'_{0}b'_{0}}^{FG}}{\langle u, \widetilde{m}_{a'_{0}b'_{0}}^{FG}\rangle}\right\rangle
	\le
	\sum_{b\in B}\left\langle g_{b},\ \frac{\widetilde{m}_{a'_{0}b'_{0}}^{FG}}{\langle u, \widetilde{m}_{a'_{0}b'_{0}}^{FG}\rangle}\right\rangle=1,
	\]
	and similarly
	\begin{align}
	\label{ineq gb epsilon}
	\sum_{b\in O_{d_{B}}(b'_{0};\,w_{2})}\left\langle g_{b},\ \frac{\widetilde{m}_{a'_{0}b'_{0}}^{FG}}{\langle u, \widetilde{m}_{a'_{0}b'_{0}}^{FG}\rangle}\right\rangle
	\ge
	1-\epsilon_{1}-\epsilon_{2}
	\end{align}
	holds for an arbitrary $w_{2}\ge\mathcal{W}_{\epsilon_{2}}(\widetilde{M}^{G}, G)$. 
	Because 
	\[
	\omega'_{0}:=\frac{\widetilde{m}_{a'_{0}b'_{0}}^{FG}}{\langle u, \widetilde{m}_{a'_{0}b'_{0}}^{FG}\rangle}
	\]
	defines a state (\eqref{eq_eigenstate0} in Lemma \ref{lem_eigenstates}), \eqref{ineq fa epsilon} and \eqref{ineq gb epsilon} together with the definition of the overall width \eqref{def_overall width} result in 
	\begin{align*}
	&w_{1}\ge W_{\epsilon_{1}+\epsilon_{2}}({\omega'}_{0}^{F}),\\
	&w_{2}\ge W_{\epsilon_{1}+\epsilon_{2}}({\omega'}_{0}^{G}).
	\end{align*}
	These equations hold for any $w_{1}\ge\mathcal{W}_{\epsilon_{1}}(\widetilde{M}^{F}, F)$ and $w_{2}\ge\mathcal{W}_{\epsilon_{2}}(\widetilde{M}^{G}, G)$, so we finally obtain
	\begin{align*}
	&\mathcal{W}_{\epsilon_{1}}(\widetilde{M}^{F}, F)\ge W_{\epsilon_{1}+\epsilon_{2}}({\omega'}_{0}^{F})\\
	&\mathcal{W}_{\epsilon_{2}}(\widetilde{M}^{G}, G)\ge W_{\epsilon_{1}+\epsilon_{2}}({\omega'}_{0}^{G}).
	\end{align*}
	\qed
\end{pf}
The next corollary results immediately from Proposition \ref{prop_error bar and Werner}. It describes a similar content to Theorem \ref{thm_error bar} in terms of another measure. 
\begin{cor}
	\label{cor_Werner measure}
	Let $\Omega$ be a transitive state space and its positive cone $V_{+}$ be self-dual with respect to $\langle\cdot,\cdot\rangle_{GL(\Omega)}$, and let $(F, G)$ be a pair of ideal observables on $\Omega$.
	For an arbitrary approximate joint observable $\widetilde{M}^{FG}$ of $(F, G)$ and $\epsilon_{1}, \epsilon_{2}\in(0,1]$ satisfying $\epsilon_{1}+\epsilon_{2}\le 1$, there exists a state $\omega\in\Omega$ such that
	\[
	\begin{aligned}
	&D_{W}(\widetilde{M}^{F}, F)\ge \frac{\epsilon_{1}}{2}\ W_{\epsilon_{1}+\epsilon_{2}}(\omega^{F}),\\
	&D_{W}(\widetilde{M}^{G}, G)\ge \frac{\epsilon_{2}}{2}\ W_{\epsilon_{1}+\epsilon_{2}}(\omega^{G}).
	\end{aligned}
	\]
\end{cor}
There is also another formulation by means of minimum localization error and $l_{\infty}$ distance.
\begin{thm}
	\label{thm_l-distance}
	Let $\Omega$ be a transitive state space and its positive cone $V_{+}$ be self-dual with respect to $\langle\cdot,\cdot\rangle_{GL(\Omega)}$, and let $(F, G)$ be a pair of ideal observables on $\Omega$.
	For an arbitrary approximate joint observable $\widetilde{M}^{FG}$ of $(F, G)$, there exists a state $\omega\in\Omega$ such that
	\[
	D_{\infty}(\widetilde{M}^{F}, F)+D_{\infty}(\widetilde{M}^{G}, G)\ge LE(\omega^{F})+LE(\omega^{G}).
	\]
\end{thm}
\begin{pf}
	We can see from \eqref{eq_eigenstate} in Lemma \ref{lem_eigenstates} and the definition of the $l_{\infty}$ distance \eqref{def_l distance} that 
	\[
	\left|
	\left\langle f_{a},\ \frac{f_{a}}{\langle u, f_{a}\rangle}\right\rangle-\left\langle \widetilde{m}^{F}_{a},\ \frac{f_{a}}{\langle u, f_{a}\rangle}\right\rangle
	\right|
	\le
	D_{\infty}(\widetilde{M}^{F}, F)
	\]
	holds for all $a\in A$, which can be rewritten as
	\begin{equation*}
	1-\sum_{b\in B}\left\langle \widetilde{m}^{FG}_{ab},\ \frac{f_{a}}{\langle u, f_{a}\rangle}\right\rangle
	\le
	D_{\infty}(\widetilde{M}^{F}, F),
	\end{equation*}
	for all $a\in A$. Multiplying both sides by $\langle u, f_{a}\rangle$ and taking the summation over $a$, we have
	\[
	1-\sum_{a\in A}\sum_{b\in B}\left\langle \widetilde{m}^{FG}_{ab}, f_{a}\right\rangle
	\le
	D_{\infty}(\widetilde{M}^{F}, F),
	\] 
	namely
	\begin{equation}
	\label{ineq_fa2}
	1-\sum_{a'\in A}\sum_{b'\in B}\langle u, \widetilde{m}_{a'b'}^{FG}\rangle\left\langle f_{a'},\ \frac{\widetilde{m}_{a'b'}^{FG}}{\langle u, \widetilde{m}_{a'b'}^{FG}\rangle}\right\rangle
	\le
	D_{\infty}(\widetilde{M}^{F}, F)
	\end{equation}
	In a similar way, we also have
	\begin{equation}
	\label{ineq_gb2}
	1-\sum_{a'\in A}\sum_{b'\in B}\langle u, \widetilde{m}_{a'b'}^{FG}\rangle\left\langle g_{b'},\ \frac{\widetilde{m}_{a'b'}^{FG}}{\langle u, \widetilde{m}_{a'b'}^{FG}\rangle}\right\rangle
	\le
	D_{\infty}(\widetilde{M}^{G}, G).
	\end{equation}
	Since $\sum_{a'\in A}\sum_{b'\in B}\langle u, \widetilde{m}_{a'b'}^{FG}\rangle=1$, \eqref{ineq_fa2} and \eqref{ineq_gb2} give
	\begin{equation*}
	\begin{aligned}
	\sum_{a'\in A}\sum_{b'\in B}\langle u, \widetilde{m}_{a'b'}^{FG}\rangle
	\left[
	\left(
	1-\left\langle f_{a'},\ \frac{\widetilde{m}_{a'b'}^{FG}}{\langle u, \widetilde{m}_{a'b'}^{FG}\rangle}\right\rangle
	\right)\right.
	+
	\left.\left(
	1-\left\langle g_{b'},\ \frac{\widetilde{m}_{a'b'}^{FG}}{\langle u, \widetilde{m}_{a'b'}^{FG}\rangle}\right\rangle
	\right)
	\right]\\
	\le D_{\infty}(\widetilde{M}^{F}, F)+D_{\infty}(\widetilde{M}^{G}, G),
	\end{aligned}
	\end{equation*}
	which indicates that there exists a $(a'_{0}, b'_{0})\in A\times B$ satisfying
	\begin{align}
	\left(
	1-\left\langle f_{a'_{0}},\ \frac{\widetilde{m}_{a'_{0}b'_{0}}^{FG}}{\langle u, \widetilde{m}_{a'_{0}b'_{0}}^{FG}\rangle}\right\rangle
	\right)
	+
	\left(
	1-\left\langle g_{b'_{0}},\ \frac{\widetilde{m}_{a'_{0}b'_{0}}^{FG}}{\langle u, \widetilde{m}_{a'_{0}b'_{0}}^{FG}\rangle}\right\rangle
	\right)\qquad\qquad\notag\\
	\le
	D_{\infty}(\widetilde{M}^{F}, F)+D_{\infty}(\widetilde{M}^{G}, G)\label{ineq_fa gb2}.
	\end{align}
	Because 
	\[
	\omega'_{0}:=\frac{\widetilde{m}_{a'_{0}b'_{0}}^{FG}}{\langle u, \widetilde{m}_{a'_{0}b'_{0}}^{FG}\rangle}
	\]
	is a state (\eqref{eq_eigenstate0} in Lemma \ref{lem_eigenstates}), we can conclude from \eqref{ineq_fa gb2} and the definition of the minimum localization error \eqref{def_localization error} that
	\[
	LE({\omega'}_{0}^{F})+LE({\omega'}_{0}^{G})\le D_{\infty}(\widetilde{M}^{F}, F)+D_{\infty}(\widetilde{M}^{G}, G),
	\]
	which proves the theorem.
	\qed\end{pf}
It is easy to see from the proofs that our theorems can be generalized to the case where three or more observables are considered.
\begin{rmk}
	It was claimed in \cite{PhysRevA.101.052104} similarly to our theorems that PURs imply MURs in GPTs. 
	However, the result in \cite{PhysRevA.101.052104} was obtained for a pair of binary (i.e. two-outcome), extreme, sharp, and {\it postprocessing clean} \cite{PhysRevA.97.062102} observables. 
	It is known that any effect of a sharp and postprocessing clean observable is pure and indecomposable, and such observables do not always exist for a GPT \cite{PhysRevA.97.062102,KIMURA2010175}. 
	The only finite-dimensional quantum theory admitting those observables is a qubit system (remember that pure and indecomposable effects correspond to rank-1 projections in finite-dimensional quantum theories). On the other hand, although our GPTs are assumed to be transitive and self-dual, or regular polygon theories, our theorems are obtained for more general forms of observables \eqref{def of ideal meas} always possible to be defined.
\end{rmk}

Theorem \ref{thm_l-distance} (and Theorem \ref{thm_for even polygon}) has an application to evaluate the {\it degree of incompatibility} \cite{Busch_2013,PhysRevA.89.022123,PhysRevA.87.052125} of a GPT.

\begin{eg}[Evaluation of degree of incompatibility]
	\label{eg_deg of inc}
	Suppose that $\Omega$ is an arbitrary state space, and $F$ and $G$ are two-outcome observables on $\Omega$, namely $F=\{f_{0}, f_{1}\}$ and $G=\{g_{0}, g_{1}\}$, and consider similarly to \eqref{def_appro Z and X} their ``fuzzy'' versions
	\begin{equation}
	\label{def_appro F and G}
	\begin{aligned}
	\widetilde{F}^{\lambda}:&=\lambda F+(1-\lambda)\left\{\frac{u}{2},\ \frac{u}{2}\right\}=\left\{\lambda f_{0}+\frac{1-\lambda}{2}u,\ \lambda f_{1}+\frac{1-\lambda}{2}u\right\},\\
	\widetilde{G}^{\lambda}:&=\lambda G+(1-\lambda)\left\{\frac{u}{2},\ \frac{u}{2}\right\}=\left\{\lambda g_{0}+\frac{1-\lambda}{2}u,\ \lambda g_{1}+\frac{1-\lambda}{2}u\right\}
	\end{aligned}
	\end{equation}
	for $\lambda\in[0, 1]$. It is known that we can find a $\lambda_{F, G}\ge\frac{1}{2}$ such that the distorted observables $\widetilde{F}^{\lambda}$ and $\widetilde{G}^{\lambda}$ in \eqref{def_appro F and G} are jointly measurable for any $\lambda\in[0, \lambda_{F, G}]$, and $\lambda_{\mathrm{opt}}:=\inf_{F, G}\lambda_{F, G}$ can be thought describing the degree of incompatibility of the theory. $\lambda_{\mathrm{opt}}$ has been calculated in various theories: for example, $\lambda_{\mathrm{opt}}=\frac{1}{\sqrt{2}}$ in finite-dimensional quantum theories \cite{PhysRevA.87.052125}, and $\lambda_{\mathrm{opt}}=\frac{1}{2}$ in the square theory (a regular polygon theory with $n=4$) \cite{PhysRevA.96.022113}.
	
	To see how Theorem \ref{thm_l-distance} contributes to the degree of incompatibility, we consider the situations in Theorem \ref{thm_l-distance} (and Theorem \ref{thm_for even polygon}) with the marginals $\widetilde{M}^{F}$ and $\widetilde{M}^{G}$ of the approximate joint observable being $\widetilde{F}^{\lambda}$ and $\widetilde{G}^{\lambda}$ in \eqref{def_appro F and G} for $\lambda\in[0, \lambda_{F, G}]$ respectively. In this case, we can represent the measurement error $D_{\infty}(\widetilde{F}^{\lambda}, F)$ in a more explicit way:
	\begin{align}
	\label{def_l distance1}
	D_{\infty}(\widetilde{F}^{\lambda}, F)
	&=\underset{\omega\in\Omega}{\sup}\ \underset{i\in \{0,1\}}{\max}\left|\left(\lambda f_{i}+\frac{1-\lambda}{2}u\right)(\omega)-f_{i}(\omega)\right|\notag\\
	&=(1-\lambda)\ \underset{\omega\in\Omega}{\sup}\ \underset{i\in \{0,1\}}{\max}\left|f_{i}(\omega)-\frac{1}{2}\right|\notag\\
	&=\frac{1-\lambda}{2},
	\end{align}
	where we use the relation
	\[
	\left|f_{0}(\omega)-\frac{1}{2}\right|=\left|(u-f_{1})(\omega)-\frac{1}{2}\right|=\left|f_{1}(\omega)-\frac{1}{2}\right|
	\]
	and the fact that there is an ``eigenstate'' $\omega_{i}$ for each ideal effect $f_{i}$ satisfying $f_{i}(\omega_{i})=1$ as we have seen in \eqref{eq_eigenstate} or \eqref{eq_pf2}. Therefore, we can conclude from Theorem \ref{thm_l-distance} (and Theorem \ref{thm_for even polygon}) that for any $\lambda\in[0, \lambda_{F, G}]$ and for some state $\omega_{0}$
	\begin{align*}
	1-\lambda\ge \left(1-\underset{i\in \{0,1\}}{\max}f_{i}(\omega_{0})\right)+\left(1-\underset{j\in \{0,1\}}{\max}g_{j}(\omega_{0})\right)
	\end{align*}
	holds, that is,
	\begin{align}
	\label{eq_deg of inc for bi 1}
	\lambda_{F, G}\le \underset{\omega\in\Omega}{\max}\left(\underset{i\in \{0,1\}}{\max}f_{i}(\omega)+\underset{j\in \{0,1\}}{\max}g_{j}(\omega)\right)-1
	\end{align}
	holds, and $\lambda_{\mathrm{opt}}$ can be evaluated by taking the infimum of both sides of \eqref{eq_deg of inc for bi 1} over all two-outcome observables. 
	We remark that the maximum value in the right hand side of \eqref{eq_deg of inc for bi 1} does exist due to the compactness of $\Omega$.
	The concrete value of the right hand side of \eqref{eq_deg of inc for bi 1} for regular polygon theories will be given in Subsection \ref{3subsec:concrete values for polygon}.
\end{eg}

\section{Entropic uncertainty relations in a class of GPTs}
\label{4sec:entropic URs}
Entropic uncertainty relations have the advantages of their compatibility with information theory and independence from the structure of the sample spaces. 
They indeed have been applied to the field of quantum information in various ways \cite{RevModPhys.89.015002}. 
In this section, we present our main results on two types of entropic uncertainty in a certain class of GPTs. 
While our results reproduce entropic uncertainty relations obtained in finite-dimensional quantum theories, they indicate that similar relations hold also in a broader class of physical theories.

\subsection{Entropic PURs}
\label{subsec:entropic PURs}
We continue following the notations in the previous section.
Let us consider a GPT with its state space $\Omega$, and two ideal observables (see \eqref{def of ideal meas}) $F=\{f_{a}\}_{a\in A}$ and $G=\{g_{b}\}_{b\in B}$ on $\Omega$.
Here we do not assume that $A$ and $B$ are metric spaces but assume that they are finite sets.
For the probability distribution $\omega^{F}=\{f_{a}(\omega)\}_{a}$ obtained in the measurement of $F$ on a state $\omega\in\Omega$  (and similarly for $\{g_{b}(\omega)\}_{b}$), its Shannon entropy is defined as 
\begin{align}
\label{def:Shannon entropy}
H\left(\omega^{F}\right)=-\sum_{a\in A}f_{a}(\omega)\log{f_{a}(\omega)}.
\end{align}
Note that $H\left(\omega^F\right)\ge0$ and $H\left(\omega^F\right)=0$ if and only if $\omega^F$ is definite, i.e. $f_{a^{*}}(\omega)=1$ for some $a^{*}$ and $f_{a}(\omega)=0$ for $a\neq a^{*}$.
If there exists a relation such as
\begin{align*}
H\left(\omega^F\right)+H\left(\omega^G\right)\ge\Gamma_{F, G}\quad\ ^\forall\omega\in\Omega
\end{align*}
with a constant $\Gamma_{F, G}>0$, then it is called an entropic PUR because it demonstrates that we cannot prepare a state which makes simultaneously $H\left(\omega^F\right)$ and $H\left(\omega^G\right)$ vanish, or $\omega^F$ and $\omega^G$ definite.
One way to obtain an entropic PUR is to consider the Landau-Pollak-type relation \cite{Uffink_PhD,PhysRevA.71.052325,PhysRevA.76.062108}:
\begin{align}
\label{def:L-P UR}
\max_{a\in A}f_{a}(\omega)+\max_{b\in B}g_{b}(\omega)\le \gamma_{F,G}\quad\ ^\forall\omega\in\Omega
\end{align}
with a constant $\gamma_{F,G}\in(0, 2]$. Remark that relations of the form \eqref{def:L-P UR} always can be found for any pair of observables. It is known \cite{PhysRevLett.60.1103, inequalities1988} that $\max_{a\in A}f_{a}(\omega)$ is related with $H\left(\omega^F\right)$ by
\[
\exp\left[-H\left(\omega^F\right)\right]\le\max_{a\in A}f_{a}(\omega),
\] 
and thus we can observe from \eqref{def:L-P UR}
\begin{align*}
\exp\left[-H\left(\omega^F\right)\right]+\exp\left[-H\left(\omega^G\right)\right]\le \gamma_{F, G}.
\end{align*}
Considering that
\begin{align*}
\exp\left[-H\left(\omega^F\right)\right]+\exp\left[-H\left(\omega^G\right)\right]
\ge 2\exp\left[\frac{-H\left(\omega^F\right)-H\left(\omega^G\right)}{2}\right]
\end{align*} 
holds, we can finally obtain an entropic relation
\begin{align}
\label{eq:entropic PUR via L-P}
H\left(\omega^F\right)+H\left(\omega^G\right)\ge -2\log\frac{\gamma_{F, G}}{2}\quad\ ^\forall\omega\in\Omega.
\end{align}
If $\gamma_{F, G}<2$, then \eqref{eq:entropic PUR via L-P} gives an entropic PUR because it indicates that it is impossible to prepare a state which makes both $H\left(\omega^F\right)$ and $H\left(\omega^G\right)$ zero, that is, there is no state preparation on which $F$ and $G$ take simultaneously definite values (note that \eqref{def:L-P UR} also gives a PUR if $\gamma_{F,G}<2$). In a finite-dimensional quantum theory with its state space $\Omega_{\mathrm{QT}}$, it can be shown that 
\begin{align}
\label{eq:quantum LP}
\max_{a}f_{a}(\omega)+\max_{b}g_{b}(\omega)\le 1+\max_{a, b}|\braket{f_{a}|g_{b}}|\quad\ ^\forall\omega\in\Omega_{\mathrm{QT}},
\end{align}
where $F=\{\ketbra{f_{a}}{f_{a}}\}_{a}$ and $B=\{\ketbra{g_{b}}{g_{b}}\}_{b}$ are rank-1 PVMs. 
In that case, \eqref{eq:entropic PUR via L-P} can be rewritten as
\begin{align}
\label{eq:Deutsch ent PUR}
H\left(\omega^F\right)+H\left(\omega^G\right)\ge2\log\frac{2}{1+\underset{a, b}{\max}|\braket{f_{a}|g_{b}}|}\quad\ ^\forall\omega\in\Omega_{\mathrm{QT}},
\end{align}
which is the entropic PUR proven by Deutsch \cite{PhysRevLett.50.631}. 
There have been studies to find a better bound \cite{PhysRevLett.60.1103} or generalization \cite{10.2307/25051432} of \eqref{eq:Deutsch ent PUR}. 

\begin{rmk}
	\label{rmk:majorization}
	Entropic PURs in quantum theory can be derived also by means of majorization \cite{PhysRevA.84.052117,PhysRevLett.111.230401,PhysRevA.89.052115,Pucha_a_2013,Pucha_a_2018,e21030270}.
	This method of majorization can be also applied to GPTs.
	To see this, let us introduce probability vectors $\mathbf{f}(\omega)$ and $\mathbf{g}(\omega)$ defined simply through $\omega^F=\{f_{a}(\omega)\}_{a}$ and $\omega^G=\{g_{b}(\omega)\}_{b}$ respectively.
	By adding outcomes to either $A$ or $B$, we can assume without loss of generality that their cardinalities are equal: $|A|=|B|=d$, and $\mathbf{f}(\omega)$ and $\mathbf{g}(\omega)$ are $d$-dimensional vectors.
	If $d$-dimensional probability vectors  $\mathbf{p}=(p_{i})_{i}$ and $\mathbf{q}=(q_{i})_{i}$ satisfy 
	\[
	\sum_{j=1}^{k}p_{j}^{\downarrow}\le
	\sum_{j=1}^{k}q_{j}^{\downarrow}\quad^\forall k=1, 2, \cdots, d,
	\]
	where $p_{j}^{\downarrow}$'s are obtained thorough ordering the components of $\mathbf{p}$ in decreasing order: $\{p_{j}^{\downarrow}\}_{j}=\{p_{i}\}_{i}$ and $p_{1}^{\downarrow}\ge p_{2}^{\downarrow}\ge p_{3}^{\downarrow}\ge \cdots$ (similarly for $q_{j}^{\downarrow}$'s),
	then $\mathbf{p}$ is called {\it majorized} by $\mathbf{q}$ and we write $\mathbf{p}\prec \mathbf{q}$.
	For $\mathbf{f}(\omega)$ and $\mathbf{g}(\omega)$, a relation of the form
	\begin{align}
	\label{eq:UR_mj}
	\mathbf{f}(\omega)\otimes\mathbf{g}(\omega)\prec\mathbf{r}\quad\ ^\forall\omega\in\Omega,
	\end{align}
	where $\mathbf{r}=(r_{i})_{i}$ is a $d^{2}$-dimensional probability vector defined below, was proven in \cite{PhysRevLett.111.230401}. 
	The vector $\mathbf{r}$ was given by 
	\begin{align*}
	\mathbf{r}=(R_1, R_2-R_1, \cdots, R_{d}-R_{d-1}, 0, 0, \cdots, 0)
	\end{align*}
	with
	\[
	\left\{
	\begin{aligned}
	&R_{k}=\max_{\mathcal{I}_{k}}\max_{\omega\in\Omega}\sum_{(x, y)\in\mathcal{I}_{k}}f_{a}(\omega)g_{b}(\omega)\\
	&\mathcal{I}_{k}=\{(a_{1}, b_{1}), \cdots, (a_{k}, b_{k})\mid
	(a_{i}, b_{i})\in A\times B,\ (a_{i}, b_{i})\neq(a_{j}, b_{j}) \ \mbox{for $i\neq j$}
	\}
	\end{aligned}
	\right.
	\]
	(thus we can see $R_{k}=1$ for $d\le k\le d^{2}$ because $F$ and $G$ are ideal).
	From \eqref{eq:UR_mj}, we can derive \cite{PhysRevA.84.052117} 
	\begin{align}
	\label{eq:entropic PUR via mj}
	H\left(\omega^F\right)+H\left(\omega^G\right)\ge H(\{r_{i}\}_i)\quad\ ^\forall\omega\in\Omega,
	\end{align}
	which gives a similar entropic relation to \eqref{eq:entropic PUR via L-P}.
	Note that when $F$ and $G$ are binary, the vector $\mathbf{r}$ is completely determined by 
	\[
	R_{1}=\max_{(a, b)}f_{a}(\omega)g_{b}(\omega).
	\]
	In \cite{PhysRevLett.111.230401}, $R_{1}$ was evaluated as
	\begin{align*}
	R_{1}=\max_{(a, b)}f_{a}(\omega)g_{b}(\omega)
	\le\frac{\gamma^2}{4}
	\end{align*}
	with
	\[
	\gamma=\max_{(a, b)}\ (f_{a}+g_{b})(\omega),
	\]
	and it was shown that in quantum theory the equality holds:
	\[
	R_{1}=\max_{(a, b)}f_{a}(\omega)g_{b}(\omega)
	=\frac{\gamma^2}{4}.
	\]
	We will consider in Subsection \ref{3subsec:concrete values for polygon} similar cases when $R_{1}=\frac{\gamma^2}{4}$ holds, and give concrete value of $\gamma$.
\end{rmk}

\subsection{Entropic MURs}
\label{subsec:entropic MURs}
Let $\Omega$ be a state space which is transitive and its positive cone $V_+$ satisfy $V_{+}=V^{*int}_{+\langle\cdot,\cdot\rangle_{GL(\Omega)}}$, and we hereafter denote the inner product $\langle\cdot,\cdot\rangle_{GL(\Omega)}$ simply by $\langle\cdot,\cdot\rangle$ as in the previous section. 
There can be defined measurement error in terms of entropy in the identical way with the quantum one by Buscemi et al. \cite{PhysRevLett.112.050401}. 
Let in the GPT $E=\{e_{x}\}_{x\in X}$ be an ideal observable and $M=\{m_{\hat{x}}\}_{\hat{x}\in \hat{X}}$ be an observable with finite outcome sets $X, \hat{X}$.
Since 
\begin{align}
\label{eq:eigenstate}
\left\langle e_{x'},\ \frac{e_{x}}{\langle u, e_{x}\rangle}\right\rangle=\delta_{x'x}
\end{align}
holds for all $x, x'\in X$, and
\begin{equation}
\begin{aligned}
\omega_{M}=u&=\sum_{x}e_{x}\\
&=\sum_{x}\langle u, e_{x}\rangle \frac{e_{x}}{\langle u, e_{x}\rangle}
\end{aligned}
\end{equation}
holds from Lemma \ref{lem_u=omegaM} and Lemma \ref{lem_eigenstates}, the joint probability distribution 
\begin{equation}
\label{def:joint dist}
\{p(x, \hat{x})\}_{x, \hat{x}}=\{\langle e_{x}, m_{\hat{x}}\rangle\}_{x, \hat{x}}=\left\{\langle u, e_{x}\rangle\left\langle\frac{e_{x}}{\langle u, e_{x}\rangle},\  m_{\hat{x}}\right\rangle\right\}_{x, \hat{x}}
\end{equation}
is considered to be obtained in the measurement of $M$ on the ``eigenstates" $\{e_{x}/\langle u, e_{x} \rangle\}_{x}$ of $E$ (see \eqref{eq:eigenstate}) with the initial distribution
\begin{align}
\label{eq:initial dist}
\left\{
p(x)
\right\}_{x}=
\left\{
\langle u, e_{x}\rangle
\right\}_{x}.
\end{align}
According to \cite{PhysRevLett.112.050401}, the conditional entropy
\begin{equation}
\label{def:entropic MN}
\begin{aligned}
\mathsf{N}(M;E):
&=H(E|M)\\
&=\sum_{\hat{x}}p(\hat{x})H\left(\{p(x|\hat{x})\}_{x}\right)\\
&=\sum_{\hat{x}}\ang{u, m_{\hat{x}}}H\left(\left\{\ang{e_{x},\  \frac{m_{\hat{x}}}{\ang{u, m_{\hat{x}}}}}\right\}_{x}\right)
\end{aligned}
\end{equation}
calculated via \eqref{def:joint dist} describes how inaccurately the actual observable $\mathcal{M}$ can estimate the input eigenstates of the ideal observable $E$. 
In fact, if we consider measuring $\mathcal{M}$ on $e_{x}/\langle u, e_{x}\rangle$ and estimating the input state from the output probability distribution 
\[
\{p(\hat{x}|x)\}_{\hat{x}}=\left\{\left\langle m_{\hat{x}},\  \frac{e_{x}}{\langle u, e_x\rangle}\right\rangle\right\}_{\hat{x}}
\]
by means of a guessing function $f:\hat{X}\to X$, then the error probability $p_{\mathrm{error}}^{f}(x)$ is given by
\[
p_{\mathrm{error}}^{f}(x)=1-\sum_{\hat{x}: f(\hat{x})=x}p(\hat{x}|x)=\sum_{\hat{x}: f(\hat{x})\neq x}p(\hat{x}|x).
\]
When similar procedures are conducted for all $x\in X$ with the probability distribution $\{p(x)\}_{x}$ in \eqref{eq:initial dist}, the total error probability $p_{\mathrm{error}}^{f}$ is
\begin{align}
p_{\mathrm{error}}^{f}=\sum_{x}p(x)\ p_{\mathrm{error}}^{f}(x)=\sum_{x\in X}\sum_{\hat{x}: f(\hat{x})\neq x}p(x, \hat{x}),
\end{align}
and it was shown in \cite{PhysRevLett.112.050401} that 
\[
\min_{f}p_{\mathrm{error}}^{f}\to0\quad\iff\quad\mathsf{N}(M;E)=H(E|M)\to0.
\]
We can conclude from the consideration above that the entropic quantity \eqref{def:entropic MN} represents the difference between $E$ to be measured ideally and $\mathcal{M}$ measured actually, and thus we can define their entropic measurement error as \eqref{def:entropic MN}.

We are now in the position to derive a similar entropic relation to \cite{PhysRevLett.112.050401} with the generalized entropic measurement error \eqref{def:entropic MN}. 
We continue focusing on a GPT with its state space $\Omega$ being transitive and $V_{+}$ being self-dual with respect to the inner product $\langle\cdot,\cdot\rangle_{GL(\Omega)}\equiv\langle\cdot,\cdot\rangle$, that is, $V_{+}=V^{*int}_{+\langle\cdot,\cdot\rangle}$. 
Let $F=\{f_{a}\}_{a\in A}$ and $G=\{g_{b}\}_{b\in B}$ be a pair of ideal observables defined in \eqref{def of ideal meas}, and consider their approximate joint observable $\widetilde{M}^{FG}:=\{\widetilde{m}_{ab}^{FG}\}_{(a, b)\in A\times B}$ and its marginals
\[
\begin{aligned}
&\widetilde{M}^{F}:=\{\widetilde{m}_{a}^{F}\}_{a},\quad \widetilde{m}_{a}^{F}:=\sum_{b\in B}\widetilde{m}_{ab}^{FG};\\
&\widetilde{M}^{G}:=\{\widetilde{m}_{b}^{G}\}_{b},\quad \widetilde{m}_{b}^{G}:=\sum_{a\in A}\widetilde{m}_{ab}^{FG}.
\end{aligned}
\]
as in the previous section.
We can prove the following theorem.
\begin{thm}
	\label{thm:entropic MUR}
	Suppose that $\Omega$ is a transitive state space with its positive cone $V_{+}$ being self-dual with respect to $\langle\cdot,\cdot\rangle_{GL(\Omega)}\equiv\langle\cdot,\cdot\rangle$, $F=\{f_{a}\}_{a}$ and $G=\{g_{b}\}_{b}$ are ideal observables on $\Omega$, and $\widetilde{M}^{FG}$ is an arbitrary approximate joint observable of $(F, G)$ with its marginals $\widetilde{M}^{F}$ and $\widetilde{M}^{G}$.
	If there exists a relation
	\begin{align*}
	H\left(\omega^{F}\right)+H\left(\omega^G\right)\ge \Gamma_{F, G}\quad\ ^\forall\omega\in\Omega
	\end{align*}
	with a constant $\Gamma_{F, G}$, then it also holds that
	\begin{align*}
	\mathsf{N}(\widetilde{M}^{F};F)+\mathsf{N}(\widetilde{M}^{G};G)\ge \Gamma_{F, G}.
	\end{align*}
\end{thm}
\begin{pf}
	Since for every $\hat{a}\in A$ and $\hat{b}\in B$ $\omega_{\hat{a}\hat{b}}:=\widetilde{m}^{FG}_{\hat{a}\hat{b}}/\langle u, \widetilde{m}_{\hat{a}\hat{b}}\rangle$ is a state due to the self-duality, it holds that
	\[
	H\left(\omega_{\hat{a}\hat{b}}^F\right)+H\left(\omega_{\hat{a}\hat{b}}^G\right)\ge \Gamma_{F, G}
	\]
	for all $\hat{a}\in A$ and $\hat{b}\in B$. 
	Therefore, taking into consideration that $\langle u, \widetilde{m}^{FG}_{\hat{a}\hat{b}}\rangle\ge0$ for all $\hat{a}, \hat{b}$ and $\sum_{\hat{a}\hat{b}}\langle u, \widetilde{m}^{FG}_{\hat{a}\hat{b}}\rangle=\langle u,u\rangle=\langle u,\omega_{M}\rangle=1$, we have
	\[
	\sum_{\hat{a}\in A}\sum_{\hat{b}\in B}\langle u, \widetilde{m}^{FG}_{\hat{a}\hat{b}}\rangle\left[H\left(\omega_{\hat{a}\hat{b}}^F\right)+H\left(\omega_{\hat{a}\hat{b}}^G\right)\right]\ge \Gamma_{A, B},
	\]
	or equivalently (see \eqref{def:entropic MN}) 
	\begin{align}
	\label{eq:proof1}
	H(A\mid\widetilde{M}^{FG})+H(B\mid\widetilde{M}^{FG})\ge \Gamma_{F, G}.
	\end{align}
	Note that the conditional entropy $H(A\mid\widetilde{M}^{FG})$ is obtained through a joint probability distribution $\{p(a, \hat{a}, \hat{b})\}_{a, \hat{a}, \hat{b}}:=\{\langle f_{a}, \widetilde{m}^{FG}_{\hat{a}\hat{b}}\rangle\}$, and we can also obtain $H(A\mid\widetilde{M}^{F})$ from its marginal distribution $\{p(a, \hat{a})\}_{a, \hat{a}}=\{\langle f_{a}, \widetilde{m}^{F}_{\hat{a}}\rangle\}$. 
	The quantity 
	\[
	H(A\mid\widetilde{M}^{F})-H(A\mid\widetilde{M}^{FG})
	\]
	defined from those two conditional entropies is called the (classical) conditional mutual information, and it is known \cite{Cover:2006:EIT:1146355} to be nonnegative:
	\[
	H(A\mid\widetilde{M}^{F})-H(A\mid\widetilde{M}^{FG})\ge 0.
	\]
	A similar relation holds also for $H(G\mid\widetilde{M}^{FG})$ and $H(G\mid\widetilde{M}^{G})$, and thus, together with \eqref{eq:proof1}, we can conclude that 
	\[
	H(F\mid\widetilde{M}^{F})+H(G\mid\widetilde{M}^{G})\ge\Gamma_{F, G}
	\]
	holds, which proves the theorem.\qed
\end{pf}
Theorem \ref{thm:entropic MUR} is a generalization of the quantum result \cite{PhysRevLett.112.050401} to a class of GPTs. 
In fact, when we consider a finite-dimensional quantum theory and a pair of rank-1 PVMs $F=\{\ketbra{f_{a}}{f_{a}}\}_{a}$ and $G=\{\ketbra{g_{b}}{g_{b}}\}_{b}$, our theorem results in the one in \cite{PhysRevLett.112.050401} with the quantum bound $\Gamma_{F, G}=-2\log\max_{a, b}|\braket{f_{a}|g_{b}}|$ by Maassen and Uffink \cite{PhysRevLett.60.1103}. Theorem \ref{thm:entropic MUR} demonstrates that if there is an entropic PUR, i.e. $\Gamma_{F, G}>0$, then there is also an entropic MUR which shows that we cannot make both $\mathsf{N}(\widetilde{M}^{F};F)$ and $\mathsf{N}(\widetilde{M}^{G};G)$ vanish. 
It is again easy to prove that this theorem holds for three or more observables.

\begin{rmk}
	\label{rmk:successive}
	There is another type of entropic uncertainty relation on successive measurements in quantum theory \cite{Srinivas2003_entropic_successive,PhysRevA.89.032108,succesive_2015_Renyi,https://doi.org/10.1002/andp.201600130}.
	With a suitable introduction of transformations associated with ideal observables, we can derive similar entropic relations also in GPTs considered above.
	For an ideal observable $E=\{e_{x}\}_{x\in X}$, we define the corresponding (Schr\"{o}dinger) channel $\Phi_{E}$, which gives the post-measurement states as
	\begin{align}
	\label{eq:Schrodinger_ideal}
	\Phi_{E}\colon\Omega\to\Omega\colon\omega\mapsto\sum_{x}\langle e_{x}, \omega\rangle\frac{e_{x}}{\langle u, e_{x}\rangle}
	\end{align}
	in analogy with the channel associated with a rank-1 projective measurement (L\"{u}ders measurement \cite{Busch_quantummeasurement} for a rank-1 PVM) in quantum theory (remember \eqref{eq:eigenstate}).
	Note that this channel is found easily to be a measure-and-prepare channel  (see Example \ref{2eg: obs as channels}).
	In the Heisenberg picture, it becomes
	\begin{align}
	\label{eq:Heisenberg_ideal}
	\Phi_{E}^{*}\colon\mathcal{E}_{\Omega}\to\mathcal{E}_{\Omega}\colon e\mapsto\sum_{x}\left\langle e, \frac{e_{x}}{\langle u, e_{x}\rangle}\right\rangle e_{x}.
	\end{align}
	
	Let $F=\{f_{a}\}_a$ and $G=\{g_{b}\}_{b}$ be ideal observables associated with the channel defined in \eqref{eq:Schrodinger_ideal} (or \eqref{eq:Heisenberg_ideal}).
	It is easy to see that 
	\begin{align*}
	H\left(\omega^{F}\right)+H\left(\omega^G\right)\ge \Gamma_{F, G}\quad\ ^\forall\omega\in\Omega
	\end{align*}
	with 
	\[
	\Gamma_{F, G}:=\inf_{\omega}\left[H\left(\omega^F\right)+H\left(\omega^G\right)\right]
	\]
	holds.
	We consider measuring successively $F$ and $G$ on a state $\omega$: measuring $F$ first, and then $G$. 
	The observed statistics are $\omega^{F}=\{f_{a}(\omega)\}_{a}$ and $\Phi_{F}(\omega)^{G}=\{g_{b}(\Phi_{F}(\omega))\}_{b}$, and we can derive
	\begin{align}
	\label{eq:successive_MUR}
	H\left(\omega^F\right)+H\left(\Phi_{F}(\omega)^{G}\right)\ge \Gamma'_{F, G}
	\end{align}
	with 
	\begin{align}
	\label{eq:successive_MUR_bound}
	\Gamma'_{F, G}:&=\inf_{\Phi_{F}(\omega)}\left[H\left(\omega^F\right)+H\left(\Phi_{F}(\omega)^{G}\right)\right]\\
	&=\inf_{\Phi_{A}(\omega)}\left[H\left(\Phi_{F}(\omega)^{F}\right)+H\left(\Phi_{F}(\omega)^{G}\right)\right]
	\end{align}
	because $f_{a}(\omega)=f_{a}(\Phi_{F}(\omega))$.
	We can see that $\Gamma'_{F, G}\ge\Gamma_{F, G}$ holds, and thus there is more uncertainty in the successive measurement than the individual measurements of $F$ and $G$.
	The entropic relation\eqref{eq:successive_MUR} together with \eqref{eq:successive_MUR_bound} can be considered as a generalization of the quantum result \cite{Srinivas2003_entropic_successive}.
	Note that similarly to \cite{Srinivas2003_entropic_successive} we can present another bound for \eqref{eq:successive_MUR} in terms of the joint entropy.
	In fact, considering that $\{f_{a}\}_{a}$ and $\{\Phi_{F}^{*}(g_b)\}$ are jointly measurable ($\left\{\left\langle g_{b}, \frac{f_{a}}{\langle u, f_{a}\rangle}\right\rangle f_{a}\right\}_{ab}$ is the joint observable), that is, the probability distributions $\{f_{a}(\omega)\}_{a}$ and $\{g_{b}(\Phi_{F}(\omega))\}_{b}$ are obtained from the joint distribution $\left\{\left\langle g_{b}, \frac{f_{a}}{\langle u, f_{a}\rangle}\right\rangle \langle f_{a}, \omega\rangle\right\}_{ab}$, 
	it can be shown \cite{Cover:2006:EIT:1146355} that 
	\[
	H\left(\omega^{F}\right)+H\left(\Phi_{F}(\omega)^G\right)\ge H\left(\left\{\left\langle g_{b}, \frac{f_{a}}{\langle u, f_{a}\rangle}\right\rangle \langle f_{a}, \omega\rangle\right\}_{ab}\right).
	\]
	It is easy to see that the right hand side is also greater than or equal to $\Gamma_{F, G}$.
\end{rmk}

\section{Uncertainty relations in regular polygon theories}
\label{4sec:eg UR}
In this section, we restrict ourselves to regular polygon theories, and consider similar situations to the previous sections.
\subsection{Extensions of previous theorems}
Our theorems in Section \ref{4sec:URs in GPTs} and Section \ref{4sec:entropic URs} have been proven only for a class of theories such as finite-dimensional classical and quantum theories, and regular polygon theories with odd sides (see Section \ref{2sec:example}).
What is essential to the proofs of the theorems is that we can see effects as states (the self-duality), and that every effect of an ideal observable is an ``eigenstate'' of itself (Lemma \ref{lem_eigenstates}). 
In fact, taking those points into consideration, although it may be a minor generalization, we can demonstrate similar theorems for even-sided regular polygon theories. 
\begin{thm}
	\label{thm_for even polygon}
	Theorem \ref{thm_error bar}, Corollary \ref{cor_Werner measure}, Theorem \ref{thm_l-distance}, and Theorem \ref{thm:entropic MUR} hold for every regular polygon theory.
\end{thm}
\begin{pf}
	We only need to prove the claim for even-sided regular polygon theories.
	The proof is done by confirming that the claim of Lemma \ref{lem_eigenstates} holds for even-sided regular polygon theories with modified parametrizations.
	We again denote the inner product $\langle \cdot ,\cdot \rangle_{GL(\Omega_{n})}$ by $\langle\cdot ,\cdot \rangle$ in this proof. 
	
	In the $n$-sided regular polygon theory with even $n$, if $F=\{f_{a}\}_{a}$ is an ideal observable, then it is of the form
	\begin{equation}
	\label{def_even_ideal}
	F=\{f_{0}, f_{1}\}
	\end{equation}
	with
	\begin{equation}
	\label{def_even_ideal2}
	f_{0}=e_{i}^{n}\quad\mbox{and}\quad f_{1}=u-e_{i}^{n}=e_{i+\frac{n}{2}}^{n}
	\end{equation}
	for some $i$ (remember that we do not consider the trivial observable $F=\{u\}$). Let us introduce an affine bijection
	\begin{align}
	\label{def_affine bijection}
	\psi:=\left(
	\begin{array}{ccc}
	r_{n} & 0 & 0 \\
	0 & r_{n} & 0 \\
	0 & 0 & 1
	\end{array}
	\right)
	\end{align}
	on $\R^{3}$. 
	Because $(e, \omega)_{E}=(\psi^{-1}(e), \psi(\omega))_{E}$ holds for any $\omega\in\Omega_{n}$ and $e\in\mathcal{E}(\Omega_{n})$, we can consider an equivalent expression of the theory with $\psi\left(\Omega_{n}\right)=:\widehat{\Omega}_{n}$ and $\psi^{-1}\left(\mathcal{E}(\Omega_{n})\right)$ being its state and effect space respectively (remember that $(\cdot,\cdot)_E$ is the standard Euclidean inner product). 
	The pure states \eqref{def_polygon pure state} and  the extreme effects \eqref{def_polygon pure effect} shown in Subsection \ref{eg_polygon} are modified as
	\begin{align}
	\omega_{i}^{n}&\ \rightarrow\ 
	\hat{\omega}_{i}^{n}:=\psi\left(\omega_{i}^{n}\right)
	=
	\left(
	\begin{array}{c}
	r_{n}^{2}\cos({\frac{2\pi i}{n}})\\
	r_{n}^{2}\sin({\frac{2\pi i}{n}})\\
	1
	\end{array}
	\right);\label{def_polygon pure state2}\\
	e_{i}^{n}&\ \rightarrow\ 
	\check{e}_{i}^{n}:=\psi^{-1}\left(e_{i}^{n}\right)
	=\frac{1}{2}
	\left(
	\begin{array}{c}
	\cos({\frac{(2i-1)\pi}{n}})\\
	\sin({\frac{(2i-1)\pi}{n}})\\
	1
	\end{array}
	\right)\label{def_polygon pure effect2}
	\end{align}
	respectively, and their conic hull (the positive cone and the internal dual cone) as 
	\begin{align*}
	V_{+}&\ \rightarrow\ \widehat{V}_{+}:=\psi\left(V_{+}\right);
	\\
	V^{*int}_{+\langle \cdot ,\cdot \rangle}&\ \rightarrow\ 
	\widecheck{V}^{*int}_{+\langle \cdot ,\cdot \rangle}:=\psi^{-1}\left(V^{*int}_{+\langle \cdot ,\cdot \rangle}\right),
	\end{align*}
	respectively. Note in the equations above that $GL(\Omega_{n})=GL(\widehat{\Omega}_{n})$ and $(\cdot ,\cdot )_{E}=\langle \cdot ,\cdot \rangle_{GL(\Omega_{n})}=\langle \cdot ,\cdot \rangle_{GL(\widehat{\Omega}_{n})}=\langle\cdot ,\cdot \rangle$ hold, and $\omega_{M}=u=\ ^{t}(0,0,1)$ is invariant for $\psi$ (and $\psi^{-1}$). We can also find that an observable $E=\{e_{a}\}_{a}$ in the original expression is rewritten as $\widecheck{E}:=\{\check{e}_{a}\}_{a}$ with $\check{e}_{a}:=\psi^{-1}(e_{a})$, and that an ideal observable $F$ in \eqref{def_even_ideal} and \eqref{def_even_ideal2} gives
	\begin{equation}
	\label{def_even_ideal3}
	\widecheck{F}=\{\check{f}_{0}, \check{f}_{1}\}
	\end{equation}
	with
	\begin{equation}
	\label{def_even_ideal4}
	\check{f}_{0}=\check{e}_{i}^{n}\quad\mbox{and}\quad \check{f}_{1}=u-\check{e}_{i}^{n}=\check{e}_{i+\frac{n}{2}}^{n}
	\end{equation}
	which is also ideal in the rewritten theory. 
	Since 
	\begin{equation}
	\label{eq_pf2}
	\left\langle \check{e}_{i}^{n},\ \frac{\check{e}_{i}^{n}}{\langle u, \check{e}_{i}^{n}\rangle}\right\rangle=1
	\end{equation}
	holds for any $i$ (see \eqref{def_polygon pure effect2}), we can conclude together with \eqref{def_even_ideal3} and \eqref{def_even_ideal4} that any ideal observables $\widecheck{F}=\{\check{f}_{k}\}_{k=0, 1}$ satisfies
	\begin{equation}
	\label{eq_pf3}
	\left\langle \check{f}_{k},\ \frac{\check{f}_{k}}{\langle u, \check{f}_{k}\rangle}\right\rangle=1.
	\end{equation}
	On the other hand, it can be seen from \eqref{def_polygon pure state2} and \eqref{def_polygon pure effect2} that $\widehat{V}_{+}$ generated by \eqref{def_polygon pure state2} includes $\widecheck{V}^{*int}_{+\langle\cdot ,\cdot \rangle}$ generated by \eqref{def_polygon pure effect2}, i.e. $\widecheck{V}^{*int}_{+\langle\cdot ,\cdot \rangle}\subset \widehat{V}_{+}$ (see FIG \ref{fig_1}).
	\begin{figure}[h]
		\centering
		\includegraphics[bb=0.000000 0.000000 757.000000 649.000000,scale=0.3]{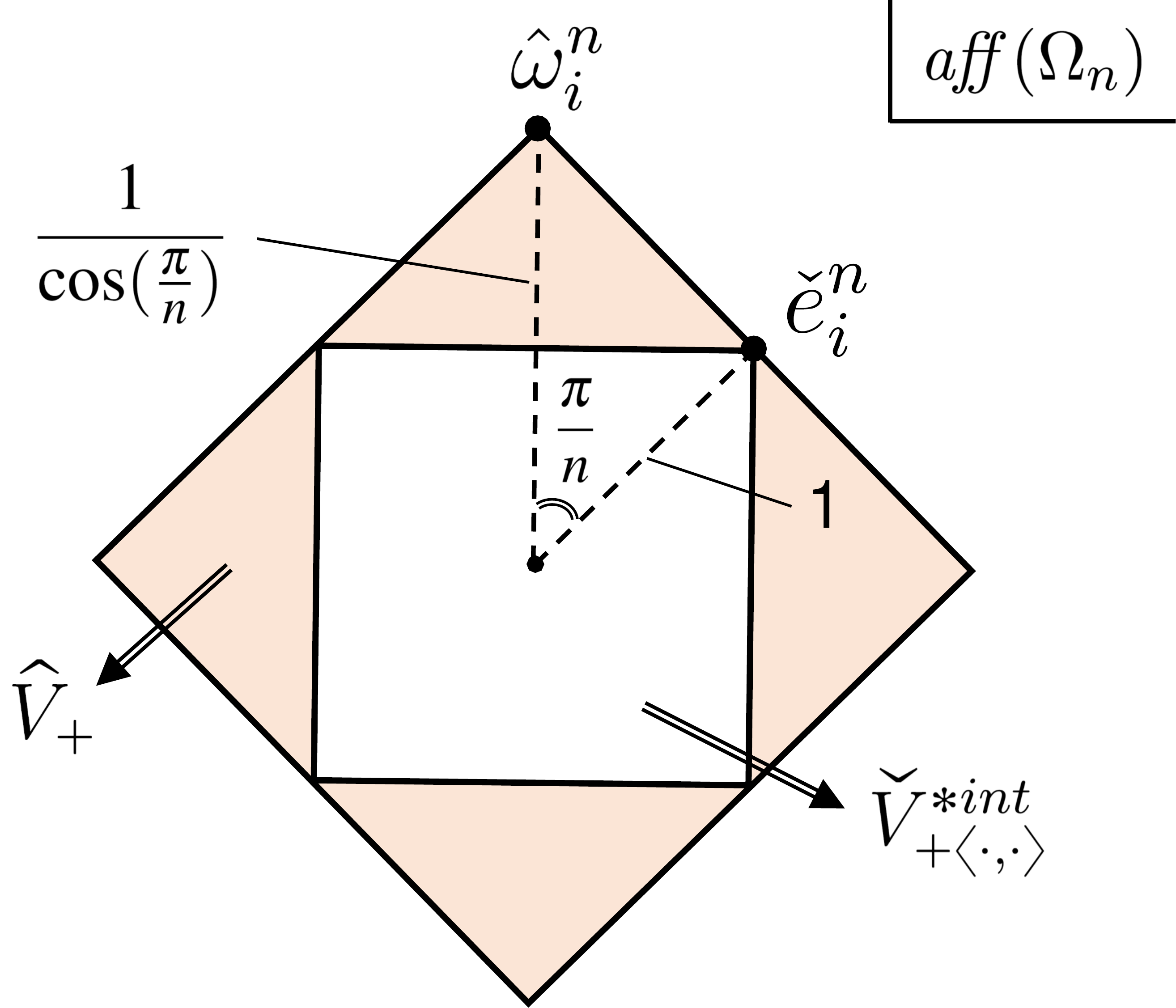}
		\caption{Illustration of $(\mathit{aff}(\Omega_{n})\cap\widehat{V}_{+})=\widehat{\Omega}_{n}$ generated by $\{\hat{\omega}_{i}^{n}\}_{i=1}^{n}$ \eqref{def_polygon pure state2}
			and $(\mathit{aff}(\Omega_{n})\cap\widecheck{V}^{*int}_{+\langle\cdot ,\cdot \rangle})$ generated by $\{2\check{e}_{i}^{n}\}_{i=1}^{n}$ \eqref{def_polygon pure effect2}
			for $n=4$.
			It is observed that
			$\widecheck{V}^{*int}_{+\langle\cdot ,\cdot \rangle}\subset \widehat{V}_{+}$, which holds also for every even $n$.
		}
		\label{fig_1}
	\end{figure}\\
	Therefore,
	\begin{equation}
	\label{eq_pf1}
	\frac{\check{e}}{\langle u,\check{e}\rangle}\in\widehat{\Omega}_{n}
	\end{equation}
	holds for any effect $\check{e}\in \widecheck{V}^{*int}_{+\langle\cdot ,\cdot \rangle}$.
	It follows from \eqref{eq_pf3} and \eqref{eq_pf1} that the claim of Lemma \ref{lem_eigenstates} holds also for even-sided regular polygon theories in a rewritten expression \eqref{def_polygon pure state2} and \eqref{def_polygon pure effect2}.

	We also need to confirm that all of our measures \eqref{def_overall width}, \eqref{def_localization error}, \eqref{def_error bar width}, \eqref{def_Werner's measure}, \eqref{def_l distance}, \eqref{def:Shannon entropy}, and \eqref{def:entropic MN} depend only on probabilities, and thus they are invariant for the modification above. For example, for a pair of observables $M=\{m_{a}\}_{a}$ and $F=\{f_{a}\}_{a}$ on the original state space $\Omega_{n}$, we can see easily from \eqref{def_localization error} and \eqref{def_l distance} that
	\begin{align*}
	LE(\omega^{F})
	&=1-\underset{a\in A}{\max}\ f_{a}(\omega)\\
	&=1-\underset{a\in A}{\max}\ \check{f}_{a}(\hat{\omega})\\
	&=LE(\hat{\omega}^{\widecheck{F}})
	\end{align*}
	and
	\begin{align*}
	D_{\infty}(M, F)
	&=\underset{\omega\in\Omega_{n}}{\sup}\ \underset{a\in A}{\max}\left|m_{a}(\omega)-f_{a}(\omega)\right|\\
	&=\underset{\hat{\omega}\in\widehat{\Omega}_{n}}{\sup}\ \underset{a\in A}{\max}\left|\check{m}_{a}(\hat{\omega})-\check{f}_{a}(\hat{\omega})\right|\\
	&=D_{\infty}(\widecheck{M}, \widecheck{F})
	\end{align*}
	respectively. 
	It results in that if Theorem \ref{thm_l-distance} holds in the modified theory, then it holds also in the original theory. 
	In fact, by virtue of \eqref{eq_pf3} and \eqref{eq_pf1} (the ``generalized version of Lemma \ref{lem_eigenstates}''), we can repeat the same calculations as in Theorem \ref{thm_l-distance}, and obtain a similar result to it in the modified theory. 
	Similar considerations can be adapted also for the other measures, and it proves Theorem \ref{thm_for even polygon}. 
	\qed
\end{pf}

\subsection{Concrete values for Landau-Pollak-type bounds}
\label{3subsec:concrete values for polygon}
In this part, we shall concentrate on the Landau-Pollak-type relation (see \eqref{eq:quantum LP}) for the $n$-sided regular polygon theory of the form
\begin{align}
\label{eq:LP}
\max_{a}f_{a}(\omega)+\max_{b}g_{b}(\omega)\le \Gamma_{F, G}(n)\quad\ ^\forall\omega\in\Omega_n,
\end{align}
where $F=\{f_a\}_a$ and $G=\{g_b\}_b$ are ideal observables as usual, and show a concrete calculation for the bound $\Gamma_{F, G}(n)$ of uncertainty.

Let us focus on the state space $\Omega_{n}$. 
Any nontrivial ideal observable is of the form $\{e_{i}^{n}, u-e_{i}^{n}\}$ (see \eqref{def_polygon pure effect}). 
Note that although $\{e_{i}^{n}\}_{i=0, 1, 2}$ is also an ideal observable when $n=3$ (a classical trit system), we focus only on ideal observables with two outcomes in this subsection.
Thus if we consider a pair of ideal observables $F$ and $G$, then we can suppose that they are binary: $F=F_{i}\equiv\{f_{i}^{0}, f_{i}^{1}\}$ and $G=G_{j}\equiv\{g_{j}^{0}, g_{j}^{1}\}$ with $f_{i}^{0}=e_{i}^{n}$ and $g_{j}^{0}=e_{j}^{n}$ for $i, j\in\{0, 1, \cdots, n-1\}$ (or $i, j\in[0, 2\pi)$ when $n=\infty$). 
On the other hand, it holds that
\begin{equation}
\begin{aligned}
\label{eq:bound for states}
\max_{x=0, 1}f_{i}^{x}(\omega)+\max_{y=0, 1}g_{j}^{y}(\omega)
&\le\sup_{\omega\in\Omega_{n}}\max_{(x, y)\in\{0, 1\}^{2}}[(f_{i}^{x}+g_{j}^{y})(\omega)]\\
&=\max_{\omega\in\Omega_{n}^{\mathrm{ext}}}\max_{(x, y)\in\{0, 1\}^{2}}[(f_{i}^{x}+g_{j}^{y})(\omega)]
\end{aligned}
\end{equation}
because $\Omega_{n}$ is a compact set and any state can be represented as a convex combination of pure states. Therefore, if we let $\omega^{n}_k$ be a pure state (\eqref{def_polygon pure state} and \eqref{def_disc pure state}), then the value
\begin{align}
\label{eq:bound for pure state1}
\gamma^{n}_{F_{i}, G_{j}}:=\max_{k}\max_{(x, y)\in\{0, 1\}^{2}}[(f_{i}^{x}+g_{j}^{y})(\omega^{n}_k)]
\end{align}
gives a Landau-Pollak-type relation
\begin{equation}
\label{eq:LP bound}
\max_{x=0, 1}f_{i}^{x}(\omega)+\max_{y=0, 1}g_{j}^{y}(\omega)\le\gamma^{n}_{A^{i}, B^{j}}\quad\ ^\forall\omega\in\Omega_{n}.
\end{equation}
From this inequality, we can derive, for example, entropic relations
\begin{align}
\label{eq: ent PUR for polygon}
H\left(\omega^{F}\right)+H\left(\omega^G\right)\ge -2\log\frac{\gamma^{n}_{F_{i}, G_{j}}}{2}\quad\ ^\forall\omega\in\Omega_{n}
\end{align}
and
\begin{align}
\label{eq: ent MUR for polygon}
\mathsf{N}(\widetilde{M}^{F};F)+\mathsf{N}(\widetilde{M}^{G};G)\ge-2\log\frac{\gamma^{n}_{F_{i}, G_{j}}}{2}.
\end{align}
\begin{table}[h]
	\centering
	\caption{The value $(f_{i}^{x}+g_{j}^{y})(\omega^{n}_k)$ when $n$ is even.}
		\begin{tabular}{|c||c|}
			\hline
			$x=0, y=0$ & $1+r_{n}^{2}\cos\left[\frac{\theta_{i}+\theta_{j}}{2}-\phi_{k}\right]\cos\left[\frac{\theta_{i}-\theta_{j}}{2}\right]$              \rule[-3.5mm]{0mm}{10.5mm}                    \\ \hline
			$x=1, y=0$ & $1+r_{n}^{2}\sin\left[\frac{\theta_{i}+\theta_{j}}{2}-\phi_{k}\right]\sin\left[\frac{\theta_{i}-\theta_{j}}{2}\right]$                \rule[-3.5mm]{0mm}{10.5mm}                 \\ \hline
			$x=0, y=1$ & ($i \longleftrightarrow j$ in the case of $x=1, y=0$)                                                                          \rule[-3.5mm]{0mm}{10.5mm}    \\ \hline
			$x=1, y=1$ & $1-r_{n}^{2}\cos\left[\frac{\theta_{i}+\theta_{j}}{2}-\phi_{k}\right]\cos\left[\frac{\theta_{i}-\theta_{j}}{2}\right]$\rule[-3.5mm]{0mm}{10.5mm} \\ \hline
			\multicolumn{2}{|c|}{$\theta_{i}=\frac{2i-1}{n}\pi,\ \theta_{j}=\frac{2j-1}{n}\pi,\ \phi_{k}=\frac{2k}{n}\pi\quad(i, j, k=0, 1, \cdots, n-1)$}  \rule[-3.5mm]{-1.3mm}{10.5mm} \\ \hline
		\end{tabular}
	\label{table:even}
\end{table}
\begin{table}[h]
	\centering
	\caption{The value $(f_{i}^{x}+g_{j}^{y})(\omega^{n}_k)$ when $n$ is odd.}
		\begin{tabular}{|c||c|}
			\hline
			$x=0, y=0$ & $\frac{2}{1+r_{n}^{2}}+\frac{2r_{n}^{2}}{1+r_{n}^{2}}\cos\left[\frac{\theta_{i}+\theta_{j}}{2}-\phi_{k}\right]\cos\left[\frac{\theta_{i}-\theta_{j}}{2}\right]$ \rule[-3.5mm]{0mm}{10.5mm}   \\ \hline
			$x=1, y=0$ & $1+\frac{2r_{n}^{2}}{1+r_{n}^{2}}\sin\left[\frac{\theta_{i}+\theta_{j}}{2}-\phi_{k}\right]\sin\left[\frac{\theta_{i}-\theta_{j}}{2}\right]$                    \rule[-3.5mm]{0mm}{10.5mm}    \\ \hline
			$x=0, y=1$ & ($i \longleftrightarrow j$ in the case of $x=1, y=0$)                                                                                                         \rule[-3.5mm]{0mm}{10.5mm}   \\ \hline
			$x=1, y=1$ & $\frac{2r_{n}^{2}}{1+r_{n}^{2}}-\frac{2r_{n}^{2}}{1+r_{n}^{2}}\cos\left[\frac{\theta_{i}+\theta_{j}}{2}-\phi_{k}\right]\cos\left[\frac{\theta_{i}-\theta_{j}}{2}\right]$    \rule[-3.5mm]{0mm}{10.5mm}    \\ \hline
			\multicolumn{2}{|c|}{$\theta_{i}=\frac{2i}{n}\pi,\ \theta_{j}=\frac{2j}{n}\pi,\ \phi_{k}=\frac{2k}{n}\pi\quad(i, j, k=0, 1, \cdots, n-1)$}  \rule[-3.5mm]{-1.3mm}{10.5mm} \\ \hline
		\end{tabular}
	\label{table:odd}
\end{table}
\begin{table}[h]
	\centering
	\caption{The value $(f_{i}^{x}+g_{j}^{y})(\omega^{n}_k)$ when $n$ is $\infty$.}
		\begin{tabular}{|c||c|}
			\hline
			$x=0, y=0$ & $1+\cos\left[\frac{\theta_{i}+\theta_{j}}{2}-\phi_{k}\right]\cos\left[\frac{\theta_{i}-\theta_{j}}{2}\right]$                   \rule[-3.5mm]{0mm}{10.5mm}                       \\ \hline
			$x=1, y=0$ & $1+\sin\left[\frac{\theta_{i}+\theta_{j}}{2}-\phi_{k}\right]\sin\left[\frac{\theta_{i}-\theta_{j}}{2}\right]$                               \rule[-3.5mm]{0mm}{10.5mm}          \\ \hline
			$x=0, y=1$ & ($i \longleftrightarrow j$ in the case of $x=1, y=0$)                                                                                            \rule[-3.5mm]{0mm}{10.5mm}       \\ \hline
			$x=1, y=1$ & $1-\cos\left[\frac{\theta_{i}+\theta_{j}}{2}-\phi_{k}\right]\cos\left[\frac{\theta_{i}-\theta_{j}}{2}\right]$ \rule[-3.5mm]{0mm}{10.5mm}\\ \hline
			\multicolumn{2}{|c|}{$\theta_{i}=i,\ \theta_{j}=j,\ \phi_{k}=k\quad(0\le i, j, k<2\pi)$}  \rule[-3.5mm]{-1.3mm}{10.5mm} \\ \hline
		\end{tabular}
	\label{table:infty}
\end{table}\\
Table \ref{table:even} - Table \ref{table:infty} show the value of $(f_{i}^{x}+g_{j}^{y})(\omega^{n}_k)$ in terms of the angles $\theta_{i}$, $\theta_{j}$, and $\phi_{k}$ between the $x$-axis and the effects $f_{i}^{0}=e^{n}_i$, $g_{j}^{0}=e^{n}_{j}$, and the state $\omega^{n}_k$ respectively when viewed from the $z$-axis (see \eqref{def_polygon pure state} - \eqref{def_polygon pure effect} in Subsection \ref{eg_polygon}). 
Maximizing the values in those tables over all pure states, we can obtain the optimal bound $\gamma^{n}_{F_{i}, G_{j}}$ in \eqref{eq:bound for pure state1} for each regular polygon theory. Note that focusing only on the case where $j=0$ and $0<i<\frac{n}{2}$ ($0<i<\pi$ when $n=\infty$) is sufficient for the universal description of $\gamma^{n}_{F_{i}, G_{j}}$ due to the geometric symmetry of the regular polygon theories.
$\gamma^{n}_{F_{i}, G_{0}}$ for the regular polygon theory with $n(<\infty)$ sides is exhibited in Table \ref{table:even2} and Table \ref{table:odd2}, 
and $\gamma^{n}_{F_{i}, G_{0}}$ for the disc theory (the regular polygon theory with $n=\infty$ sides) can be calculated from Table \ref{table:infty} as
\begin{equation}
\label{eq: infty bound}
\gamma^{n}_{F_{i}, G_{0}}=\max\left\{1+\cos\frac{\theta'_{i}}{2},\ 1+\sin\frac{\theta'_{i}}{2}\right\},
\end{equation}
where $\theta'_{i}=\theta_{i}-\theta_{0}=\theta_{i}$ similarly to Table \ref{table:even2} and Table \ref{table:odd2}. 
\eqref{eq: infty bound} can be regarded as giving the quantum bound in \eqref{eq:quantum LP} for a qubit system in terms of the usual Bloch representation. 
Note that when $n$ is even or $\infty$, due to the geometric symmetry, $(f_{i}^{x}+g_{j}^{y})(\omega^{n}_k)$ takes its maximum where $\omega^{n}_k$ lies just ``halfway" between the effects $f_{i}^{x}$ and $g_{j}^{y}$, that is, $f_{i}^{x}(\omega)=g_{j}^{y}(\omega)$ and thus $f_{i}^{x}(\omega)g_{j}^{y}(\omega)=\frac{1}{4}(f_{i}^{x}(\omega)+g_{j}^{y}(\omega))^{2}$ holds (see Remark \ref{rmk:majorization}), while this does not hold generally when $n$ is odd.
From Table \ref{table:even2}, Table \ref{table:odd2} and \eqref{eq: infty bound}, we can obtain the corresponding entropic inequalities \eqref{eq: ent PUR for polygon} (also \eqref{eq:entropic PUR via mj}) and \eqref{eq: ent MUR for polygon} for an arbitrary regular polygon theory.
We should recall that the value $\gamma^{n}_{F_{i}, G_{0}}$ can be used also to evaluate the nonlocality of the theory via its degree of incompatibility (see Example \ref{eg_deg of inc}).
\begin{table}[h]
	\centering
	\caption{The value $\gamma^{n}_{F_{i}, G_{0}}$ when $n$ is even.}
		\begin{tabular}{|c||c|}
			\hline
			$n\equiv0$ (mod 4), $i$: even & $\max\left\{1+\cos\frac{\theta'_{i}}{2},\ 1+\sin\frac{\theta'_{i}}{2}\right\}$ \rule[-3.5mm]{0mm}{10.5mm} \\ \hline
			$n\equiv0$ (mod 4), $i$: odd\ \  & $\max\left\{1+r_{n}^{2}\cos\frac{\theta'_{i}}{2},\ 1+r_{n}^{2}\sin\frac{\theta'_{i}}{2}\right\}$
			\rule[-3.5mm]{0mm}{10.5mm} \\ \hline
			$n\equiv2$ (mod 4), $i$: even & $\max\left\{1+\cos\frac{\theta'_{i}}{2},\ 1+r_{n}^{2}\sin\frac{\theta'_{i}}{2}\right\}$ \rule[-3.5mm]{0mm}{10.5mm} \\ \hline
			$n\equiv2$ (mod 4), $i$: odd\ \   & $\max\left\{1+r_{n}^{2}\cos\frac{\theta'_{i}}{2},\ 1+\sin\frac{\theta'_{i}}{2}\right\}$ 
			\rule[-3.5mm]{0mm}{10.5mm} \\ \hline
			\multicolumn{2}{|c|}{$\theta'_{i}=\frac{2i}{n}\pi=\theta_{i}-\theta_{0}$}  \rule[-3.5mm]{-1.3mm}{10.5mm} \\ \hline
		\end{tabular}
	\label{table:even2}
\end{table}

\begin{table}[h]
	\centering
	\caption{The value $\gamma^{n}_{F_{i}, G_{0}}$ when $n$ is odd.}
		\begin{tabular}{|c||c|}
			\hline
			$i$: even & 
			$\max\left\{
			\frac{2r_{n}^{2}}{1+r_{n}^{2}}+\frac{2}{1+r_{n}^{2}}\cos\frac{\theta'_{i}}{2},\ 
			1+\frac{1}{\cos\frac{\pi}{2n}}\sin\frac{\theta'_{i}}{2}
			\right\}$ \rule[-3.5mm]{0mm}{10.5mm} \\ \hline
			$i$: odd\ \   & $\max\left\{
			\frac{2r_{n}^{2}}{1+r_{n}^{2}}+\frac{2r_{n}^{2}}{1+r_{n}^{2}}\cos\frac{\theta'_{i}}{2},\ 
			1+\frac{1}{\cos\frac{\pi}{2n}}\sin\frac{\theta'_{i}}{2}
			\right\}$ 
			\rule[-3.5mm]{0mm}{10.5mm} \\ \hline
			\multicolumn{2}{|c|}{$\theta'_{i}=\frac{2i}{n}\pi=\theta_{i}-\theta_{0}=\theta_{i}$}  \rule[-3.5mm]{-1.3mm}{10.5mm} \\ \hline
		\end{tabular}
	\label{table:odd2}
\end{table}
\begin{rmk}
	\label{rmk}
	With the angle $\theta'_{i}$ 
	fixed, we can see from Table \ref{table:even2}, Table \ref{table:odd2}, and \eqref{eq: infty bound} that $\gamma^{n}_{F_{i}, G_{0}}\ge\gamma^{\infty}_{F_{i}, G_{0}}$ holds for all $n$.
	In fact, if we assume, for example, $n$ is odd and $i$ is even, then 
	\[
	\gamma^{n}_{F_{i}, G_{0}}=\max\left\{
	\frac{2r_{n}^{2}}{1+r_{n}^{2}}+\frac{2}{1+r_{n}^{2}}\cos\frac{\theta'_{i}}{2},\ 
	1+\frac{1}{\cos\frac{\pi}{2n}}\sin\frac{\theta'_{i}}{2}
	\right\}
	\]
	(see Table \ref{table:odd2}), and it can be easily shown that
	\begin{align*}
	\frac{2r_{n}^{2}}{1+r_{n}^{2}}+\frac{2}{1+r_{n}^{2}}\cos\frac{\theta'_{i}}{2}
	&\ge
	1+\cos\frac{\theta'_{i}}{2},
	\\
	1+\frac{1}{\cos\frac{\pi}{2n}}\sin\frac{\theta'_{i}}{2}
	&\ge
	1+\sin\frac{\theta'_{i}}{2}
	\end{align*}
	hold for $0<i<\frac{n}{2}$ (or $0<\theta'_{i}<\frac{\pi}{2}$).
	\begin{figure}[h]
		\centering
		\includegraphics[bb=0.000000 0.000000 923.000000 501.000000, scale=0.413]{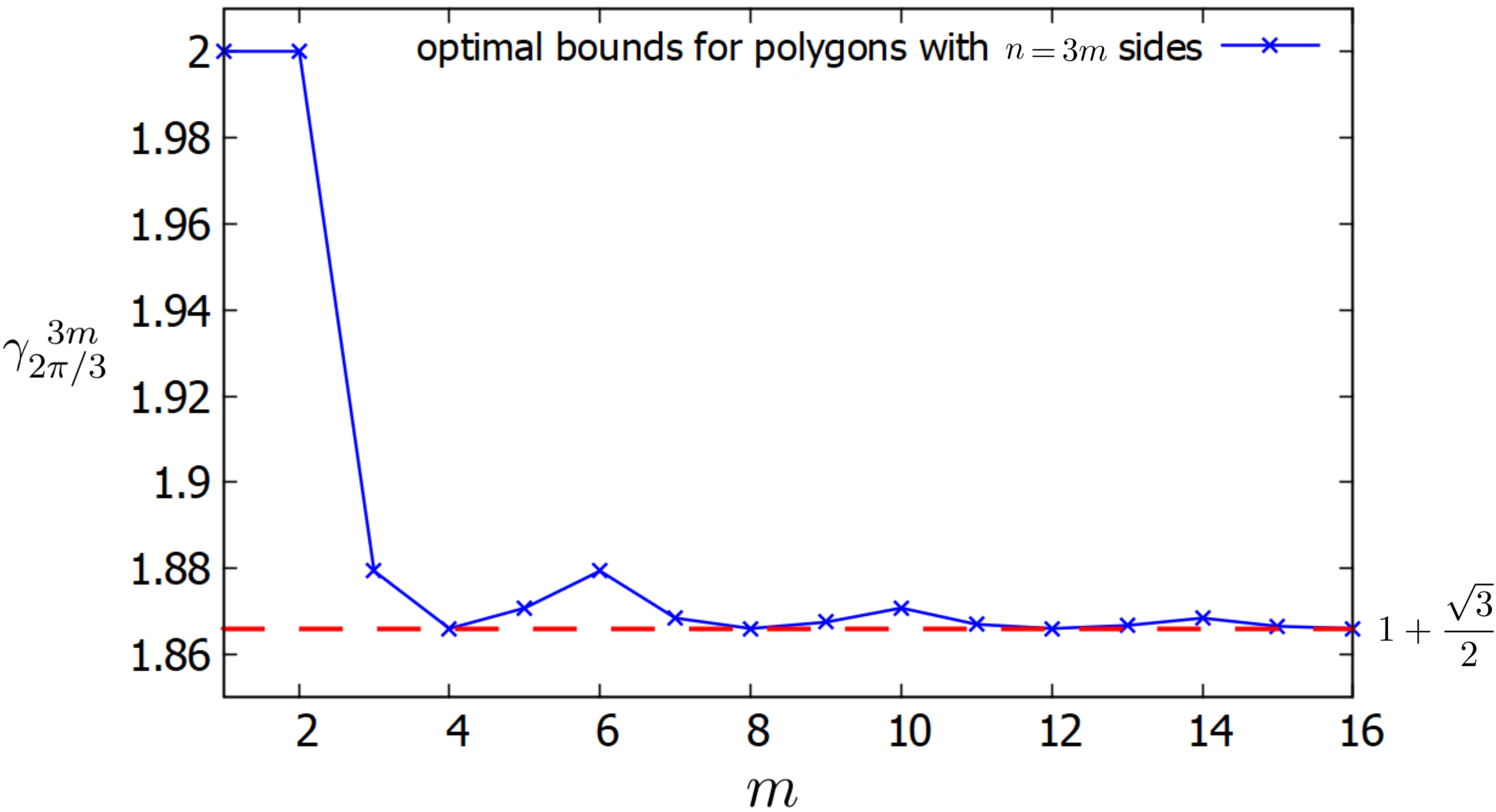}
		\caption{The optimal bound $\gamma^{\ 3m}_{2\pi/3}$ for the Landau-Pollak-type inequality on a pair of observables $(F_{m}, G_{0})$ in the regular polygon theory with $n=3m$.}
		\label{fig_11}
	\end{figure}
	Thus we can conclude 
	$\gamma^{n}_{F_{i}, G_{0}}\ge\gamma^{\infty}_{F_{i}, G_{0}}$.
	
	To see this in a more explicit way, let us consider, as an illustration, regular polygon theories with $n=3m$ ($m=1, 2, \cdots$), and let the angle $\theta'_{i}$ be $\theta'_{i}=\frac{2\pi}{3}$ (i.e. $i=m$). 
	We can calculate the corresponding optimal bound $\gamma^{\ n}_{2\pi/3}\equiv\gamma^{\ 3m}_{2\pi/3}$ for any $m$ from Table \ref{table:even2}, Table \ref{table:odd2} and \eqref{eq: infty bound}, and describe its behavior as a function of $m$ in Figure \ref{fig_11}. 
	There can be observed that theories with $m=1, 2$ ($n=3, 6$) admit $\gamma^{\ 3m}_{2\pi/3}=2$, that is, there is a state on which both $F_{i}=F_{m}$ and $G_{0}$ take simultaneously exact values when $m=1, 2$.
	It exhibits that when $m\ge3$, there exists preparation uncertainty for this $(F_{i}, G_{0})$.
	Hence it follows from our theorems that there also exists measurement uncertainty for $(F_{i}, G_{0})$, and their entropic representations (entropic PUR and MUR) are given by similar inequalities with the same bound.
	Also, it can be observed that $\gamma^{\ 3m}_{2\pi/3}\ge\gamma^{\ \infty}_{2\pi/3}=1+\frac{\sqrt{3}}{2}$ holds for all $m$, which has been shown in the argument above.
	Note that we can derive easily an observable-independent relation
	\[
	\min_{i}\gamma^{n}_{F_{i}, G_{0}}\ge\min_{i}\gamma^{\infty}_{F_{i}, G_{0}}.
	\]
	In other words, the disc theory shows the ``maximum uncertainty'' in terms of the Landau-Pollak-type formulation.
\end{rmk}

%% file: chap4.tex
\chapter{Testing incompatibility of quantum devices with few states}
\label{chap:incomp dim}
Quantum information processing, including the exciting fields of quantum communication and quantum computation, is ultimately based on the fact that there are new types of resources that can be utilized in carefully designed information processing protocols. 
The best-known feature of quantum information is that quantum systems can be in superposition and entangled states, and these resources lead to applications such as superdense coding and quantum teleportation.
While superposition and entanglement are attributes of quantum states,  quantum measurements have also features that can power a new type of applications. 
The best known and most studied property is the incompatibility of pairs (or collections) of quantum measurements \cite{Heinosaari_2016}.
It is crucial e.g. in the BB84 quantum key distribution protocol \cite{BB84} that the used measurements are incompatible.

From the resource perspective, it is important to 
quantify the incompatibility. 
There have been several studies on incompatibility robustness, i.e., how incompatibility is affected by noise.
This is motivated by the fact that noise is unavoidable in any actual implementation of quantum devices and similar to other quantum properties (e.g. entanglement), large amount of noise destroys incompatibility.  
Earlier studies have mostly focused on quantifying noise \cite{Designolle_2019} and finding those pairs or collections of measurements that are most robust to certain types of noise \cite{HEINOSAARI20141695}, or to find conditions under which all incompatibility is completely erased \cite{incomp_break_2015}.
In this work, we introduce quantifications of incompatibility which are motivated by an operational aspect of testing whether a collection of devices is incompatible or not. 
We focus on two integer valued quantifications of incompatibility, called \emph{compatibility dimension} and \emph{incompatibilility dimension}. 
We formulate these concepts for arbitrary collections of devices. 
Roughly speaking, the first one quantifies how many states we minimally need to use to detect incompatibility if we choose the test states carefully, whereas the second one quantifies how many (affinely independent) states we may have to use if we cannot control their choice.
We study some of the basic properties of these quantifications of incompatibility and we present several examples to demonstrate their behaviour.

This part is organized as follows.
In Section \ref{4sec:S0-compatibility}, we introduce the notion of compatibility and incompatibility dimension, which reflects operationally how easy it is to detect the incompatibility of quantum devices considered. 
We also give brief reviews on related studies recently reported in \cite{PhysRevA.100.042308,Kiukas_2020,doi:10.1063/5.0028658,PhysRevA.103.022203} for the case of quantum observables, and explain the interconnections of these studies to ours.
In Section \ref{4sec:witness}, we show that incompatibility dimension is related with the concept of {\itshape incompatibility witness} \cite{doi:10.1063/1.5126496,PhysRevLett.122.130402,PhysRevA.98.012133}.
We also derive a useful bound for incompatibility dimension by means from the relation between them.
In Section \ref{4sec:qubit}, we give a particular analysis for compatibility and incompatibility dimensions of a pair of mutually unbiased qubit observables.
We show that, remarkably, even for the standard example of noisy orthogonal qubit observables the incompatibility dimension has a jump in a point where all noise robustness measures are continuous and indicate nothing special to happen.
More precisely, the noise parameter has a threshold value where the number of needed test states to reveal incompatibility shifts from 2 to 3. 
This means that even in this simple class of incompatible pairs of qubit observables there is a qualitative difference in the incompatibility of less noisy and more noisy pairs of observables. 
An interesting additional fact is that the compatibility dimension of these pairs of observables does not depend on the noise parameter.

For simplicity and clarity, we will restrict to finite-dimensional Hilbert spaces and observables with a finite number of outcomes. 
Our definitions apply not only to quantum theory but also to any GPT.
However, for the sake of concreteness, we keep the discussion in the realm of quantum theory. 
The main definitions work in any GPT without any changes.
We expect that similar findings as the aforementioned result on noisy orthogonal qubit observables can be made in subsequent studies on other collections of devices.

\section{(In)compatibility on a subset of states}
\label{4sec:S0-compatibility}
In this section, we introduce the notion of incompatibility dimension and compatibility dimension as quantifications of incompatibility.
We again mention that we focus on compatibility and incompatibility in quantum theory in this chapter, but those quantities can be defined naturally also in GPTs.

\subsection{(In)compatibility for quantum devices}
We start with presenting explicit descriptions of compatibility and incompatibility for quantum observables, although we have already given their definitions in the general framework of GPTs (see Definition \ref{2def:compatibility} and Proposition \ref{2prop:compatibility for obs}). 
A quantum observable is mathematically described as a positive operator valued measure (POVM) \cite{heinosaari_ziman_2011}. 
A quantum observable with finite number of outcomes is hence a map $x \mapsto \A(x)$ from the outcome set to the set of linear operators on a Hilbert space.
The compatibility of quantum observables $\A_1,\ldots,\A_n$ with outcome sets $X_1,\ldots,X_n$ means that there exists an observable $\G$, called \emph{joint observable}, defined on the product outcome set $X_1 \times \cdots \times X_n$ such that from an outcome $(x_1,\ldots,x_n)$ of $\G$, one can infer outcomes for every $\A_1,\ldots,\A_n$ by ignoring the other outcomes.
More precisely, the requirement is that
\begin{equation}\label{eq:pp}
	\begin{split}
		& \A_1(x_1) = \sum_{x_2,\ldots,x_n} \G(x_1,x_2,\ldots,x_n), \\
		& \A_2(x_2) = \sum_{x_1,x_3\ldots,x_n} \G(x_1,x_2,\ldots,x_n), \\
		& \quad\vdots \\
		& \A_n(x_n) = \sum_{x_1,\ldots,x_{n-1}} \G(x_1,x_2,\ldots,x_n). \\
	\end{split}
\end{equation}
If $\A_1,\ldots,\A_n$ are not compatible, then they are called \emph{incompatible}.

\begin{eg}(\emph{Unbiased qubit observables})\label{ex:qubit-1}
	We recall a standard example to fix the notation that we will use in later examples.
	An unbiased qubit observable is a dichotomic observable with outcomes $\pm$ and
	determined by a vector $\va\in\R^3$, $|\va| \leq 1$ via 
	\begin{equation*}
		\A^{\va}(\pm) = \half ( \id \pm \va \cdot \vsigma ) \,,
	\end{equation*}
	where $\va \cdot \vsigma = a_1\sigma_1 + a_2\sigma_2 + a_3\sigma_3$ and $\sigma_i$, $i=1,2,3$, are the Pauli matrices.
	The Euclidean norm $|\va|$ of $\va$ reflects the noise in $\A^{\va}$; in the extreme case of $|\va|=1$ the operators $\A^{\va}(\pm)$ are projections and the observable is called sharp.
	As shown in \cite{PhysRevD.33.2253}, two unbiased qubit observables $\A^{\va}$ and $\A^{\vb}$ are compatible if and only if 
	\begin{equation}\label{eq:paul}
		|\va + \vb| + |\va - \vb| \leq 2 \, .
	\end{equation}
	There are two extreme cases.
	Firstly, if $\A^{\va}$ is sharp then it is compatible with some $\A^{\vb}$ if and only if $\vb=r\va$ for some $-1\leq r \leq 1$.
	Secondly, if $|\va|=0$, then $\A^{\va}(\pm) = \half \id$ and it is called a trivial qubit observable, in which case it is compatible with all other qubit observables. 
\end{eg}

How can we test if a given family of observables is compatible or incompatible?
From the operational point of view, the existence of 
an observable $\G$ satisfying (\ref{eq:pp}) is equivalent to 
the existence of $\G$ such that for any state $\varrho$ the equation
\begin{equation}\label{eq:naive}
	\Tr[\varrho \A_1(x_1)]=  \sum_{x_2,\ldots,x_n} \Tr[\varrho \G(x_1,x_2,\ldots,x_n)]
\end{equation}
holds.
Before contemplating into these questions, we recall that analogous definitions of quantum compatibility and incompatibility make sense for other types of quantum devices, in particular, for instruments and channels \cite{Heinosaari_2016,HeMiRe2014,Haapasalo_2015,Heinosaari_2017,PhysRevA.97.022112, doi:10.1063/1.5008300, Haapasalo2019}. 
We denote by $\state(\hi)$ the set of all density operators on a Hilbert space $\hi$.
The input space of all types of devices must be $\state(\hin)$ on the same Hilbert space $\hin$
as the devices operate on a same system.
We denote $\state(\hin)$ simply by $\state$.
A device is a completely positive map and the ``type'' of the device is characterized by its output space.
Output spaces for the three basic types of devices are:
\begin{itemize}
	\item observable: $P(X):=\{p=\{p(x)\}_{x\in X}\mid0 \leq p(x)\leq 1,\ 
	\sum_{x}p(x)=1\}$,
	\item channel: $\mathcal{S}(\hi_{out})$, 
	\item instrument: $\mathcal{S}(\hi_{out})\otimes 
	P(X)$. 
\end{itemize}
In this classification, an observable $\A$ is identified with a map $\varrho\mapsto\Tr[\varrho \A(x)]$ from $\state(\hin)$ to $P(X)$. 
We limit our investigation to the cases where the number of outcomes in $X$ is finite and the output Hilbert space $\hi_{out}$ is finite-dimensional.
Regarding $P(X) \subset \state(\C^{|X|})$ as the set of all diagonal density operators, we can summarize that quantum devices are normalized completely positive maps to different type of output spaces.  

Devices $\D_1, \ldots, \D_n$ are \emph{compatible} if there exists a device $\D$ that can simulate $\D_1, \ldots, \D_n$ simultaneously, meaning that by ignoring disjoint parts of the output of $\D$ we get the same actions as $\D_1, \ldots, \D_n$ (see \cite{Heinosaari_2016}).  
This kind of device is called a \emph{joint device} of $\D_1, \ldots, \D_n$. 
The input space of $\D$ is the same as for $\D_1, \ldots, \D_n$, but the output space is the tensor product of their output spaces.
As an illustration, let $\D_j\colon\state(\hi_{in})\to\state(\hi_j)$ $(j=1, \ldots, n)$ be quantum channels. 
They are compatible iff there exists a channel $\D\colon\state(\hi_{in})\to\state(\bigotimes_{j=1}^{n}\hi_j)$ satisfying
\begin{equation*}
	\begin{split}
		& \D_1(\varrho) = \mathrm{tr}_{\hi_2,\ldots,\hi_n} \D(\varrho), \\
		& \D_2(\varrho) = \mathrm{tr}_{\hi_1, \hi_3, \ldots,\hi_n} \D(\varrho), \\
		& \quad\vdots \\
		& \D_n(\varrho) = \mathrm{tr}_{\hi_1,\ldots,\hi_{n-1}} \D(\varrho) \\
	\end{split}
\end{equation*}
for all $\varrho\in\state(\hi_{in})$ (see \eqref{eq:pp}).
If $\D_1, \ldots, \D_n$ are not compatible, then they are \emph{incompatible}.
We recall a qubit example to exemplify the general definition. 
\begin{eg}(\emph{Unbiased qubit observable and partially depolarizing noise})\label{ex:pauli}
	A measurement of an unbiased qubit observable $\A^{\va}$ necessarily disturbs the system.
	This trade-off is mathematically described by the compatibility relation between observables and channels. 
	Let us consider partially depolarizing qubit channels, which have the form
	\begin{equation}
		\Gamma_{p}(\varrho) = p \varrho + (1-p) \tfrac{1}{2} \id 
	\end{equation}
	for $0 \leq p \leq 1$.
	A joint device for a channel and observable is an instrument. 
	Hence, $\A^{\va}$ and $\Gamma_p$ are compatible if there exists an instrument $x \mapsto \Phi_x$ such that
	\begin{equation*}
		\sum_x \Phi_x(\varrho) = \Gamma_p(\varrho) \quad \textrm{and} \quad \Tr[\Phi_x(\varrho)] = \Tr[\varrho \A^{\va}(x)]
	\end{equation*}
	for all states $\varrho$ and outcomes $x$.
	It has been proven in \cite{PhysRevA.97.022112} that $\A^{\va}$ and $\Gamma_p$ are compatible if and only if 
	\begin{equation}
		|\va | \leq \frac{1}{2} \left( 1- p + \sqrt{(1-p)(1+3p)} \right)  \, .
	\end{equation}
	This shows that higher is the norm $|\va |$, smaller must $p$ be.
\end{eg}

\subsection{(In)compatibility dimension of devices}
To test the incompatibility we should hence check the validity of \eqref{eq:naive} in a subset of states that spans the whole state space.
An obvious question is then if we really need all those states, or if a smaller number of test states is enough.
Further, does the number of needed test states depend on the given family of observables? How does noise affect the number of needed test states?
The earlier discussion motivates the following definition, which is central to our investigation. 
\begin{defi}\label{def:comp}
	Let $\state_0\subset\state$.
	Devices $\D_1,\ldots,\D_n$ are \emph{$\state_0$-compatible} if there exist compatible devices $\D'_1,\ldots,\D'_n$ of the same type such that
	\begin{equation}\label{eq:comp-same}
		\D'_j(\varrho) = \D_j(\varrho)
	\end{equation}
	for all $j=1,\ldots,n$ and states $\varrho\in\state_0$.
	Otherwise, $\D_1,\ldots,\D_n$ are \emph{$\state_0$-incompatible}.
\end{defi}
The definition is obviously interesting only when $\D_1,\ldots,\D_n$ are incompatible in the usual sense, i.e., with respect to the full state space.
In that case the definition means that if devices $\D_1,\ldots,\D_n$ are $\state_0$-compatible, their incompatibility cannot be verified by taking test states from $\state_0$ only, and vice versa, if devices $\D_1,\ldots,\D_n$ are $\state_0$-incompatible, their actions on $\state_0$ cannot be simulated by any collection of compatible devices and therefore their incompatibility should be able to be observed in some way.

The $\state_0$-compatibility depends not only on the size of $\state_0$ but also on its structure.
We start with a simple example showing that 
there exist sets $\state_0$ such that 
an arbitrary family of devices is $\state_0$-compatible. 
\begin{eg}\label{ex:distinguishable}
	Any set of devices $\D_1,\ldots, \D_n$ is $\state_0$-compatible if 
	$\state_0=\{\varrho_1, \ldots, \varrho_k\}$ 
	consists of perfectly distinguishable states. 
	In fact, one may construct a device 
	$\D_k'$ which outputs $\D_k(\varrho_j)$ after 
	confirming an input state is $\varrho_j$ by measuring 
	an observable that distinguishes the states in $\state_0$. 
	It is easy to see that the devices $\D_1', \ldots, \D_n'$ are compatible. 
	The same argument works for devices in general probabilistic theories and one can use the same reasoning for a subset $\state_0$ that is broadcastable \cite{PhysRevLett.99.240501}.
	(We recall that a subset $\state_0$ is broadcastable if there exists a channel $B:\state \to \state\otimes\state$ such that the bipartite state $B(\varrho)$ has marginals equal to $\varrho$ for all $\varrho\in\state_0$.) 
	For instance, two qubit states $\id/2$ and $|0\rangle \langle 0|$ are broadcastable even though not distinguishable.
	Any pair of qubit channels $\Lambda_1$ and $\Lambda_2$ is $\state_0$-compatible for $\state_0=\{\id/2,|0\rangle \langle 0| \}$ as we can define 
	$\Lambda'_j(\varrho) = \sum_{i=0}^1 \langle i | 
	\varrho |i\rangle \Lambda_j(|i\rangle\langle i |)$ for $j=1,2$.
	The channel $\Lambda'_j$ has clearly the same action as $\Lambda_j$ on $\state_0$.
	A joint channel $\Lambda$ for $\Lambda'_1$ and $\Lambda'_2$ is given as
	\begin{align*}
		\Lambda(\varrho) =  \sum_{i=0}^1 \langle i | 
		\varrho |i\rangle \, \Lambda_1(|i\rangle\langle i |) \otimes  \Lambda_2(|i\rangle\langle i |),
	\end{align*}
	and it is clear that, in fact, $\pTr{2}{\Lambda(\varrho)}=\Lambda_1(\varrho)$ and $\pTr{1}{\Lambda(\varrho)}=\Lambda_2(\varrho)$.
\end{eg}

For a subset $\state_0 \subset \state$, we denote by $\bar{\state}_0$ the intersection of the linear hull of $\state_0$ with $\state$, i.e.,
\begin{align*}
	\bar{\state}_0 = \{ \varrho \in \state \mid \textrm{$\varrho = \sum_{i=1}^l c_i \varrho_i$ for some $c_i \in \C$ and $\varrho_i\in\state_0$} \}
\end{align*}
In this definition we can assume without restriction that $c_i\in\R$ and $\sum_i c_i = 1$ as they follow from the positivity and unit-trace of states.
Since the condition \eqref{eq:comp-same} is linear in $\varrho$, we conclude that devices $\D_1,\ldots,\D_n$ are $\state_0$-compatible if and only if they are $\bar{\state}_0$-compatible.
This makes sense: if we can simulate the action of devices for states in $\state_0$, we can simply calculate the action for all states that are linear combinations of those states.
This observation also shows that a reasonable way to quantify the size of a subset $\state_0$ for the task in question is the number of affinely independent states. 
We consider the following questions. 
Given a collection of incompatible devices $\D_1,\ldots,\D_n$,
\begin{itemize}
	\item[(a)] what is the smallest subset $\state_0$ such that $\D_1,\ldots,\D_n$ are $\state_0$-incompatible?
	\item[(b)] what is the largest subset $\state_0$ such that $\D_1,\ldots,\D_n$ are $\state_0$-compatible?
\end{itemize}
Smallest and largest here mean the number of affinely independent states in $\bar{\state}_0$.
It agrees with the linear dimension of the linear hull of $\state_0$, or $\mathrm{dim}\mathit{aff}\state_{0}+1$, where $\mathrm{dim}\mathit{aff}\state_{0}$ is the affine dimension of the affine hull $\mathit{aff}\state_{0}$ of $\state_0$ \cite{Rockafellar+2015,boyd_vandenberghe_2004}.
The answer to (a) quantifies how many states we need to use to detect incompatibility if we choose them carefully, whereas the answer to (b) quantifies how many (affinely independent) states we may have to use if we cannot control their choice. 
Hence for both of these quantities lower number means more incompatibility in the sense of easier detection.
The precise mathematical definitions read as follows.

\begin{defi}
	For a collection of incompatible devices $\D_1,\ldots,\D_n$, we denote
	\begin{align*}
		\chi_{incomp}(\D_1,\ldots,\D_n)=\min_{\state_{0}\subset\state}\{\mathrm{dim}\mathit{aff}\state_{0}+1
		\mid\mbox{$\D_1,\ldots,\D_n$: $\state_{0}$-incompatible}\}
	\end{align*}
	and
	\begin{align*}
		\chi_{comp}(\D_1,\ldots,\D_n)=\max_{\state_{0}\subset\state}\{\mathrm{dim}\mathit{aff}\state_{0}+1
		\mid\mbox{$\D_1,\ldots,\D_n$: $\state_{0}$-compatible}\}.
	\end{align*}
	We call these numbers the \emph{incompatibility dimension} and \emph{compatibility dimension} of $\D_1,\ldots,\D_n$ respectively.
\end{defi}
From Example \ref{ex:distinguishable} and the fact that the linear dimension of the linear hull of $\state$ is $d^2$ we conclude that
\begin{equation}\label{eq:incomp-bounds}
	2 \leq \chi_{incomp}(\D_1,\ldots,\D_n) \leq d^2 
\end{equation}
and
\begin{equation}\label{eq:comp-bounds}
	d \leq \chi_{comp}(\D_1,\ldots,\D_n) \leq d^2-1 \, .
\end{equation}
Further, from the definitions of these quantities it directly follows that
\begin{equation}
	\chi_{incomp}(\D_1,\ldots,\D_n) \leq \chi_{comp}(\D_1,\ldots,\D_n) +1 \, .
\end{equation}
We note that based on their definitions, both $\chi_{incomp}$ and $\chi_{comp}$ are expected to be smaller for collections of devices that are more incompatible.
The following monotonicity property of $\chi_{incomp}$ and $\chi_{comp}$ under pre-processing is a basic property that any quantification of incompatibility is expected to satisfy.
\begin{prop}
	Let $\Lambda:\state\to\state$ be a quantum channel and let $\widetilde{\D}_j$ be a pre-processing of $\D_j$ with $\Lambda$ for each $j=1,\ldots,n$, i.e., $\widetilde{\D}_j(\varrho)=\D_j(\Lambda(\varrho)))$.
	If $\widetilde{\D}_j$'s are 
	incompatible, then also $\D_j$'s are incompatible and 
	\begin{equation}
		\chi_{incomp}(\widetilde{\D}_1,\ldots,\widetilde{\D}_n) \geq \chi_{incomp}(\D_1,\ldots,\D_n) 
	\end{equation}
	and
	\begin{equation}
		\chi_{comp}(\widetilde{\D}_1,\ldots,\widetilde{\D}_n) \geq \chi_{comp}(\D_1,\ldots,\D_n) \, .
	\end{equation}
\end{prop}

\begin{pf}
	Suppose that $\D_1, \ldots, \D_n$ are $\state_0$-compatible for some subset $\state_0$. 
	Let $\D'$ be a device that gives  devices $\D'_1, \ldots, \D'_n$ as marginals and these marginals satisfy \eqref{eq:comp-same} in $\state_0$.
	Then the pre-processing of $\D'$ with $\Lambda$ gives $\widetilde{\D}_1,\ldots,\widetilde{\D}_n$ as marginals in $\state_0$.
	The claimed inequalities then follow.\qed
\end{pf}
The post-processing map of a device $\D$ depends on type of the device. 
For instance, the output set of an observable is $P(X)$ and post-processing is then described as a stochastic matrix \cite{Martens1990}.
We formulate and prove the following monotonicity property of $\chi_{incomp}$ and $\chi_{comp}$ under post-processing only for observables. The formulation is analogous for other types of devices.
\begin{prop}
	\label{prop:post-processing}
	Let $\widetilde{\A}_j$ be a post-processing of $\A_j$ (i.e. $\widetilde{\A}_j(x')=\sum_x \nu_j(x',x)\A_j(x)$ for some stochastic matrix $\nu_j$) for each $j=1,\ldots,n$.
	If $\widetilde{\A}_j$'s are $\state_0$-incompatible, then also $\A_j$'s are $\state_0$-incompatible and  
	\begin{equation}
		\chi_{incomp}(\widetilde{\A}_1,\ldots,\widetilde{\A}_n) \geq \chi_{incomp}(\A_1,\ldots,\A_n) 
	\end{equation}
	and
	\begin{equation}
		\chi_{comp}(\widetilde{\A}_1,\ldots,\widetilde{\A}_n) \geq \chi_{comp}(\A_1,\ldots,\A_n) \, .
	\end{equation}
\end{prop}

\begin{pf}
	Suppose that $\A_1, \ldots, \A_n$ are 
	$\state_0$-compatible for some subset $\state_0$. 
	This means that there exists an observable 
	$\G$ satisfying for all $\varrho\in\state_0$, any $j$ and $x_j$, 
	\begin{equation}\label{eq:pp-1}
		\Tr[\varrho \A_j(x_j)]
		= \sum_{l\neq j} \sum_{x_l} 
		\Tr[\varrho \G(x_1, \ldots, x_n)] \, .
	\end{equation}
	We define an observable 
	$\widetilde{\G}$ 
	\[
	\widetilde{\G}(x'_1, \ldots, x'_n)
	= \sum_{x_1, \ldots, x_n}
	\nu(x'_1|x_1)\cdots \nu(x'_n|x_n)
	\G(x_1, \ldots, x_n),
	\] 
	and it then satisfies 
	\begin{eqnarray}
		\Tr[\varrho \widetilde{\A}_j(x'_j)]
		= \sum_{l \neq j} \sum_{x'_l}
		\Tr[\varrho \widetilde{\G}(x'_1, \ldots, x'_n)]
	\end{eqnarray}
	for all $\varrho\in\state_0$, any $j$ and $x'_j$.
	This shows that $\widetilde{\A}_1,\ldots,\widetilde{\A}_n$ are $\state_0$-compatible.
	The claimed inequalities then follow.\qed
\end{pf}
We will now have some examples to demonstrate the values of $\chi_{incomp}$ and $\chi_{comp}$ in some standard cases.
\begin{eg}\label{ex:identity}
	Let us consider the identity channel $\mbox{id}: \mathcal{S}(\mathbb{C}^d)
	\to \mathcal{S}(\mathbb{C}^d)$. It follows from the definitions that two identity channels are $\state_0$-compatible if and only if $\state_0$ is a broadcastable set. 
	It is known that a subset of states is broadcastable only if the states commute with each other \cite{PhysRevLett.76.2818}, and for this reason the pair of two identity channels is $\state_0$-incompatible whenever $\state_0$ contains two noncommuting states. 
	Therefore, we have $\chi_{incomp}(\mbox{id}, \mbox{id}) =2$.
	On the other hand, $\state_0$ consisting of 
	distinguishable states makes the identity channels
	$\state_0$-compatible. 
	As $\state_0$ consisting of commutative states 
	has at most $d$ affinely independent states, 
	we conclude that $\chi_{comp}(\mbox{id}, \mbox{id})=d$. 
\end{eg}
A comparison of the results of Example \ref{ex:identity} to the bounds \eqref{eq:incomp-bounds} and \eqref{eq:comp-bounds} shows that the pair of identity channels has the smallest possible incompatibility and compatibility dimensions. 
This is quite expectable as that pair is consider to be the most incompatible pair - any device can be post-processed from the identity channel. 
Perhaps surprisingly, the lower bound of $\chi_{incomp}$ can be attained already with a pair of dichotomic observables; this is shown in the next example.

\begin{eg}\label{ex:pq}
	Let $P$ and $Q$ be two noncommuting one-dimensional projections in a $d$-dimensional Hilbert space $\hi$.
	We define two dichotomic observables $\A$ and $\B$
	as 
	\begin{align*}
		\A(1)=P \, , \A(0)=\id-P \, , \quad \B(1)=Q \, , \B(0)=\id-Q \, .
	\end{align*} 
	Let us then consider a subset consisting of two states, 
	$$
	\state_0=\{\varrho^P, \varrho^Q\}:=
	\{\tfrac{1}{d-1}(\id-P), \tfrac{1}{d-1}(\id-Q)\} \, .
	$$
	We find that the dichotomic observables $\A$ and $\B$ are $\state_0$-incompatible. 
	To see this, let us make a counter assumption that  $\A$ and $\B$ are $\state_0$-compatible, in which case there exists $\G$ such that the marginal condition \eqref{eq:naive} holds for both observables and for all $\varrho\in\state_0$.
	We have $\Tr[\varrho^P \A(1)]=0$ and therefore
	\begin{align*}
		0=\Tr[(\id-P) \G(1,1)] = \Tr[(\id-P) \G(1,0)].
	\end{align*}
	It follows that $\G(1,1)=\alpha P$ and $\G(1,0) =\beta P$.
	Further, $\Tr[P \A(1)] =1$ and hence $\alpha+\beta=1$.
	In a similar way we obtain $\G(1,1)=\gamma Q$ and $\G(0,1) =\delta Q$ with $\gamma+\delta=1$.
	It follows that $\alpha=\gamma=0$ and $\beta=\delta=1$.
	But $\G(1,0) + \G(0,1) = P + Q$ contradicts $\G(1,0) + \G(0,1) \leq \id$.
	Thus we conclude $\chi_{incomp}(\A, \B)
	=2$. 
\end{eg}
For two incompatible sharp qubit observables (Example \ref{ex:qubit-1}) the previous example gives a concrete subset of two states such that the observables are incompatible and proves that $\chi_{incomp}(\A^{\va},\A^{\vb})=2$ for such a pair.
The incompatibility dimension for unsharp qubit observables is more complicated and will be treated in Section \ref{4sec:qubit}.
\begin{eg}\label{ex:fix}
	Let us consider two observables $\A$ and $\B$.
	Fix a state $\varrho_0\in\state$ and define 
	\begin{equation*}
		\state_0 = \{ \varrho\in\state : \Tr[\varrho\A(x)] = \Tr[\varrho_0\A(x)] \  \forall x \} \, .
	\end{equation*}
	Then $\A$ and $\B$ are $\state_0$-compatible. 
	To see this, we define an observable $\G$ as
	\begin{equation*}
		\G(x,y) =  \Tr[\varrho_0\A(x)] \B(y) \, .
	\end{equation*}
	It is then straightforward to verify that \eqref{eq:naive} holds for all $\varrho \in \state_0$.
	As a special instance of this construction, let $\A^{\va}$ be a qubit observable and $\va\neq 0$ (see Example \ref{ex:qubit-1}).
	We choose $\state_0=\{ \varrho\in\state\mid \Tr[\varrho\A^{\va}(+)]=\half \}$.
	We then have $\state_0 = \{ \half(\id + \vr\cdot\vsigma)\mid \vr\cdot\va=0 \}$ and hence
	$\mathrm{dim}\mathit{aff}\state_0=2$.
	Based on the previous argument, $\A^{\va}$ is $\state_0$-compatible with any $\A^{\vb}$.
	Therefore, $\chi_{comp}(\A^{\va},\A^{\vb})=3$ for all incompatible qubit observables $\A^{\va}$ and $\A^{\vb}$.
\end{eg}

\subsection{Remarks on other formulations of incompatibility dimension}

The notion of $\state_0$-compatibility for quantum observables has  been introduced in \cite{PhysRevA.100.042308} and in that particular case (i.e. quantum observables) it is equivalent to Definition \ref{def:comp}. 
In the current investigation, our focus is on the largest or smallest $\state_0$ on which devices $\D_1,\ldots,\D_n$ are compatible or incompatible, and this has some differences to the earlier approaches. In \cite{doi:10.1063/5.0028658}, the term ``compatibility dimension" was introduced and
for observables $\A_1,\ldots,\A_n$ on a $d$-dimensional Hilbert space $\hi=\C^{d}$: it is given by
\begin{align*}
	R(\A_1,\ldots,\A_n)&=\max\{r\le d\mid
	\exists V\colon\C^{r}\to\C^{d}\ isometry\ \\
	&\qquad\qquad\qquad s.t.\  V^{*}\A_1V, ,\ldots,V^{*}\A_n V\  are\  compatible
	\},
\end{align*}
Evaluations of $R(\A_1,\ldots,\A_n)$ in various cases such as $n=2$ and $\A_{1}$ and $\A_{2}$ are rank-1 were presented in \cite{doi:10.1063/5.0028658}.
To describe it in our notions, let us denote $\C^{r}$ by $\hik$, and define $\state_{\hi}$ and $\state_{\hik}$ as the set of all density operator on $\hi$ and $\hik$ respectively. 
We also introduce $\state_{V\hik}$ as
\begin{align*}
	\state_{V\hik}:=\{\varrho\in\state\mid \mathrm{supp}\varrho\subset V\hik\}=
	V\state_{\hik}V^{*}\subset\state_{\hi}.
\end{align*}
Then we can see that the $\state_{\hik}$-compatibility of $V^{*}\A_1V, ,\ldots,V^{*}\A_n V$ is equivalent to the $\state_{V\hik}$-compatibility of $\A_1,\ldots,\A_n$.
Therefore, if we focus only on sets of states such as $\state_{V\hik}$ (i.e. states with fixed support), then there is no essential difference between our compatibility dimension and the previous one: $R(\A_1,\ldots,\A_n)=r$ iff $\chi_{comp}(\A_1,\ldots,\A_n)=r^{2}$. 
In \cite{doi:10.1063/5.0028658} also the concept of ``strong compatibility dimension" was defined as
\begin{align*}
	\overline{R}(\A_1,\ldots,\A_n)=&\max\{r\le d\mid
	\forall V\colon\C^{r}\to\C^{d}\ isometry\ \\
	&\qquad\qquad\qquad s.t.\  V^{*}\A_1V, ,\ldots,V^{*}\A_n V\  are\  compatible
	\}.
\end{align*}
It is related to our notion of incompatibility dimension.
In fact, if we only admit sets of states such as $\state_{V\hik}$, then $\overline{R}(\A_1,\ldots,\A_n)$ and $\chi_{incomp}(\A_1,\ldots,\A_n)$ are essentially the same: $\overline{R}(\A_1,\ldots,\A_n)=r$ iff $\chi_{incomp}(\A_1,\ldots,\A_n)=(r+1)^{2}$.

Similar notions have been introduced and investigated also in \cite{Kiukas_2020,PhysRevA.103.022203}.
As in \cite{doi:10.1063/5.0028658}, these works focus on quantum observables and on subsets of states that are lower dimensional subspaces of the original state space.
Therefore, the notions are not directly applicable in GPTs.
In \cite{PhysRevA.103.022203} incompatibility is classified into three types.
They are explained exactly in terms of \cite{doi:10.1063/5.0028658} as\\
(i) incompressive incompatibility: $(\A_1,\ldots,\A_n)$ are $\state_{V\hik}$-compatible for all $\hik$ and $V$\\
(ii) fully compressive incompatibility: $(\A_1,\ldots,\A_n)$ are $\state_{V\hik}$-incompatible for all nontrivial $\hik$ and $V$\\
(iii) partly compressive incompatibility: there is a $V$ and $\hik$ such that $(\A_1,\ldots,\A_n)$ are $\state_{V\hik}$-compatible, and some $V'$ and $\hik'$ such that $(\A_1,\ldots,\A_n)$ are $\state_{V'\hik'}$-incompatible.\\
In \cite{PhysRevA.103.022203} concrete constructions of these three types of incompatible observables were given.

\section{Incompatibility dimension and incompatibility witness}
\label{4sec:witness}
In this section we show how the notion of incompatibility dimension is related to the notion of incompatibility witness.
\subsection{Relation between incompatibility dimension and incompatibility witness for observables}
An \emph{incompatibility witness} is an affine functional $\xi$ defined on $n$-tuples of observables such that $\xi$ takes non-negative values on all compatible $n$-tuples and a negative value at least for some incompatible $n$-tuple \cite{doi:10.1063/1.5126496,PhysRevLett.122.130402,PhysRevA.98.012133}.
Every incompatibility witness $\xi$ is of the form 
\begin{align}
	\label{eq:standard-form0}
	\xi(\oplus_{j=1}^{n} \A_j)=\delta - 
	f(\oplus_{j=1}^{n} \A_j),
\end{align}
where $\delta\in \mathbb{R}$ and $f$ is a linear functional on $\oplus_{j=1}^n \mathcal{L}_s(\hi)^{m_j}$ with $\mathcal{L}_s(\hi)$ being the set of all 
self-adjoint operators on $\hi$ and $m_j$ the number of outcomes of $\A_j$.
It can be written also in the form
\begin{equation}
	\label{eq:standard-form}
	\xi(\A_1,\ldots,\A_n) = \delta - \sum_{j=1}^n \sum_{x_j=1}^{m_j} c_{j,x_j} \Tr[\varrho_{j,x_j}\A_j(x_j)],
\end{equation}
where $c_{j,x_j}$'s are real numbers, and $\varrho_{j,x_j}$'s are states.
This result has been proven in \cite{PhysRevLett.122.130402} for incompatibility witnesses acting on pairs of observables and the generalization to $n$-tuples is straightforward.
A witness $\xi$ \emph{detects} the incompatibility of observables $\A_1,\ldots,\A_n$ if $\xi(\A_1,\ldots,\A_n) <0$.
The following proposition gives a simple relation between incompatibility dimension and incompatibility witness.
\begin{prop}
	\label{prop:dim-witness}
	Assume that an incompatibility witness $\xi$ has the form \eqref{eq:standard-form} and it detects the incompatibility of observables $\A_1,\ldots,\A_n$.
	Then $\A_1,\ldots,\A_n$ are $\state_0$-incompatible for $\state_0 = \{ \varrho_{j,x_j}\mid j=1,\ldots,n, x_j=1,\ldots,m_j\}$.
\end{prop}
\begin{pf}
	Let $\A_1,\ldots,\A_n$ be $\state_0$-compatible.
	Then we have compatible observables $\widetilde{\A}_1,\ldots,\widetilde{\A}_n$ such that $\Tr[\varrho\A_j(x_j)]=\Tr[\varrho\widetilde{\A}_j(x_j)]$ for all $\varrho\in\state_0$.
	This implies that 
	\begin{equation*}
		\xi(\A_1,\ldots,\A_n) = \xi(\widetilde{\A}_1,\ldots,\widetilde{\A}_n) \geq 0 \, , 
	\end{equation*}
	which contradicts the assumption that $\xi$ detects the incompatibility of observables $\A_1,\ldots,\A_n$.\qed
\end{pf}
It has been shown in \cite{PhysRevLett.122.130402} that any incompatible pair of observables is detected by some incompatibility witness of the form \eqref{eq:standard-form}.
The proof is straightforward to generalize to $n$-tuples of observables, and thus, together with Proposition \ref{prop:dim-witness}, we can obtain
\begin{equation}
	\label{eq:upper bound}
	\chi_{incomp}(\A_1,\ldots,\A_n) \leq m_1 + \cdots + m_n.
\end{equation}
That is, the incompatibility dimension of $\A_1,\ldots,\A_n$ can be evaluated via their incompatibility witness (we will derive a better upper bound later in this section).
We can further prove the following proposition.

\begin{prop}
	\label{prop:dim-witness0}
	The statements (i) and (ii) for a set of incompatible observables $\{\A_1, \ldots, \A_n\}$ are equivalent:
	\begin{itemize}
		\item[(i)]$\chi_{incomp}(\A_1, \ldots, \A_n)\leq N$ 
		\item[(ii)]There exist a family of linearly independent 
		states $\{\varrho_1, \ldots, \varrho_N\}$ and real numbers $\delta$ and $\{c_{l,j,x_j}\}_{l, j, x_{j}}$ 
		$(l=1,\ldots, N, j=1,\ldots,n, x_j=1,\ldots,m_j)$ such that the incompatibility witness $\xi$ defined by
		\begin{align*}
			\xi(\B_1, \ldots, \B_n)
			=\delta- \sum_{l=1}^{N}\sum_{j=1}^{n} \sum_{x_j=1}^{m_j} 
			c_{l,j,x_j}\mbox{tr}[\varrho_l \B_j(x_j)]
		\end{align*}
		detects the incompatibility of $\{\A_1, \ldots, \A_n\}$.
	\end{itemize}
\end{prop}
The claim $\mathit{(i)}\Rightarrow\mathit{(ii)}$ may be regarded as the converse of the previous argument to obtain \eqref{eq:upper bound}.
It manifests that we can find an incompatibility witness detecting the incompatibility of $\{\A_1, \ldots, \A_n\}$ reflecting their incompatibility dimension.
\begin{pf}
	$\mathit{(ii)}\Rightarrow\mathit{(i)}$ can be proven in the same way as Proposition \ref{prop:dim-witness}.
	Thus we focus on proving $\mathit{(i)}\Rightarrow\mathit{(ii)}$.
Suppose that a family of observables $\{\A_1, \ldots, \A_n\}$ satisfies $\chi_{incomp}(\A_1, \ldots, \A_n)= N$.
	Then there exists a family of linearly independent states $\{\varrho_1, \varrho_2, \ldots, \varrho_N\}$ in $\mathcal{L}_s(\hi)$ on which $\{\A_1, \ldots, \A_n\}$ are incompatible.
	We can regard the family $\{\A_1, \ldots, \A_n\}$ as an element of a vector space $\mathcal{L}$ defined as $\mathcal{L}:=\oplus_{j=1}^n \mathcal{L}_s(\hi)^{m_j}$, that is, $\A:=\oplus_{j=1}^n\A_j\in \mathcal{L}$.
	For each $l=1,\ldots,N$, $j=1, \ldots, n$, and $x_j=1, \ldots, m_j$, let us define a subset $K(\A, \varrho_l, j,x_j)$ of 
	$\mathcal{L}$ as  
	\begin{align}
		\label{eq:def of K}
		\begin{aligned}
			K(\A, \varrho_l, j,x_j):=\{\B\in\mathcal{L}\mid\langle \varrho_l | \B_j(x_j)\rangle_{HS}
			= \langle \varrho_l | \A_j(x_j)\rangle_{HS}\}, 
		\end{aligned} 
	\end{align}
	where $\langle \varrho_l | \A_j(x_j)\rangle_{HS}
	:=\mbox{tr}[\varrho_l\A_j(x_j)]$ is the Hilbert-Schmidt inner 
	product on $\mathcal{L}_s(\hi)$.
	Note that this inner product can be naturally extended to an inner product $\langle\langle\cdot | \cdot\rangle\rangle$ on $\mathcal{L}$: 
	\begin{align*}
		\langle\langle\A | \B\rangle\rangle=\sum_{j=1}^{n}\sum_{x_{j}=1}^{m_j}\langle \A_j(x_j)|\B_j(x_j)\rangle_{HS} \, .
	\end{align*}
	Embedding $\varrho_{l}$ into $\mathcal{L}$ by $\hat{\varrho}_l^{j,x_j}=
	\oplus_{i=1}^{n} \oplus_{y=1}^{m_{i}} \delta_{ij} \delta_{yx_j} 
	\varrho_l$ for each $j,x_j$ and $l$, we obtain another representation 
	of \eqref{eq:def of K} as 
	\begin{align}
		\label{eq:def of K2}
		K(\A, \varrho_l, j,x_j)
		=\{\B\mid\langle \langle \hat{\varrho}_l^{j,x}| \B\rangle\rangle
		= \langle \langle \hat{\varrho}_l^{j,x_j}|\A\rangle\rangle\} \, .
	\end{align}
	Thus this set is a hyperplane in $\mathcal{L}$. 
	Note that $\{\hat{\varrho}_l^{j,x}\}_{l,j,x_j}$ is a linearly independent set in $\mathcal{L}$.
	Consider an affine set 
	$K:=\cap_{l=1}^N \cap_{j=1}^{n}\cap_{x_{j}=1}^{m_j} K(\A,\varrho_l,j,x_j)$. 
	Because $\{\A_1, \ldots, \A_n\}$ is incompatible in $\{\varrho_{1},\cdots,\varrho_N\}$, it satisfies 
	\begin{align}
		\label{eq:KC-empty} 
		K\cap C=\emptyset,
	\end{align}
	where $C:=\{\CC\in\mathcal{L}\mid\mbox{$\{\CC_{1},\ldots\CC_n\}$ is compatible} \}$.
	Thus, by the separating hyperplane theorem \cite{Rockafellar+2015}, there exists a hyperplane in $\mathcal{L}$ which separates strongly the (closed) convex sets $K$ and $C$.
	In the following, we will show that one of those separating hyperplanes can be constructed from $\{\hat{\varrho}_l^{j,x}\}_{l,j,x_j}$.

	Let us extend a family of linearly 
	independent vectors 
	$\{\hat{\varrho}_l^{j,x_j}\}_{l, j, x_j}$ to form a basis of $\mathcal{L}$.  
	That is, we introduce a 
	basis 
	$\{v_b\}_{b=1,\ldots, 
		\dim \mathcal{L}}$  of $\mathcal{L}$ satisfying  
	$\{v_a\}_{a=1,\ldots, 
		N (\sum_j m_j)}= \{\hat{\varrho}_l^{j,x_j}\}_{l, j, x_j}$. 
	We introduce its dual basis $\{w_b\}_{
		b=1,2,\ldots, \dim \mathcal{L}}$
	satisfying
	$\langle \langle v_a|
	w_b\rangle \rangle=\delta_{ab}$. 
	Because $K$ can be written as
	\begin{align*}
		K=\{\B\mid
		\langle \langle
		\hat{\varrho}_l^{j,x_j} |(\B-\A)\rangle\rangle
		=0, \forall l, j,x_j \},
	\end{align*}
	it is represented in terms this (dual) basis as 
	\begin{align*}
		K=\A+K_0, 
	\end{align*}
	where $K_0$ is an affine set defined by 
	\begin{align}
		K_0:&=
		\{ \sum_{a=N(\sum_j m_j)+1}^{\dim\mathcal{L}}c_a w_a \mid c_a\in\R\}
		\label{eq:expresison_of_K_0}
	\end{align}
	Now we can construct a hyperplane separating $K$ and $C$.
	To do this, let us focus on the convex sets $K_0$ and $C':=C-\A$ instead of $K$ and $C$, which satisfy
	$K_0\cap C'=\emptyset$ because of \eqref{eq:KC-empty}.
	We can apply the separating hyperplane theorem (Theorem 11.2 in \cite{Rockafellar+2015}) for the affine set $K_0$ and convex set $C'$.
	There exists a hyperplane $H_0$ in $\mathcal{L}$ such that $K_0$ and $C'$ are contained by $H_0$ and one of its associating open half-spaces respectively.
	That is, there exists $h\in\mathcal{L}$ satisfying
	\begin{align*}
		\label{eq:hyperplane2}
		H_0=\{\B\in \mathcal{L}\mid\langle \langle
		\B |h\rangle\rangle=0
		\}
	\end{align*}
	with $K_0\subset H_0$, and $\langle \langle
	\CC' |h\rangle\rangle<0$ for all $\CC'\in C'$.
	Let us examine the vector $h$.
	It satisfies 
	\[
	\langle \langle
	w_{a} |h\rangle\rangle=0\ \ \mbox{for all $a=N(\sum_j m_j)+1,\ldots, \dim\mathcal{L}$}
	\]
	because $K_0\subset H_0$ (see \eqref{eq:expresison_of_K_0}).
	Thus if we write $h$ as $h=\sum_{a=1}^{\dim\mathcal{L}}c_a v_{a}$, then we can find that $c_a=0$ holds for all $a=N(\sum_j m_j)+1,\ldots, \dim\mathcal{L}$.
	It follows that
	\begin{align*}
		h=\sum_{a=1}^{N(\sum_j m_j)}c_a v_a=\sum_l\sum_j\sum_{x_j}c_{l, j, x_j}\hat{\varrho}_l^{j,x_j}
	\end{align*}
	holds, and the hyperplane $H_0$ can be written as
	\begin{align*}
		\label{eq:hyperplane}
		H_0=\{\B\in \mathcal{L}\mid\sum_l\sum_j\sum_{x_j}c_{l, j, x_j}\Tr[\varrho_l \B_j(x_j)]=0
		\}.
	\end{align*}
	Then the hyperplane $H':=\A+H_0$, a translation of $H_0$, of the form
	\[
	H'=\{\B\in \mathcal{L}\mid\sum_l\sum_j\sum_{x_j}c_{l, j, x_j}\Tr[\varrho_l \B_j(x_j)]=\delta'
	\}
	\]
	contains the original sets $K$, and satisfy
	\[
	\sum_l\sum_j\sum_{x_j}c_{l, j, x_j}\Tr[\varrho_l \CC_j(x_j)]<\delta'
	\] 
	for all $\CC\in C$.
	We can displace $H'$ slightly in the direction of $C$ to obtain a hyperplane $H$ defined as
	\[
	H=\{\B\in \mathcal{L}\mid\sum_l\sum_j\sum_{x_j}c_{l, j, x_j}\Tr[\rho_l \B_j(x_j)]=\delta
	\},
	\]
	which (strongly) separates $H'$ (in particular $K$) and $C$ because $H'$ is closed and $C$ is compact (see Corollary 11.4.2 in \cite{Rockafellar+2015}).
	The claim now follows as $\A\in K$.\qed
\end{pf}

\subsection{An upper bound on the incompatibility dimension of observables via incompatibility witness}

We can give a better upper bound than \eqref{eq:upper bound} for the incompatibiliy dimension by slightly modifing the previous argument in \cite{PhysRevLett.122.130402} on incompatibility witness.
\begin{prop}
	\label{prop:upper-bound}
	Let $\A_1,\ldots,\A_n$ be incompatible observables with $m_1,\ldots,m_n$ outcomes, respectively. 
	Then
	\begin{align*}
		\chi_{incomp}(\A_1,\ldots,\A_n) \leq 
		\sum_{j=1}^n m_j -n+1.
	\end{align*}
\end{prop}
\begin{pf}
	We continue following the same notations as the proof of Proposition \ref{prop:dim-witness0}.
	Let us assume that the incompatibility of $\A_1,\ldots,\A_n$ is detected by an incompatibility witness $\xi$.
	The functional $\xi$ is of the form 
	\begin{align*}
		\xi(\A)=\delta - 
		f(\A)
	\end{align*}
	with a real number $\delta$ and a functional $f$ on $\mathcal{L}$ (see \eqref{eq:standard-form0}).
	Then Riesz representation theorem shows that 
	the functional $f$ can be represented as 
	\begin{align*}
		f(\A) 
		&=\sum_{j=1}^n \sum_{x_j}^{m_j}
		\langle F_j(x_j) | \A_j(x_j)\rangle_{HS}
	\end{align*}
	with some $F_j(x_j) \in \mathcal{L}_s(\hi)$ $(j=1, \ldots, n,\ x_j=1, \ldots, m_{j})$.
	If we define $F'_j(x_j) = F_j(x_j) +\epsilon_j \id$, then we find 
	\begin{align*}
		\xi(\A)=\delta + d \sum_j \epsilon_j 
		- \sum_{j=1}^n \sum_{x_j=1}^{m_j} 
		\langle F'_j(x_j) | \A_j(x_j)\rangle_{HS}.  
	\end{align*}
	We choose $\epsilon_j$ so that 
	$$
	\sum_{x_j}\mbox{tr}[F'_j(x_j)] =\sum_{x_j} \langle F'_j (x_j)| \id\rangle_{HS}=0
	$$
	holds.  
	The choice of $\{F'_j(x_j)\}_{j, x_j}$ has still some 
	freedom. 
	Each $F'_j(x_j)$ can be replaced with 
	$F''_j(x_j)=F'_j(x_j) + T_j$, where $T_j
	\in \mathcal{L}_s(\hi)$ satisfies $\mbox{tr}[T_j]
	=\langle T_j |\id\rangle_{HS}=0$. 
	In fact, it holds that
	\begin{align*}
		\sum_{x_j} \langle F''_j(x_j)|\A_j(x_j)\rangle_{HS} 
		&=\sum_{x_j} \langle F'_j (x_j) |\A_j(x_j)\rangle_{HS} 
		+ \sum_{x_j} \langle T_j|\A_j(x_j)\rangle_{HS}\\
		&= \sum_{x_j} \langle F'_j(x_j) | \A_j(x_j)\rangle_{HS} 
		+ \langle T_j |\id\rangle_{HS}\\
		&=\sum_{x_j} \langle F'_j(x_j) | \A_j(x_j)\rangle_{HS}.  
	\end{align*}
	We choose $T_j$ as $m_j T_j = - \sum_{x_j=1}^{m_j} F'_j(x_j)$ which indeed satisfies 
	\[
		m_j \langle T_j | \id\rangle_{HS}= -\sum_{x_j=1}^{m_j} \langle F'_j(x_j)|\id\rangle_{HS}=0, 
	\]
	i.e., $\Tr [T_{j}]=0$,
	to obtain 
	\begin{align*}
		\sum_{x_j} F''_j(x_j) =0. 
	\end{align*}
	We further choose 
	large numbers $\alpha_j \geq 0$ 
	so that $G_j(x_j):=F''_j(x_j) + \alpha_j 
	\id \geq 0$ for all $j$ and $x_j$.
	Now we obtain a representation of 
	the witness which is equivalent to $\xi$ for $n$-tuples of observables
	as 
	\begin{align*}
		\xi^*(\A)
		= \delta + d \sum_j (\epsilon_j +\alpha_j) 
		- \sum_j \sum_{x_j} \langle G_j(x_j) |\A_j(x_j)\rangle_{HS}, 
	\end{align*} 
	where positive operators 
	$G_j(x_j)$'s satisfy $\sum_{x_j} 
	G_j (x_j) = m_j \alpha_j \id$. 
	Defining density operators 
	$\varrho_j(x_j)$ by 
	$\varrho_j(x_j) = \frac{G_j(x_j)}{\mbox{tr}[G_j(x_j)]}$, 
	we obtain yet another representation 
	\begin{align*}
		\xi^*(\A) 
		= 
		\delta + d\sum_j (\epsilon_j + \alpha_j)
		- \sum_j \sum_{x_j} 
		\mbox{tr}[G_j(x_j)] \mbox{tr}[\varrho_j(x_j)
		\A_j(x_j)]
	\end{align*} 
	with $\varrho_j(x_j)$'s satisfying 
	constraints  
	\begin{eqnarray}
		\label{eq:witness constraint}
		\sum_{x_j} \mbox{tr}[G_j(x_j)] \varrho_j(x_j)
		= m_j \alpha_j \id. 
	\end{eqnarray}
	Thus, according to Proposition \ref{prop:dim-witness}, $\A_1,\ldots,\A_n$ are $\state_{0}$-incompatible with $\state_{0}=\{\varrho_j(x_j)\}_{j, x_{j}}$.
	To evaluate $\mathrm{dim}\mathit{aff}\state_{0}$, we focus on the condition \eqref{eq:witness constraint}.
	Introducing parameters $p_j(x_{j}):=\mbox{tr}[G_j(x_j)] /dm_j \alpha_j$ such that $\sum_{x_j}p_j(x_{j})=1$, we obtain
	\[
	\sum_{x_j} p_j(x_{j}) \varrho_j(x_j)
	= \frac{1}{d}\id,
	\]
	or
	\[
	\sum_{x_j} p_j(x_{j}) \tilde{\varrho_j}(x_j)=0,
	\]
	where $\tilde{\varrho_j}(x_j):=\varrho_j(x_j)-\frac{1}{d}\id$.
	It follows that $\{\tilde{\varrho_j}(x_j)\}_{x_{j}}$  are linearly dependent, and thus
	\[
	\mathrm{dim}\mathit{span}\{\tilde{\varrho_j}(x_j)\}_{x_{j}}\le m_{1}-1.
	\]
	Similar arguments for the other $j$'s result in
	\[
	\mathrm{dim}\mathit{span}\{\tilde{\varrho_j}(x_j)\}_{j, x_{j}}\le \sum_{j}(m_{j}-1)=\sum_{j}m_{j}-n.
	\]
	Considering that 
	\[
	\mathrm{dim}\mathit{span}\{\tilde{\varrho_j}(x_j)\}_{j, x_{j}} =\mathrm{dim}\mathit{aff}\{\varrho_j(x_j)\}_{j, x_{j}}
	\]
	holds, we can obtain the claim of the proposition. \qed
\end{pf}
The bound in Proposition \ref{prop:upper-bound} is not tight in general since the right-hand side of the inequality can exceed the bound obtained in \eqref{eq:incomp-bounds}.  
However, for small $n$ and $m_j$'s, the bound can be tight. 
In fact, while for $n=2$ and $m_1=m_2=2$ it gives $\chi_{incomp}(\A_1, \A_2)\leq 3$, we will construct an example which attains this upper bound in the next section.  

\section{(In)compatibility dimension for mutually unbiased qubit observables}
\label{4sec:qubit}

In this section we study the incompatibility dimension of pairs of unbiased qubit observables introduced in Example \ref{ex:qubit-1}.
We concentrate on pairs that are mutually unbiased, i.e., $\Tr[\A^\va(\pm) \A^\vb(\pm)] = 1/2$ (this terminology originates from the fact that if the observables are sharp, then the respective orthonormal bases are mutually unbiased. In the previously written form the definition makes sense also for unsharp observables \cite{PhysRevA.88.032312}).
The condition of mutual unbiasedness is invariant under a global unitary transformation, hence it is enough to fix the basis $\vx=(1,0,0)$, $\vy=(0,1,0)$, $\vz=(0,0,1)$ in $\R^3$ and choose two of these unit vectors. 
We will study the observables $\A^{t\vx}$ and $\A^{t\vy}$, where $0\leq t \leq 1$.
The observables are written explicitly as
\begin{align*}
	\A^{t\vx}(\pm)=
	\frac{1}{2}(\id \pm t \sigma_1) \, , \quad 
	\A^{t\vy}(\pm)=
	\frac{1}{2}(\id \pm t \sigma_2).
\end{align*}
The condition \eqref{eq:paul} shows that $\A^{t\vx}$ and $\A^{t\vy}$ 
are incompatible if and only if $1/\sqrt{2} < t \leq 1$. 
The choice of having mutually unbiased observables as well as using a single noise parameter instead of two is to simplify the calculations. 

We have seen in Example \ref{ex:fix} that $\chi_{comp}(\A^{t\vx},\A^{t\vy})=3$ for all values $t$ for which the pair is incompatible.
We have further seen (discussion after Example \ref{ex:pq}) that $\chi_{incomp}(\A^{\vx},\A^{\vy})=2$, 
and from Prop. \ref{prop:upper-bound} follows that $\chi_{incomp}(\A^{t\vx},\A^{t\vy}) \leq 3$ for all $1/\sqrt{2} < t \leq 1$.
The remaining question is then about the exact value of $\chi_{incomp}(\A^{t\vx},\A^{t\vy})$, which can depend on the noise parameter $t$ and will be in our focus in this section (see Table \ref{table_incomp}). 

\begin{table}[h]
	\centering
	\begin{tabular}{c||c|c}
		& $\chi_{incomp}(\A^{t\vx},\A^{t\vy})$ & $\chi_{comp}(\A^{t\vx},\A^{t\vy})$ \\ \hline
		$t\le\frac{1}{\sqrt{2}}$ & - & - \\ \hline
		$\frac{1}{\sqrt{2}}<t<1$ 
		& \begin{tabular}{c}
			2 or 3\\
			(Proposition \ref{prop:qubit-threshold})
		\end{tabular}	 
		& \multirow{2}{*}{
			\begin{tabular}{c}
				3\\
				(Example \ref{ex:fix})
			\end{tabular}	
		} \\ \cline{1-2}
		$t=1$ 
		& \begin{tabular}{c}
			2\\
			(Example \ref{ex:pq})
		\end{tabular}	 
		&  \\ \hline
	\end{tabular}
	\caption{$\chi_{incomp}$ and $\chi_{comp}$ for $(\A^{t\vx},\A^{t\vy})$ with $0\le t\le1$. For $t\le1/\sqrt{2}$ the observables $\A^{t\vx}$ and $\A^{t\vy}$ are compatible and $\chi_{incomp}$ and $\chi_{comp}$ are not defined.}
	\label{table_incomp}
\end{table}

Let us first make a simple observation that follows from Prop. \ref{prop:post-processing}. 
Considering that $\A^{s\vx}$ is obtained as a post-processing of $\A^{t\vx}$ if and only if $s\leq t$, we conclude that
\begin{align*}
	\chi_{incomp}(\A^{s\vx},\A^{s\vy})=2\ 
	\Rightarrow\ \chi_{incomp}(\A^{t\vx},\A^{t\vy})=2 \quad \textrm{for $\frac{1}{\sqrt{2}}<s \leq t$} \, , 
\end{align*}
and 
\begin{align*}
	\chi_{incomp}(\A^{s' \vx},\A^{s' \vy})=3\ 
	\Rightarrow\ \chi_{incomp}(\A^{t'\vx},\A^{t'\vy})=3 \quad \textrm{for $s'\geq t'>\frac{1}{\sqrt{2}}$} \, .
\end{align*}
Interestingly, there is a threshold value $t_0$ where the value of $\chi_{incomp}(\A^{t\vx},\A^{t\vy})$ changes; this is the content of the following proposition.

\begin{prop}\label{prop:qubit-threshold}
	There exists $1/\sqrt{2}<t_0<1$ such that $\chi_{incomp}(\A^{t\vx},\A^{t\vy})=3$ for $1/\sqrt{2}<t\le t_0$ and $\chi_{incomp}(\A^{t\vx},\A^{t\vy})=2$ for $t_0<t \leq 1$.
\end{prop}
The main line of the lengthy proof of Proposition \ref{prop:qubit-threshold} is the following. 
Defining two subsets $L$ and $M$ of $(\frac{1}{\sqrt{2}}, 1]$ as
\begin{align}
	\label{eq: sets of xi=2,3}
	\begin{aligned}
		L:=\{t\mid\chi_{incomp}(\A^{t\vx},\A^{t\vy})=2\},\ \ 
		M:=\{t\mid\chi_{incomp}(\A^{t\vx},\A^{t\vy})=3\},
	\end{aligned}
\end{align}
we see that
\begin{align}
	\inf L=\sup M(=:t_{0}')
\end{align}
holds unless $L$ and $M$ are empty.
By its definition, the number $t_{0}'$ satisfies
\begin{align*}
	&\chi_{incomp}(\A^{t\vx},\A^{t\vy})=2\ \ \mbox{for\  $t>t_{0}'$},\ \ 
	\chi_{incomp}(\A^{t\vx},\A^{t\vy})=3\ \ \mbox{for\  $t<t_{0}'$}.
\end{align*}
Based on the considerations above, the proof of Proposition \ref{prop:qubit-threshold} proceeds as follows.
First, in Part 1 - 3 (Subsection \ref{4subsec:proof part 1} - \ref{4subsec:proof part 3}), we prove that $M$ is nonempty while $L$ has already been shown to be nonempty as $t=1\in L$.
It will be found that $\chi_{incomp}(\A^{t\vx},\A^{t\vy})=3$ for $t$ sufficiently close to $\frac{1}{\sqrt{2}}$, and thus $t_{0}'$ introduced above can be defined successfully.
Then we demonstrate in Part 4 (Subsection \ref{4subsec:proof part 4}) that $\sup M=\max M$, i.e. $t_{0}'$ is equal to $t_{0}$ in the claim of Prop. \ref{prop:qubit-threshold}.

\begin{rmk}
	In \cite{PhysRevA.100.042308} a similar problem to ours was considered.
	While in that work the focus was on several affine sets, and a threshold value $t_{0}$ was given for each of them by means of their semidefinite programs where observables $\{\A^{t\vx}, \A^{t\vy}, \A^{t\vz}\}$ become compatible, we are considereding \emph{all} affine sets with dimension 2.
\end{rmk}

\subsection{Proof of Proposition \ref{prop:qubit-threshold} : Part 1}
\label{4subsec:proof part 1}
In order to prove that
$M$ is nonempty, let us introduce some relevant notions: 
\begin{align*}
	&D:= \{\mathbf{v}\mid |\mathbf{v}| \leq 1,\ v_z=0\}\subset B:= \{\mathbf{v}\mid|\mathbf{v}| \leq 1\},\\
	&\state_D:= \{\varrho^{\mathbf{v}}\mid
	\mathbf{v} \in D\}\subset\state=\{\varrho^{\mathbf{v}}\mid
	\mathbf{v} \in B\},
\end{align*}
where $\mathbf{v}=v_{\x}\vx+v_{\y}\vy+v_{\z}\vz\in\R^3$, and $\varrho^{\mathbf{v}}:=\frac{1}{2}(\id+\mathbf{v}\cdot\sigma)$.
Since $\state_D$ is a convex set, we can treat $\state_D$ almost like a quantum system.
In the following, we will do it without giving precise definitions because they are obvious. 
For an observable $\E$ on $\mathcal{S}$ with effects $\{\E(x)\}_{x}$, 
we write its restriction 
to $\state_D$ as $\E|_D$ with effects $\{\E(x)|_D\}_{x}$, which is an observable on $\state_D$.
It is easy to obtain the following Lemma. 
\begin{lem}
	The followings are equivalent:
	\begin{itemize}
		\item[(i)]$\A^{t\vx}$ and $\A^{t\vy}$ are 
		incompatible (thus $\frac{1}{\sqrt{2}}<t\leq 1$).
		\item[(ii)] 
		$\A^{t\vx}$ and $\A^{t\vy}$
		are $\state_D$-incompatible.
		\item[(iii)]
		$\A^{t\vx}|_D$ and $\A^{t\vy}|_D$ 
		are incompatible as observables on $\state_D$. 
	\end{itemize}
\end{lem}
\begin{pf}
	(i) $\Rightarrow$ (iii). 
	Suppose that $\A^{t\vx}|_D$ and $\A^{t\vy}|_D$ are compatible in $\state_D$. 
	There exists an observable $\M$ on $\state_D$ 
	whose marginals coincide with 
	$\A^{t\vx}|_D$ and $\A^{t\vy}|_D$. 
	One can extend this $\M$ to 
	the whole $\state$ so that it does not 
	depend on $\z$ (for example, one can simply regard its effect $c_{0}\id+c_{1}\sigma_{1}+c_{2}\sigma_{2}$ as an effect on $\state$). Since both $\A^{t\vx}|_D$ and $\A^{t\vy}|_D$ also do not depend on $\z$, the extension of $\M$ gives a joint observable of $\A^{t\vx}$ and $\A^{t\vx}$. 
	\\
	(iii) $\Rightarrow $ (ii). 
	Suppose that $\A^{t\vx}$ and 
	$\A^{t\vy}$ are $\state_D$-compatible. 
	There exists an observable $\M$ on $\state$ 
	whose marginals coincide with $\A^{t\vx}$ 
	and $\A^{t\vy}$ in $\state_D$. 
	The restriction of $\M$ on $\state_D$ 
	proves that (iii) is false. 
	\\
	(ii) $\Rightarrow $ (i). 
	Suppose that $\A^{t\vx}$ and 
	$\A^{t\vy}$ are compatible,
	then they are $\state_D$-compatible. \qed
\end{pf}
This lemma demonstrates that the incompatibility of $\A^{t\vx}$ and $\A^{t\vy}$ means the incompatibility of $\A^{t\vx}|_{D}$ and $\A^{t\vy}|_{D}$.
We can present further observations.
\begin{lem}
	Let us consider two pure states 
	$\varrho^{\mathbf{r}_1}$ and $\varrho^{\mathbf{r}_2}$ 
	($\mathbf{r}_1, \mathbf{r}_2 \in \partial B$, $\mathbf{r}_1\neq\mathbf{r}_2$), 
	and a convex subset $\mathcal{S}_0$ of $\state$ generated by them:
	$\mathcal{S}_0:= \{p\varrho^{\mathbf{r}_1}
	+ (1-p)\varrho^{\mathbf{r}_2}\mid0 \leq p \leq 1\}$.
	We also introduce an affine projection $P$ by $P\varrho^{\mathbf{v}}=\varrho_{\mathsf{P} \mathbf{v}}$, where $\varrho^{\mathbf{v}} \in \state$ with $\mathbf{v}=v_x \vx
	+ v_y \vy + v_z \vz$ and $\mathsf{P}\mathbf{v}=v_x \vx
	+ v_y \vy$, and extend it affinely. 
	The affine hull of $\mathcal{S}_0$ is projected to $\state_{D}$ as
	\begin{eqnarray}
		P\state_0:= 
		\{\lambda P \varrho^{\mathbf{r}_{1}} + 
		(1-\lambda) P \varrho^{\mathbf{r}_{2}}\mid\lambda \in \mathbf{R}\}
		\cap \state_D. 
	\end{eqnarray}
	If $\A^{t\vx}$ and $\A^{t\vy}$ are $\mathcal{S}_0$-incompatible, 
	then their restrictions
	$\A^{t\vx}|_D$ and $\A^{t\vy}|_D$ are 
	$P\mathcal{S}_0$-incompatible.  
\end{lem}
\begin{pf}
	Suppose that $\A^{t\vx}$ and $\A^{t\vy}$ are $\mathcal{S}_0$-incompatible. 
	It implies $\mathsf{P} \mathbf{r}_{1}\neq\mathsf{P} \mathbf{r}_{2}$, i.e., $P \varrho^{\mathbf{r}_{1}}\neq P \varrho^{\mathbf{r}_{2}}$ (see Example \ref{ex:fix}), and thus $P\state_0$ is a segment in $\state_{D}$.
	If $\A^{t\vx}|_D$ and $\A^{t\vy}|_D$ 
	are $P\state_0$-compatible, then there exists a joint observable $\M$ 
	on $\state_D$ such that 
	its marginals coincide with $\A^{t\vx}|_D$ 
	and $\A^{t\vy}|_D$
	on $P\state_0 \subset \state_D$. 
	This $\M$ can be extended to an observable on
	$\state$ so that the extension does not depend on 
	$\z$.
	Because 
	\begin{align*}
		&\Tr[\A^{t\vx}(\pm)P\varrho^{\mathbf{r}_{1}}]=\Tr[\A^{t\vx}(\pm)\varrho^{\mathbf{r}_{1}}],\\
		&\Tr[\A^{t\vx}(\pm)P\varrho^{\mathbf{r}_{2}}]=\Tr[\A^{t\vx}(\pm)\varrho^{\mathbf{r}_{2}}]
	\end{align*}
	(and their $\vy$-counterparts) hold due to the independence of $\A^{t\vx}(\pm)$ from $\sigma_3$, the marginals of $\M$ coincide with $\A^{t\vx}$ and $\A^{t\vy}$ on $\state_0$.
	It results in the $\mathcal{S}_0$-compatibility of $\A^{t\vx}$ 
	and $\A^{t\vy}$,
	which is a contradiction.\qed
\end{pf}
It follows from this lemma that $\chi_{incomp}(\A^{t\vx}|_{D},\A^{t\vy}|_{D})$ is two when $\chi_{incomp}(\A^{t\vx},\A^{t\vy})$ is two, equivalently $\chi_{incomp}(\A^{t\vx},\A^{t\vy})$ is three when $\chi_{incomp}(\A^{t\vx}|_{D},\A^{t\vy}|_{D})$ is three (remember that $\chi_{incomp}(\A^{t\vx},\A^{t\vy})\le3$).
In fact, the converse also holds.
\begin{lem}
	$\chi_{incomp}(\A^{t\vx}|_{D},\A^{t\vy}|_{D})$ is three
	when $\chi_{incomp}(\A^{t\vx}, \A^{t\vy})$ is three.
\end{lem}
\begin{pf}
	Let $\chi_{incomp}(\A^{t\vx},\A^{t\vy})=3$.
	It follows that $\A^{t\vx}$ and $\A^{t\vy}$ are $S$-compatible for any line $S\subset\state$.
	In particular, $\A^{t\vx}$ and $\A^{t\vy}$ are $S'$-compatible for any line $S'$ in $\state_{D}$, and thus there is an observable $\M$ such that its marginals coincide with $\A^{t\vx}$ and $\A^{t\vy}$ on $S'$.
	It is easy to see that the marginals of $\M|_{D}$ coincide with $\A^{t\vx}|_{D}$ and $\A^{t\vy}|_{D}$ on $S'$, which results in the $S'$-compatibility of $\A^{t\vx}|_{D}$ and $\A^{t\vy}|_{D}$.
	Because $S'$ is arbitrary, we can conclude $\chi_{incomp}(\A^{t\vx}|_{D},\A^{t\vy}|_{D})=3$.\qed
\end{pf}
The lemmas above manifest that if $\A^{t\vx}$ and $\A^{t\vy}$ are incompatible, then $\A^{t\vx}|_{D}$ and $\A^{t\vy}|_{D}$ are also incompatible and 
\[
\chi_{incomp}(\A^{t\vx},\A^{t\vy})=\chi_{incomp}(\A^{t\vx}|_{D},\A^{t\vy}|_{D}).
\]
Therefore, in the following, we denote $\A^{t\vx}|_{D}$ and $\A^{t\vy}|_{D}$ simply by $\A^{t\vx}_{D}$ and $\A^{t\vy}_{D}$ respectively, and focus on the quantity $\chi_{incomp}(\A^{t\vx}_{D},\A^{t\vy}_{D})$ instead of the original $\chi_{incomp}(\A^{t\vx}, \A^{t\vy})$.

Before proceeding to the next step, let us confirm our strategy in the following parts.
In Part 2 (Subsection \ref{4subsec:proof part 2}), we will consider a line (segment) $\state_{1}$ in $\state_{D}$, and consider for $0<t<1$ all pairs of observables $(\widetilde{\A}_{1}^{t}, \widetilde{\A}_{2}^{t})$ on $\state_{D}$ that coincide with $(\A^{t\vx}_{D}, \A^{t\vy}_{D})$ on $\state_{1}$.
Then we will investigate the (in)compatibility of those $\widetilde{\A}_{1}^{t}$ and $\widetilde{\A}_{2}^{t}$ in order to obtain $\chi_{incomp}(\A^{t\vx}_{D},\A^{t\vy}_{D})$ in Part 3 (Subsection \ref{4subsec:proof part 3}). 
It will be shown that when $t$ is sufficiently small, there exists a compatible pair $(\widetilde{\A}_{1}^{t}, \widetilde{\A}_{2}^{t})$ for any $\state_{1}$, that is, $\A^{t\vx}_{D}$ and $\A^{t\vy}_{D}$ are $\state_{1}$-compatible for any line $\state_{1}$.
It results in $\chi_{incomp}(\A^{t\vx}_{D},\A^{t\vy}_{D})=3$, and thus $M\neq \emptyset$.

\subsection{Proof of Proposition \ref{prop:qubit-threshold}
	: Part 2}
\label{4subsec:proof part 2}
\begin{figure}[h]
	\centering
	\includegraphics[bb=0.000000 0.000000 786.000000 630.000000, scale=0.345]{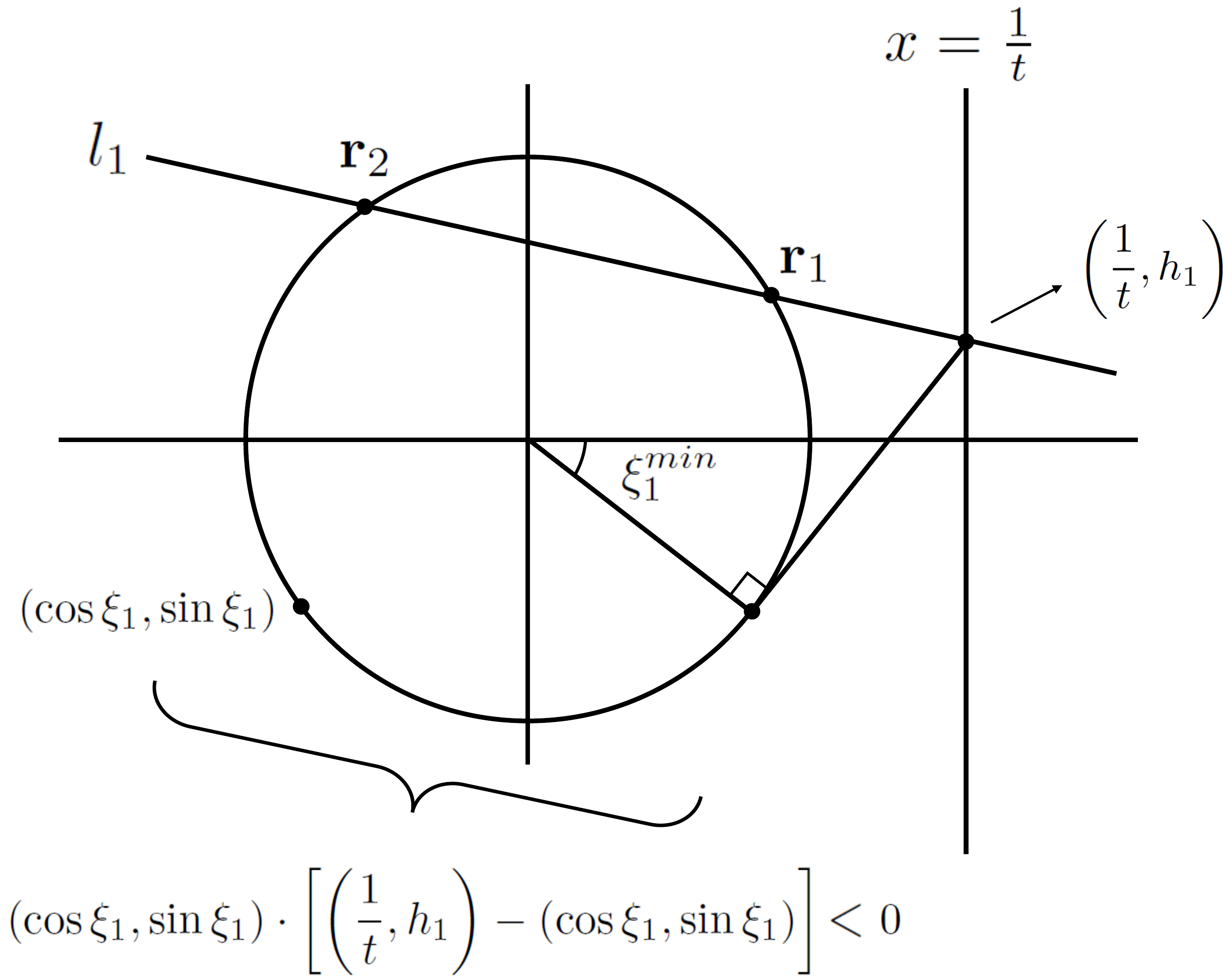}
	\caption{Geometric description of determining $\xi_{1}^{min}$.}
	\label{Fig:xi_min}
\end{figure}

\begin{figure}[h]
	\centering
	\includegraphics[bb=0.000000 0.000000 660.000000 507.000000, scale=0.335]{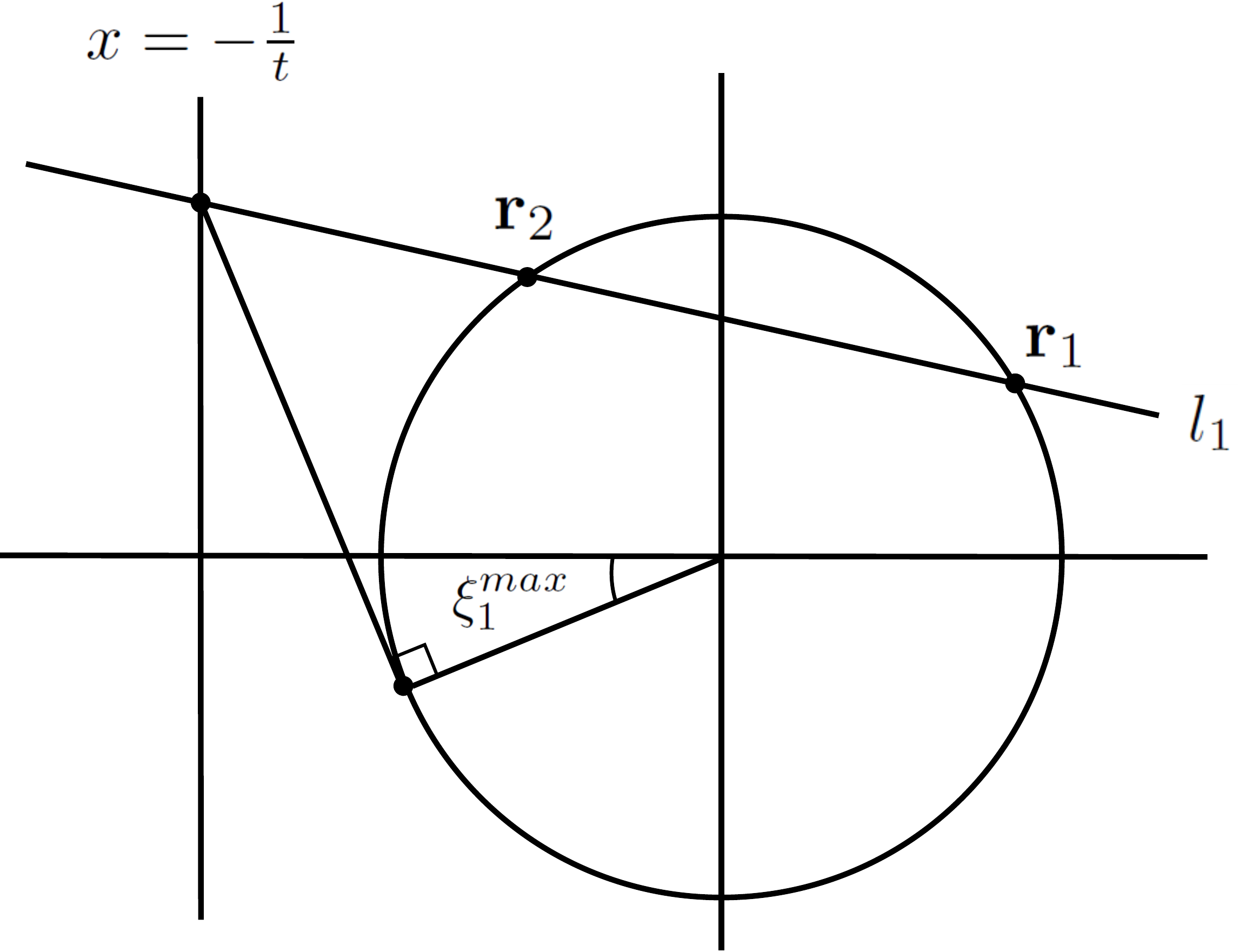}
	\caption{Geometric description of determining $\xi_{1}^{max}$.}
	\label{Fig:xi_max}
\end{figure}

Let us consider two pure states 
$\varrho^{\mathbf{r}_1}$ and $\varrho^{\mathbf{r}_{2}}$ 
with $\mathbf{r}_1, \mathbf{r}_2 \in \partial D$ ($\mathbf{r}_{1}\neq\mathbf{r}_{2}$), 
and a convex set 
$\state_1:= \{p \varrho^{\mathbf{r}_1}
+ (1-p) \varrho^{\mathbf{r}_2}\mid 0 \leq p \leq 1\}$. 
We set parameters $\varphi_1$ and $\varphi_2$ 
as 
\begin{align}
	&\mathbf{r}_1= \cos \varphi_1 \vx
	+ \sin \varphi_1 \vy, \\
	&\mathbf{r}_2 = \cos \varphi_2 \vx 
	+ \sin \varphi_2 \vy, 
\end{align}
where $-\pi\le\varphi_1<\varphi_2<\pi$.
By exchanging $\pm$ properly, without loss of 
generality
we can assume 
the line connecting $\mathbf{r}_1$ and 
$\mathbf{r}_2$ passes through above the origin
(instead of below).
In this case, from geometric consideration, we have 
\begin{equation}
	\label{eq:bound for varphi 0}
	\begin{aligned}
		&0<\varphi_2 - \varphi_1 \leq \pi, \\
		&0 \le \frac{\varphi_1 + \varphi_2}{2} \le \frac{\pi}{2}. 
	\end{aligned}
\end{equation}
Note that when $\varphi_2 - \varphi_1=\pi$, the states $\varrho^{\mathbf{r}_{1}}$ and $\varrho^{\mathbf{r}_{2}}$ are perfectly distinguishable, which results in the $\state_{1}$-compatibility of $\A^{t\vx}_{D}$ and $\A^{t\vy}_{D}$ (see Example \ref{ex:distinguishable}).
On the other hand, when $\frac{\varphi_1 + \varphi_2}{2}=0$ or $\frac{\pi}{2}$,  $\Tr[\varrho\A^{t\vx}_{D}
	(+)]$ or $\Tr[\varrho\A^{t\vy}_{D}
	(+)]$ is constant for $\varrho\in\state_1$ respectively, so $\A^{t\vx}_D$ and $\A^{t\vy}_D$ are $\state_1$-compatible (see Example \ref{ex:fix}). 
Thus, instead of \eqref{eq:bound for varphi 0}, we hereafter assume 
\begin{equation}
	\label{eq:bound for varphi }
	\begin{aligned}
		&0<\frac{\varphi_2 - \varphi_1}{2} < \frac{\pi}{2}, \\
		&0 < \frac{\varphi_1 + \varphi_2}{2} < \frac{\pi}{2}.
	\end{aligned}
\end{equation}
Next, we consider a binary observable $\widetilde{\A}_{1}^{t}$
on $\state_D$ that coincides with $\A^{t\vx}_{D}$ on $\state_1\subset\state_{D}$.
There are many possible $\widetilde{\A}_{1}^{t}$, and each $\widetilde{\A}_{1}^{t}$ is determined completely by its effect $\widetilde{\A}_{1}^{t}(+)$ corresponding to the outcome `+' because it is binary.
The effect $\widetilde{\A}_{1}^{t}(+)$ is associated with a vector $\mathbf{v}_1\in D$ defined as
\begin{eqnarray}
	\mathbf{v}_1:=
	argmax_{\mathbf{v}\in D} \mbox{tr}[\varrho_{\mathbf{v}} 
	\widetilde{\A}_{1}^{t}(+)].
\end{eqnarray} 
Let us introduce a parameter $\xi_1\in[-\pi, \pi)$ by
\begin{eqnarray}
	\label{eq:E1_v1}
	\mathbf{v}_1 = \cos \xi_{1}
	\vx + \sin \xi_{1} \vy,
\end{eqnarray}
and express $\widetilde{\A}_{1}^{t}(+)$ as
\begin{eqnarray}
	\label{eq:E1_Bloch}
	\widetilde{\A}_{1}^{t}(+) 
	= \frac{1}{2}\left(
	(1+w(\xi_{1})) \id + \mathbf{m}_1(\xi_1)
	\cdot \mathbf{\sigma}\right),
\end{eqnarray}
where we set
\begin{equation}
	\label{eq:E1_A1}
	\mathbf{m}_1(\xi_1) = C_1(\xi_1) \mathbf{v}_1\quad\mbox{with}\quad0\le C_1(\xi_1)\le1.
\end{equation}
Because 
\begin{align*}
	&\Tr[\varrho^{\mathbf{r}_{1}}\A^{t\vx}_{D} (+)]=\Tr[\varrho^{\mathbf{r}_{1}}\widetilde{\A}_{1}^{t}(+)],\\
	&\Tr[\varrho^{\mathbf{r}_{2}}\A^{t\vy}_{D} (+)]=\Tr[\varrho^{\mathbf{r}_{2}}\widetilde{\A}_{1}^{t}(+)],
\end{align*}
namely
\begin{equation}
	\label{eq:E1}
	\begin{aligned}
		&\frac{1}{2}+\frac{t}{2}\cos\varphi_{1}=\frac{1+w_1(\xi_1)}{2}+\frac{C_{1}(\xi_{1})}{2}\cos(\varphi_{1}-\xi_{1}),\\
		&\frac{1}{2}+\frac{t}{2}\cos\varphi_{2}=\frac{1+w_1(\xi_1)}{2}+\frac{C_{1}(\xi_{1})}{2}\cos(\varphi_{2}-\xi_{1}),
	\end{aligned}
\end{equation}
hold, we can obtain
\begin{align}
	\label{eq:A_1explicit}
	C_1(\xi_1)
	&= \frac{t(\cos \varphi_1 
		- \cos \varphi_2)}{\cos (\varphi_1 - \xi_1)
		-\cos (\varphi_2 - \xi_1)} = \frac{t\sin\varphi_{0}}{
		\sin (\varphi_{0}- \xi_1)},\\
	\label{eq:x_1explicit}
	w_1(\xi_1) 
	&= - t \left(
	\frac{\sin (\varphi_1 - \varphi_2)}{
		2 \sin(\frac{\varphi_1 - \varphi_2}{2})}\right)
	\cdot \left(\frac{\sin \xi_1}{\sin 
		(\varphi_{0} -\xi_1)} \right)=\frac{-t\cos\psi_{0}\sin \xi_1}{\sin 
		(\varphi_{0} -\xi_1)},
\end{align}
where we set $\varphi_{0}:=\frac{\varphi_{1}+\varphi_{2}}{2}$ and $\psi_{0}:=\frac{\varphi_{2}-\varphi_{1}}{2}$ ($0<\varphi_{0}<\frac{\pi}{2}$, $0<\psi_{0}<\frac{\pi}{2}$).
Note that if $\sin (\varphi_{0}- \xi_1)=0$ or $\cos (\varphi_1 - \xi_1)
-\cos (\varphi_2 - \xi_1)=0$ holds, then $\cos \varphi_1 
- \cos \varphi_2=0$ holds (see \eqref{eq:A_1explicit}). It means $\varphi_{0}=0$, which is a contradiction, and thus $\sin (\varphi_{0}- \xi_1)\neq0$ (that is, $C_{1}(\xi_{1})$ and $w_{1}(\xi_{1})$ in \eqref{eq:A_1explicit}, \eqref{eq:x_1explicit} are well-defined).
Moreover, because $C_{1}(\xi_{1})\ge0$, we can see from \eqref{eq:A_1explicit} that $\sin (\varphi_{0}- \xi_1)>0$ holds, which results in 
\begin{align}
	0\le\xi_{1}<\varphi_{0}\label{eq:xi_1max},
\end{align}
or
\begin{align}
	-\pi+\varphi_{0}<\xi_{1}\le0\label{eq:xi_1min}.
\end{align}
In addition, $\xi_1$ is restricted also by the condition that $\widetilde{\A}_{1}^{t}(\pm)$ are positive.
Since the eigenvalues of $\widetilde{\A}_{1}^{t}(\pm)$ are $\frac{1}{2}((1+w_1(\xi_1)) \pm C_1(\xi_1))$, the restriction comes from both 
\begin{equation}
	\begin{aligned}
		&1+w_1(\xi_1) +C_1(\xi_1) \leq 2,
		\\
		&1+w_1(\xi_1) - C_1(\xi_1) \geq 0, 
	\end{aligned}
\end{equation}
equivalently
\begin{align}
	&1-w_1(\xi_1) \geq C_1(\xi_1)\label{condition:min}, 
	\\
	&1+w_1(\xi_1) \geq C_1(\xi_1)\label{condition:max}.
\end{align}
When \eqref{eq:xi_1min} (i.e. $\sin\xi_{1}\le0$) holds, $w_{1}(\xi_{1})\ge0$ holds, and thus \eqref{condition:min} is sufficient. 
It is written explicitly as
\begin{align*}
	\sin\left(\varphi_{0}-\xi_{1}\right)+t\sin\xi_{1}\cos\psi_{0}
	\ge t\sin\varphi_{0},
\end{align*}
or
\begin{align}
	\label{eq:E1_innerproduct1}
	\frac{1}{t}\cos\xi_{1}+	\frac{1}{t\sin\varphi_{0}}\left(t\cos\psi_{0}-\cos\varphi_{0}\right)\sin\xi_{1}
	\ge
	1.
\end{align}
In order to investigate \eqref{eq:E1_innerproduct1}, we adopt a geometric method here while it can be solved in an analytic way.
Let us define 
\begin{equation}
	\label{eq:def of h1}
	h_{1}(t, \varphi_{0}, \psi_{0})=\frac{1}{t\sin\varphi_{0}}\left(t\cos\psi_{0}-\cos\varphi_{0}\right).
\end{equation}
Then we can rewrite \eqref{eq:E1_innerproduct1} as
\begin{equation}
	\label{eq:E1_innerproduct2}
	(\cos\xi_{1}, \sin\xi_{1})
	\cdot
	\left[\left(\frac{1}{t}, h_{1}\right)-(\cos\xi_{1}, \sin\xi_{1})\right]\ge0.
\end{equation}
In fact, it can be verified easily that $\left(\frac{1}{t}, h_{1}\right)$ is the intersection of the line $l_{1}:=\{\lambda\mathbf{r}_1+(1-\lambda)\mathbf{r}_2\mid \lambda\in\R\}$ and the line $x=\frac{1}{t}$ in $\R^{2}$.
Considering this fact, we can find that $\xi_{1}$ satisfies \eqref{eq:E1_innerproduct2} if and only if
\begin{equation}
	\xi_{1}^{min}(t, \varphi_{0}, \psi_{0})\le\xi_{1}\le0,
\end{equation}
where $\xi_{1}^{min}(t, \varphi_{0}, \psi_{0})$ is determined by the condition 
\begin{align}
	\left[\left(\frac{1}{t}, h_{1}\right)-(\cos\xi_{1}^{min}, \sin\xi_{1}^{min})\right]\perp(\cos\xi_{1}^{min}, \sin\xi_{1}^{min})
\end{align}
(see FIG. \ref{Fig:xi_min}). 
Analytically, it corresponds to the case where the equality of \eqref{eq:E1_innerproduct1} holds:
\begin{align}
	\label{eq:xi_1min_identity}
	\frac{1}{t}\cos\xi_{1}^{min}+	\frac{1}{t\sin\varphi_{0}}\left(t\cos\psi_{0}-\cos\varphi_{0}\right)\sin\xi_{1}^{min}
	=
	1,
\end{align}
or
\[
1-w_{1}(\xi_{1}^{min})=C_{1}(\xi_{1}^{min}).
\]
It can be represented explicitly as
\begin{equation}
	\label{eq:xi1_equation}
	\begin{aligned}
		&\left( 
		t^2 \cos^2\psi_{0}
		-2t \cos \varphi_{0}\cos \psi_{0}
		+1\right)
		\sin^2\xi_1^{min}\\
		&\qquad\qquad\qquad- 2t \sin\varphi_{0}
		\left( t \cos\psi_{0}
		-\cos \varphi_{0}\right) \sin \xi_1^{min}
		+(t^2 -1) \sin^2 \varphi_{0}=0,
	\end{aligned}
\end{equation}
and $\sin\xi_{1}^{min}$ is obtained as its negative solution.
Note that since the coefficient $(t^2 \cos^2\psi_{0}-
2t \cos \varphi_{0}\cos \psi_{0}
+1)$ is strictly positive, the solutions do not show any singular behavior.
In summary, we have obtained
\begin{equation}
	\xi_{1}^{min}(t, \varphi_{0}, \psi_{0})\le\xi_{1}\le0
\end{equation}
with $\xi_{1}^{min}(t, \varphi_{0}, \psi_{0})$ uniquely determined for $t$, $\varphi_{0}$, and $\psi_{0}$ by 
\begin{align}
	\left\{
	\begin{aligned}
		&-\pi+\varphi_{0}<\xi_{1}^{min}(t, \varphi_{0}, \psi_{0})\le0,\\
		&1-w_{1}(\xi_{1}^{min}(t, \varphi_{0}, \psi_{0}))=C_{1}(\xi_{1}^{min}(t, \varphi_{0}, \psi_{0})).
	\end{aligned}
	\right.
	\label{eq:xi_1min_detailed}
\end{align}
On the other hand, when \eqref{eq:xi_1max} (i.e. $\sin\xi_{1}\ge0$) holds, \eqref{condition:max} is sufficient. 
It results in a tight condition for $\xi_{1}$:
\begin{align}
	0\le\xi_{1}\le\xi_{1}^{max}(t, \varphi_{0}, \psi_{0}),
\end{align}
where $\xi_{1}^{max}(t, \varphi_{0}, \psi_{0})$ is a constant uniquely determined for $\varphi_{0}$ and $\psi_{0}$ by 
\begin{align}
	\left\{
	\begin{aligned}
		&0\le\xi_{1}^{max}(t, \varphi_{0}, \psi_{0})<\varphi_{0}\\
		&1+w_{1}(\xi_{1}^{max}(t, \varphi_{0}, \psi_{0}))=C_{1}(\xi_{1}^{max}(t, \varphi_{0}, \psi_{0})).
	\end{aligned}
	\right.
	\label{eq:xi_1max_detailed}
\end{align}
We remark that this can be obtained by a similar geometric method to the previous case: consider the intersection of the line $l_{1}$ and the line $x=-\frac{1}{t}$ in turn (see FIG. \ref{Fig:xi_max}).
Overall, we have demonstrated that $\xi_{1}$ for $\widetilde{\A}_{1}^{t}$ satisfies
\begin{align}
	\xi_{1}^{min}(t, \varphi_{0}, \psi_{0})\le\xi_{1}\le\xi_{1}^{max}(t, \varphi_{0}, \psi_{0}),
\end{align}
where $\xi_{1}^{min}(t, \varphi_{0}, \psi_{0})$ and $\xi_{1}^{max}(t, \varphi_{0}, \psi_{0})$ are obtained thorough \eqref{eq:xi_1min_detailed} and \eqref{eq:xi_1max_detailed} respectively.
Note that $\xi_{1}^{min}(t, \varphi_{0}, \psi_{0})$ and $\xi_{1}^{max}(t, \varphi_{0}, \psi_{0})$ depend
continuously on $t$ (and $\varphi_1, \varphi_2$ through 
$\varphi_0$ and $\psi_{0}$).

Similarly, we consider a binary observable $\widetilde{\A}_{2}^{t}$ on $\state_D$ which coincides with $\A_{D}^{t\vy}$ in $\state_1$, and focus on its effect $\widetilde{\A}_{2}^{t}(+)$.
We define parameters $\mathbf{v}_2\in D$ and $\xi_2\in[-\pi, \pi)$ as
\begin{equation}
	\begin{aligned}
		\mathbf{v}_2 
		= \sin \xi_2 \vx +\cos \xi_2 \vy
		= argmax_{\mathbf{v}\in D} \mbox{tr}
		[\widetilde{\A}_{2}^{t} (+) \varrho_{\mathbf{v}}]. 
	\end{aligned}
\end{equation}
$\widetilde{\A}_{2}^{t}(+)$ is represented as
\begin{eqnarray}
	\label{eq:E2_Bloch}
	\widetilde{\A}_{2}^{t}(+) = \frac{1}{2}
	\left( (1+w_2(\xi_2))\id + \mathbf{m}_2(\xi_2)\right)
\end{eqnarray}
with
\[
\mathbf{m}_2(\xi_2) = C_2(\xi_2) \mathbf{v}_2\quad(0\le C_2(\xi_2)\le1).
\]
\eqref{eq:E1} becomes
\begin{equation}
	\begin{aligned}
		&\frac{1}{2}+\frac{t}{2}\cos\left(\frac{\pi}{2}-\varphi_{1}\right)=\frac{1+x_2(\xi_2)}{2}+\frac{C_{2}(\xi_{2})}{2}\cos\left(\frac{\pi}{2}-\varphi_{1}-\xi_{1}\right),\\
		&\frac{1}{2}+\frac{t}{2}\cos\left(\frac{\pi}{2}-\varphi_{2}\right)=\frac{1+x_2(\xi_2)}{2}+\frac{C_{2}(\xi_{2})}{2}\cos\left(\frac{\pi}{2}-\varphi_{2}-\xi_{1}\right),
	\end{aligned}
\end{equation}
so defining $\overline{\varphi_{1}}:=\frac{\pi}{2}-\varphi_{1}$ and $\overline{\varphi_{2}}:=\frac{\pi}{2}-\varphi_{2}$, we can obtain similarly to \eqref{eq:A_1explicit} and \eqref{eq:x_1explicit}
\begin{align}
	\label{eq:A_2explicit}
	&C_2(\xi_2)
	= \frac{t\sin\overline{\varphi_{0}}}{
		\sin (\overline{\varphi_{0}}- \xi_1)}
	, \\
	\label{eq:x_2explicit}
	&w_2(\xi_2) = \frac{-t\cos\psi_{0}\sin \xi_2}{\sin 
		(\overline{\varphi_{0}}-\xi_2)},
\end{align}
where $\overline{\varphi_{0}}:=\frac{\overline{\varphi_{1}}+\overline{\varphi_{2}}}{2}=\frac{\pi}{2}-\varphi_{0}$.
It follows that properties of $\widetilde{\A}_{2}^{t}$ can be obtained just by replacing $\xi_{1}$ and $\varphi_{0}$ exhibited in the argument for $\widetilde{\A}_{1}^{t}$ by $\xi_{2}$ and $\overline{\varphi_{0}}$ respectively.  
Remark that $0<\overline{\varphi_{0}}<\frac{\pi}{2}$ holds similarly to $\varphi_{0}$, and that the change $\psi_{0}\rightarrow\overline{\psi_{0}}:=\frac{\overline{\varphi_{2}}-\overline{\varphi_{1}}}{2}=-\psi_{0}$ does not affect the equations above, so we dismiss it.
From \eqref{eq:A_2explicit} and \eqref{eq:x_2explicit}, we have 
\begin{align}
	\xi_{2}^{min}(t, \varphi_{0}, \psi_{0})
	\le
	\xi_{2}
	\le
	\xi_{2}^{max}(t, \varphi_{0}, \psi_{0}),
\end{align}
where 
\begin{equation}
	\begin{aligned}
		\xi_{2}^{min}(t, \varphi_{0}, \psi_{0})
		=\xi_{1}^{min}(t, \overline{\varphi_{0}}, \psi_{0})=\xi_{1}^{min}\left(t, \frac{\pi}{2}-\varphi_{0}, \psi_{0}\right),
	\end{aligned}
\end{equation}
and
\begin{equation}
	\begin{aligned}
		\xi_{2}^{max}(t, \varphi_{0}, \psi_{0})
		=\xi_{1}^{max}(t, \overline{\varphi_{0}}, \psi_{0})
		=\xi_{1}^{max}\left(t, \frac{\pi}{2}-\varphi_{0}, \psi_{0}\right),
	\end{aligned}
\end{equation}
which satisfy
\begin{align}
	\left\{
	\begin{aligned}
		&-\frac{\pi}{2}+\varphi_{0}<\xi_{2}^{min}(t, \varphi_{0}, \psi_{0})
		\le0\\
		&1-w_{2}(\xi_{2}^{min}(t, \varphi_{0}, \psi_{0}))=C_{2}(\xi_{2}^{min}(t, \varphi_{0}, \psi_{0}))
	\end{aligned}
	\right.
	\label{eq:xi_2min_detailed}
\end{align}
and
\begin{align}
	\left\{
	\begin{aligned}
		&0\le
		\xi_{2}^{max}(t, \varphi_{0}, \psi_{0})<
		\frac{\pi}{2}-\varphi_{0}\\
		&1+w_{2}(\xi_{2}^{max}(t, \varphi_{0}, \psi_{0}))=C_{2}(\xi_{2}^{max}(t, \varphi_{0}, \psi_{0}))
	\end{aligned}
	\right.
	\label{eq:xi_2max_detailed}
\end{align}
respectively.

\subsection{Proof of Proposition \ref{prop:qubit-threshold}
	: Part 3}
\label{4subsec:proof part 3}
In this part, we shall consider the (in)compatibility of the observables $\widetilde{\A}_{1}^{t}$ and $\widetilde{\A}_{2}^{t}$ defined in $\textbf{(a)}$ for $t$ close to $\frac{1}{\sqrt{2}}$ ($t\sim\frac{1}{\sqrt{2}}$).
It is related directly with the $\state_{1}$-(in)compatibility of $\A^{t\vx}_{D}$ and $\A^{t\vy}_{D}$ as we have shown in the beginning of this section.
Let us examine the behavior of $\xi_{1}^{min}(t, \varphi_{0}, \psi_{0})$ for $t\sim\frac{1}{\sqrt{2}}$.
We denote $\xi_{1}^{min}(t=\frac{1}{\sqrt{2}}, \varphi_{0}, \psi_{0})$ and $h_{1}(t=\frac{1}{\sqrt{2}}, \varphi_{0}, \psi_{0})$ simply by $\widehat{\xi}_{1}^{min}(\varphi_{0}, \psi_{0})$ and $\widehat{h}_{1}(\varphi_{0}, \psi_{0})$ respectively.
The following lemma is useful.
\begin{lem}
	\label{lem:decrease}
	With $\varphi_{0}$ fixed, $\widehat{\xi}_{1}^{min}$ is a strictly decreasing function of $\psi_{0}$.
\end{lem}
\begin{pf}
	The claim can be observed to hold by a geometric consideration in terms of FIG. \ref{Fig:xi_min}. 
	In fact, increasing $\psi_0$ with $\varphi_0$ fixed corresponds to moving the line $l_1$ down with its inclination fixed. 
	The movement makes $h_1$ (or $\widehat{h}_{1}$) and hence $\xi_1^{min}$ (or $\widehat{\xi}_{1}^{min}$) smaller, which proves the claim.
	Here we show an analytic proof of this fact.
	We can see from \eqref{eq:def of h1} and \eqref{eq:xi_1min_identity} that 
	\begin{align}
		\label{eq:xi1_hat}
		\sqrt{2}\cos\widehat{\xi}_{1}^{min}+	\widehat{h}_{1}\sin\widehat{\xi}_{1}^{min}
		=
		1,
	\end{align}
	i.e.
	\[
	\widehat{h}_{1}=\frac{1}{\sin\widehat{\xi}_{1}^{min}}\left(1-\sqrt{2}\cos\widehat{\xi}_{1}^{min}\right)
	\]
	holds  
	(note that $\sin\widehat{\xi}_{1}^{min}\neq0$ because $\sin\widehat{\xi}_{1}^{min}=0$ contradicts \eqref{eq:xi1_hat}).
	Then the claim follows from the observation that
	\[
	\frac{d\widehat{h}_{1}}{d\widehat{\xi}_{1}^{min}}=\frac{1}{(\sin\widehat{\xi}_{1}^{min})^{2}}\left(\sqrt{2}-\cos\widehat{\xi}_{1}^{min}\right)>0,
	\]
	and $\widehat{h}_{1}=\frac{1}{\sin\varphi_{0}}\left(\cos\psi_{0}-\sqrt{2}\cos\varphi_{0}\right)$ is a decreasing function of $\psi_{0}$.\qed
\end{pf}
From this lemma, it follows that 
\begin{align}
	\label{eq:Xi1_def}
	\widehat{\xi}_{1}^{min}(\varphi_{0}, \psi_{0})<
	\lim_{\psi_{0}\rightarrow +0}\widehat{\xi}_{1}^{min}(\varphi_{0}, \psi_{0})=:{\Xi}_{1}^{min}(\varphi_{0}),
\end{align} 
and
\begin{align}
	\label{eq:Xi2_def}
	\widehat{\xi}_{2}^{min}(\varphi_{0}, \psi_{0})<{\Xi}_{2}^{min}(\varphi_{0})
\end{align}
hold for all $\varphi_{0}\in (0, \frac{\pi}{2})$ and $\psi_{0}\in (0, \frac{\pi}{2})$, where
\begin{equation}
	\label{eq:Xi2_def0}
	\begin{aligned}
		&\widehat{\xi}_{2}^{min}(\varphi_{0}, \psi_{0}):=\xi_{2}^{min}\left(t=\frac{1}{\sqrt{2}},\varphi_{0}, \psi_{0}\right) \left(=\widehat{\xi}_{1}^{min}\left(\frac{\pi}{2}-\varphi_{0}, \psi_{0}\right)\right),\\
		&{\Xi}_{2}^{min}(\varphi_{0}):=\lim_{\psi_{0}\rightarrow +0}\widehat{\xi}_{2}^{min}(\varphi_{0}, \psi_{0})\left(=\Xi_{1}^{min}\left(\frac{\pi}{2}-\varphi_{0}\right)\right).
	\end{aligned}
\end{equation}
We can prove the following lemma.
\begin{lem}
	\label{lem:Xi_bounds}
	\[
	\Xi_{1}^{min}(\varphi_{0})+\Xi_{2}^{min}(\varphi_{0})\le-\frac{\pi}{2}
	\]
	holds for all $0<\varphi_{0}<\frac{\pi}{2}$.
\end{lem}
\begin{pf}
	Let us define 
	\begin{align*}
		H_{1}(\varphi_{0}):=\lim_{\psi_{0}\rightarrow +0}\widehat{h}_{1}(\varphi_{0}, \psi_{0})
		&=\lim_{\psi_{0}\rightarrow +0}h_{1}\left(t=\frac{1}{\sqrt{2}}, \varphi_{0}, \psi_{0}\right)\\
		&=\frac{1}{\sin\varphi_{0}}\left(1-\sqrt{2}\cos\varphi_{0}\right).
	\end{align*}
	It holds similarly to \eqref{eq:xi1_hat} that
	\begin{align}
		\label{eq:Xi1_eq}
		\sqrt{2}\cos\Xi_{1}^{min}+	H_{1}\sin\Xi_{1}^{min}
		=
		1.
	\end{align}
	Hence, together with $\sin^{2}\Xi_{1}^{min}+\cos^{2}\Xi_{1}^{min}=1$, we can obtain 
	\begin{align}
		\label{eq:Xi1=H1}
		\cos\Xi_{1}^{min}=\frac{1}{\sqrt{2}}\cdot\frac{2+H_{1}\sqrt{2H_{1}^{2}+2}}{H_{1}^{2}+2},
	\end{align}
	or its more explicit form 
	\begin{align}
		\cos\Xi_{1}^{min}=\frac{1}{\sqrt{2}}\cdot\frac{4-3\sqrt{2}\cos\varphi_{0}}{3-2\sqrt{2}\cos\varphi_{0}}.
	\end{align}
	It results in
	\begin{align}
		\label{eq:Xi_arccos}
		\Xi_{1}^{min}(\varphi_{0})=-\arccos\left(\frac{1}{\sqrt{2}}\cdot\frac{4-3\sqrt{2}\cos\varphi_{0}}{3-2\sqrt{2}\cos\varphi_{0}}\right),
	\end{align}
	where we follow the convention that $\arccos\colon[-1, 1]\to[0, \pi]$, and thus ${\Xi}_{1}^{min}\in(-\pi+\varphi_{0}, 0]$ is obtained through $-\arccos\colon[-1, 1]\to[-\pi, 0]$.
	Because
	\[
	\frac{d}{d\varphi_{0}}\left(\frac{1}{\sqrt{2}}\cdot\frac{4-3\sqrt{2}\cos\varphi_{0}}{3-2\sqrt{2}\cos\varphi_{0}}\right)
	=\frac{\sin\varphi_{0}}{(3-2\sqrt{2}\cos\varphi_{0})^{2}},
	\]
	and
	\begin{align*}
		\sqrt{1-\left(\frac{1}{\sqrt{2}}\cdot\frac{4-3\sqrt{2}\cos\varphi_{0}}{3-2\sqrt{2}\cos\varphi_{0}}\right)^{2}}
		=\sqrt{\left(\frac{\sin\varphi_{0}}{3-2\sqrt{2}\cos\varphi_{0}}\right)^{2}}=\frac{\sin\varphi_{0}}{3-2\sqrt{2}\cos\varphi_{0}},
	\end{align*}
	we can observe that
	\begin{align*}
		\frac{d\Xi_{1}^{min}}{d\varphi_{0}}
		=\left(\frac{\sin\varphi_{0}}{3-2\sqrt{2}\cos\varphi_{0}}\right)^{-1}\cdot\frac{\sin\varphi_{0}}{(3-2\sqrt{2}\cos\varphi_{0})^{2}}=\frac{1}{3-2\sqrt{2}\cos\varphi_{0}},
	\end{align*}
	and
	\begin{align}
		\frac{d^{2}\Xi_{1}^{min}}{d\varphi_{0}^{2}}=\frac{-2\sqrt{2}\sin\varphi_{0}}{(3-2\sqrt{2}\cos\varphi_{0})^{2}}<0,
	\end{align}
	which means $\Xi_{1}^{min}$ is concave.
	Therefore, for any $\varphi_{0}\in(0, \frac{\pi}{2})$, the concavity results in 
	\begin{align*}
		\frac{1}{2}\Xi_{1}^{min}(\varphi_{0})+\frac{1}{2}\Xi_{2}^{min}(\varphi_{0})
		&=\frac{1}{2}\Xi_{1}^{min}(\varphi_{0})+\frac{1}{2}\Xi_{1}^{min}\left(\frac{\pi}{2}-\varphi_{0}\right)\\
		&\le\Xi_{1}^{min}\left(\frac{1}{2}\varphi_{0}+\frac{1}{2}\left(\frac{\pi}{2}-\varphi_{0}\right)\right)\\
		&=\Xi_{1}^{min}\left(\frac{\pi}{4}\right).
	\end{align*}
	Since we can see form \eqref{eq:Xi_arccos} that $\Xi_{1}^{min}\left(\frac{\pi}{4}\right)=-\frac{\pi}{4}$,
	\[
	\Xi_{1}^{min}(\varphi_{0})+\Xi_{2}^{min}(\varphi_{0})\le-\frac{\pi}{2}
	\]
	holds for any $\varphi_{0}\in(0, \frac{\pi}{2})$.\qed
\end{pf}
According to Lemma \ref{lem:decrease} and Lemma \ref{lem:Xi_bounds},
\begin{align*}
	\widehat{\xi}_{1}^{min}(\varphi_{0}, \psi_{0})+\widehat{\xi}_{2}^{min}(\varphi_{0}, \psi_{0})
	&<\Xi_{1}^{min}(\varphi_{0})+\Xi_{2}^{min}(\varphi_{0})\le-\frac{\pi}{2},
\end{align*}
that is,
\begin{align*}
	\xi_{1}^{min}\left(\hspace{-0.15mm}t\hspace{-0.15mm}=\hspace{-0.15mm}\frac{1}{\sqrt{2}}, \varphi_{0}, \psi_{0}\right)+\xi_{2}^{min}\left(\hspace{-0.15mm}t\hspace{-0.15mm}=\hspace{-0.15mm}\frac{1}{\sqrt{2}}, \varphi_{0}, \psi_{0}\right)
	<-\frac{\pi}{2}
\end{align*}
holds for any $\varphi_{0}$ and $\psi_{0}$ (i.e. for any $\varphi_{1}$ and $\varphi_{2}$).
However, we cannot conclude that 
\begin{align}
	\label{eq:suff small t}
	\xi_{1}^{min}\left(t, \varphi_{0}, \psi_{0}\right)+\xi_{2}^{min}\left(t, \varphi_{0}, \psi_{0}\right)
	\le-\frac{\pi}{2}
\end{align}
holds for $t\sim\frac{1}{\sqrt{2}}$: it may fail when 
\begin{align*}
	\sup_{\varphi_{0}, \psi_{0}}	\left[\xi_{1}^{min}\left(\hspace{-0.15mm}t\hspace{-0.15mm}=\hspace{-0.15mm}\frac{1}{\sqrt{2}}, \varphi_{0}, \psi_{0}\right)
	+\xi_{2}^{min}\left(\hspace{-0.15mm}t\hspace{-0.15mm}=\hspace{-0.15mm}\frac{1}{\sqrt{2}}, \varphi_{0}, \psi_{0}\right)\right]=-\frac{\pi}{2}.
\end{align*}
On the other hand, because we can observe similarly to Lemma \ref{lem:decrease} that $\xi_{1}^{min}$ is a strictly decreasing function of $\psi_{0}$, 
it is anticipated that \eqref{eq:suff small t} holds for $t\sim\frac{1}{\sqrt{2}}$ and for $\psi_{0}$ sufficiently close to $\frac{\pi}{2}$.
In fact, for $\psi_{0}\in[\frac{\pi}{4}, \frac{\pi}{2})$, we can prove the following proposition.

\begin{prop}
	\label{prop:xi1_detail_1}
	There exists a constant $C<-\frac{\pi}{2}$ such that 
	\begin{align*}
		\widehat{\xi}_{1}^{min}(\varphi_{0}, \psi_{0})+\widehat{\xi}_{2}^{min}(\varphi_{0}, \psi_{0})
		<C,
	\end{align*}
	i.e.
	\begin{align*}
		\xi_{1}^{min}\left(t=\frac{1}{\sqrt{2}}, \varphi_{0}, \psi_{0}\right)+\xi_{2}^{min}\left(t=\frac{1}{\sqrt{2}}, \varphi_{0}, \psi_{0}\right)
		<C,
	\end{align*}
	holds for all $\psi_{0}\in[\frac{\pi}{4}, \frac{\pi}{2})$ and $\varphi_{0}\in(0, \frac{\pi}{2})$.
\end{prop}
\begin{pf}
	Because 
	\begin{align*}
		\widehat{\xi}_{1}^{min}(\varphi_{0}, \psi_{0})+\widehat{\xi}_{2}^{min}(\varphi_{0}, \psi_{0})=
		\widehat{\xi}_{1}^{min}(\varphi_{0}, \psi_{0})+\widehat{\xi}_{1}^{min}\left(\frac{\pi}{2}-\varphi_{0}, \psi_{0}\right),
	\end{align*}
	we can assume without loss of generality that $0<\varphi_{0}\le\frac{\pi}{4}$.
	Due to Lemma \ref{lem:decrease}, it holds for any $\psi_{0}\in[\frac{\pi}{4}, \frac{\pi}{2})$ that
	\begin{equation}
		\begin{aligned}
			&\widehat{\xi}_{1}^{min}(\varphi_{0}, \psi_{0})
			\le\widehat{\xi}_{1}^{min}\left(\varphi_{0}, \psi_{0}=\frac{\pi}{4}\right),\\
			&\widehat{\xi}_{1}^{min}\left(\frac{\pi}{2}-\varphi_{0}, \psi_{0}\right)
			\le\widehat{\xi}_{1}^{min}\left(\frac{\pi}{2}-\varphi_{0}, \psi_{0}=\frac{\pi}{4}\right).
		\end{aligned}
	\end{equation}
	Let us denote $\widehat{\xi}_{1}^{min}\left(\varphi_{0}, \psi_{0}=\frac{\pi}{4}\right)$ 
	simply by $\widetilde{{\Xi}}_{1}^{min}\left(\varphi_{0}\right)$. 
	In order to investigate $\widetilde{{\Xi}}_{1}^{min}\left(\varphi_{0}\right)$ and $\widetilde{{\Xi}}_{1}^{min}\left(\frac{\pi}{2}-\varphi_{0}\right)$, we have to recall \eqref{eq:xi1_hat}. 
	Similarly to \eqref{eq:Xi1_eq} and \eqref{eq:Xi1=H1} in the proof of Lemma \ref{lem:Xi_bounds}, it results in
	\begin{equation}
		\label{eq:xi1_hat_pi/4}
		\begin{aligned}
			\cos\widetilde{{\Xi}}_{1}^{min}
			=\frac{1}{\sqrt{2}}\cdot\frac{2+\widetilde{H}_{1}\sqrt{2\widetilde{H}_{1}^{2}+2}}{\widetilde{H}_{1}^{2}+2},
		\end{aligned}
	\end{equation}
	where
	\begin{equation}
		\label{eq:h1_hat_pi/4}
		\begin{aligned}
			\widetilde{H}_{1}\left(\varphi_{0}\right)
			=\widehat{h}_{1}\left(\varphi_{0}, \psi_{0}=\frac{\pi}{4}
			\right)=\frac{1}{\sin\varphi_{0}}\left(\frac{1}{\sqrt{2}}-\sqrt{2}\cos\varphi_{0}\right).
		\end{aligned}
	\end{equation}
	Note that  in this case we cannot apply a similar method to the one in Lemma \ref{lem:Xi_bounds} because $\widetilde{\Xi}_{1}^{min}$ does not have a clear form like \eqref{eq:Xi_arccos}.
	Alternatively, we focus on the following monotone relations 
	between $\widetilde{\Xi}_{1}^{min}$, $\widetilde{H}_{1}$, and $\varphi_{0}$ (referring to the proof of Lemma \ref{lem:decrease} may be helpful):
	\begin{align}
		\label{eq:monotones}
		\frac{d\widetilde{\Xi}_{1}^{min}}{d\widetilde{H}_{1}}>0,
		\quad
		\frac{d\widetilde{H}_{1}}{d\varphi_{0}}>0
		\quad
		\left(\mbox{thus}\ \ \frac{d\widetilde{\Xi}_{1}^{min}}{d\varphi_{0}}>0\right).
	\end{align}
	From these relations, it can be seen that our restriction $0<\varphi_{0}\le\frac{\pi}{4}$ is equivalent to the condition $\widetilde{H}_{1}\le1-\sqrt{2}$ since $\widetilde{H}_{1}\left(0\right)=-\infty$ and $\widetilde{H}_{1}\left(\frac{\pi}{4}\right)=1-\sqrt{2}$.
	The claim of the proposition can be shown easily when $\widetilde{H}_{1}\le-1$ (or $0<\varphi_{0}\le\varphi^{*}:=\arccos\frac{2+\sqrt{10}}{6}$, where $\widetilde{H}_{1}(\varphi^{*})=-1$).
	\begin{figure}[h]
		\centering
		\includegraphics[bb=0.000000 0.000000 702.000000 519.000000, scale=0.34]{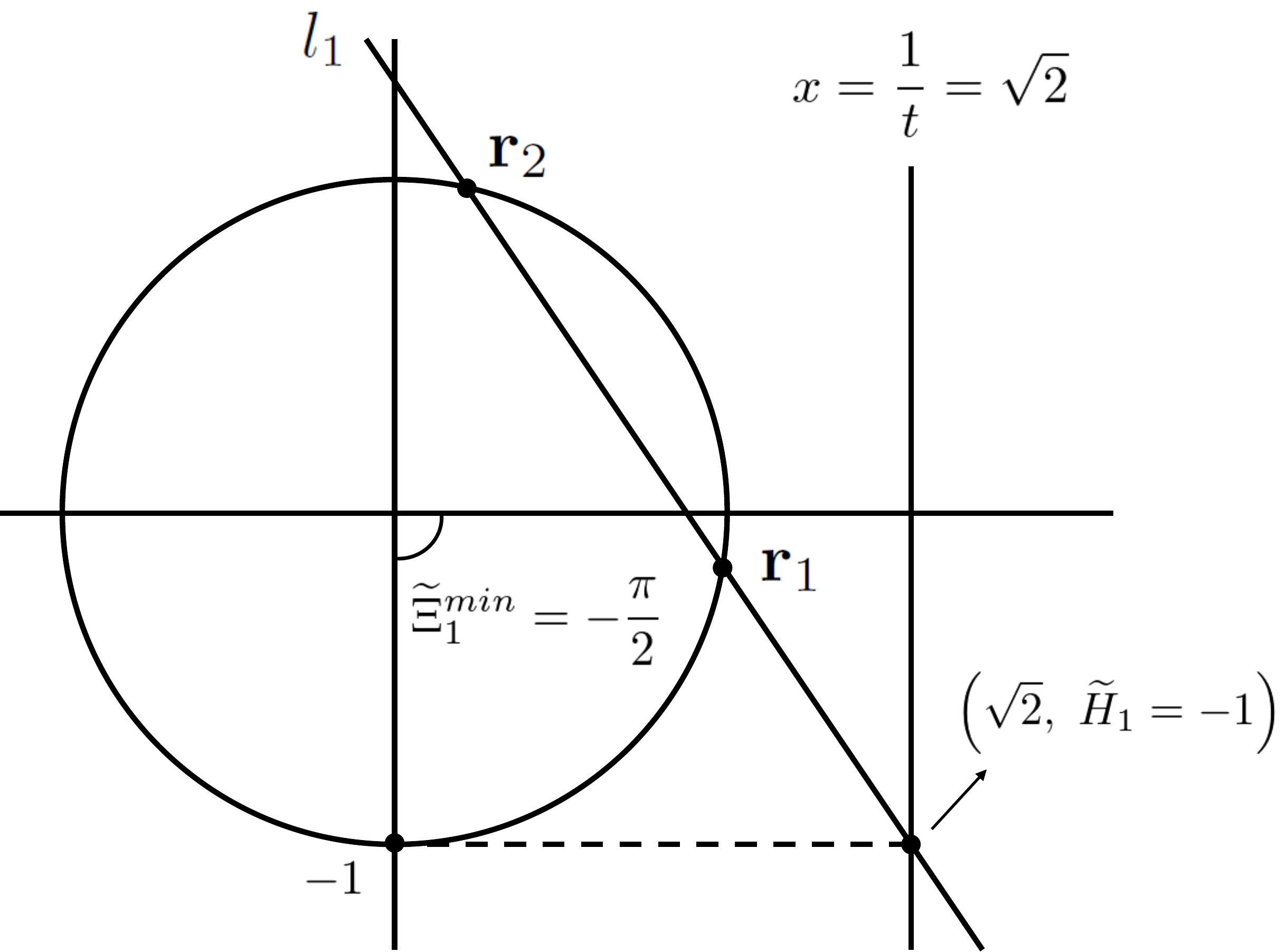}
		\caption{Geometric description of $\widetilde{\Xi}_{1}^{min}$.
			It can be observed that $\widetilde{\Xi}_{1}^{min}=-\frac{\pi}{2}$ when $\widetilde{H}_{1}=-1$.}
		\label{Fig:Xi=-pi/2}
	\end{figure}
	In fact, 
	\begin{align*}
		\widetilde{\Xi}_{1}^{min}(\varphi_{0})\le\widetilde{\Xi}_{1}^{min}\left(\varphi^{*}\right)=-\frac{\pi}{2}
	\end{align*}
	and
	\begin{align*}
		\widetilde{\Xi}_{1}^{min}\left(\frac{\pi}{2}-\varphi_{0}\right)<\widetilde{\Xi}_{1}^{min}\left(\frac{\pi}{2}\right)=-\arccos\frac{2\sqrt{2}+\sqrt{3}}{5}
	\end{align*}
	hold (see FIG. \ref{Fig:Xi=-pi/2} and \eqref{eq:xi1_hat_pi/4}), and thus we can conclude
	\[
	\widetilde{\Xi}_{1}^{min}(\varphi_{0})+\widetilde{\Xi}_{1}^{min}\left(\frac{\pi}{2}-\varphi_{0}\right)<C_{1},
	\]
	where
	\[
	C_{1}=-\frac{\pi}{2}-\arccos\frac{2\sqrt{2}+\sqrt{3}}{5}\left(<-\frac{\pi}{2}\right).
	\]
	When $-1<\widetilde{H}_{1}\le1-\sqrt{2}$ (or $\varphi^{*}<\varphi_{0}\le\frac{\pi}{4}$), we need a bit complicated evaluations.
	It holds similarly to the previous calculations that
	\begin{align*}
		&\widetilde{\Xi}_{1}^{min}(\varphi_{0})\le\widetilde{\Xi}_{1}^{min}\left(\frac{\pi}{4}\right),
		\\
		&\widetilde{\Xi}_{1}^{min}\left(\frac{\pi}{2}-\varphi_{0}\right)<\widetilde{\Xi}_{1}^{min}\left(\frac{\pi}{2}-\varphi^{*}\right).
	\end{align*}
	Since \[\widetilde{\Xi}_{1}^{min}=-\frac{\pi}{4}\iff\widetilde{H}_{1}=0\iff\varphi_{0}=\frac{\pi}{3},
	\]  $\widetilde{\Xi}_{1}^{min}(\frac{\pi}{4})<-\frac{\pi}{4}=\widetilde{\Xi}_{1}^{min}(\frac{\pi}{3})$ holds due to the monotone relations \eqref{eq:monotones}.
	On the other hand, we have
	\begin{align*}
		\cos\varphi^{*}-\cos\frac{\pi}{6}
		=\frac{2+\sqrt{10}}{6}-\frac{\sqrt{3}}{2}
		=-0.0056...<0,
	\end{align*}
	that is,
	\[
	\varphi^{*}>\frac{\pi}{6}.
	\]
	It follows that $\frac{\pi}{2}-\varphi^{*}<\frac{\pi}{3}$, and thus $\widetilde{\Xi}_{1}^{min}\left(\frac{\pi}{2}-\varphi_{0}\right)<-\frac{\pi}{4}$.
	Therefore, we can conclude also in this case
	\[
	\widetilde{\Xi}_{1}^{min}(\varphi_{0})+\widetilde{\Xi}_{1}^{min}\left(\frac{\pi}{2}-\varphi_{0}\right)<C_{2},
	\]
	where
	\[
	C_{2}=\widetilde{\Xi}_{1}^{min}\left(\frac{\pi}{4}\right)+\widetilde{\Xi}_{1}^{min}\left(\frac{\pi}{2}-\varphi^{*}\right)\left(<-\frac{\pi}{2}\right).
	\]
	Overall, we have obtained
	\[
	\widetilde{\Xi}_{1}^{min}(\varphi_{0}, \psi_{0})+\widetilde{\Xi}_{2}^{min}\left(\varphi_{0}, \psi_{0}\right)<\max\{C_{1}, C_{2}\}\left(<-\frac{\pi}{2}\right)
	\]
	for all $\varphi_{0}\in(0, \frac{\pi}{2})$ and $\psi_{0}\in[\frac{\pi}{4}, \frac{\pi}{2})$.\qed
\end{pf}
By virtue of this proposition, for $t$ sufficiently close to $\frac{1}{\sqrt{2}}$, 
\begin{align*}
	\xi_{1}^{min}\left(t, \varphi_{0}, \psi_{0}\right)+\xi_{2}^{min}\left(t, \varphi_{0}, \psi_{0}\right)
	\le-\frac{\pi}{2}
\end{align*}
follows from the continuity of $\xi_{1}^{min}$ and $\xi_{2}^{min}$ with respect to $t$ when $\frac{\pi}{4}\le\psi_{0}<\frac{\pi}{2}$.
It means that there always exist $\xi_{1}^{\star}\ge\xi_{1}^{min}$ and $\xi_{2}^{\star}\ge\xi_{2}^{min}$ for such $t$ and for any $\varphi_{1}$ and $\varphi_{2}$ satisfying $\xi_{1}^{\star}+\xi_{2}^{\star}=-\frac{\pi}{2}.$
For these $\xi_{1}^{\star}$ and $\xi_{2}^{\star}$, it holds that $\mathbf{v}_1=-\mathbf{v}_2$, and thus $\widetilde{\A}_{1}^{t}$ and $\widetilde{\A}_{2}^{t}$ are compatible, i.e. $\A_{D}^{t\vx}$ and $\A_{D}^{t\vy}$ 
are $\state_1$-compatible.

On the other hand, when $0<\psi_{0}<\frac{\pi}{4}$, it may not hold for $t\sim\frac{1}{\sqrt{2}}$ that $\xi_{1}^{min}\left(t, \varphi_{0}, \psi_{0}\right)+\xi_{2}^{min}\left(t, \varphi_{0}, \psi_{0}\right)
\le-\frac{\pi}{2}$, and thus we cannot apply the same argument.
Nevertheless, we can demonstrate that there exist $\xi_1$ and $\xi_2$ 
such that $\widetilde{\A}^{t}_1$ and $\widetilde{\A}^{t}_2$ are compatible even when $0<\psi_{0}<\frac{\pi}{4}$.
To see this, let us assume $0<\psi_{0}<\frac{\pi}{4}$ and apply the necessary and sufficient condition 
for (in)compatibility. 
According to the result proven in \cite{PhysRevA.78.012315,Busch_Coexistence,PhysRevA.81.062116}, $\widetilde{\A}^{t}_1$ and $\widetilde{\A}^{t}_2$ with \eqref{eq:E1_Bloch} and \eqref{eq:E2_Bloch} respectively are 
compatible if and only if
\begin{align}
	\label{eq:compatible iff}
	\begin{aligned}
		\left( 1- F_1^2 - F_2^2\right)
		\left( 1- \frac{w_1^2}{F_1^2} - \frac{w_2^2}{F_2^2}
		\right)
		\leq \left( \mathbf{m}_1 \cdot \mathbf{m}_2 
		- w_1 w_2\right)^2
	\end{aligned} 
\end{align}
holds, where 
\begin{align}
	F_1 &:= \frac{1}{2}\hspace{-0.25mm}
	\left(
	\sqrt{(1+w_1)^2 - C_1^2}
	+ \sqrt{(1-w_1)^2 - C_1^2}
	\right),
	\\
	F_2 &:= \frac{1}{2}\hspace{-0.25mm}
	\left(
	\sqrt{(1+w_2)^2 - C_2^2}
	+ \sqrt{(1-w_2)^2 - C_2^2}
	\right). 
\end{align}
For $\xi_1^{min}$ and $\xi_2^{min}$, 
since it holds that 
\begin{align}
	1-w_1(\xi_1^{min}) &= C_1 (\xi_1^{min}),\\
	1-w_2(\xi_2^{min}) &= C_2(\xi_2^{min}), 
\end{align}
they become 
\begin{align}
	F_1 = \sqrt{w_1(\xi_1^{min})}, \quad
	F_2= \sqrt{w_2(\xi_2^{min}) }. 
\end{align}
Therefore, \eqref{eq:compatible iff} can be rewritten as
	\begin{equation}
		\label{eq:compatible iff for min}
		\begin{aligned}
			&[
			(1\hspace{-0.12em}-\hspace{-0.12em} \sin (\xi_1^{min}\hspace{-0.12em}+\hspace{-0.12em} \xi_2^{min}))
			w_1(\xi_1^{min})w_2(\xi_2^{min})\\
			&\quad\qquad-(1\hspace{-0.12em}+\hspace{-0.12em}\sin(\xi_1^{min} + \xi_2^{min}))
			(1\hspace{-0.12em}-\hspace{-0.12em} w_1(\xi_1^{min})\hspace{-0.12em} - \hspace{-0.12em}w_2(\xi_2^{min}))
			]\\
			&\quad\cdot\left[
			(1-w_1(\xi_1^{min}))(1-w_2(\xi_2^{min}))(1- \sin (\xi_1^{min}+ \xi_2^{min}))
			\right]
			\ge 0. 
		\end{aligned}
	\end{equation}
If $1-\sin(\xi_1^{min} + \xi_2^{min})=0$, then \eqref{eq:compatible iff for min} holds, that is, $\widetilde{\A}^{t}_1$ and $\widetilde{\A}^{t}_2$ for $\xi_1^{min}$ and $\xi_2^{min}$ respectively are compatible.
Therefore, we hereafter assume $1-\sin(\xi_1^{min} + \xi_2^{min})>0$, and rewrite \eqref{eq:compatible iff for min} as (note that $0<w_1(\xi_1^{min})<1$, $0<w_2(\xi_2^{min})<1$)
\begin{align}
	\begin{aligned}
		(1+\sin(\xi_1^{min} + \xi_2^{min}))
		(1- w_1(\xi_1^{min}) - w_2(\xi_2^{min}))\qquad\qquad\qquad\\
		\leq (1- \sin (\xi_1^{min}+ \xi_2^{min}))
		w_1(\xi_1^{min})w_2(\xi_2^{min}). 
	\end{aligned}
\end{align}
In other words, 
$\widetilde{\A}^{t}_1$ and 
$\widetilde{\A}^{t}_2$ with respect to 
$\xi^{min}_1$ and $\xi^{min}_2$ are incompatible 
if and only if
\begin{align}
	\label{eq:incompatible iff for min}
	\begin{aligned}
		(1+\sin(\xi_1^{min} + \xi_2^{min}))
		(1- w_1(\xi_1^{min}) - w_2(\xi_2^{min}))\qquad\qquad\qquad\\
		> (1- \sin (\xi_1^{min}+ \xi_2^{min}))
		w_1(\xi_1^{min})w_2(\xi_2^{min})
	\end{aligned}
\end{align}
holds.
In order to investigate whether \eqref{eq:incompatible iff for min} holds, it is helpful to introduce a function $Z$ defined as 
	\begin{equation}
		\label{eq:def of Z}
		\begin{aligned}
			Z(t, \varphi_{0}, \psi_{0}):= &\left[1\hspace{-0.12em}+\hspace{-0.12em}\sin(\xi_1^{min}(t, \varphi_{0}, \psi_{0})\hspace{-0.12em}+\hspace{-0.12em}\xi_2^{min}(t, \varphi_{0}, \psi_{0}))\right]\\
			&\qquad\qquad\quad\left[1\hspace{-0.12em}+ \hspace{-0.12em}w_1(\xi_1^{min}(t, \varphi_{0}, \psi_{0})) \hspace{-0.12em}+\hspace{-0.12em} w_2(\xi_2^{min}(t, \varphi_{0}, \psi_{0}))\right]\\
			&-\left[1- \sin (\xi_1^{min}(t, \varphi_{0}, \psi_{0})+\xi_2^{min}(t, \varphi_{0}, \psi_{0}))\right]\\
			&\qquad\qquad\qquad w_1(\xi_1^{min}(t, \varphi_{0}, \psi_{0}))
			w_2(\xi_2^{min}(t, \varphi_{0}, \psi_{0})). 
		\end{aligned}
	\end{equation}
Because
\begin{align*}
	(1+\sin(\xi_1^{min} + \xi_2^{min}))
	(1- w_1(\xi_1^{min}) - w_2(\xi_2^{min}))\qquad\qquad\qquad\qquad\\
	<(1+\sin(\xi_1^{min} + \xi_2^{min}))
	(1+ w_1(\xi_1^{min}) + w_2(\xi_2^{min})),
\end{align*}
\begin{eqnarray}
	\label{eq:Z>0}
	Z(t, \varphi_{0}, \psi_{0}) >0
\end{eqnarray}
holds if $\widetilde{\A}^{t}_1$ and 
$\widetilde{\A}^{t}_2$ with respect to 
$\xi^{min}_1$ and $\xi^{min}_2$ are incompatible.
Let us focus on the case where $t=\frac{1}{\sqrt{2}}$ (i.e. $\xi_{1}^{min}=\widehat{\xi}_{1}^{min}$).
If a pair $(\varphi_{0}, \psi_{0})$ satisfies $\widehat{\xi}_{1}^{min}(\varphi_{0}, \psi_{0})\le-\frac{\pi}{2}$ or $\widehat{\xi}_{2}^{min}(\varphi_{0}, \psi_{0})\le-\frac{\pi}{2}$, then 
\begin{align*}
	\widehat{\xi}_{1}^{min}(\varphi_{0}, \psi_{0})+\widehat{\xi}_{2}^{min}(\varphi_{0}, \psi_{0})
	<C
\end{align*}
with
\begin{align*}
	C&=-\frac{\pi}{2}+
	\lim_{\substack{\varphi_{0}\rightarrow \frac{\pi}{2}-0\\ 
			\psi_{0}\rightarrow+0}}
	\widehat{\xi}_{1}^{min}\left(\varphi_{0}, \psi_{0}\right)
	=-\frac{\pi}{2}-\arccos\left(\frac{2\sqrt{2}}{3}\right)
	<-\frac{\pi}{2}
\end{align*}
holds due to similar monotone relations to \eqref{eq:monotones} between $\varphi_{0}, \psi_{0},$ and $\widehat{\xi}_{1}^{min}$ (remember that $\widehat{\xi}_{2}^{min}(\varphi_{0}, \psi_{0})=\widehat{\xi}_{1}^{min}\left(\frac{\pi}{2}-\varphi_{0}, \psi_{0}\right)$).
Therefore, in this case, we can apply the same argument as Proposition \ref{prop:xi1_detail_1}, which results in the compatibility of $\widetilde{\A}^{t}_1$ and 
$\widetilde{\A}^{t}_2$ for $t\sim\frac{1}{\sqrt{2}}$.
On the other hand, let us examine the case where  $(\varphi_{0}, \psi_{0})$ satisfies $\psi_{0}\in(0, \frac{\pi}{4})$, and $\widehat{\xi}_{1}^{min}(\varphi_{0}, \psi_{0})>-\frac{\pi}{2}$ and $\widehat{\xi}_{2}^{min}(\varphi_{0}, \psi_{0})>-\frac{\pi}{2}$.
Because $\psi_{0}\in(0, \frac{\pi}{4})$, we obtain for general $t$ (see \eqref{eq:x_1explicit})
\begin{align}
	\begin{aligned}
		w_1(\xi^{min}_1) 
		> - \frac{t}{\sqrt{2}} \frac{\sin \xi_1^{min}}{
			\sin (\varphi_0- \xi_1^{min})}\geq \frac{t}{\sqrt{2}} (-\sin \xi_1^{min})
		. 
	\end{aligned}
\end{align}
For $t=\frac{1}{\sqrt{2}}$, since 
\[
-\frac{\pi}{2}<\widehat{\xi}_{1}^{min}(\varphi_{0}, \psi_{0})<\lim_{\substack{\varphi_{0}\rightarrow \frac{\pi}{2}-0\\ 
		\psi_{0}\rightarrow+0}}
\widehat{\xi}_{1}^{min}\left(\varphi_{0}, \psi_{0}\right),
\] 
it gives a bound 
\begin{eqnarray}
	w_1(\xi_1^{min}) >  \frac{1}{2} \sin \widehat{\xi}_0, 
\end{eqnarray}
where we define
\begin{align*}
	\widehat{\xi}_0 = -\lim_{\substack{\varphi_{0}\rightarrow \frac{\pi}{2}-0\\ 
			\psi_{0}\rightarrow+0}}
	\widehat{\xi}_{1}^{min}\left(\varphi_{0}, \psi_{0}\right)=\arccos \left(\frac{2\sqrt{2}}{3}\right).
\end{align*}
Let $\varepsilon$ be a positive constant satisfying $\varepsilon<\frac{1}{16} (\sin\widehat{\xi}_0)^{2}$.
Due to the continuity of sine, there exists a positive constant $\delta$ such that $\sin x\in(-1, -1+\varepsilon)$ whenever $x\in\left(-\frac{\pi}{2}-\delta, -\frac{\pi}{2}\right)$.
If $(\varphi_{0}, \psi_{0})$ satisfies $\widehat{\xi}_{1}^{min}(\varphi_{0}, \psi_{0})+\widehat{\xi}_{2}^{min}(\varphi_{0}, \psi_{0})\le-\frac{\pi}{2}-\delta$, then it again leads to the same argument as Proposition \ref{prop:xi1_detail_1}, and we can see that $\widetilde{\A}^{t}_1$ and $\widetilde{\A}^{t}_2$ for this $(\varphi_{0}, \psi_{0})$ are compatible.
Conversely, if $(\varphi_{0}, \psi_{0})$ satisfies $-\frac{\pi}{2}-\delta<\widehat{\xi}_{1}^{min}(\varphi_{0}, \psi_{0})+\widehat{\xi}_{2}^{min}(\varphi_{0}, \psi_{0})<-\frac{\pi}{2}$ (remember Lemma \ref{lem:Xi_bounds}), then 
\[
-1<\sin(
\widehat{\xi}_{1}^{min}(\varphi_{0}, \psi_{0})+\widehat{\xi}_{2}^{min}(\varphi_{0}, \psi_{0})
)<-1+\varepsilon
\]
follows from the definition of $\delta$.
Therefore, by virtue of \eqref{eq:def of Z}, we have
	\begin{align*}
		Z\left(t=\frac{1}{\sqrt{2}}, \varphi_{0}, \psi_{0}\right)
		&=\left[1+\sin(\widehat{\xi}_1^{min}(\varphi_{0}, \psi_{0})+\widehat{\xi}_2^{min}(\varphi_{0}, \psi_{0}))\right]\\
		&\qquad\qquad\qquad\left[1+ w_1(\widehat{\xi}_1^{min}(\varphi_{0}, \psi_{0})) + w_2(\widehat{\xi}_2^{min}(\varphi_{0}, \psi_{0}))\right]\\
		&\qquad-\left[1- \sin (\widehat{\xi}_1^{min}(\varphi_{0}, \psi_{0})+\widehat{\xi}_2^{min}(\varphi_{0}, \psi_{0}))\right]\\
		&\qquad\qquad\qquad\quad w_1(\widehat{\xi}_1^{min}(\varphi_{0}, \psi_{0}))
		w_2(\widehat{\xi}_2^{min}(\varphi_{0}, \psi_{0}))
		\\
		&<\varepsilon\left[1+ w_1(\widehat{\xi}_1^{min}(\varphi_{0}, \psi_{0})) + w_2(\widehat{\xi}_2^{min}(\varphi_{0}, \psi_{0}))\right]\\
		&\qquad\qquad\quad\quad\ -(2-\varepsilon)w_1(\widehat{\xi}_1^{min}(\varphi_{0}, \psi_{0}))
		w_2(\widehat{\xi}_2^{min}(\varphi_{0}, \psi_{0}))
		\\
		&=\varepsilon\left[1+w_1(\widehat{\xi}_1^{min}(\varphi_{0}, \psi_{0}))\right]\left[1+w_2(\widehat{\xi}_1^{min}(\varphi_{0}, \psi_{0}))\right]
		\\&\qquad\qquad\qquad\ -2w_1(\widehat{\xi}_1^{min}(\varphi_{0}, \psi_{0}))
		w_2(\widehat{\xi}_2^{min}(\varphi_{0}, \psi_{0}))
		\\
		&<4\varepsilon-2w_1(\widehat{\xi}_1^{min}(\varphi_{0}, \psi_{0}))
		w_2(\widehat{\xi}_2^{min}(\varphi_{0}, \psi_{0})).
	\end{align*}
Because 
\begin{align*}
	4\varepsilon-2w_1(\widehat{\xi}_1^{min}(\varphi_{0}, \psi_{0}))
	w_2(\widehat{\xi}_2^{min}(\varphi_{0}, \psi_{0}))
	&<\frac{1}{4} (\sin\widehat{\xi}_0)^{2}-\frac{1}{2} (\sin\widehat{\xi}_0)^{2}\\
	&=-\frac{1}{4} (\sin\widehat{\xi}_0)^{2},
\end{align*}
it follows that
\[
Z\left(t=\frac{1}{\sqrt{2}}, \varphi_{0}, \psi_{0}\right)
<-\frac{1}{4} (\sin\widehat{\xi}_0)^{2}<0.
\]
Therefore, for $t\sim\frac{1}{\sqrt{2}}$, it holds that $Z\left(t, \varphi_{0}, \psi_{0}\right)\le0$, that is, $\widetilde{\A}_1^{t}$ and 
$\widetilde{\A}_2^{t}$ with respect to 
$\xi^{min}_1$ and $\xi^{min}_2$ are compatible.
Overall, we have demonstrated that when $t\sim\frac{1}{\sqrt{2}}$, there exist compatible observables $\widetilde{\A}_1^{t}$ and 
$\widetilde{\A}_2^{t}$ for any line $\state_{1}\subset\state_{D}$ such that they agree with $\A^{t\vx}_{D}$ and $\A^{t\vy}_{D}$ on $\state_{1}$ respectively.
That is, when $t\sim\frac{1}{\sqrt{2}}$, the observables $\A^{t\vx}_{D}$ and $\A^{t\vy}_{D}$ are $\state_{1}$-compatible for any line $\state_{1}\subset\state_{D}$.
Therefore, we can conclude that $\chi_{incomp}(\A^{t\vx}_{D},\A^{t\vy}_{D})=3$ for $t\sim\frac{1}{\sqrt{2}}$, and thus the set $M$ in \eqref{eq: sets of xi=2,3} is nonempty.

\subsection{Proof of Proposition \ref{prop:qubit-threshold}
	: Part 4}
\label{4subsec:proof part 4}
In this part, we shall show that 
\[
t_{0}':=\inf L=\sup M\in M,
\]
where $L$ and $M$ are defined in \eqref{eq: sets of xi=2,3}.
In order to prove this, we will see that if $t\in L$, then $t-\delta\in L$ for sufficiently 
small $\delta>0$, that is, $t_{0}'\notin L$.

Let us focus again on a system described by 
a two-dimensional disk
state space $\mathcal{S}_D$. 
It is useful to identify this system with the system of a quantum bit with real coefficients by replacing 
$\{\sigma_1, \sigma_2\}$ with 
$\{\sigma_3, \sigma_1\}$.
Then, defining $\mathcal{E}_{D}$ as the set of all effects on $\mathcal{S}_D$, we can see that any $E\in\mathcal{E}_{D}$  can be expressed as a real-coefficient positive matrix smaller than $\id$.
We also define $O_{D}(2)
\subset \mathcal{E}_{D}\times \mathcal{E}_{D}$ as the set of all binary observables  on $\state_{D}$, 
which is isomorphic naturally to $\mathcal{E}_{D}$
since a binary observable $\A$ is completely specified by its effect $\A(+) \in \mathcal{E}_{D}$. 
With introducing a topology (e.g. norm topology) on $\mathcal{E}_{D}$, it also can be observed that $O_{D}(2)$ is homeomorphic to $\mathcal{E}_{D}$.
Note that because the system is described by finite-dimensional matrices, any (natural) topology (norm topology, weak topology, etc.) coincides with each other.  
For a pair of states $\{\varrho^{\mathbf{r}_{1}}, \varrho^{\mathbf{r}_2}\}$ in $\state_{D}$, and a binary observable $\A
\in O_{D}(2)$, we define a set of observables $C(\A: \varrho^{\mathbf{r}_{1}}, 
\varrho^{\mathbf{r}_{2}})$ as the set of all binary observables $\widetilde{\A}\in O_{D}(2)$ such that 
\begin{align*}
	&\Tr[\varrho^{\mathbf{r}_{1}}\widetilde{\A}(\pm)]=\Tr[\varrho^{\mathbf{r}_{1}}\A(\pm)],\\ &\Tr[\varrho^{\mathbf{r}_{2}}\widetilde{\A}(\pm)]=\Tr[\varrho^{\mathbf{r}_{2}}\A(\pm)].
\end{align*}
It can be confirmed easily that  
$C(\A: \varrho^{\mathbf{r}_{1}}, 
\varrho^{\mathbf{r}_{2}})$ is closed in $O_{D}(2)\simeq \mathcal{E}_{D}$. 
Let us denote by $O_{D}(4)$ the set of all observables 
with four outcomes, which is a compact (i.e. bounded and closed) subset of 
$\mathcal{E}_{D}^4$.  
For each $\M=\{\M(x,y)\}\in O_{D}(4)$, 
we can introduce a pair of binary observables 
by 
\[
\pi_1(\M)= \left\{\sum_{y}\M(x,y)\right\}_x,\  
\pi_2(\M)=\left\{\sum_x \M(x,y)\right\}_y.
\]
Since $\pi_j\colon O_{D}(4)\to O_{D}(2)$ is continuous, 
the set of all compatible binary observables 
denoted by
\begin{align*}
	JM(2,2):=\{(\pi_1(\M), \pi_2(\M))\mid\M \in O_{D}(4)\}
\end{align*}
is compact in $O_{D}(2)\times O_{D}(2)
\simeq \mathcal{E}_{D} \times \mathcal{E}_D$ as well. 
As we have seen in the previous part, $\chi_{incomp}(\A^{t\vx}, \A^{t\vy})=2$ (i.e. $\chi_{incomp}(\A_{D}^{t\vx}, \A_{D}^{t\vy})=2$) if and only if 
there exists a pair of vectors
$\mathbf{r}_{1}, \mathbf{r}_{2} \in \partial D$ such that 
\begin{align*}
	\left(C(\A_{D}^{t\vx}: \varrho^{\mathbf{r}_{1}}, \varrho^{\mathbf{r}_{2}})
	\times C(\A^{t\vy}_{D}: \varrho^{\mathbf{r}_{1}}, 
	\varrho^{\mathbf{r}_{2}}) \right)
	\cap JM(2:2) =\emptyset. 
\end{align*} 

\par
Let us examine concrete representations of the 
sets. 
Each effect $E\in\mathcal{E}_{D}$ is written as 
$E=\frac{1}{2}(e_0 \id + \mathbf{e}\cdot\sigma)=\frac{1}{2}(e_0 \id + e_1 \sigma_1 + e_2\sigma_2)$ 
with $(e_0, \mathbf{e})=(e_0, e_1, e_2) 
\in \R^3$ satisfying $0 \leq  
e_0 \pm |\mathbf{e}| \leq 2$.

\begin{figure}[h]
	\centering
	\includegraphics[bb=0.000000 0.000000 876.000000 499.000000, scale=0.28]{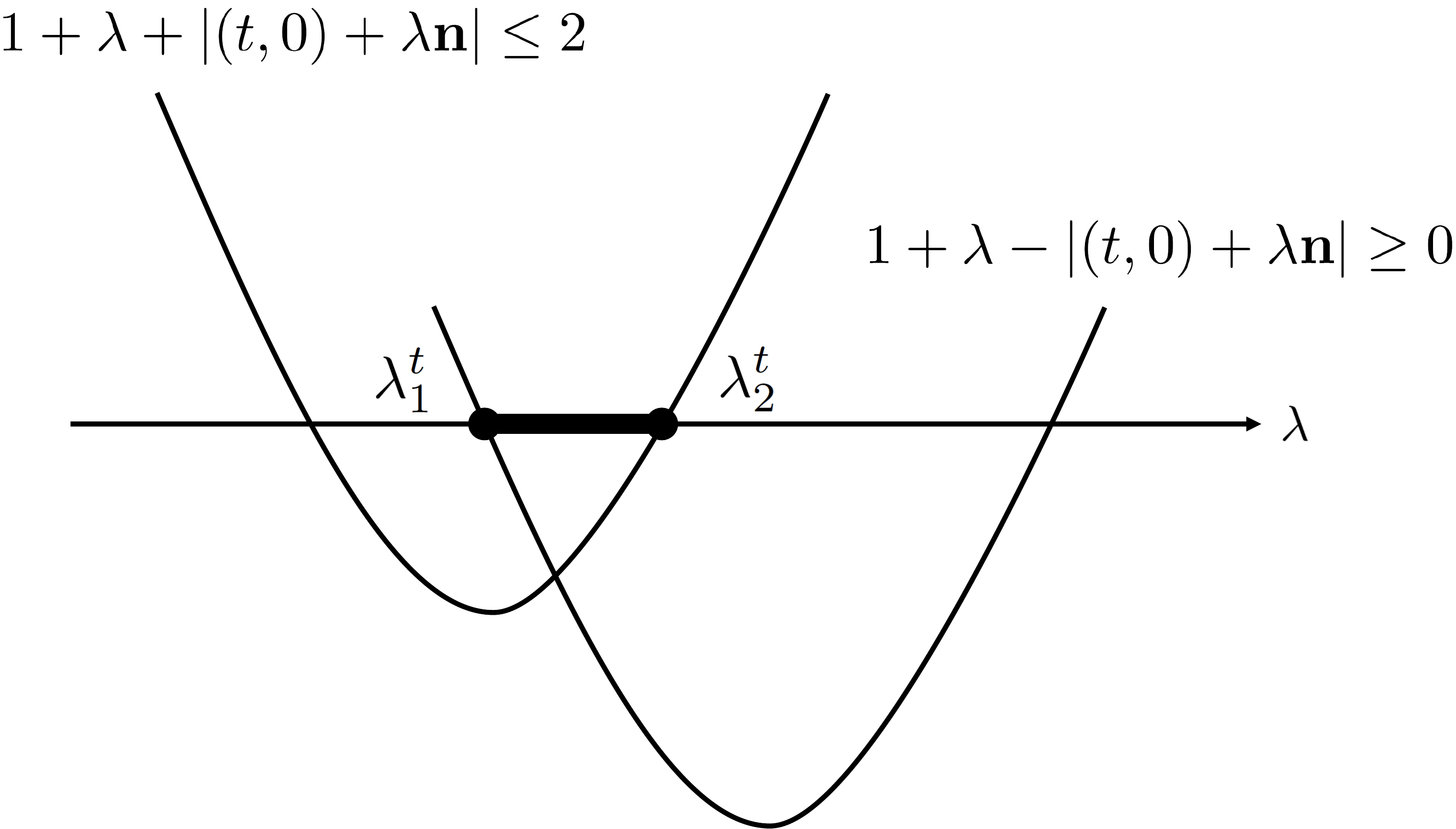}
	\caption{Solutions for $\lambda$.}
	\label{Fig:lambda}
\end{figure}

If we consider another effect $F=\frac{1}{2}(f_0 \id + \mathbf{f}\cdot\sigma)$, the operator norm of $E-F$ 
is calculated as 
\begin{align}
	\label{eq:norm for qubit effect}
	\Vert E-F\Vert =\frac{1}{2}\left( |e_0 - f_0 | 
	+ |\mathbf{e} - \mathbf{f}|\right). 
\end{align}
We may employ this norm to define 
a topology on $\mathcal{E}_{D}$ and $O_{D}(2)
\simeq \mathcal{E}_{D}$. 
On the other hand, each state in $\state_{D}$ is parameterized as 
$\varrho^{\mathbf{r}_{1}} = \frac{1}{2}(\id + x_1 \sigma_1 + 
y_1\sigma_2)$, where $\mathbf{r}_{1}=(x_1, y_1)$ 
satisfies $|\mathbf{r}_{1}|\leq 1$. 
For an effect $E$ and a state $\varrho^{\mathbf{r}_{1}}$, we have 
$\mbox{tr}[\varrho^{\mathbf{r}_{1}}E]
=\frac{1}{2}(e_0 + \mathbf{r}_{1}\cdot \mathbf{e})
$. 
In particular, when considering $\A^{t\vx}(\pm)=\frac{1}{2}(\id \pm 
t \sigma_1)$, 
a binary observable $\CC$ determined by the effect $\CC(+)=\frac{1}{2}(c_0 \id  + \mathbf{c}\cdot 
\mathbf{\sigma})=\frac{1}{2}(c_0 \id  + c_{1}\sigma_1+c_{2}\sigma_{2})$ satisfies 
$\CC \in 
C(\A^{t\vx}: \varrho^{\mathbf{r}_{1}}, \varrho^{\mathbf{r}_{2}})$ if and only if 
\begin{align*}
	&\Tr[\varrho^{\mathbf{r}_{1}}\A^{t\vx}(+)]=\Tr[\varrho^{\mathbf{r}_{1}}\CC(+)],\\ &\Tr[\varrho^{\mathbf{r}_{2}}\A^{t\vx}(+)]=\Tr[\varrho^{\mathbf{r}_{2}}\CC(+)],
\end{align*}
i.e.
\begin{align*}
	1 +t x_1 &=c_0 + \mathbf{r}_{1}\cdot \mathbf{c}
	= c_0 + x_1 c_1+y_1 c_2, \\
	1 + t x_2 &=c_0 + \mathbf{r}_{2}\cdot \mathbf{c}
	= c_0 + x_2 c_1 +y_2 c_2. 
\end{align*}
hold, where we set $\mathbf{r}_{2}=(x_{2}, y_{2})$.
The set of their solutions for $(c_0, \mathbf{c})$ is represented as 
\begin{align*}
	(c_0, \mathbf{c})=(1, t, 0)+\lambda'
	\left(
	-\frac{x_{1}y_{2}-y_{1}x_{2}}{x_{1}-x_{2}},\  
	-\frac{y_{1}-y_{2}}{x_{1}-x_{2}},\ 
	1
	\right)
\end{align*}
with $\lambda'\in\R$.
Let us define a vector $\mathbf{n}\in\R^{2}$ such that 
\[
(1, \mathbf{n})\cdot(1, \mathbf{r}_{1})=(1, \mathbf{n})\cdot(1, \mathbf{r}_{2})=0
\]
(i.e. $\mathbf{n}\cdot\mathbf{r}_{1}=\mathbf{n}\cdot\mathbf{r}_{2}=-1$).
It is easy to see that 
\[
\left(
-\frac{x_{1}y_{2}-y_{1}x_{2}}{x_{1}-x_{2}},\  
-\frac{y_{1}-y_{2}}{x_{1}-x_{2}},\ 
1
\right)
\propto
(1, \mathbf{n}),
\]
and thus the set of solutions can be rewritten as
\begin{align}
	\label{eq:solution for C}
	(c_0, \mathbf{c})=(1, t, 0)+\lambda(1, \mathbf{n})
\end{align}
with $\lambda\in\R$.
Note that because we are interested in the case where $\A_{D}^{t\vx}$ and $\A_{D}^{t\vy}$ are $\state_{1}$-incompatible, we do not consider the case where $\mathbf{r}_{1}$ and $\mathbf{r}_{2}$ are parallel or when $x_{1}=x_{2}$ corresponding to $\psi_{0}=\frac{\pi}{2}$ or $\varphi_{0}=0$ in Part 1 respectively.
Therefore, the vector $\mathbf{n}=(n_{x}, n_{y})$ can be defined successfully, and it is easy to verify that $|\mathbf{n}|=\sqrt{n_{x}^{2}+n_{y}^{2}}>1$.
\begin{figure}[h]
	\centering
	\includegraphics[bb=0.000000 0.000000 539.000000 496.000000, scale=0.33]{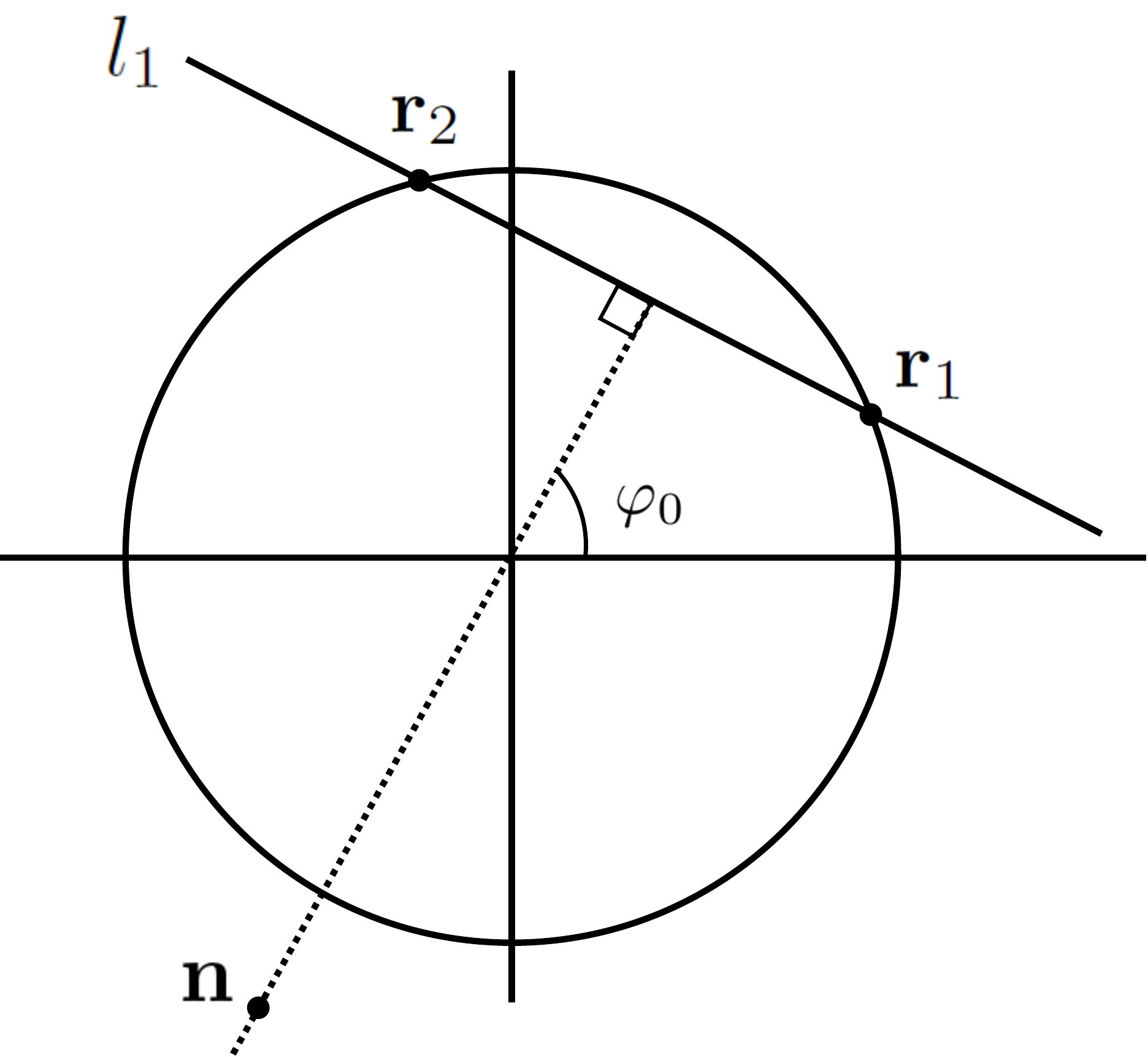}
	\caption{Geometric description of $\mathbf{n}$: we can observe that it lies in the third quadrant.}
	\label{Fig:vector n}
\end{figure}
Moreover, because $\varphi_{0}$ is supposed to be $0<\varphi_{0}<\frac{\pi}{2}$ as shown in Part 1, we can assume without loss of generality that its components $n_{x}$ and $n_{y}$ are negative (see FIG. \ref{Fig:vector n}).
In order for $\CC$ to be an element of $C(\A^{t\vx}: \varrho^{\mathbf{r}_{1}}, \varrho^{\mathbf{r}_{2}})$, \eqref{eq:solution for C} should also satisfy
\[
0\leq 1+\lambda \pm |(t, 0) + \lambda\mathbf{n}|
\leq 2,
\]
i.e.
\begin{align*}
	1+\lambda - |(t, 0) + \lambda\mathbf{n}| \ge 0,\quad
	1+\lambda + |(t, 0) + \lambda\mathbf{n}| \le 2.
\end{align*}
It can be reduced to 
\begin{align}
	\label{eq:lambda}
	\lambda_{1}^{t}\le \lambda \le \lambda_{2}^{t}
\end{align}
with
\begin{equation}
	\label{eq:lambda2}
	\begin{aligned}
		&\lambda_{1}^{t}=\frac{1-n_{x}t-\sqrt{(1-n_{x}t)^{2}+(|\mathbf{n}|^{2}-1)(1-t^2)}}{|\mathbf{n}|^{2}-1},\\
		&\lambda_{2}^{t}=\min\left\{1,\ \frac{-1-n_{x}t+\sqrt{(1+n_{x}t)^{2}+(|\mathbf{n}|^{2}-1)(1-t^2)}}{|\mathbf{n}|^{2}-1}\right\},
	\end{aligned}
\end{equation}
where we used $|\mathbf{n}|>1$ and $n_{x}<0$ (see FIG. \ref{Fig:lambda}).
Overall, $C(\A^{t\vx}_{D}: \varrho^{\mathbf{r}_{1}}, \varrho^{\mathbf{r}_{2}})$
is isomorphic to the set parameterized as 
\begin{equation}
	\label{eq:explicit C}
	\begin{aligned}
		\{(1, t, 0) + \lambda(1, \mathbf{n})\mid\lambda_{1}^{t}\le \lambda \le \lambda_{2}^{t}\},
	\end{aligned}
\end{equation}
where $\lambda_{1}^{t}$ and $\lambda_{2}^{t}$ are shown in \eqref{eq:lambda2}.
Remark that the same argument can be applied for $C(\A^{t\vy}: \varrho^{\mathbf{r}_{1}}, \varrho^{\mathbf{r}_{2}})$.

We shall now prove $t_{0}'=\inf L\notin L$.
Suppose that $t\in L$, i.e. $\chi_{incomp}(\A^{t\vx}_{D}, \A^{t\vy}_{D})=2$.
It follows that there exist $\mathbf{r}_{1}$ and $\mathbf{r}_{2}$ in $\partial D$ such that
\begin{align*}
	\left(C(\A_{D}^{t\vx}: \varrho^{\mathbf{r}_{1}}, \varrho^{\mathbf{r}_{2}})
	\times C(\A^{t\vy}_{D}: \varrho^{\mathbf{r}_{1}}, 
	\varrho^{\mathbf{r}_{2}}) \right)
	\cap JM(2:2) =\emptyset. 
\end{align*} 
Denoting $C(\A_{D}^{t\vx}: \varrho^{\mathbf{r}_{1}}, \varrho^{\mathbf{r}_{2}})$ and $C(\A_{D}^{t\vy}: \varrho^{\mathbf{r}_{1}}, \varrho^{\mathbf{r}_{2}})$ simply by $X^{t}$ and $Y^{t}$ respectively, we can rewrite it as
\[
X^{t}\times Y^{t} \cap JM(2:2) =\emptyset.
\]
We need the following lemma.
\begin{lem}
	\label{lem: metric}
	Let $\delta>0$.
	There exists $\Delta>0$ such that 
	for all $\tau\in[0, \Delta]$ and for all $\CC\in X^{t-\tau}$, there exists $\A\in X^{t}$ satisfying 
	\[
	d(\CC, \A):=\|\CC(+)-\A(+)\|<\delta
	\]
	where $d$ is a metric on $O_{D}(2)$ defined through the operator norm $\|\cdot\|$ on $\mathcal{E}_{D}\simeq O_{D}(2)$.
\end{lem}

\begin{pf}
	By its definition, $X^{t}$ is a convex set of $O_{D}(2)$, and thus  for all $\E\in O_{D}(2)$ we can define successfully the distance between $\E$ and $X^{t}$: 
	\[
	d(\E, X^{t})=\min_{\F\in X^{t}}d(\E, \F).
	\]
	In particular, for $\E'\in X^{t-\Delta'}\subset O_{D}(2)$ with $\Delta'>0$ and $\E'(+)=\frac{1}{2}(e'_{0}\id+\mathbf{e}'\cdot\sigma)$, it becomes
	\begin{align}
		\label{eq:d(E', X^t)}
		\begin{aligned}
			d(\E', X^{t})
			=\min_{\F\in X^{t}}d(\E', \F)
			=\min_{\F\in X^{t}}\frac{1}{2}\left( |e'_0 - f_0 | 
			+ |\mathbf{e} - \mathbf{f}|\right),
		\end{aligned}
	\end{align}
	where $\F(+)=\frac{1}{2}(f_{0}\id+\mathbf{f}\cdot\sigma)$ (see \eqref{eq:norm for qubit effect}).
	Since, in terms of \eqref{eq:explicit C}, $\E'\in X^{t-\Delta'}$ and $\F\in X^{t}$ imply 
	\[
	(e_{0}', \mathbf{e}')=(1, t-\Delta', 0) + \lambda'(1, \mathbf{n})
	\]
	with $\lambda_{1}^{t-\Delta'}\le \lambda' \le \lambda_{2}^{t-\Delta'}$ and 
	\[
	(f_{0}, \mathbf{f})=(1, t, 0) + \lambda(1, \mathbf{n})
	\]
	with $\lambda_{1}^{t}\le \lambda \le \lambda_{2}^{t}$ respectively, \eqref{eq:d(E', X^t)} can be rewritten as
	\begin{align*}
		\begin{aligned}
			2d(\E', X^{t})=\min_{\lambda\in[\lambda_{1}^{t}, \lambda_{2}^{t}]}
			\left(
			|\lambda'-\lambda|+|(-\Delta', 0)+(\lambda'-\lambda)\mathbf{n}|
			\right).
		\end{aligned}
	\end{align*}
	It follows that 
	\begin{align}
		\label{eq:d(E', X^t) bound}
		2d(\E', X^{t})
		\le
		\Delta'+\min_{\lambda\in[\lambda_{1}^{t}, \lambda_{2}^{t}]}|\lambda'-\lambda|(1+|\mathbf{n}|).
	\end{align}
	Let us evaluate its right hand side.
	It is easy to see that 
	\[
	\min_{\lambda\in[\lambda_{1}^{t}, \lambda_{2}^{t}]}|\lambda'-\lambda|
	=\left\{
	\begin{aligned}
		\lambda_{1}^{t}&-\lambda' & \quad&(\lambda'<\lambda_{1}^{t})\\
		&0 & \quad&(\lambda_{1}^{t}\le\lambda'\le\lambda_{2}^{t})\\
		\lambda'&-\lambda_{2}^{t}
		& \quad &(\lambda'>\lambda_{2}^{t})
	\end{aligned}
	\right.
	.
	\]
	Suppose that $\lambda'<\lambda_{1}^{t}$ holds, for example.
	In this case, because $\lambda_{1}^{t-\Delta'}\le\lambda'$, we can obtain
	\[
	\lambda_{1}^{t}-\lambda'
	\le
	\lambda_{1}^{t}-\lambda_{1}^{t-\Delta'}.
	\]
	In a similar way, it can be demonstrated that
	\begin{align*}
		\sup_{\lambda'\in[\lambda_{1}^{t-\Delta'}, \lambda_{2}^{t-\Delta'}]}\min_{\lambda\in[\lambda_{1}^{t}, \lambda_{2}^{t}]}|\lambda'-\lambda|
		=\max\left\{
		\lambda_{1}^{t}-\lambda_{1}^{t-\Delta'},\ 0,\ 
		\lambda_{2}^{t-\Delta'}-\lambda_{2}^{t}
		\right\}.
	\end{align*}
	By virtue of \eqref{eq:lambda2}, the right hand side converges to 0 as $\Delta'\rightarrow0$, and thus we can see from \eqref{eq:d(E', X^t) bound} that
	\[
	\sup_{\E'\in X^{t-\Delta'}}
	d(\E', X^{t})\underset{\Delta'\to0}{\longrightarrow}0
	\]
	It results in that there exists $\Delta>0$ such that for all $\tau\in[0, \Delta]$, 
	\[
	\sup_{\E'\in X^{t-\tau}}
	d(\E', X^{t})<\delta
	\]
	holds, that is, $d(\CC, X^{t})<\delta$ holds for any $\CC\in X^{t-\tau}$.
	Moreover, because $X^{t}$ is convex, there exists $\A\in X^{t}$ satisfying $d(\CC, X^{t})=d(\CC, \A)$, which proves the claim of the lemma.\qed
\end{pf}
Note that a similar statement also holds for $Y^{t}$: there exists $\widetilde{\Delta}>0$ such that 
for all $\widetilde{\tau}\in[0, \widetilde{\Delta}]$ and for all $\D\in Y^{t-\widetilde{\tau}}$, there exists $\B\in Y^{t}$ satisfying $d(\D, \B)<\delta$.
Let $V:=O_{D}(2)\times O_{D}(2)(\simeq\mathcal{E}_{D}\times\mathcal{E}_{D})$ and 
let $d_{V}$ be a product metric on $V$ defined as 
\begin{align*}
	d_{V}\left((\A,\B), (\CC,\D)\right) 
	= \max\{d(\A, \CC), d(\B, \D)\}.
\end{align*}
According to Lemma \ref{lem: metric} and its $Y^{t}$-counterpart, if we take $\Delta_{0}=\min\{\Delta, \widetilde{\Delta}\}(>0)$, then 
there exists $(\A, \B)\in X^{t}\times Y^{t}$ for all $(\CC, \D)\in X^{t-\Delta_{0}}\times Y^{t-\Delta_{0}}$ such that $d_{V}((\A,\B), (\CC,\D))<\delta$.
On the other hand, as we have seen, it holds that
\[
X^{t}\times Y^{t} \cap JM(2:2) =\emptyset.
\]
Since $X^{t}\times Y^{t}$ and $JM(2:2)$ are closed in $V$, and $V$ is a metric space, we can apply Urysohn's Lemma \cite{Kelley1975}.
It follows that there exists a continuous (in fact uniformly continuous since $V$ is compact) function $f\colon V\to [0,1]$ satisfying $f(U)=0$ for any $U\in X^{t}\times Y^{t}$ and $f(W)=1$ for any $W\in JM(2:2)$.
The uniform continuity of $f$ implies that for some $\varepsilon\in(0,1)$, there is $\delta>0$ such that
\begin{align}
	\label{eq:continuous f}
	\begin{aligned}
		d_{V}\left((\E', \F'), (\E, \F)\right)<\delta\ \Rightarrow
		\ \left|f\left((\E', \F')\right)-f\left((\E, \F)\right)\right|<\varepsilon
	\end{aligned}
\end{align}
holds for any $(\E, \F)\in V$.
For this $\delta$, we can apply the argument above:
we can take $\Delta_{0}>0$ such that for any $(\CC, \D)\in X^{t-\Delta_{0}}\times Y^{t-\Delta_{0}}$, there exists $(\A, \B)\in X^{t}\times Y^{t}$ satisfying $d_{V}((\A,\B), (\CC,\D))<\delta$.
Because $f((\A,\B))=0$, we have $f((\CC,\D))<\varepsilon<1$ (see \eqref{eq:continuous f}), and thus $(\CC,\D)\notin JM(2:2)$ .
It indicates that $X^{t-\Delta_{0}}\times Y^{t-\Delta_{0}}\cap JM(2:2)=\emptyset$, that is, there is $\Delta_{0}>0$ for any $t\in L$ satisfying $t-\Delta_{0}\in L$.
Therefore, $t_{0}'=\inf L\notin L$ can be concluded.

%% file: chap5.tex
\chapter{Thermodynamical entropy of mixing in regular polygon theories}
\label{chap:TD entropy in GPT}
The concept of entropy plays an important role in thermodynamics \cite{Callen_thermo,zemansky_thermo}. 
It is possible to calculate the thermodynamical entropy of a mixture of classically different kinds of particles (such as a mixture of nitrogens and oxygens), and similar ideas were applied by von Neumann to the case where the system was composed of particles with different quantum internal states \cite{von1955mathematical}.
Similarly to the previous parts, it is expected that generalizing the notion of entropy to GPTs will help us to understand how entropy can affect our world.
In fact, 
there have been researches which aim to introduce and investigate the concept of entropy in GPTs from informational perspectives \cite{KIMURA2010175,1367-2630-12-3-033024,Short_2010,KimuraEntropiesinGeneralProbabilisticTheoriesandTheirApplicationtotheHolevoBound}. 
In those researches, some kinds of entropy were defined in all theories of GPTs and their information-theoretical properties were investigated. 
Meanwhile, there have been also researches referring to the thermodynamical entropy in terms of the microcanonical or canonical formulation in GPTs \cite{Chiribella_2017,chiribella2016entanglement}, and researches referring to the thermodynamical entropy of mixing in GPTs \cite{1367-2630-19-4-043025,EPTCS195.4}. 
However, in those works, the entropy was only defined in or applied to some restricted theories of GPTs with special assumptions such as the existence of a spectral decomposition for any state into perfectly distinguishable pure states.
In particular, it can be found that regular polygon theories do not always satisfy those assumptions, and thus entropy in regular polygon has never been investigated although they can be regarded as intermediate theories of a classical trit system and a qubit-like system, where entropy is defined successfully.
It seems natural to ask how entropy of mixing behaves in regular polygon theories.

In this part, we consider thermodynamical entropy of mixing in regular polygon theories. 
It is proven that the operationally natural thermodynamical entropy of a mixture of ideal particles with different internal states described by a regular polygon theory exists if and only if the state space of theory is triangle-shaped or disc-shaped, i.e., the theory is either classical or quantum-like. 
More precisely, we demonstrate that the thermodynamical entropy of mixing satisfying conditions imposed in \cite{1367-2630-19-4-043025}, where the concrete operational construction of the entropy was given as von Neumann did under the assumption of the existence of semipermeable membranes, does not exist in all the regular polygon theories except for classical and quantum-like ones. 

This part is organized as follows. 
In Section \ref{5sec:entropy}, we present a generalization of thermodynamically natural entropy of mixing in GPTs.
We will see that the notion of perfect distinguishability plays an important role to define entropy also in GPTs.
Then we demonstrate our main theorem and its proof in Section \ref{5sec:entropy main result}.

\section{Entropy of mixing in GPTs}
\label{5sec:entropy}
In this section, we introduce the thermodynamically consistent definition of entropy of mixing in GPTs based on the notion of perfect distinguishability.

\subsection{Perfect distinguishablity for regular polygon theories}
\begin{figure}[h]
	\hfill
	\begin{minipage}[b]{0.32\linewidth}
		\centering
		\includegraphics[scale=0.42]{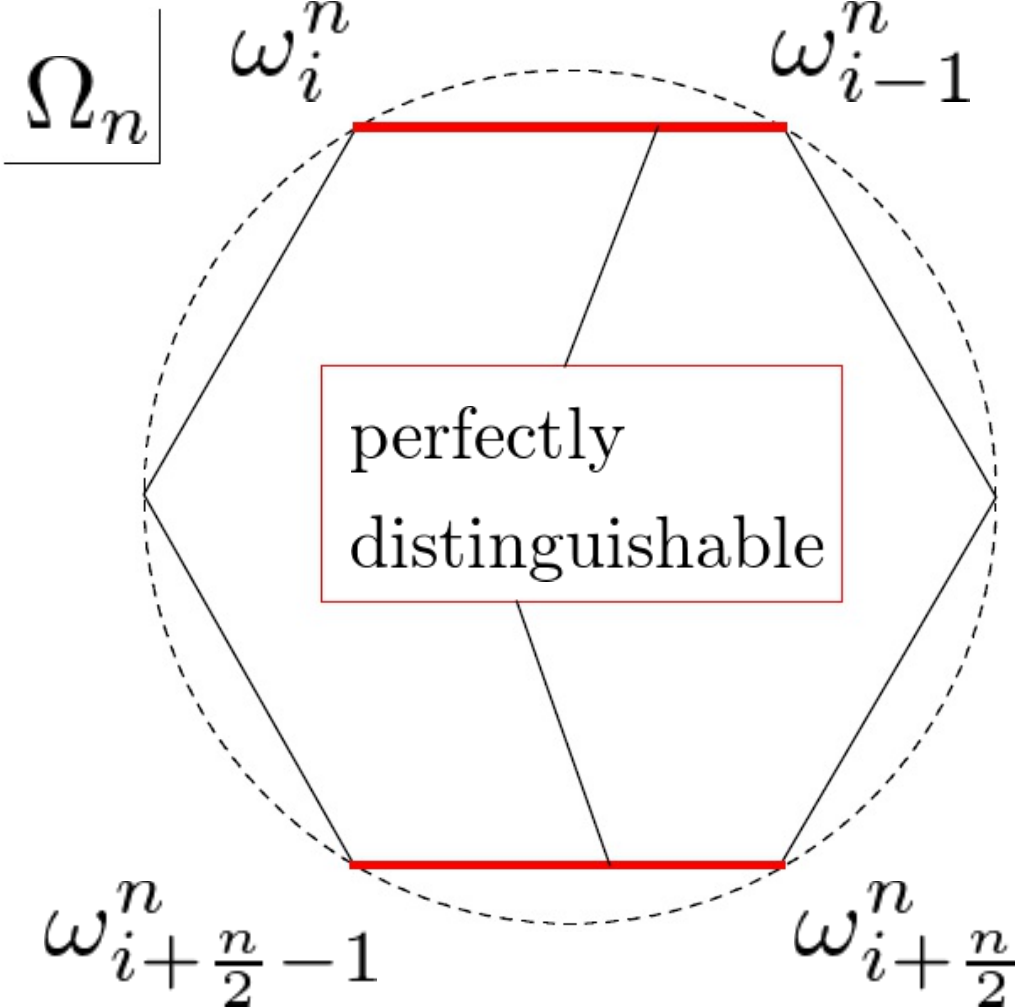}
		\subcaption{$n$ is an even number.}
	\end{minipage}
	\hfill
	\begin{minipage}[b]{0.32\linewidth}
		\centering
		\includegraphics[scale=0.42]{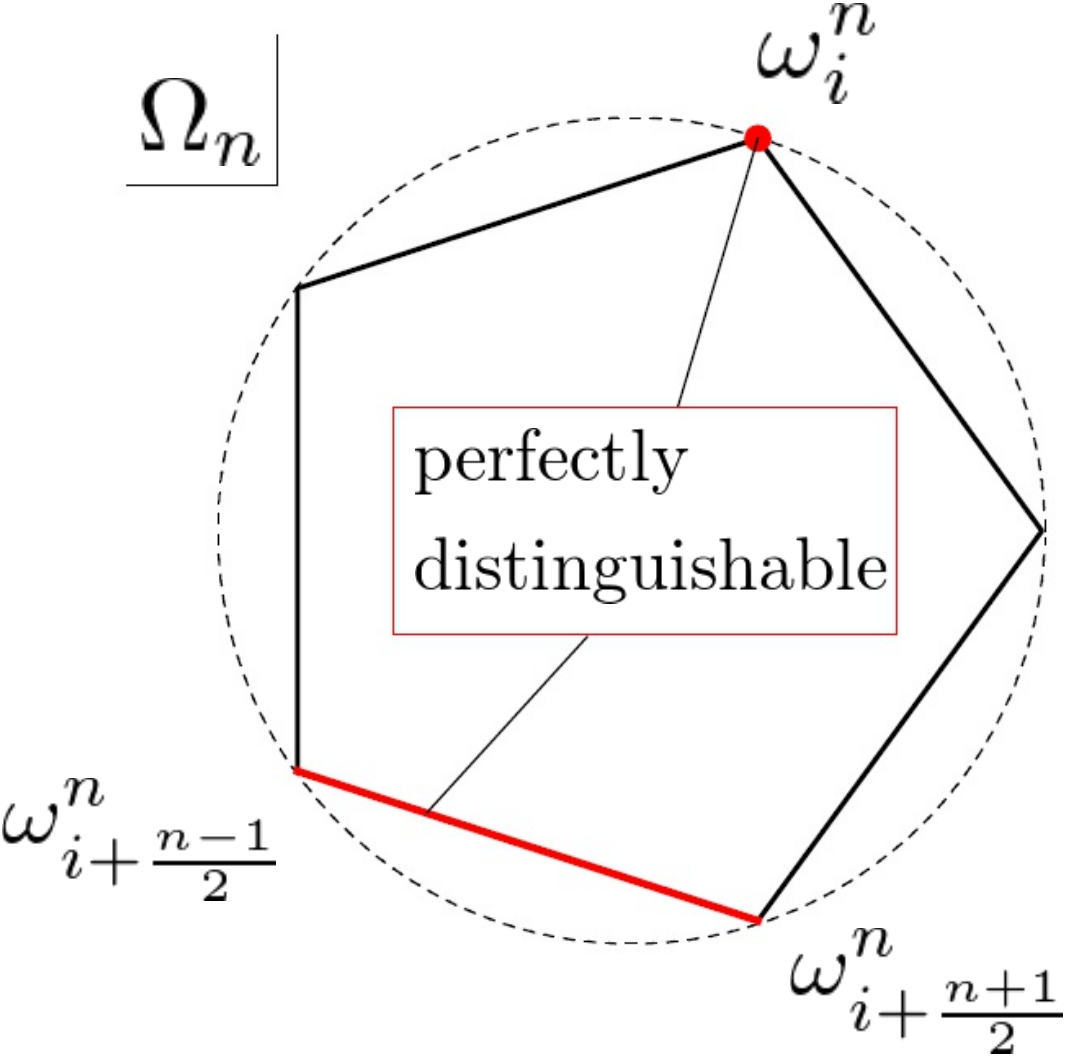}
		\subcaption{$n$ is an odd number.}
	\end{minipage}
	\hfill
	\begin{minipage}[b]{0.32\linewidth}
		\centering
		\includegraphics[scale=0.42]{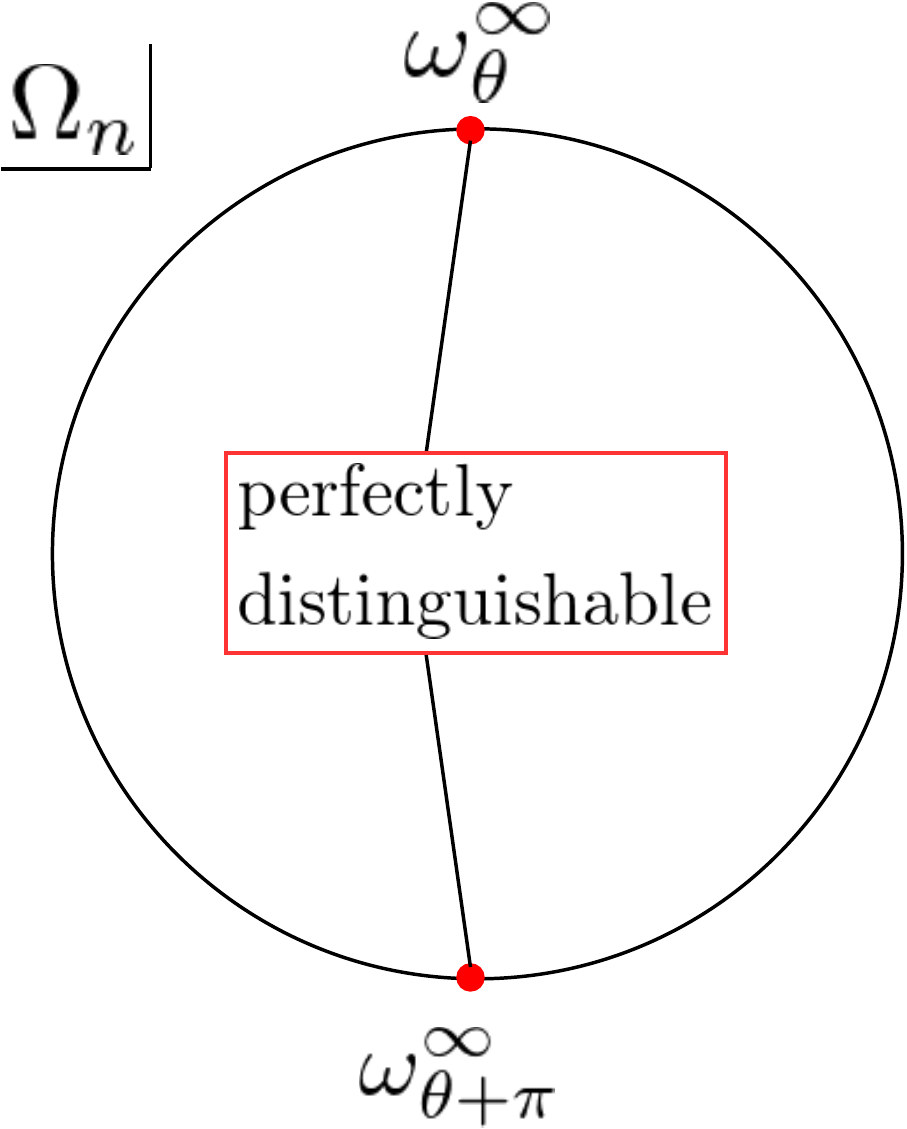}
		\subcaption{$n=\infty$.}
	\end{minipage}
	\hfill
	\null
	\caption{Pairs of perfectly distinguishable states in the $n$-gon state space.}
	\label{fig:n-gon}
\end{figure}

We recall that a family of states $\{\omega_i\}_i$ is called perfectly distinguishable if there exists an observable $\{e_i\}_i$ such that $e_{i}(\omega_j)=\delta_{ij}$.
Let us characterize perfectly distinguishable states in regular polygon theories. 
We first consider the regular polygon theory with $n$ sides, where $n$ is an even number greater than two.
Calculating the Euclidean inner product (denoted by $(\cdot, \cdot)$ here) of pure effects and pure states in Subsection \ref{eg_polygon}, we obtain 
\begin{align*}
	\left(e^n_i,\  \omega^n_i\right)=\left(e^n_i,\  \omega^{n}_{i-1}\right)=1,\ \ (e^n_i,\  \omega^{n}_{i+\frac{n}{2}-1})=(e^n_i,\ \omega^{n}_{i+\frac{n}{2}})=0.
\end{align*}
These equations indicate that any state in $\Omega_{n}^{[i-1,\ i]}$ is perfectly distinguishable from any state in $\Omega_{n}^{[i+\frac{n}{2}-1,\ i+\frac{n}{2}]}$, where we define
\begin{align*}
	\Omega_{n}^{[k-1,\ k]}=\{\omega\in\Omega_{n}\mid \omega=p\omega^{n}_{k-1}+(1-p)\omega^{n}_{k},\ 0\le p\le1\},
\end{align*}
since the measurement $\{e^{n}_{i},\ u-e^{n}_{i}\}$ distinguishes perfectly those two states. 
For odd $n\ (\ge3)$, we obtain
\begin{align*}
		\left(e^n_i,\  \omega^n_i\right)=1,\ \ 
		(e^n_i,\  \omega^{n}_{i+\frac{n-1}{2}})=(e^n_i,\ \omega^{n}_{i+\frac{n+1}{2}})=0.
\end{align*}
Hence $\omega^{n}_{i}$ and an arbitrary state in $\Omega_{n}^{[i+\frac{n-1}{2},\ i+\frac{n+1}{2}]}$ are perfectly distinguishable.
Finally, when $n=\infty$, 
\begin{align*}
	\left(e^\infty_\theta,\  \omega^\infty_\theta\right)=1,\ \ 
	(e^\infty_\theta,\  \omega^{\infty}_{\theta+\pi})=0.
\end{align*}
hold, so there is only one perfectly distinguishable state for each pure state (see Figure \ref{fig:n-gon}). 

\subsection{Entropy of mixing in GPTs}
\label{2c}
In this part, we consider the thermodynamical entropy of mixing in a system composed of ideal gases with different internal degrees of freedom described by a GPT. In thermodynamics, it is well known that a mixture of several classically distinct ideal gases, such like a mixture of ideal hydrogens and nitrogens, causes an increase of entropy. The amount of increase by the mixture can be calculated under the assumption of the existence of semipermeable membranes which distinguish perfectly those particles. We assume in a similar way that if the internal states $\omega_{1}, \omega_{2}, \cdots, \omega_{l}$ described by a GPT are perfectly distinguishable, then there exist semipermeable membranes which can identify completely a state among them without disturbing every $\omega_{j}$ $(j=1, 2, \cdots, l)$. 

We consider ideal gases in thermal equilibrium with its temperature $T$, volume $V$, and $N$ particles, and do not focus on the mechanical part of the particles in the following. All of these $N$ particles are in the same internal state $\omega=\sum_{i=1}^{l}p_{i}\omega_{i}$, where $\{\omega_{1}, \omega_{2}, \cdots, \omega_{l}\}$ is a perfectly distinguishable set of states, and $\forall i,\  p_{i}\ge0,\ \mbox{and}\  \sum_{i=1}^{l}p_{i}=1$, meaning that this system is composed of the mixture of $l$ different kinds of particles whose internal states are $\omega_{1}, \omega_{2}, \cdots, \omega_{l}$ with a probability weight $\{p_{1}, p_{2}, \cdots, p_{l}\}$. We note again that in this chapter, classical species of particles are also regarded as the internal states of them. In classical thermodynamics, thermodynamical entropy is calculated by constructing concrete thermodynamical operations such as isothermal or adiabatic quasistatic operations. We follow this doctrine of thermodynamics also in GPTs that thermodynamical entropy, especially thermodynamical entropy of mixing, should be operationally-derived quantity. In fact, as shown in \cite{1367-2630-19-4-043025}, our assumption of the existence of semipermeable membranes makes it possible to realize concrete thermodynamical operations to calculate the thermodynamical entropy of mixing of the system mentioned above in the same way as von Neumann did when the internal degrees of freedom were quantum \cite{von1955mathematical}. Strictly speaking, it has been demonstrated operationally in \cite{1367-2630-19-4-043025} that the thermodynamical entropy of mixing in the system is
\begin{equation}
	\label{eq1}
	S(\omega)=\sum_{i=1}^{l}p_{i}S(\omega_{i})-\sum_{i=1}^{l}p_{i}\log p_{i}\ ,
\end{equation}
where $S(\sigma)$ means the per-particle thermodynamical entropy of mixing in the system which consists of particles in the same state $\sigma$, and we set the Boltzmann constant $\kB=1$ (also $0\log0=0$). In the process of deriving $\eqref{5eq:TD}$, the additivity and extensivity of the thermodynamical entropy, and the continuity of $S$ with respect to states are assumed. 
The latter one is needed in order to apply $\eqref{5eq:TD}$ to arbitrary states with an arbitrary probability weight, while its operational derivation has been given only when each $p_{i}N$ is the number of particles in the state $\omega_{i}$ and thus each $p_{i}$ is rational.
We impose additional assumption that the entropy of any pure state is equal to zero, that is, $S(\sigma)=0$ whenever $\sigma$ is a pure state.

\section{Main result}
\label{5sec:entropy main result}
Our main result is in the following form.
\begin{thm*}
	Consider a system in thermal equilibrium composed of ideal gases whose internal states are described an element of the state space $\Omega_{n}$ of the regular polygon theory with $n$ sides ($n\ge3$).
	The (per-particle) thermodynamical entropy of mixing $S:\Omega_{n}\rightarrow\mathbb{R}$ satisfying
	\begin{align}
		\label{5eq:TD}
S(\omega)=\sum_{i=1}^{l}p_{i}S(\omega_{i})-\sum_{i=1}^{l}p_{i}\log p_{i},
	\end{align}
where $\{\omega_{i}\}_i$ is a family of perfectly distinguishable states, exists if and only if $n=3\ \mbox{or}\ \infty$, that is, the state space is classical or quantum-like.
\end{thm*}
\begin{pf}
	For $n=3$, because it is a classical system, any $\omega\in\Omega_{3}$ is decomposed uniquely into perfectly distinguishable pure states as $\omega=p\omega_{0}^{3}+q\omega_{1}^{3}+(1-p-q)\omega_{2}^{3}$, where $\omega_{i}^{3}\ (i=0, 1, 2)$ are the three pure states in $\Omega_{3}$ and $\{p, q, 1-p-q\}$ is a probability weight.
	In this settings, we define $S$ as 
	\begin{align*}
		S(\omega)=-p\log p-q\log q-(1-p-q)\log (1-p-q).
	\end{align*}
	This $S$ gives the well-defined entropy satisfying \eqref{5eq:TD}. Similarly, when $n=\infty$, any state has only one decomposition into perfectly distinguishable (pure) states except for the central state of $\Omega_{\infty}$ (the maximally mixed state).
	For states that are not maximally mixed, we define $S$ as 
	\begin{align*}
		S(\omega)=H(p),
	\end{align*}
	where we decompose a non-maximally-mixed $\omega\in\Omega_{\infty}$ as $\omega=p\omega_{\theta}^{\infty}+(1-p)\omega_{\theta+\pi}^{\infty}\ (0\le p\le1)$ and $H(p)=-p\log p-(1-p)\log(1-p)$ is the {\it 1-bit Shannon entropy}. We can apply this $S$ to the maximally mixed state, for the probability weight does not depend on the way of decompositions and they are always $\{\frac{1}{2},\frac{1}{2}\}$. Therefore, we can define successfully the thermodynamical entropy $S$ which meets \eqref{5eq:TD} for $n=3, \infty$. In the following, we prove the only if part.

\begin{figure}[h]
	\hfill
	\begin{minipage}[b]{0.47\linewidth}
		\centering
		\includegraphics[scale=0.45]{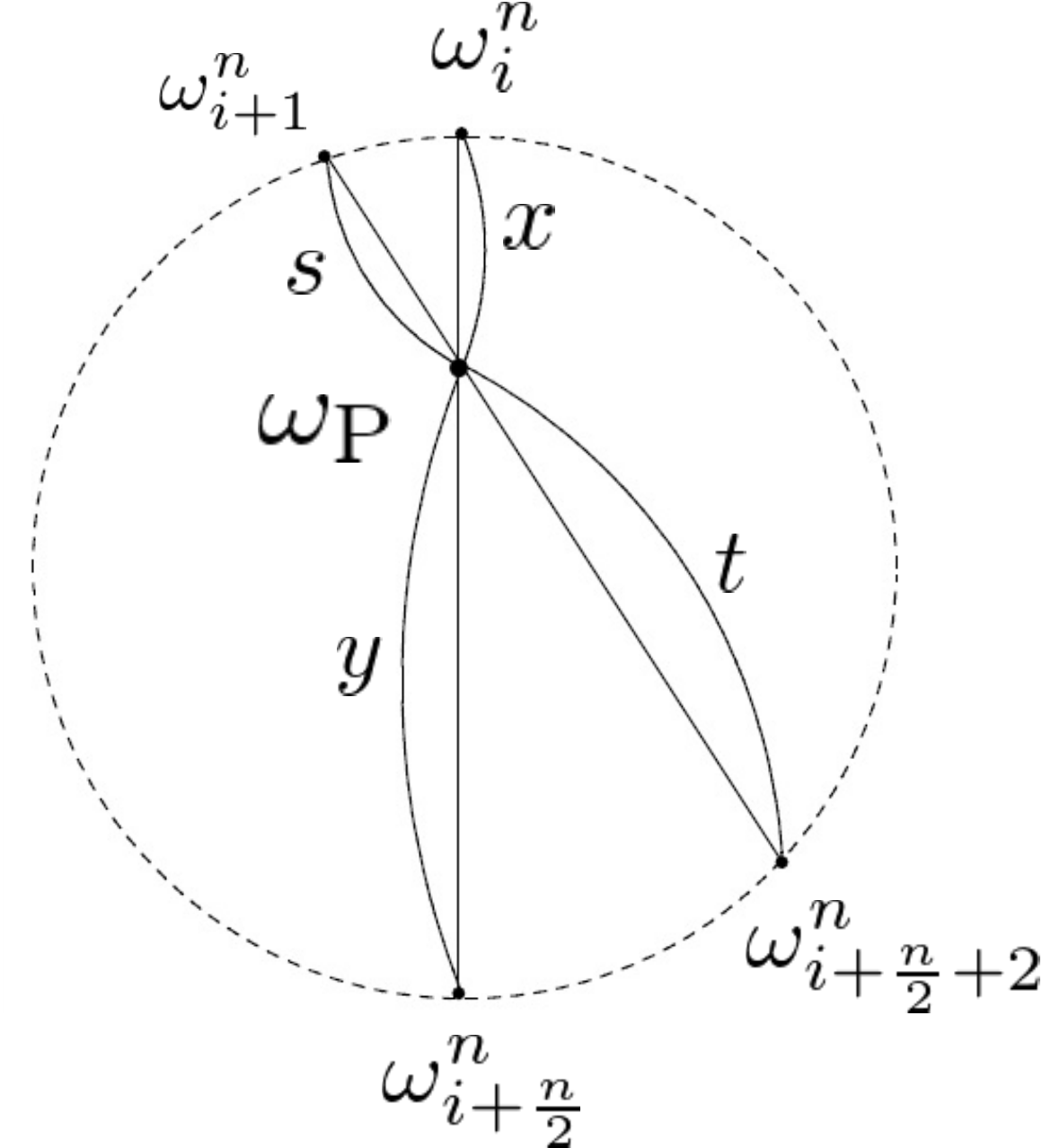}
	\end{minipage}
	\hfill
	\begin{minipage}[b]{0.48\linewidth}
		\centering
		\includegraphics[scale=0.45]{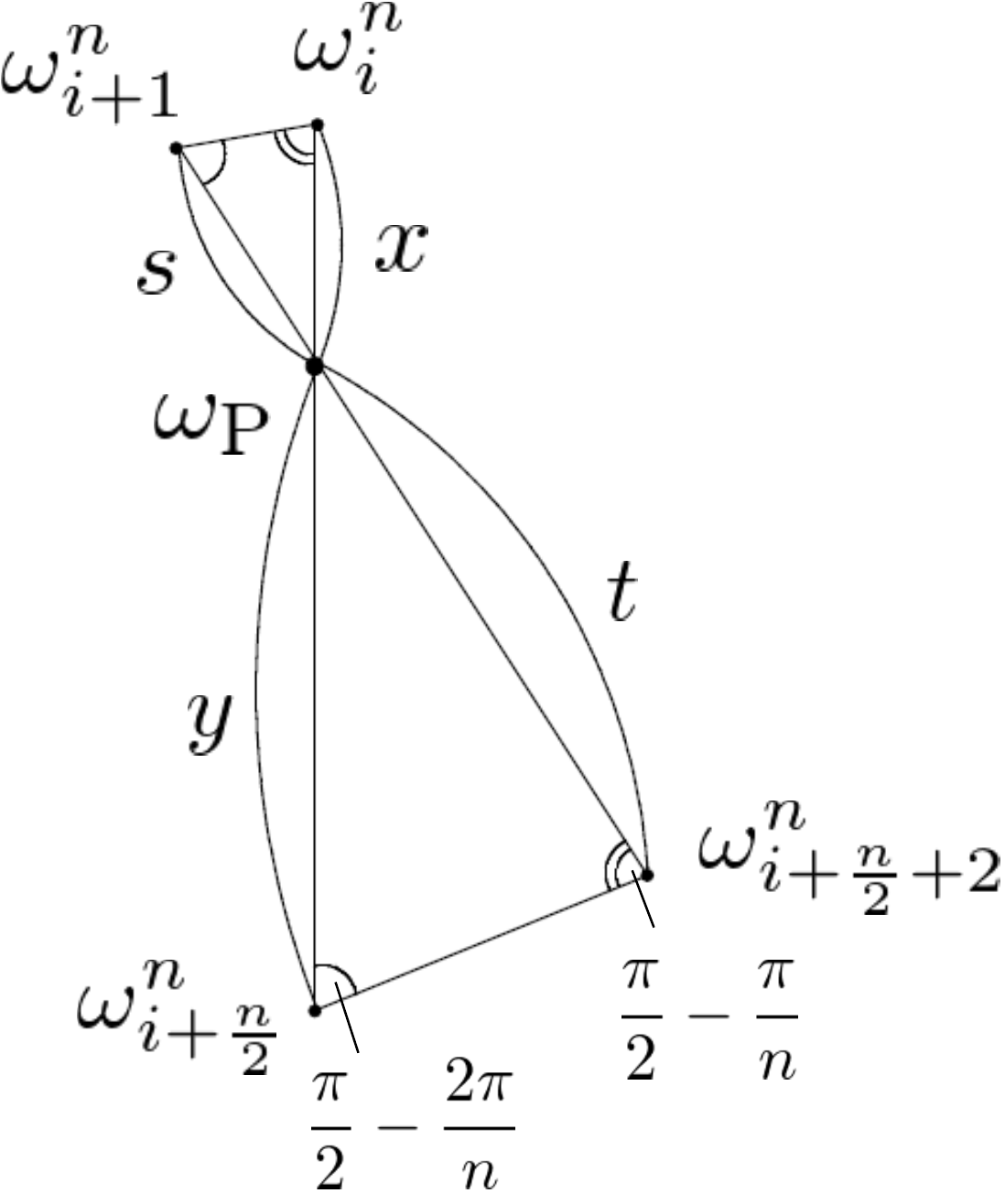}
	\end{minipage}
	\hfill\null
	\caption{Illustration of the state $\omega_{\mathrm{P}}$.}
	\label{fig:proof_even}
\end{figure}
The case when $n=4$ was proven in \cite{1367-2630-19-4-043025}, so we only consider $n\ge5$. At first, we assume $n$ is an even number, and consider the state $\omega_{\mathrm{P}}$ represented in Figure \ref{fig:proof_even}, 
	that is,
	\begin{align*}
		\omega_{\mathrm{P}}=\frac{y}{x+y}\omega_{i}^{n}+\frac{x}{x+y}\omega_{i+\frac{n}{2}}^{n}=\frac{t}{s+t}\omega_{i+1}^{n}+\frac{s}{s+t}\omega_{i+\frac{n}{2}+2}^{n},
	\end{align*}
	where $x, y, s, t$ are all nonnegative, and $x\le y$ and $s\le t$ as shown in Figure \ref{fig:proof_even}.
	Note that $\{\omega_{i}^{n},\ \omega_{i+\frac{n}{2}}^{n}\}$ and $\{\omega_{i+1}^{n},\ \omega_{i+\frac{n}{2}+2}^{n}\}$ are two perfectly distinguishable pairs of pure states. 
	From the observations in the previous section, we obtain two forms of the thermodynamical entropy of mixing:
	\begin{equation}
		\label{eq2}
		S(\omega_{\mathrm{P}})=H\left(\frac{x}{x+y}\right)=H\left(\frac{s}{s+t}\right),
	\end{equation}
	which means
	\begin{align*}
		\frac{x}{x+y}=\frac{s}{s+t}
	\end{align*}
	because $x\le y$ and $s\le t$. 
	On the other hand, applying sine theorem to Figure \ref{fig:proof_even} we can see that
	\begin{align*}
		\frac{x}{\sin(\frac{\pi}{2}-\frac{2\pi}{n})}=\frac{s}{\sin(\frac{\pi}{2}-\frac{\pi}{n})},\ \ 
		\frac{y}{\sin(\frac{\pi}{2}-\frac{\pi}{n})}=\frac{t}{\sin(\frac{\pi}{2}-\frac{2\pi}{n})},
	\end{align*}
	namely
	\begin{align*}
		\frac{s}{s+t}=\frac{x}{x+\left(\frac{\cos\frac{2\pi}{n}}{\cos\frac{\pi}{n}}\right)^{2}\ y}
	\end{align*}
holds.
	It follows from these two equations that
	\[
	\left(\frac{\cos\frac{2\pi}{n}}{\cos\frac{\pi}{n}}\right)^{2}=1,
	\] and because for even $n$, this equation holds if and only if $n=\infty$ ($\cos(\frac{\pi}{n})=1$), the entropy in \eqref{eq2} has been proven to be ill-defined.

	Next, we consider the case where $n$ is an odd number greater than three. We define the state $\omega_{\mathrm{A}}$ as $\omega_{\mathrm{A}}=\frac{1}{2}(\omega_{i+\frac{n-1}{2}}^{n}+\omega_{i+\frac{n+1}{2}}^{n})$,
	and consider two states $\omega_{\mathrm{Q}}$ and $\omega_{\mathrm{R}}$ shown in Figure \ref{fig:proof_odd}, where $j=\frac{n+1}{4}\ \mbox{or}\ \frac{n-1}{4}$ corresponding to the case where $n\equiv3$ or $n\equiv1$ (mod 4) respectively.
	\begin{figure}[!h]
		\hfill
		\begin{minipage}[b]{0.49\linewidth}
			\centering
			\includegraphics[scale=0.44]{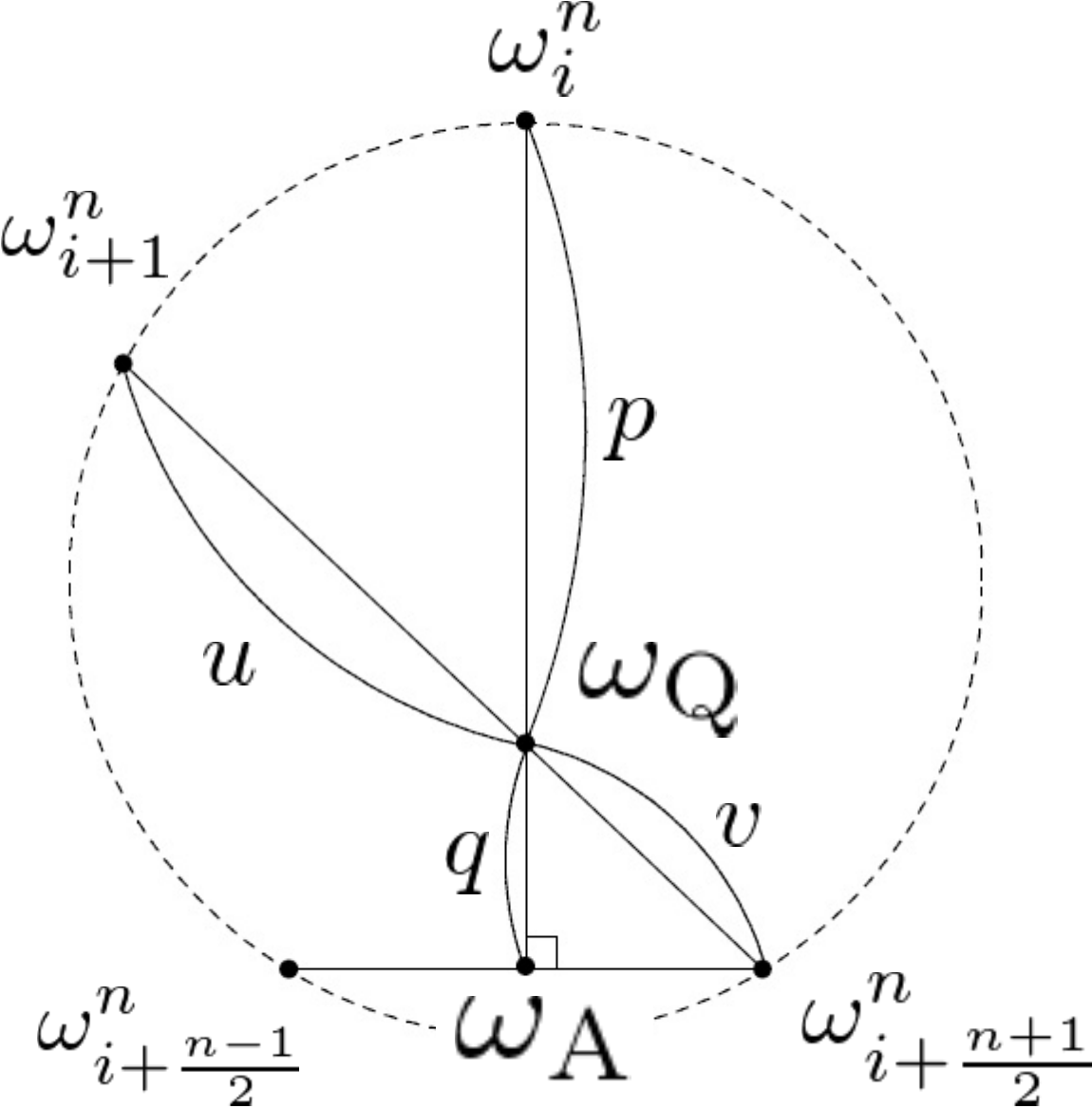}
			\subcaption{Illustration of the state $\omega_{\mathrm{Q}}$.}
			\label{fig:proof_odd1}
		\end{minipage}
		\hfill
		\begin{minipage}[b]{0.49\linewidth}
			\centering
			\includegraphics[scale=0.43]{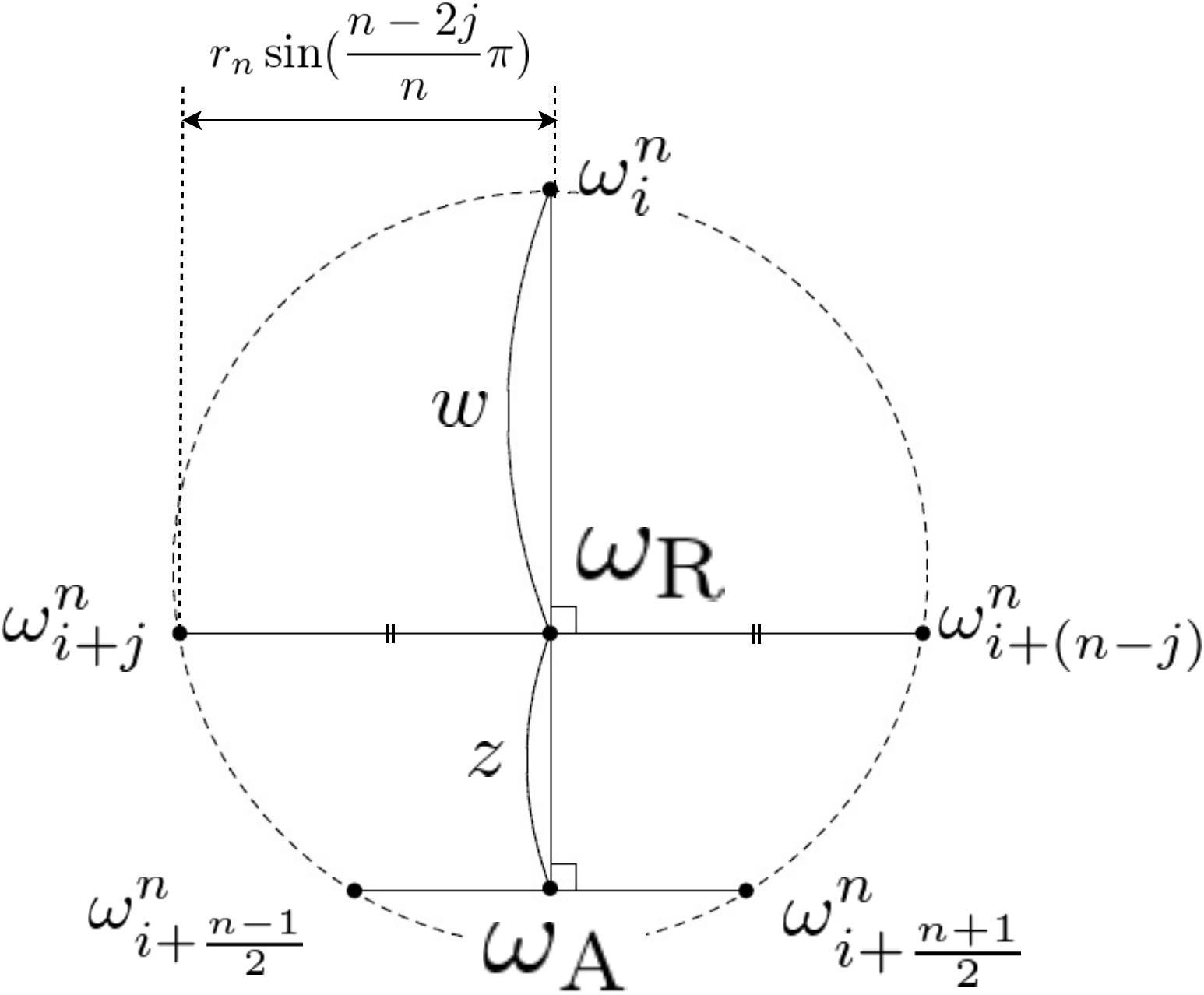}
			\subcaption{Illustration of the state $\omega_{\mathrm{R}}$.}
			\label{fig:proof_odd2}
		\end{minipage}
		\hfill
		\null
		\caption{Illustration of the states $\omega_{\mathrm{Q}}$ and $\omega_{\mathrm{R}}$.}
		\label{fig:proof_odd}
	\end{figure}
	Note that $\{\omega_{i}^{n},\ \omega_{\mathrm{A}}\}$ and $\{\omega_{i+1}^{n},\ \omega_{i+\frac{n+1}{2}}^{n}\}$ in Figure \ref{fig:proof_odd1}, and $\{\omega_{i+j}^{n},\ \omega_{i+(n-j)}^{n}\}$ in Figure \ref{fig:proof_odd2} are perfectly distinguishable pairs of states. 
	Then
	\begin{align*}
		S(\omega_{\mathrm{Q}})=\frac{p}{p+q}S(\omega_{\mathrm{A}})+H\left(\frac{p}{p+q}\right)=H\left(\frac{u}{u+v}\right),
	\end{align*}
	and
	\begin{align*}
		S(\omega_{\mathrm{R}})=\frac{w}{w+z}S(\omega_{\mathrm{A}})+H\left(\frac{w}{w+z}\right)=H\left(\frac{1}{2}\right)
	\end{align*}
	hold. 
	We assume that the entropies of the two states $\omega_{\mathrm{Q}},\ \omega_{\mathrm{R}}$ are well-defined (so is $\omega_{\mathrm{A}}$). 
	Then
	\begin{align}
		S(\omega_{\mathrm{A}})&=\frac{p+q}{p}\left\{H\left(\frac{u}{u+v}\right)-H\left(\frac{p}{p+q}\right)\right\}\label{eq:3} \\
		&=\frac{w+z}{w}\left\{H\left(\frac{1}{2}\right)-H\left(\frac{w}{w+z}\right)\right\}\label{eq:4}
	\end{align}
	holds. 
	Let us give the explicit expressions of \eqref{eq:4}.
	From Figure \ref{fig:proof_odd1_app}, we obtain
	\begin{figure}[h]
		\hfill
		\begin{minipage}[b]{0.47\linewidth}
			\centering
			\includegraphics[scale=0.45]{proof_odd1.pdf}
		\end{minipage}
		\hfill
		\begin{minipage}[b]{0.45\linewidth}
			\centering
			\includegraphics[scale=0.47]{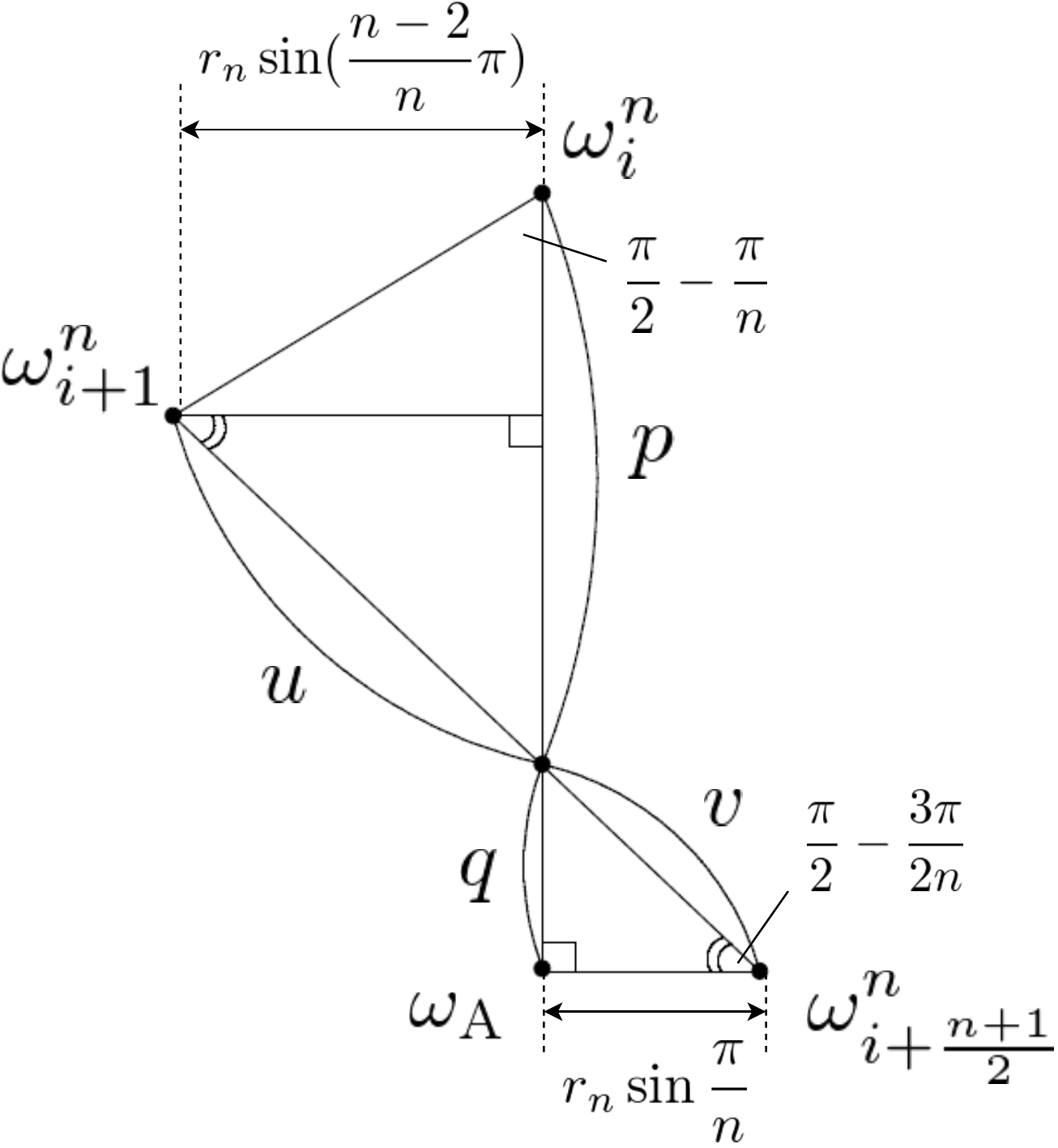}
		\end{minipage}
		\hfill\null
		\caption{The decomposition for $\omega_Q$.}
		\label{fig:proof_odd1_app}
	\end{figure}
	\begin{align*}
		\frac{u}{v}=\frac{r_{n}\sin(\frac{n-2}{n}\pi)}{r_{n}\sin\frac{\pi}{n}}
		=\frac{\sin\frac{2\pi}{n}}{\sin\frac{\pi}{n}}=2\cos\frac{\pi}{n},
	\end{align*}
	and by sine theorem, 
	\begin{align*}
		\frac{p}{\sin(\frac{\pi}{2}-\frac{\pi}{2n})}=\frac{u}{\sin(\frac{\pi}{2}-\frac{\pi}{n})},\ \ 
		q=v\cos\frac{3\pi}{2n}
	\end{align*}
	hold. 
	Therefore,
	\begin{align*}
		\frac{q}{p}&=\frac{v}{u}\cdot\frac{\cos\frac{\pi}{n}\cdot\cos\frac{3\pi}{2n}}{\cos\frac{\pi}{2n}}\\
		&=\frac{\cos\frac{3\pi}{2n}}{2\cos\frac{\pi}{2n}}\\
		&=\frac{1}{2}\left(4\cos^{2}\frac{\pi}{2n}-3\right)\\
		&=\frac{1}{2}\left(2\cos\frac{\pi}{n}-1\right).
	\end{align*}
	On the other hand, from Figure \ref{fig:proof_odd2}, we obtain
	\begin{align}
		\label{eq:app3}
		\begin{aligned}
			\frac{z}{w}&=\frac{2r_{n}\cos^{2}\frac{\pi}{2n}-2r_{n}\cos^{2}(\frac{n-2j}{2n}\pi)}{2r_{n}\cos^{2}(\frac{n-2j}{2n}\pi)}\\
			&=\frac{\cos^{2}\frac{\pi}{2n}-\sin^{2}\frac{j\pi}{n}}{\sin^{2}\frac{j\pi}{n}}\ \ \ \ \mbox{for}\ \ j=\frac{n\pm1}{4}.
		\end{aligned}
	\end{align}
	Since
	\begin{equation}
		\label{eq:app4}
		\begin{aligned}
			\sin^{2}\frac{j\pi}{n}&=\frac{1}{2}\left(1-\cos\frac{2j\pi}{n}\right)\\
			&=\frac{1}{2}\left(1-\cos\frac{(n\pm1)\pi}{2n}\right)\\
			&=\frac{1}{2}\left(1\pm\sin\frac{\pi}{2n}\right),
		\end{aligned}
	\end{equation}
	the equation above can be written as
	\begin{align*}
		\frac{z}{w}&=\frac{2(1+\sin\frac{\pi}{2n})(1-\sin\frac{\pi}{2n})-(1\pm\sin\frac{\pi}{2n})}{1\pm\sin\frac{\pi}{2n}}\\
		&=1\mp2\sin\frac{\pi}{2n},
	\end{align*}
	where the double sign corresponds to the ones in \eqref{eq:app3} and \eqref{eq:app4}, and the upper and lower sign correspond to the case of $n\equiv3$ and $n\equiv1$ (mod 4) respectively. 
	Substituting these results to \eqref{eq:3} and \eqref{eq:4}, we obtain
	\begin{align*}
		S(\omega_{\mathrm{A}})
		&=\left(\frac{2\cos\frac{\pi}{n}+1}{2}\right)\left\{H\left(\frac{1}{2\cos\frac{\pi}{n}+1}\right)-H\left(\frac{2}{2\cos\frac{\pi}{n}+1}\right)\right\}\\
		&=\left(2\mp2\sin\frac{\pi}{2n}\right)\left\{H\left(\frac{1}{2}\right)-H\left(\frac{1}{2\mp2\sin\frac{\pi}{2n}}\right)\right\}.
	\end{align*}
\begin{figure}[h]
	\hfill
	\begin{minipage}[b]{0.49\linewidth}
		\includegraphics[scale=0.4]{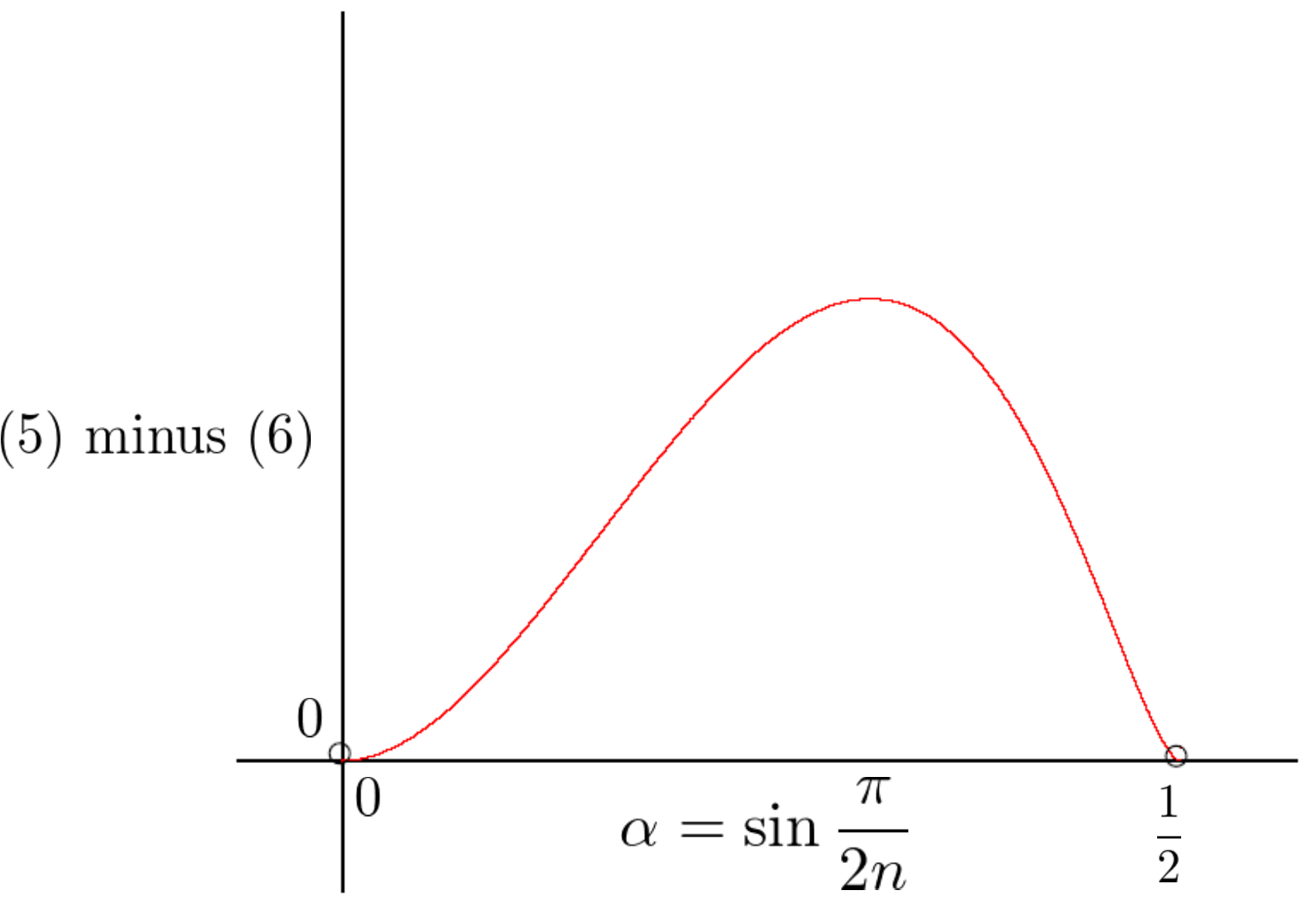}
		\subcaption{$n\equiv3$.}
		\label{fig:graph_equiv3}
	\end{minipage}
	\hfill
	\begin{minipage}[b]{0.49\linewidth}
		\centering
		\includegraphics[scale=0.4]{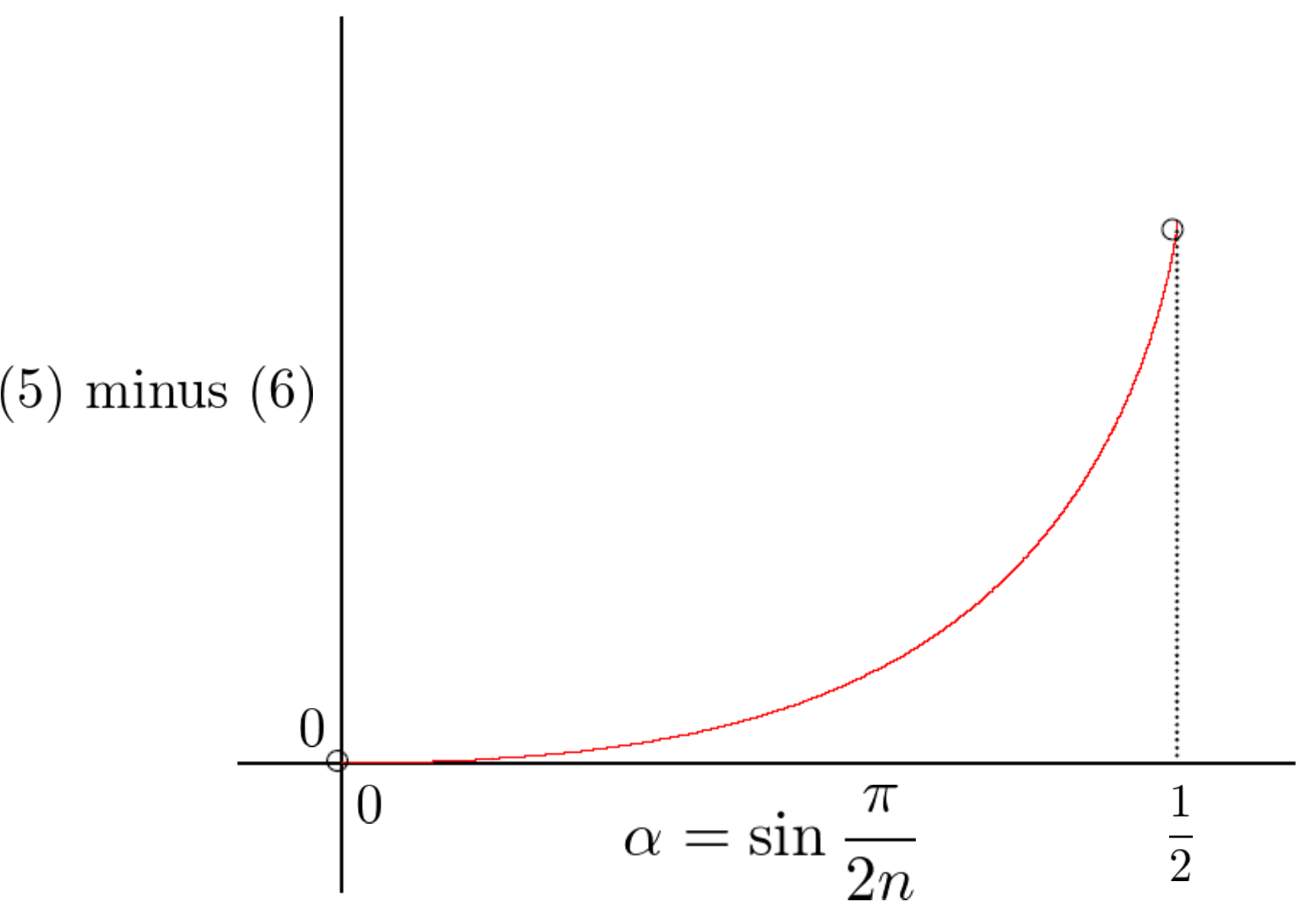}
		\subcaption{$n\equiv1$.}
		\label{fig:graph_equiv1}
	\end{minipage}
	\hfill\null
	\caption{The difference between \eqref{eq:5} and \eqref{eq:6}.}
	\label{fig:graph}
\end{figure}\\
	By letting $\alpha=\sin\frac{\pi}{2n}$, it can be rewritten as
	\begin{align}
		S(\omega_{\mathrm{A}})&=2\alpha^{2}\log2+\frac{1-4\alpha^{2}}{2}\log(1-4\alpha^{2})-(1-2\alpha^{2})\log(1-2\alpha^{2})\label{eq:5}\\
		&=(1\mp2\alpha)\log(1\mp2\alpha)-(2\mp2\alpha)\log(1\mp\alpha)\label{eq:6},
	\end{align}
	where the upper and lower signs correspond to the case of $n\equiv3$ and $n\equiv1$ (mod 4) respectively.
	The differences between \eqref{eq:5} and \eqref{eq:6} in the case of $n\equiv3$ and $n\equiv1$ are displayed in Figures \ref{fig:graph_equiv3} and \ref{fig:graph_equiv1} respectively, and we can see that the two forms of $S(\omega_{\mathrm{A}})$ shown in \eqref{eq:5} and \eqref{eq:6} do not agree with each other. 
	In conclusion, it has been proven that if $n\neq3,\ \infty$, then there exists some state whose thermodynamical entropy of mixing is ill-defined.\qed
\end{pf}
We can see the ill-defined values of entropy become well-defined if $n=3,\ \infty$ in our proof. 
For example, when $n$ is an odd number, $\alpha=\sin\frac{\pi}{2n}$ is equal to $\frac{1}{2}$ or $0$ if $n$ is equal to three or infinite, respectively, and two values \eqref{eq:5} and \eqref{eq:6} coincide with each other in these cases (see Figure \ref{fig:graph}). 

\begin{rmk}
Similar results were obtained in \cite{1367-2630-19-4-043025}, where it was assumed that any state could be represented as a convex combination of perfectly distinguishable pure states. 
However, a state of a regular polygon theory is not always represented by a convex combination of perfectly distinguishable pure states. 
For instance, we can see from Figure \ref{fig:n-gon} that the state $\omega_{\mathrm{A}}$ in Figure \ref{fig:proof_odd1} or Figure \ref{fig:proof_odd2} can not be decomposed into perfectly distinguishable pure states. 
Thus regular polygon theories generally do not satisfy the assumption in the previous study \cite{1367-2630-19-4-043025}, and our result is the one about the exsistence of well-defined thermodynamical entropy in such a broader class of theories where ``spectral decompositions" of states are not generally possible.
\end{rmk}

%% file: Takakura_PhD.bbl
\begin{thebibliography}{100}

\bibitem{Takakura2020}
R.~Takakura and T.~Miyadera, \href {https://doi.org/10.1063/5.0017854}
  {``Preparation uncertainty implies measurement uncertainty in a class of
  generalized probabilistic theories,''} {\em Journal of Mathematical Physics},
  vol.~61, no.~8, p.~082203, 2020.

\bibitem{Takakura_entropic_2021}
R.~Takakura and T.~Miyadera, \href {https://doi.org/10.1088/1751-8121/ac0c5c}
  {``Entropic uncertainty relations in a class of generalized probabilistic
  theories,''} {\em Journal of Physics A: Mathematical and Theoretical},
  vol.~54, p.~315302, July 2021.

\bibitem{PhysRevA.104.032228}
T.~Heinosaari, T.~Miyadera, and R.~Takakura, \href
  {https://doi.org/10.1103/PhysRevA.104.032228} {``Testing incompatibility of
  quantum devices with few states,''} {\em Physical Review A}, vol.~104,
  p.~032228, Sept. 2021.

\bibitem{Takakura_2019}
R.~Takakura, \href {https://doi.org/10.1088/1751-8121/ab4a2e} {``Entropy of
  mixing exists only for classical and quantum-like theories among the regular
  polygon theories,''} {\em Journal of Physics A: Mathematical and
  Theoretical}, vol.~52, p.~465302, Oct. 2019.

\bibitem{von1955mathematical}
J.~Von~Neumann, {\em Mathematical Foundations of Quantum Mechanics}.
\newblock Princeton: Princeton University Press, 1955.

\bibitem{Hardy_Spekken_foundation}
L.~Hardy and R.~Spekkens, \href {https://arxiv.org/abs/2103.07469} {``Why
  physics needs quantum foundations,''} {\em Physics in Canada}, vol.~66,
  no.~2, pp.~73--76, 2010.

\bibitem{Chiribella_Spekken_2016}
G.~Chiribella and R.~W. Spekkens, eds., \href
  {https://doi.org/https://doi.org/10.1007/978-94-017-7303-4} {{\em Quantum
  Theory: Informational Foundations and Foils}}.
\newblock No.~181 in Fundamental Theories of Physics, Springer, Dordrecht,
  1st~ed., 2016.

\bibitem{Heisenberg1927}
W.~Heisenberg, \href {https://doi.org/10.1007/BF01397280} {``{\"U}ber den
  anschaulichen inhalt der quantentheoretischen {K}inematik und {M}echanik,''}
  {\em Zeitschrift f{\"u}r Physik}, vol.~43, pp.~172--198, Mar. 1927.

\bibitem{PhysRev.47.777}
A.~Einstein, B.~Podolsky, and N.~Rosen, \href
  {https://doi.org/10.1103/PhysRev.47.777} {``Can quantum-mechanical
  description of physical reality be considered complete?,''} {\em Physical
  Review}, vol.~47, pp.~777--780, May 1935.

\bibitem{Bell_inequality}
J.~S. Bell, \href {https://doi.org/10.1103/PhysicsPhysiqueFizika.1.195} {``On
  the {E}instein {P}odolsky {R}osen paradox,''} {\em Physics Physique Fizika},
  vol.~1, pp.~195--200, Nov. 1964.

\bibitem{BB84}
C.~H. Bennett and G.~Brassard, \href {https://arxiv.org/abs/2003.06557}
  {``Quantum cryptography: Public key distribution and coin tossing,''} in {\em
  International Conference on Computers, Systems \& Signal Processing},
  pp.~175--179, 1984.

\bibitem{E91}
A.~K. Ekert, \href {https://doi.org/10.1103/PhysRevLett.67.661} {``Quantum
  cryptography based on {B}ell's theorem,''} {\em Physical Review Letters},
  vol.~67, pp.~661--663, Aug. 1991.

\bibitem{Lahti1985}
P.~J. Lahti and S.~Bugajski, \href
  {https://doi.org/https://doi.org/10.1007/BF00671306} {``Fundamental
  principles of quantum theory. ii. from a convexity scheme to the {DHB}
  theory,''} {\em International Journal of Theoretical Physics}, vol.~24,
  pp.~1051--1080, 1985.

\bibitem{HORODECKI19961}
M.~Horodecki, P.~Horodecki, and R.~Horodecki, \href
  {https://doi.org/https://doi.org/10.1016/S0375-9601(96)00706-2}
  {``Separability of mixed states: necessary and sufficient conditions,''} {\em
  Physics Letters A}, vol.~223, no.~1, pp.~1--8, 1996.

\bibitem{TERHAL2000319}
B.~M. Terhal, \href
  {https://doi.org/https://doi.org/10.1016/S0375-9601(00)00401-1} {``{B}ell
  inequalities and the separability criterion,''} {\em Physics Letters A},
  vol.~271, no.~5, pp.~319--326, 2000.

\bibitem{doi:10.1063/1.5126496}
C.~Carmeli, T.~Heinosaari, T.~Miyadera, and A.~Toigo, \href
  {https://doi.org/10.1063/1.5126496} {``Witnessing incompatibility of quantum
  channels,''} {\em Journal of Mathematical Physics}, vol.~60, no.~12,
  p.~122202, 2019.

\bibitem{PhysRevLett.122.130402}
C.~Carmeli, T.~Heinosaari, and A.~Toigo, \href
  {https://doi.org/10.1103/PhysRevLett.122.130402} {``Quantum incompatibility
  witnesses,''} {\em Physical Review Letters}, vol.~122, p.~130402, Apr. 2019.

\bibitem{doi:10.1063/1.3614503}
T.~Miyadera, \href {https://doi.org/10.1063/1.3614503} {``Uncertainty relations
  for joint localizability and joint measurability in finite-dimensional
  systems,''} {\em Journal of Mathematical Physics}, vol.~52, no.~7, p.~072105,
  2011.

\bibitem{PhysRevLett.112.050401}
F.~Buscemi, M.~J.~W. Hall, M.~Ozawa, and M.~M. Wilde, \href
  {https://doi.org/10.1103/PhysRevLett.112.050401} {``Noise and disturbance in
  quantum measurements: An information-theoretic approach,''} {\em Physical
  Review Letters}, vol.~112, p.~050401, Feb. 2014.

\bibitem{Wootters_Zurek_no-cloning}
W.~K. Wootters and W.~H. Zurek, \href
  {https://doi.org/doi.org/10.1038/299802a0} {``A single quantum cannot be
  cloned,''} {\em Nature}, vol.~299, pp.~802--803, 1982.

\bibitem{Heinosaari_2016}
T.~Heinosaari, T.~Miyadera, and M.~Ziman, \href
  {https://doi.org/10.1088/1751-8113/49/12/123001} {``An invitation to quantum
  incompatibility,''} {\em Journal of Physics A: Mathematical and Theoretical},
  vol.~49, p.~123001, Feb. 2016.

\bibitem{hardy2001quantum}
L.~Hardy, \href {http://arxiv.org/abs/arXiv:quant-ph/0101012} {``Quantum theory
  from five reasonable axioms,''} 2001, arXiv:quant-ph/0101012.

\bibitem{Barnum_nocloning}
H.~Barnum, J.~Barrett, M.~Leifer, and A.~Wilce, \href
  {http://arxiv.org/abs/arXiv:quant-ph/0611295} {``Cloning and broadcasting in
  generic probabilistic theories,''} 2006, arXiv:quant-ph/0611295.

\bibitem{PhysRevA.75.032304}
J.~Barrett, \href {https://doi.org/10.1103/PhysRevA.75.032304} {``Information
  processing in generalized probabilistic theories,''} {\em Physical Review A},
  vol.~75, p.~032304, Mar. 2007.

\bibitem{PhysRevLett.99.240501}
H.~Barnum, J.~Barrett, M.~Leifer, and A.~Wilce, \href
  {https://doi.org/10.1103/PhysRevLett.99.240501} {``Generalized
  no-broadcasting theorem,''} {\em Physical Review Letters}, vol.~99,
  p.~240501, Dec. 2007.

\bibitem{PhysRevA.81.062348}
G.~Chiribella, G.~M. D'Ariano, and P.~Perinotti, \href
  {https://doi.org/10.1103/PhysRevA.81.062348} {``Probabilistic theories with
  purification,''} {\em Physical Review A}, vol.~81, p.~062348, June 2010.

\bibitem{PhysRevA.84.012311}
G.~Chiribella, G.~M. D'Ariano, and P.~Perinotti, \href
  {https://doi.org/10.1103/PhysRevA.84.012311} {``Informational derivation of
  quantum theory,''} {\em Physical Review A}, vol.~84, p.~012311, July 2011.

\bibitem{Masanes_Muller_2011}
L.~Masanes and M.~P. M\"uller, \href
  {https://doi.org/10.1088/1367-2630/13/6/063001} {``A derivation of quantum
  theory from physical requirements,''} {\em New Journal of Physics}, vol.~13,
  p.~063001, June 2011.

\bibitem{barnum2012teleportation}
H.~Barnum, J.~Barrett, M.~Leifer, and A.~Wilce, \href
  {https://doi.org/https://doi.org/10.1090/psapm/071} {``Teleportation in
  general probabilistic theories,''} in {\em Proceedings of Symposia in Applied
  Mathematics}, vol.~71, pp.~25--48, 2012.

\bibitem{Lami_PhD}
L.~Lami, \href {https://ddd.uab.cat/record/187745} {{\em Non-classical
  correlations in quantum mechanics and beyond}}.
\newblock PhD thesis, Universitat Aut\`onoma de Barcelona, 2017.

\bibitem{Plavala_2021_GPTs}
M.~Pl\'avala, \href {http://arxiv.org/abs/arXiv:2103.07469} {``General
  probabilistic theories: An introduction,''} 2021, arXiv:2103.07469.

\bibitem{Mackey1963}
G.~W. Mackey, {\em The mathematical foundations of quantum mechanics: a
  lecture-note volume}.
\newblock New York ; Amsterdam: W.A. Benjamin, Inc., 1963.

\bibitem{Ludwig1964}
G.~Ludwig, \href {https://doi.org/10.1007/BF01418533} {``{V}ersuch einer
  axiomatischen {G}rundlegung der {Q}uantenmechanik und allgemeinerer
  physikalischer {T}heorien,''} {\em Zeitschrift f\"ur Physik}, vol.~181,
  pp.~233--260, 1964.

\bibitem{Ludwig1967}
G.~Ludwig, \href {https://doi.org/10.1007/BF01653647} {``Attempt of an
  axiomatic foundation of quantum mechanics and more general theories, ii,''}
  {\em Communications in Mathematical Physics}, vol.~4, pp.~331--348, 1967.

\bibitem{Davies1970}
E.~B. Davies and J.~T. Lewis, \href {https://doi.org/10.1007/BF01647093} {``An
  operational approach to quantum probability,''} {\em Communications in
  Mathematical Physics}, vol.~17, pp.~239--260, Sept. 1970.

\bibitem{cmp/1103858551}
S.~Gudder, \href {https://doi.org/https://doi.org/10.1007/BF01645250} {``Convex
  structures and operational quantum mechanics,''} {\em Communications in
  Mathematical Physics}, vol.~29, no.~3, pp.~249 -- 264, 1973.

\bibitem{Hartkaemper_Neumann_1974}
A.~Hartk\"amper and H.~Neumann, eds., \href
  {https://doi.org/10.1007/3-540-06725-6} {{\em Foundations of Quantum
  Mechanics and Ordered Linear Spaces: Advanced Study Institute held in Marburg
  1973}}, vol.~29 of {\em Lecture Notes in Physics}.
\newblock Springer-Verlag Berlin Heidelberg, 1st~ed., 1974.

\bibitem{Cassinelli2016}
G.~Cassinelli and P.~Lahti, \href {https://doi.org/10.1007/s10701-016-0022-y}
  {``An axiomatic basis for quantum mechanics,''} {\em Foundations of Physics},
  vol.~46, pp.~1341--1373, 2016.

\bibitem{Kraus1983}
K.~Kraus, \href {https://doi.org/https://doi.org/10.1007/3-540-12732-1} {{\em
  States, Effects, and Operations: Fundamental Notions of Quantum Theory}},
  vol.~190 of {\em Lecture Notes in Physics}.
\newblock Springer-Verlag Berlin Heidelberg, 1st~ed., 1983.

\bibitem{Araki_mathematical}
H.~Araki, {\em Mathematical Theory of Quantum Fields}.
\newblock International Series of Monographs on Physics, Oxford: Oxford
  University Press, 1999.

\bibitem{Busch_quantummeasurement}
P.~Busch, P.~J. Lahti, J.-P. Pellonp{\"a}{\"a}, and K.~Ylinen, \href
  {https://doi.org/https://doi.org/10.1007/978-3-319-43389-9} {{\em Quantum
  Measurement}}.
\newblock Theoretical and Mathematical Physics, Springer International
  Publishing, 2016.

\bibitem{Holevo1982}
A.~S. Holevo, {\em Probabilistic and Statistical Aspects of Quantum Theory},
  vol.~1 of {\em North-Holland series in statistics and probability}.
\newblock Amsterdam ; New York: North-Holland Pub. Co., 1982.

\bibitem{Gudder_stochastic}
S.~P. Gudder, {\em Stochastic Methods in Quantum Mechanics}.
\newblock New York: Dover, 1979.

\bibitem{Kuramochi_compactconvex_2020}
Y.~Kuramochi, \href {http://arxiv.org/abs/arXiv:2002.03504} {``Compact convex
  structure of measurements and its applications to simulability,
  incompatibility, and convex resource theory of continuous-outcome
  measurements,''} 2020, arXiv:2002.03504.

\bibitem{kimura2010physical}
G.~Kimura, K.~Nuida, and H.~Imai, \href {http://arxiv.org/abs/arXiv:1012.5361}
  {``Physical equivalence of pure states and derivation of qubit in general
  probabilistic theories,''} 2010, arXiv:1012.5361.

\bibitem{PhysRevA.71.052108}
R.~W. Spekkens, \href {https://doi.org/10.1103/PhysRevA.71.052108}
  {``Contextuality for preparations, transformations, and unsharp
  measurements,''} {\em Physical Review A}, vol.~71, p.~052108, May 2005.

\bibitem{Filippov2020_restriction}
S.~N. Filippov, S.~Gudder, T.~Heinosaari, and L.~Lepp\"aj\"arvi, \href
  {https://doi.org/10.1007/s10701-020-00352-6} {``Operational restrictions in
  general probabilistic theories,''} {\em Foundations of Physics}, vol.~50,
  p.~850–876, 2020.

\bibitem{PhysRevA.87.052131}
P.~Janotta and R.~Lal, \href {https://doi.org/10.1103/PhysRevA.87.052131}
  {``Generalized probabilistic theories without the no-restriction
  hypothesis,''} {\em Physical Review A}, vol.~87, p.~052131, May 2013.

\bibitem{Rockafellar+2015}
R.~T. Rockafellar, \href {https://doi.org/doi:10.1515/9781400873173} {{\em
  Convex Analysis}}, vol.~28 of {\em Princeton Mathematical Series}.
\newblock Princeton University Press, 1970.

\bibitem{schaefer_TopVecSp}
H.~H. Schaefer, \href {https://doi.org/10.1007/978-1-4612-1468-7} {{\em
  Topological Vector Spaces}}, vol.~3 of {\em Graduate Texts in Mathematics}.
\newblock Springer-Verlag New York, 2nd~ed., 1999.

\bibitem{Barnum_2014}
H.~Barnum, M.~P. M{\"u}ller, and C.~Ududec, \href
  {https://doi.org/10.1088/1367-2630/16/12/123029} {``Higher-order interference
  and single-system postulates characterizing quantum theory,''} {\em New
  Journal of Physics}, vol.~16, p.~123029, Dec. 2014.

\bibitem{PhysRevLett.120.200402}
A.~B. Sainz, Y.~Guryanova, A.~Ac\'{\i}n, and M.~Navascu\'es, \href
  {https://doi.org/10.1103/PhysRevLett.120.200402} {``Almost-quantum
  correlations violate the no-restriction hypothesis,''} {\em Physical Review
  Letters}, vol.~120, p.~200402, May 2018.

\bibitem{Bourbaki_topvec}
N.~Bourbaki, \href {https://doi.org/10.1007/978-3-642-61715-7} {{\em
  Topological Vector Spaces}}.
\newblock Elements of Mathematics, Springer-Verlag Berlin Heidelberg, 2003.
\newblock Original French edition published by Masson, Paris, France, 1981.

\bibitem{boyd_vandenberghe_2004}
S.~Boyd and L.~Vandenberghe, \href {https://doi.org/10.1017/CBO9780511804441}
  {{\em Convex Optimization}}.
\newblock Cambridge University Press, 2004.

\bibitem{Alfsen_compact_convex}
E.~M. Alfsen, \href {https://doi.org/10.1007/978-3-642-65009-3} {{\em Compact
  Convex Sets and Boundary Integrals}}, vol.~57 of {\em Ergebnisse der
  Mathematik und ihrer Grenzgebiete}.
\newblock Springer-Verlag Berlin Heidelberg, 1971.

\bibitem{Ellis_duality_1964}
A.~J. Ellis, \href {https://doi.org/10.1112/jlms/s1-39.1.730} {``{The Duality
  of Partially Ordered Normed Linear Spaces},''} {\em Journal of the London
  Mathematical Society}, vol.~s1-39, no.~1, pp.~730--744, 1964.

\bibitem{Kelley_Namioka_topological}
J.~L. Kelley and I.~Namioka, \href {https://doi.org/10.1007/978-3-662-41914-4}
  {{\em Linear Topological Spaces}}.
\newblock Springer-Verlag Berlin Heidelberg, 2nd~ed., 1963.

\bibitem{Conway_functionalanalysis}
J.~B. Conway, \href {https://doi.org/10.1007/978-1-4757-3828-5} {{\em A Course
  in Functional Analysis}}, vol.~96 of {\em Graduate Texts in Mathematics}.
\newblock Springer-Verlag New York, 1st~ed., 1985.

\bibitem{Guler_optimization_2010}
O.~G\"uler, \href {https://doi.org/https://doi.org/10.1007/978-0-387-68407-9}
  {{\em Foundations of Optimization}}, vol.~258 of {\em Graduate Texts in
  Mathematics book series (GTM, volume 258) Graduate Texts in Mathematics book
  series}.
\newblock Springer New York, 1st~ed., 2010.

\bibitem{Olubummo1999}
Y.~Olubummo and T.~A. Cook, \href
  {https://doi.org/https://doi.org/10.1023/A:1026646602561} {``The predual of
  an order-unit banach space,''} {\em International Journal of Theoretical
  Physics}, vol.~38, pp.~3301--3303, 1999.

\bibitem{Aliprantis_cone_2007}
C.~D. Aliprantis and R.~Tourky, {\em Cones and Duality}, vol.~84 of {\em
  Graduate Studies in Mathematics}.
\newblock American Mathematica l Society, 2007.

\bibitem{KIMURA2010175}
G.~Kimura, K.~Nuida, and H.~Imai, \href
  {https://doi.org/https://doi.org/10.1016/S0034-4877(10)00025-X}
  {``Distinguishability measures and entropies for general probabilistic
  theories,''} {\em Reports on Mathematical Physics}, vol.~66, no.~2, pp.~175
  -- 206, 2010.

\bibitem{Bratteli_Robinson}
O.~Bratteli and D.~W. Robinson, \href
  {https://doi.org/10.1007/978-3-662-02520-8} {{\em Operator Algebras and
  Quantum Statistical Mechanics 1: C*- and W*-Algebras Symmetry Groups
  Decomposition of States}}.
\newblock Texts and Monographs in Physics, Springer-Verlag Berlin Heidelberg,
  2nd~ed., 1987.

\bibitem{Haag_localquantum_1996}
R.~Haag, \href {https://doi.org/https://doi.org/10.1007/978-3-642-61458-3}
  {{\em Local Quantum Physics}}.
\newblock Texts and Monographs in Physics, Springer-Verlag Berlin Heidelberg,
  2nd~ed., 1996.

\bibitem{Fewster_algebraic_2019}
C.~J. Fewster and K.~Rejzner, \href {http://arxiv.org/abs/arXiv:1904.04051}
  {``Algebraic quantum field theory -- an introduction,''} 2019,
  arXiv:1904.04051.

\bibitem{Takesaki_operatoralgebra}
M.~Takesaki, \href {https://doi.org/https://doi.org/10.1007/978-1-4612-6188-9}
  {{\em Theory of Operator Algebras I}}.
\newblock New York: Springer, 1st~ed.

\bibitem{Sakai1998}
S.~Sakai, \href {https://doi.org/https://doi.org/10.1007/978-3-642-61993-9}
  {{\em C*-Algebras and W*-Algebras}}.
\newblock Classics in Mathematics, Springer-Verlag Berlin Heidelberg, 1st~ed.,
  1998.

\bibitem{AlfsenShultz2003}
E.~M. Alfsen and F.~W. Shultz, \href
  {https://doi.org/https://doi.org/10.1007/978-1-4612-0019-2} {{\em Geometry of
  State Spaces of Operator Algebras}}.
\newblock Mathematics: Theory \& Applications, Boston: Birkh\"auser, 1st~ed.,
  2003.

\bibitem{doi:10.1063/1.532031}
E.~G. Beltrametti and S.~Bugajski, \href {https://doi.org/10.1063/1.532031}
  {``Effect algebras and statistical physical theories,''} {\em Journal of
  Mathematical Physics}, vol.~38, no.~6, pp.~3020--3030, 1997.

\bibitem{Gudder_Pulmannova_1998}
S.~Gudder and S.~Pulmannov\'a, \href {http://eudml.org/doc/248280}
  {``Representation theorem for convex effect algebras,''} {\em Commentationes
  Mathematicae Universitatis Carolinae}, vol.~39, no.~4, pp.~645--659, 1998.

\bibitem{Gudder_EffectAlgebras}
S.~Gudder, \href {https://doi.org/10.1023/A:1026678114856} {``Convex structures
  and effect algebras,''} {\em International Journal of Theoretical Physics},
  vol.~38, pp.~3179--3187, 1999.

\bibitem{Roman_advanced_linear_alge}
S.~Roman, \href {https://doi.org/https://doi.org/10.1007/978-1-4757-2178-2}
  {{\em Advanced Linear Algebra}}, vol.~135 of {\em Graduate Texts in
  Mathematics}.
\newblock Springer-Verlag New York, 1st~ed., 1992.

\bibitem{PR-box}
S.~Popescu and D.~Rohrlich, \href
  {https://doi.org/https://doi.org/10.1007/BF02058098} {``Quantum nonlocality
  as an axiom,''} {\em Foundations of Physics}, vol.~24, pp.~379--385, 1994.

\bibitem{Barnum_post-classical}
H.~Barnum and A.~Wilce, \href {http://arxiv.org/abs/arXiv:1205.3833}
  {``Post-classical probability theory,''} 2012, arXiv:1205.3833.

\bibitem{Hardy2011_quantum}
L.~Hardy, \href {http://arxiv.org/abs/arXiv:1104.2066} {``Reformulating and
  reconstructing quantum theory,''} 2011, arXiv:1104.2066.

\bibitem{HULANICKI1968177}
A.~Hulanicki and R.~Phelps, \href
  {https://doi.org/https://doi.org/10.1016/0022-1236(68)90016-5} {``Some
  applications of tensor products of partially-ordered linear spaces,''} {\em
  Journal of Functional Analysis}, vol.~2, no.~2, pp.~177--201, 1968.

\bibitem{https://doi.org/10.1112/plms/s3-19.1.177}
A.~L. Peressini and D.~R. Sherbert, \href
  {https://doi.org/https://doi.org/10.1112/plms/s3-19.1.177} {``Ordered
  topological tensor products,''} {\em Proceedings of the London Mathematical
  Society}, vol.~s3-19, no.~1, pp.~177--190, 1969,

\bibitem{Aubrun2021}
G.~Aubrun, L.~Lami, C.~Palazuelos, and M.~Pl\'avala, \href
  {https://doi.org/https://doi.org/10.1007/s00039-021-00565-5}
  {``Entangleability of cones,''} {\em Geometric and Functional Analysis},
  vol.~31, pp.~181--205, May 2021.

\bibitem{Araki_composite_1980}
H.~Araki, \href {https://doi.org/https://doi.org/10.1007/BF01962588} {``On a
  characterization of the state space of quantum mechanics,''} {\em
  Communications in Mathematical Physics}, vol.~75, pp.~1--24, 1980.

\bibitem{heinosaari_ziman_2011}
T.~Heinosaari and M.~Ziman, \href {https://doi.org/10.1017/CBO9781139031103}
  {{\em The Mathematical Language of Quantum Theory: From Uncertainty to
  Entanglement}}.
\newblock Cambridge: Cambridge University Press, 2011.

\bibitem{Heinosaari2019nofreeinformation}
T.~Heinosaari, L.~Lepp{\"{a}}j{\"{a}}rvi, and M.~Pl{\'{a}}vala, \href
  {https://doi.org/10.22331/q-2019-07-08-157} {``No-free-information principle
  in general probabilistic theories,''} {\em {Quantum}}, vol.~3, p.~157, July
  2019.

\bibitem{Haag_Kastler_1964}
R.~Haag and D.~Kastler, \href {https://doi.org/10.1063/1.1704187} {``An
  algebraic approach to quantum field theory,''} {\em Journal of Mathematical
  Physics}, vol.~5, no.~7, pp.~848--861, 1964.

\bibitem{Kraus_CP_1971}
K.~Kraus, \href {https://doi.org/https://doi.org/10.1016/0003-4916(71)90108-4}
  {``General state changes in quantum theory,''} {\em Annals of Physics},
  vol.~64, no.~2, pp.~311--335, 1971.

\bibitem{Hayashi_etal_2015}
M.~Hayashi, S.~Ishizaka, A.~Kawachi, G.~Kimura, and T.~Ogawa, \href
  {https://doi.org/https://doi.org/10.1007/978-3-662-43502-1} {{\em
  Introduction to Quantum Information Science}}.
\newblock Graduate Texts in Physics, Springer-Verlag Berlin Heidelberg,
  1st~ed., 2015.

\bibitem{PhysRevA.94.042108}
M.~Pl\'avala, \href {https://doi.org/10.1103/PhysRevA.94.042108} {``All
  measurements in a probabilistic theory are compatible if and only if the
  state space is a simplex,''} {\em Physical Review A}, vol.~94, p.~042108,
  Oct. 2016.

\bibitem{Davies_compactconvex}
E.~B. Davies, \href {https://doi.org/10.1093/qmath/25.1.323} {``Symmetries of
  compact convex sets,''} {\em The Quarterly Journal of Mathematics}, vol.~25,
  pp.~323--328, Jan. 1974.

\bibitem{muller2012unifying}
M.~P. M{\"u}ller, O.~C.~O. Dahlsten, and V.~Vedral, \href
  {https://doi.org/https://doi.org/10.1007/s00220-012-1605-x} {``Unifying
  typical entanglement and coin tossing: on randomization in probabilistic
  theories,''} {\em Communications in Mathematical Physics}, vol.~316,
  pp.~441--487, 2012.

\bibitem{PhysRevLett.108.130401}
M.~P. M\"uller and C.~Ududec, \href
  {https://doi.org/10.1103/PhysRevLett.108.130401} {``Structure of reversible
  computation determines the self-duality of quantum theory,''} {\em Physical
  Review Letters}, vol.~108, p.~130401, Mar. 2012.

\bibitem{KIMURA2003339}
G.~Kimura, \href
  {https://doi.org/https://doi.org/10.1016/S0375-9601(03)00941-1} {``The
  {B}loch vector for n-level systems,''} {\em Physics Letters A}, vol.~314,
  no.~5, pp.~339--349, 2003.

\bibitem{Bengtsson2013}
I.~Bengtsson, S.~Weis, and K.~{\.{Z}}yczkowski, \href
  {https://doi.org/10.1007/978-3-0348-0448-6_15} {{\em Geometry of the Set of
  Mixed Quantum States: An Apophatic Approach}}, pp.~175--197.
\newblock Basel: Springer Basel, 2013.

\bibitem{1367-2630-13-6-063024}
P.~Janotta, C.~Gogolin, J.~Barrett, and N.~Brunner, \href
  {http://stacks.iop.org/1367-2630/13/i=6/a=063024} {``Limits on nonlocal
  correlations from the structure of the local state space,''} {\em New Journal
  of Physics}, vol.~13, no.~6, p.~063024, 2011.

\bibitem{Dummit_abstractalgebra}
D.~S. Dummit and R.~M. Foote, {\em Abstract Algebra}.
\newblock Hoboken, New Jersey: John Wiley \& Sons, Inc., 3rd~ed., 2003.

\bibitem{PhysRevA.96.022113}
A.~Jen\ifmmode~\check{c}\else \v{c}\fi{}ov\'a and M.~Pl\'avala, \href
  {https://doi.org/10.1103/PhysRevA.96.022113} {``Conditions on the existence
  of maximally incompatible two-outcome measurements in general probabilistic
  theory,''} {\em Physical Review A}, vol.~96, p.~022113, Aug. 2017.

\bibitem{Koashi_2006}
M.~Koashi, \href {https://doi.org/10.1088/1742-6596/36/1/016} {``Unconditional
  security of quantum key distribution and the uncertainty principle,''} {\em
  Journal of Physics: Conference Series}, vol.~36, pp.~98--102, Apr. 2006.

\bibitem{PhysRev.34.163}
H.~P. Robertson, \href {https://doi.org/10.1103/PhysRev.34.163} {``The
  uncertainty principle,''} {\em Physical Review}, vol.~34, pp.~163--164, July
  1929.

\bibitem{Uffink_PhD}
J.~B.~M. Uffink, {\em Measures of uncertainty and the uncertainty principle}.
\newblock PhD thesis, University of Utrecht, Utrecht, 1990.

\bibitem{PhysRevA.71.052325}
J.~I. de~Vicente and J.~S\'anchez-Ruiz, \href
  {https://doi.org/10.1103/PhysRevA.71.052325} {``Separability conditions from
  the {L}andau-{P}ollak uncertainty relation,''} {\em Physical Review A},
  vol.~71, p.~052325, May 2005.

\bibitem{PhysRevA.76.062108}
T.~Miyadera and H.~Imai, \href {https://doi.org/10.1103/PhysRevA.76.062108}
  {``Generalized {L}andau-{P}ollak uncertainty relation,''} {\em Physical
  Review A}, vol.~76, p.~062108, Dec. 2007.

\bibitem{Hirschman1957_entropy}
I.~I. Hirschman, \href {https://doi.org/doi:10.2307/2372390} {``A note on
  entropy,''} {\em American Journal of Mathematics}, vol.~79, no.~1,
  pp.~152--156, 1957.

\bibitem{Beckner1975_entropy}
W.~Beckner, \href {https://doi.org/doi:10.2307/1970980} {``Inequalities in
  fourier analysis,''} {\em Annals of Mathematics}, vol.~102, no.~1,
  pp.~159--182, 1975.

\bibitem{UncertaintyRelationsForInformationEntropy}
I.~Bia{\l}ynicki-Birula and J.~Mycielski, \href
  {https://doi.org/10.1007/BF01608825} {``Uncertainty relations for information
  entropy in wave mechanics,''} {\em Communications in Mathematical Physics},
  vol.~44, pp.~129--132, June 1975.

\bibitem{PhysRevLett.50.631}
D.~Deutsch, \href {https://doi.org/10.1103/PhysRevLett.50.631} {``Uncertainty
  in quantum measurements,''} {\em Physical Review Letters}, vol.~50,
  pp.~631--633, Feb. 1983.

\bibitem{PhysRevLett.60.1103}
H.~Maassen and J.~B.~M. Uffink, \href
  {https://doi.org/10.1103/PhysRevLett.60.1103} {``Generalized entropic
  uncertainty relations,''} {\em Physical Review Letters}, vol.~60,
  pp.~1103--1106, Mar. 1988.

\bibitem{10.2307/25051432}
M.~Krishna and K.~R. Parthasarathy, \href
  {http://www.jstor.org/stable/25051432} {``An entropic uncertainty principle
  for quantum measurements,''} {\em Sankhy$\bar{{\rm a}}$: The Indian Journal
  of Statistics, Series A (1961-2002)}, vol.~64, no.~3, pp.~842--851, 2002.

\bibitem{PhysRevLett.60.2447}
E.~Arthurs and M.~S. Goodman, \href
  {https://doi.org/10.1103/PhysRevLett.60.2447} {``Quantum correlations: A
  generalized {H}eisenberg uncertainty relation,''} {\em Physical Review
  Letters}, vol.~60, pp.~2447--2449, June 1988.

\bibitem{doi:10.1002/j.1538-7305.1965.tb01684.x}
E.~Arthurs and J.~L. Kelly~Jr., \href
  {https://doi.org/10.1002/j.1538-7305.1965.tb01684.x} {``On the simultaneous
  measurement of a pair of conjugate observables,''} {\em Bell System Technical
  Journal}, vol.~44, no.~4, pp.~725--729, 1965,

\bibitem{PhysRevA.67.042105}
M.~Ozawa, \href {https://doi.org/10.1103/PhysRevA.67.042105} {``Universally
  valid reformulation of the {H}eisenberg uncertainty principle on noise and
  disturbance in measurement,''} {\em Physical Review A}, vol.~67, p.~042105,
  Apr. 2003.

\bibitem{doi:10.1063/1.2759831}
P.~Busch and D.~B. Pearson, \href {https://doi.org/10.1063/1.2759831}
  {``Universal joint-measurement uncertainty relation for error bars,''} {\em
  Journal of Mathematical Physics}, vol.~48, no.~8, p.~082103, 2007,

\bibitem{PhysRevA.78.052119}
T.~Miyadera and H.~Imai, \href {https://doi.org/10.1103/PhysRevA.78.052119}
  {``{H}eisenberg's uncertainty principle for simultaneous measurement of
  positive-operator-valued measures,''} {\em Physical Review A}, vol.~78,
  p.~052119, Nov. 2008.

\bibitem{10.5555/2017011.2017020}
P.~Busch and T.~Heinosaari, ``Approximate joint measurements of qubit
  observables,'' {\em Quantum Information \& Computation}, vol.~8,
  p.~797–818, Sept. 2008.

\bibitem{10.5555/2011593.2011606}
R.~F. Werner, ``The uncertainty relation for joint measurement of postion and
  momentum,'' {\em Quantum Information \& Computation}, vol.~4, p.~546–562,
  Dec. 2004.

\bibitem{PhysRevA.101.052104}
D.~Saha, M.~Oszmaniec, L.~Czekaj, M.~Horodecki, and R.~Horodecki, \href
  {https://doi.org/10.1103/PhysRevA.101.052104} {``Operational foundations for
  complementarity and uncertainty relations,''} {\em Physical Review A},
  vol.~101, p.~052104, May 2020.

\bibitem{PhysRevLett.103.230402}
M.~M. Wolf, D.~Perez-Garcia, and C.~Fernandez, \href
  {https://doi.org/10.1103/PhysRevLett.103.230402} {``Measurements incompatible
  in quantum theory cannot be measured jointly in any other no-signaling
  theory,''} {\em Physical Review Letters}, vol.~103, p.~230402, Dec. 2009.

\bibitem{Busch_2013}
P.~Busch, T.~Heinosaari, J.~Schultz, and N.~Stevens, \href
  {https://doi.org/10.1209/0295-5075/103/10002} {``Comparing the degrees of
  incompatibility inherent in probabilistic physical theories,''} {\em
  Europhysics Letters}, vol.~103, p.~10002, July 2013.

\bibitem{PhysRevA.89.022123}
N.~Stevens and P.~Busch, \href {https://doi.org/10.1103/PhysRevA.89.022123}
  {``Steering, incompatibility, and {B}ell-inequality violations in a class of
  probabilistic theories,''} {\em Physical Review A}, vol.~89, p.~022123, Feb.
  2014.

\bibitem{PhysRevA.98.012133}
A.~Jen\ifmmode~\check{c}\else \v{c}\fi{}ov\'a, \href
  {https://doi.org/10.1103/PhysRevA.98.012133} {``Incompatible measurements in
  a class of general probabilistic theories,''} {\em Physical Review A},
  vol.~98, p.~012133, July 2018.

\bibitem{PhysRevA.87.052125}
M.~Banik, M.~R. Gazi, S.~Ghosh, and G.~Kar, \href
  {https://doi.org/10.1103/PhysRevA.87.052125} {``Degree of complementarity
  determines the nonlocality in quantum mechanics,''} {\em Physical Review A},
  vol.~87, p.~052125, May 2013.

\bibitem{1367-2630-19-4-043025}
M.~Krumm, H.~Barnum, J.~Barrett, and M.~P. M{\"u}ller, \href
  {http://stacks.iop.org/1367-2630/19/i=4/a=043025} {``Thermodynamics and the
  structure of quantum theory,''} {\em New Journal of Physics}, vol.~19, no.~4,
  p.~043025, 2017.

\bibitem{PhysRevA.97.062102}
S.~N. Filippov, T.~Heinosaari, and L.~Lepp\"aj\"arvi, \href
  {https://doi.org/10.1103/PhysRevA.97.062102} {``Simulability of observables
  in general probabilistic theories,''} {\em Physical Review A}, vol.~97,
  p.~062102, June 2018.

\bibitem{RevModPhys.89.015002}
P.~J. Coles, M.~Berta, M.~Tomamichel, and S.~Wehner, \href
  {https://doi.org/10.1103/RevModPhys.89.015002} {``Entropic uncertainty
  relations and their applications,''} {\em Reviews of Modern Physics},
  vol.~89, p.~015002, Feb. 2017.

\bibitem{inequalities1988}
G.~H. Hardy, J.~E. Littlewood, and G.~P\'olya, {\em Inequalities}.
\newblock Cambridge: Cambridge University Press, 2nd~ed., 1988.

\bibitem{PhysRevA.84.052117}
M.~H. Partovi, \href {https://doi.org/10.1103/PhysRevA.84.052117}
  {``Majorization formulation of uncertainty in quantum mechanics,''} {\em
  Physical Review A}, vol.~84, p.~052117, Nov. 2011.

\bibitem{PhysRevLett.111.230401}
S.~Friedland, V.~Gheorghiu, and G.~Gour, \href
  {https://doi.org/10.1103/PhysRevLett.111.230401} {``Universal uncertainty
  relations,''} {\em Physical Review Letters}, vol.~111, p.~230401, Dec. 2013.

\bibitem{PhysRevA.89.052115}
L.~Rudnicki, Z.~Pucha\l{}a, and K.~\ifmmode~\dot{Z}\else \.{Z}\fi{}yczkowski,
  \href {https://doi.org/10.1103/PhysRevA.89.052115} {``Strong majorization
  entropic uncertainty relations,''} {\em Physical Review A}, vol.~89,
  p.~052115, May 2014.

\bibitem{Pucha_a_2013}
Z.~Pucha{\l}a, {\L}.~Rudnicki, and K.~{\.{Z}}yczkowski, \href
  {https://doi.org/10.1088/1751-8113/46/27/272002} {``Majorization entropic
  uncertainty relations,''} {\em Journal of Physics A: Mathematical and
  Theoretical}, vol.~46, p.~272002, June 2013.

\bibitem{Pucha_a_2018}
Z.~Pucha{\l}a, {\L}.~Rudnicki, A.~Krawiec, and K.~{\.{Z}}yczkowski, \href
  {https://doi.org/10.1088/1751-8121/aab66c} {``Majorization uncertainty
  relations for mixed quantum states,''} {\em Journal of Physics A:
  Mathematical and Theoretical}, vol.~51, p.~175306, Apr. 2018.

\bibitem{e21030270}
K.~Baek, H.~Nha, and W.~Son, \href {https://doi.org/10.3390/e21030270}
  {``Entropic uncertainty relations via direct-sum majorization relation for
  generalized measurements,''} {\em Entropy}, vol.~21, no.~3, 2019.

\bibitem{Cover:2006:EIT:1146355}
T.~M. Cover and J.~A. Thomas, \href {https://doi.org/10.1002/047174882X} {{\em
  Elements of Information Theory}}.
\newblock Hoboken, New Jersey: John Wiley \& Sons, Inc., 2nd~ed., 2006.

\bibitem{Srinivas2003_entropic_successive}
M.~Srinivas, \href {https://doi.org/https://doi.org/10.1007/BF02704281}
  {``Optimal entropic uncertainty relation for successive measurements in
  quantum information theory,''} {\em Pramana}, vol.~60, pp.~1137--1152, 2003.

\bibitem{PhysRevA.89.032108}
K.~Baek, T.~Farrow, and W.~Son, \href
  {https://doi.org/10.1103/PhysRevA.89.032108} {``Optimized entropic
  uncertainty for successive projective measurements,''} {\em Physical Review
  A}, vol.~89, p.~032108, Mar. 2014.

\bibitem{succesive_2015_Renyi}
J.~Zhang, Y.~Zhang, and C.-s. Yu, \href
  {https://doi.org/https://doi.org/10.1007/s11128-015-0950-z} {``{R}\'enyi
  entropy uncertainty relation for successive projective measurements,''} {\em
  Quantum Information Processing}, vol.~14, pp.~2239--2253, 2015.

\bibitem{https://doi.org/10.1002/andp.201600130}
A.~E. Rastegin, \href {https://doi.org/https://doi.org/10.1002/andp.201600130}
  {``Entropic uncertainty relations for successive measurements of canonically
  conjugate observables,''} {\em Annalen der Physik}, vol.~528, no.~11-12,
  pp.~835--844, 2016.

\bibitem{Designolle_2019}
S.~Designolle, M.~Farkas, and J.~Kaniewski, \href
  {https://doi.org/10.1088/1367-2630/ab5020} {``Incompatibility robustness of
  quantum measurements: a unified framework,''} {\em New Journal of Physics},
  vol.~21, p.~113053, Nov. 2019.

\bibitem{HEINOSAARI20141695}
T.~Heinosaari, J.~Schultz, A.~Toigo, and M.~Ziman, \href
  {https://doi.org/https://doi.org/10.1016/j.physleta.2014.04.026} {``Maximally
  incompatible quantum observables,''} {\em Physics Letters A}, vol.~378,
  no.~24, pp.~1695--1699, 2014.

\bibitem{incomp_break_2015}
T.~Heinosaari, J.~Kiukas, D.~Reitzner, and J.~Schultz, \href
  {https://doi.org/10.1088/1751-8113/48/43/435301} {``Incompatibility breaking
  quantum channels,''} {\em Journal of Physics A: Mathematical and
  Theoretical}, vol.~48, p.~435301, Oct. 2015.

\bibitem{PhysRevA.100.042308}
L.~Guerini, M.~T. Quintino, and L.~Aolita, \href
  {https://doi.org/10.1103/PhysRevA.100.042308} {``Distributed sampling,
  quantum communication witnesses, and measurement incompatibility,''} {\em
  Physical Review A}, vol.~100, p.~042308, Oct. 2019.

\bibitem{Kiukas_2020}
J.~Kiukas, \href {https://doi.org/10.1088/1742-6596/1638/1/012003} {``Subspace
  constraints for joint measurability,''} {\em Journal of Physics: Conference
  Series}, vol.~1638, p.~012003, Oct. 2020.

\bibitem{doi:10.1063/5.0028658}
F.~Loulidi and I.~Nechita, \href {https://doi.org/10.1063/5.0028658} {``The
  compatibility dimension of quantum measurements,''} {\em Journal of
  Mathematical Physics}, vol.~62, no.~4, p.~042205, 2021,

\bibitem{PhysRevA.103.022203}
R.~Uola, T.~Kraft, S.~Designolle, N.~Miklin, A.~Tavakoli, J.-P. Pellonp\"a\"a,
  O.~G\"uhne, and N.~Brunner, \href
  {https://doi.org/10.1103/PhysRevA.103.022203} {``Quantum measurement
  incompatibility in subspaces,''} {\em Physical Review A}, vol.~103,
  p.~022203, Feb. 2021.

\bibitem{PhysRevD.33.2253}
P.~Busch, \href {https://doi.org/10.1103/PhysRevD.33.2253} {``Unsharp reality
  and joint measurements for spin observables,''} {\em Physical Review D},
  vol.~33, pp.~2253--2261, Apr. 1986.

\bibitem{HeMiRe2014}
T.~Heinosaari, T.~Miyadera, and D.~Reitzner, \href
  {https://doi.org/10.1007/s10701-013-9761-1} {``Strongly incompatible quantum
  devices,''} {\em Foundations of Physics}, vol.~44, pp.~34--57, 2014.

\bibitem{Haapasalo_2015}
E.~Haapasalo, \href {https://doi.org/10.1088/1751-8113/48/25/255303}
  {``Robustness of incompatibility for quantum devices,''} {\em Journal of
  Physics A: Mathematical and Theoretical}, vol.~48, p.~255303, June 2015.

\bibitem{Heinosaari_2017}
T.~Heinosaari and T.~Miyadera, \href {https://doi.org/10.1088/1751-8121/aa5f6b}
  {``Incompatibility of quantum channels,''} {\em Journal of Physics A:
  Mathematical and Theoretical}, vol.~50, p.~135302, Mar. 2017.

\bibitem{PhysRevA.97.022112}
T.~Heinosaari, D.~Reitzner, T.~c.~v. Ryb\'ar, and M.~Ziman, \href
  {https://doi.org/10.1103/PhysRevA.97.022112} {``Incompatibility of unbiased
  qubit observables and pauli channels,''} {\em Physical Review A}, vol.~97,
  p.~022112, Feb. 2018.

\bibitem{doi:10.1063/1.5008300}
Y.~Kuramochi, \href {https://doi.org/10.1063/1.5008300} {``Quantum
  incompatibility of channels with general outcome operator algebras,''} {\em
  Journal of Mathematical Physics}, vol.~59, no.~4, p.~042203, 2018,

\bibitem{Haapasalo2019}
E.~Haapasalo, \href {https://doi.org/10.1007/s00023-019-00827-x}
  {``Compatibility of covariant quantum channels with emphasis on {W}eyl
  symmetry,''} {\em Annales Henri Poincar\'e}, vol.~20, p.~3163–3195, 2019.

\bibitem{Martens1990}
H.~Martens and W.~M. de~Muynck, \href {https://doi.org/10.1007/BF00731693}
  {``Nonideal quantum measurements,''} {\em Foundations of Physics}, vol.~20,
  pp.~255--281, 1990.

\bibitem{PhysRevLett.76.2818}
H.~Barnum, C.~M. Caves, C.~A. Fuchs, R.~Jozsa, and B.~Schumacher, \href
  {https://doi.org/10.1103/PhysRevLett.76.2818} {``Noncommuting mixed states
  cannot be broadcast,''} {\em Physical Review Letters}, vol.~76,
  pp.~2818--2821, Apr. 1996.

\bibitem{PhysRevA.88.032312}
R.~Beneduci, T.~J. Bullock, P.~Busch, C.~Carmeli, T.~Heinosaari, and A.~Toigo,
  \href {https://doi.org/10.1103/PhysRevA.88.032312} {``Operational link
  between mutually unbiased bases and symmetric informationally complete
  positive operator-valued measures,''} {\em Physical Review A}, vol.~88,
  p.~032312, Sept. 2013.

\bibitem{PhysRevA.78.012315}
P.~Stano, D.~Reitzner, and T.~Heinosaari, \href
  {https://doi.org/10.1103/PhysRevA.78.012315} {``Coexistence of qubit
  effects,''} {\em Physical Review A}, vol.~78, p.~012315, July 2008.

\bibitem{Busch_Coexistence}
P.~Busch and H.-J. Schmidt, \href {https://doi.org/10.1007/s11128-009-0109-x}
  {``Coexistence of qubit effects,''} {\em Quantum Information Processing},
  vol.~9, p.~143–169, 2010.

\bibitem{PhysRevA.81.062116}
S.~Yu, N.-l. Liu, L.~Li, and C.~H. Oh, \href
  {https://doi.org/10.1103/PhysRevA.81.062116} {``Joint measurement of two
  unsharp observables of a qubit,''} {\em Physical Review A}, vol.~81,
  p.~062116, June 2010.

\bibitem{Kelley1975}
J.~L. Kelley, {\em General Topology}, vol.~27 of {\em Graduate Texts in
  Mathematics}.
\newblock Springer-Verlag New York, 1st~ed., 1975.

\bibitem{Callen_thermo}
H.~B. Callen, {\em Thermodynamics and an Introduction to Thermostatistics}.
\newblock Hoboken, New Jersey: John Wiley \& Sons, Inc., 2nd~ed., 1985.

\bibitem{zemansky_thermo}
M.~W. Zemansky and R.~H. Dittman, {\em Heat and Thermodynamics}.
\newblock New York: McGraw-Hill Companies, Inc., 7th~ed., 1997.

\bibitem{1367-2630-12-3-033024}
H.~Barnum, J.~Barrett, L.~O. Clark, M.~Leifer, R.~Spekkens, N.~Stepanik,
  A.~Wilce, and R.~Wilke, \href
  {http://stacks.iop.org/1367-2630/12/i=3/a=033024} {``Entropy and information
  causality in general probabilistic theories,''} {\em New Journal of Physics},
  vol.~12, no.~3, p.~033024, 2010.

\bibitem{Short_2010}
A.~J. Short and S.~Wehner, \href
  {https://doi.org/10.1088/1367-2630/12/3/033023} {``Entropy in general
  physical theories,''} {\em New Journal of Physics}, vol.~12, p.~033023, Mar.
  2010.

\bibitem{KimuraEntropiesinGeneralProbabilisticTheoriesandTheirApplicationtotheHolevoBound}
G.~Kimura, J.~Ishiguro, and M.~Fukui, \href
  {https://doi.org/10.1103/PhysRevA.94.042113} {``Entropies in general
  probabilistic theories and their application to the holevo bound,''} {\em
  Physical Review A}, vol.~94, p.~042113, Oct. 2016.

\bibitem{Chiribella_2017}
G.~Chiribella and C.~M. Scandolo, \href
  {https://doi.org/10.1088/1367-2630/aa91c7} {``Microcanonical thermodynamics
  in general physical theories,''} {\em New Journal of Physics}, vol.~19,
  p.~123043, Dec. 2017.

\bibitem{chiribella2016entanglement}
G.~Chiribella and C.~M. Scandolo, \href {http://arxiv.org/abs/arxiv:1608.04459}
  {``Entanglement as an axiomatic foundation for statistical mechanics,''}
  2016, arxiv:1608.04459.

\bibitem{EPTCS195.4}
H.~Barnum, J.~Barrett, M.~Krumm, and M.~P. M\"uller, \href
  {https://doi.org/10.4204/EPTCS.195.4} {``Entropy, majorization and
  thermodynamics in general probabilistic theories,''} in {\em {\rm Proceedings
  of the 12th International Workshop on} Quantum Physics and Logic, {\rm
  Oxford, U.K., July 15-17, 2015}}, vol.~195, pp.~43--58, Open Publishing
  Association, 2015.

\bibitem{KIMURA20141}
G.~Kimura and K.~Nuida, \href
  {https://doi.org/https://doi.org/10.1016/j.geomphys.2014.06.004} {``On affine
  maps on non-compact convex sets and some characterizations of
  finite-dimensional solid ellipsoids,''} {\em Journal of Geometry and
  Physics}, vol.~86, pp.~1--18, 2014.

\end{thebibliography}
